\documentclass{jpp}

\usepackage{graphicx}
\usepackage{natbib}
\usepackage[colorlinks,urlcolor=blue,citecolor=blue,linkcolor=blue]{hyperref}
\usepackage{amsmath}

\ifCUPmtlplainloaded \else
  \checkfont{eurm10}
  \iffontfound
    \IfFileExists{upmath.sty}
      {\typeout{^^JFound AMS Euler Roman fonts on the system,
                   using the 'upmath' package.^^J}%
       \usepackage{upmath}}
      {\typeout{^^JFound AMS Euler Roman fonts on the system, but you
                   dont seem to have the}%
       \typeout{'upmath' package installed. JPP.cls can take advantage
                 of these fonts, if you use 'upmath' package.^^J}%
      }
  \else
  \fi
\fi

\ifCUPmtlplainloaded \else
  \checkfont{msam10}
  \iffontfound
    \IfFileExists{amssymb.sty}
      {\typeout{^^JFound AMS Symbol fonts on the system, using the
                'amssymb' package.^^J}%
       \usepackage{amssymb}%
       \let\le=\leqslant  \let\leq=\leqslant
         \let\geq=\geqslant
      }{}
  \fi
\fi

\ifCUPmtlplainloaded \else
  \IfFileExists{amsbsy.sty}
    {\typeout{^^JFound the 'amsbsy' package on the system, using it.^^J}%
     \usepackage{amsbsy}}
    {\providecommand\boldsymbol[1]{\mbox{\boldmath $##1$}}}
\fi

\newcommand{\bmath}[1]{{\bf {#1}}}

\newcommand{\bB}{\bmath{B}}

\newcommand{\vpr}[2]{\bmath{#1} \!\times\! \bmath{#2}}

\newcommand{\eq}[1]{Eq.~(\ref{eq:#1})}
\newcommand{\fign}[1]{\ref{fig:#1}}

\def\comp{\,c/\omega_{\rm p}}

\newcommand{\fig}[1]{Fig.~\ref{fig:#1}}
\newcommand{\gammamax}{\gamma_{\rm max}}

\newcommand{\be}{\begin{equation}} 
\newcommand{\ee}{\end{equation}}
\newcommand{\nn}{\mbox{} \nonumber \\ \mbox{} }
\newcommand{\ba}{\begin{eqnarray}}
\newcommand{\ea}{\end{eqnarray}}

\newcommand{\Alfven}{Alfv\'{e}n }
\newcommand{\sigmain}{\sigma_{\rm in}}
\newcommand{\rhot}{r_{\rm L,hot}}
\newcommand{\vpush}{v_{\rm push}}
\newcommand{\rj}{\,r_{\rm j}}
\newcommand{\Bf}{{magnetic field}}
\newcommand{\Bfs}{{magnetic fields}}
\newcommand{\Ef}{{electric  field}}
\newcommand{\Efs}{{electric fields}}

\newcommand{\NSs}{{neutron stars}}
\newcommand{\EM}{electromagnetic}

\newcommand{\curl}{{\rm curl\, }}
\newcommand{\E}{{\bf E}}
\newcommand{\B}{{\bf B}}

\newcommand{\aap}{    {\it Astron. Astrophys.}}
\newcommand{\aaps}{   {\it Astron. Astrophys. Suppl.}}
\newcommand{\aapr}{   {\it Astron. Astrophys. Rev.}}

\newcommand{\aj}{     {\it Astron. J.}}
\newcommand{\apj}{    {\it Astrophys. J.}}
\newcommand{\apjl}{    {\it Astrophys. J. Lett.}}
\newcommand{\apss}{   {\it Astrophys. Space Sci.}}

\newcommand{\fcp}{    {\it Fundamenals Cosm. Phys.}}

\newcommand{\grl}{    {\it Geophys. Res. Lett.}}

\newcommand{\jgr}{    {\it J. Geophys. Res.}}
\newcommand{\mnras}{  {\it Mon. Not. Roy. Astron. Soc.}}
\newcommand{\nat}{    {\it Nature}}
\newcommand{\na}{    {\it Nature}}
\newcommand{\pasp}{   {\it Pub. Astron. Soc. Pac.}}
\newcommand{\pasj}{   {\it Pub. Astron. Soc. Japan}}
\newcommand{\prd}{    {\it Phys. Rev. D}}
\newcommand{\pre}{    {\it Phys. Rev. E}}
\newcommand{\solphys}{{\it Solar Phys.}}
\newcommand{\sovast}{ {\it Sov. Astron.}}
\newcommand{\ssr}{    {\it Space Sci. Rev.}}
\def\apjs{ApJS}               % Astrophysical Journal, Supplement

\title{Particle acceleration in  relativistic  magnetic flux-merging events}

\author{Maxim Lyutikov,$^1$  Lorenzo Sironi$^{2}$, Serguei S. Komissarov$^{1,3}$,  Oliver Porth$^{3,4}$}
\affiliation{$^1$ Department of Physics, Purdue University, 
 525 Northwestern Avenue,
West Lafayette, IN
47907-2036, USA; lyutikov@purdue.edu
\\
$^2$  Department of Astronomy, Columbia University, 550 W 120th St, New York, NY 10027, USA; lsironi@astro.columbia.edu
\\
$^3$  School of Mathematics,
University of Leeds, LS29JT
Leeds, UK; s.s.komissarov@leeds.ac.uk
\\
$^4$  Institut f\"{u}r Theoretische Physik, 
J. W. Goethe-Universit\"{a}t, 
D-60438, Frankfurt am Main, Germany
porth@th.physik.uni-frankfurt.de
}

\begin{document}

\maketitle

%%%%%%%%%%%

\begin{abstract} 

 Using analytical and numerical methods (fluid and particle-in-cell simulations) we study a number of model problems involving merger of magnetic flux tubes in relativistic magnetically-dominated plasma. Mergers of current-carrying  flux tubes  (exemplified by the two dimensional ``ABC'' structures)
  and zero total current magnetic flux tubes  are considered.  In all cases  regimes of spontaneous and driven evolution are investigated.

We identify two stages of particle acceleration during flux mergers: (i)  fast explosive prompt X-point collapse and (ii)  ensuing  island merger. 
The fastest acceleration occurs during the initial catastrophic  X-point collapse, with the reconnection \Ef\ of the order
of the \Bf.   During the X-point collapse particles are accelerated by charge-starved \Efs, which can reach (and even exceed) values of  the local \Bf. The explosive stage of reconnection 
 produces non-thermal power-law tails with slopes
that depend on the average magnetization $\sigma$. For plasma magnetization $\sigma \leq 10^2$ the  spectrum  power law index is  $p> 2$; in this case   the maximal energy depends linearly on the size of the reconnecting islands. For  higher magnetization, $\sigma \geq 10^2$, the spectra are hard, $p< 2$, yet the maximal energy $\gamma_{max}$ can still exceed the average magnetic energy per particle,   $ \sim \sigma$,  by  orders of magnitude (if $p$ is not too close to unity).  
 The X-point collapse stage is followed by  magnetic island merger  that dissipates a large
fraction of the initial magnetic energy in a regime of forced magnetic
reconnection, further accelerating the particles, but proceeds at a slower reconnection rate. 

\end{abstract}

% \clearpage
%%%%%%%%%

%\tableofcontents

%sssssssssssssssssssssssssssssssssssssssssssss
\section{Introduction}
\label{intro}
%sssssssssssssssssssssssssssssssssssssssssssss

In many astrophysical settings the magnetic field controls the overall dynamics
of a plasma, while the dissipation of magnetic energy may power the high energy
emission.  
The relevant  astrophysical settings include magnetars (strongly magnetized \NSs\ possessing
super-strong \Bfs), pulsars and pulsar wind nebulae, jets of Active Galactic
Nuclei, and Gamma-Ray Bursters.  All these objects are efficient emitters of X-rays
and $\gamma$-rays, and in the past two decades they have been the subjects of 
intensive observational studies using a number of very successful high-energy
satellites.  These objects seem to 
share one important property -- they include relativistic 
magnetized plasma, and often their plasma is magnetically dominated, i.e.,  
the energy density of this plasma is dominated not by the rest mass-energy of 
matter, but by the energy of the magnetic field. This is dramatically different  
from laboratory plasmas, the magnetospheres of planets, and the interplanetary 
plasma. 

Recently, these topics came to the front of astrophysical and plasma physical research. Particularly important was the
the discovery of flares  from  Crab Nebula \citep{2011Sci...331..736T,2011Sci...331..739A,2012ApJ...749...26B}, exhibiting  unusually short durations, high luminosities, and high photon energies. Exceptionally high peak energy  of the flare spectral peak {\it excludes stochastic shock acceleration} as a mechanism of acceleration of flare-producing particles. 
 An alternative possibility is that  magnetic reconnection in {\it highly magnetized plasma} is responsible for the acceleration of  the wind leptons    \citep{2010MNRAS.405.1809L,2012MNRAS.426.1374C,2014PhPl...21e6501C,2013ApJ...770..147C}. These events  challenge our understanding of particle acceleration in  PWNe,  and, possibly in other high-energy  astrophysical sources.

In conventional laboratory/space plasma the merging of magnetic islands have been invoked as an important, and perhaps dominant mechanism of particle acceleration \citep{2010ApJ...714..915O,2010PhPl...17j2902T,2012SSRv..172..227D,2014ApJ...797...28Z}. In this paper we investigate particle acceleration during magnetic island merger  in relativistic highly magnetized plasmas.   In order to describe the level of magnetization it is
convenient to use the so-called magnetization parameter 
\be
\sigma = 2  \frac{u_B}{ u_{p}} = \frac{B^2}{4 \pi \rho c^2}
\ee
where $u_B=B^2/8\pi$ is the magnetic energy density, $u_p= \rho c^2$ is the
rest mass-energy density. 

In Lyutikov \etal\ (2016a, Paper I) we studied the plasma dynamics and  particle acceleration during explosive collapse of a stressed $X$-point. In this paper we extends the work to include the large scale dynamics. The key point of our approach is that the reconnection and particle acceleration is driven by {\it macroscopic} large scale stresses of the \Bf\ (and not, \eg\ by effects on  skin depth scales  during the development of the tearing mode in collisionless plasma). 

To include the  effects of large scale magnetic stresses we consider a number magnetic configurations: (i) 2D force-free magnetic flux tubes (ABC) configuration 
 including the driven regimes \S \ref{driven}; (ii) colliding magnetic flux tubes carrying zero  total current, \S \ref{fluxtubes}.  The principal difference between these configurations is that in the case of  2D force-free magnetic flux tubes each flux tube carries a non-zero total electric current - thus even far away flux tube experience Lorentz attraction or repulsion.

%%%%%%%Lorenzo%%%%%%%
 
%ssssssssssssssssssssssssssssssssssssssssssssssssssssssssssssssssssssssssssssssssssss
\section{Collapse of a system of magnetic islands}
\label{unstr-latt}
%ssssssssssssssssssssssssssssssssssssssssssssssssssssssssssssssssssssssssssssssssssss

\subsection{2D magnetic ABC structures}
\begin{figure}
\centering
\includegraphics[width=.49\textwidth]{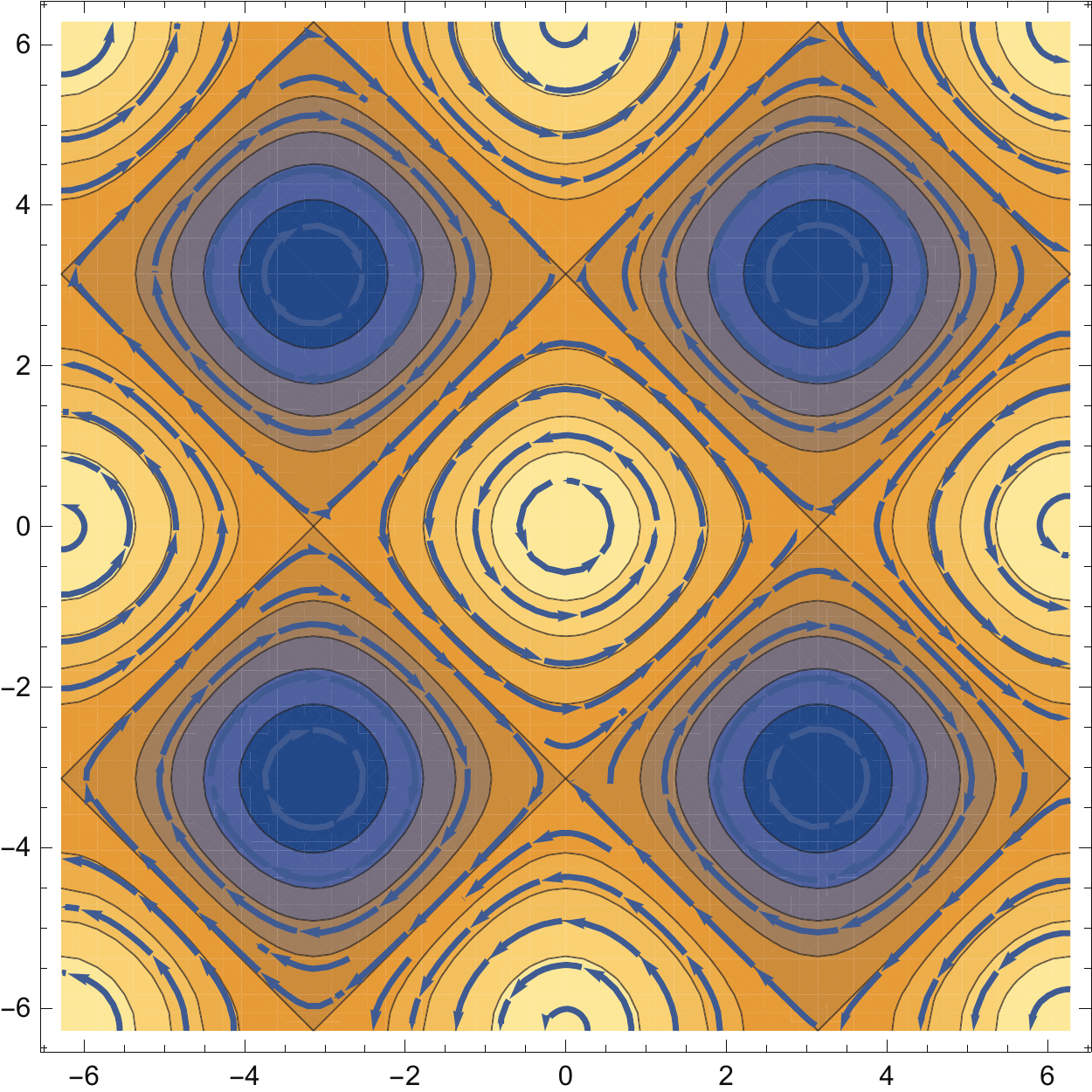}
\caption{A 2D force-free system of magnetic islands (magnetic ABC structures). Color indicates out-of-the-plane \Bf. Both types of islands have the same helicity. At the X-point  the \Bf\ and current are exactly zero. 
 }
\label{ABC}
\end{figure}

In this section we consider  large scale dynamics that can lead to the $X$-point collapse described above.
 As an initial pre-flare state of plasma we consider a 2D force-fee lattice of magnetic flux tubes 
 \ba &&
B_x = -\sin(2\pi \alpha y)B_0\,,
\nn &&
B_y = \sin(2\pi \alpha x) B_0 \,,
\nn &&
B_z =\left(\cos(2\pi \alpha x)+\cos(2\pi \alpha y)\right) B_0  \, ,
\label{Bb}
\ea
Fig. \ref{ABC}.  This constitutes  a lattice of force-free magnetic islands separated by 
$90^o$ X-points in equilibrium. Islands have alternating out-of-the-plane poloidal  fields and  alternating toroidal fields. Each magnetic flux tube  carries a magnetic flux
$ \propto  B_0/\alpha^2$, energy per unit length $ \propto   B_0^2/\alpha^2$,  helicity per unit length $  \propto  B_0^2/(\alpha^3 )$ and axial current $  \propto  B_0/( \alpha)$. Helicity of both types of flux tubes is of the same sign.

Previously,  this configuration has been considered by \cite{1983ApJ...264..635P} in the context of Solar \Bfs; he suggested that this force-free configuration is unstable. Later \cite{1994ApJ...437..851L} considered the evolution of the instability. In case of force-free plasma the instability has been recently considered by \cite{2015PhRvL.115i5002E}. 
 The configuration (\ref{Bb}) belong to a family of ABC flows  \citep[Arnold-Beltrami-Childress, \eg][]{1972RSPTA.271..411R}. ABC magnetic structures are known to be stable to ideal perturbations \cite{Arnold74,1974SoPh...39..393M,1986JFM...166..359M} {\it provided} that all three corresponding coefficients are non-zero
 \citep[though see][for a claim to the contrary]{2015PhRvL.115i5002E}. 
 Two-dimensional 
 structures considered here  do not satisfy this condition - in this case the islands can move with respect to each other, reducing their interaction energy, as we show below. The  stable full 3D ABC structures do not allow for such motion \citep[see though][who claimed the instability of 3D ABC configurations as well]{2015PhRvL.115i5002E}. Full 3D ABC  structures have all fields linked, while in case of 2D the neighboring cylinders are not linked together. The difference between stable 3D and unstable 2D ABC configurations illustrates a general principle that lower dimension magnetic structures are typically less stable.

We also point out that we start with configuration stable to internal kinks. For sufficiently high currents the non-constant-$\alpha$ Lundquist-type configurations are known to show internal kink instabilities \cite{1962PhRv..128.2016V}.

%We foresee that the configuration (\ref{Bb}) is created by {\it large scale} turbulent processes described in \S \ref{formation}, and {\it not}  through slow evolution from smaller scales,  via, \eg  an inverse cascade-type process  \citep{2014ApJ...794L..26Z} -  such cascade is expected to be highly dissipative, see Appendix \ref{inverse}.

%In addition, in appendix \ref{Shock-triggered} we consider induced collapse of such configuration. 

\subsection{The nature of the instability}
\label{Thenature}

The  instability of the 2D ABC configuration is of the kind ``parallel currents attract". In the initial configuration the attraction of parallel currents is balanced by the repulsion of anti-parallel ones.  Small amplitude fluctuations lead to fluctuating forces between the currents, that eventually lead to the disruption of the system. To identify the dominant instability mode let us consider a simplified model problem replacing each island by a solid tube carrying a given current. 
Such incompressible-type approximation is expected to be valid at early times, when the resulting motions are slow and the amount of the dissipated  magnetic energy is small. 

\begin{figure}
\includegraphics[width=0.59\linewidth]{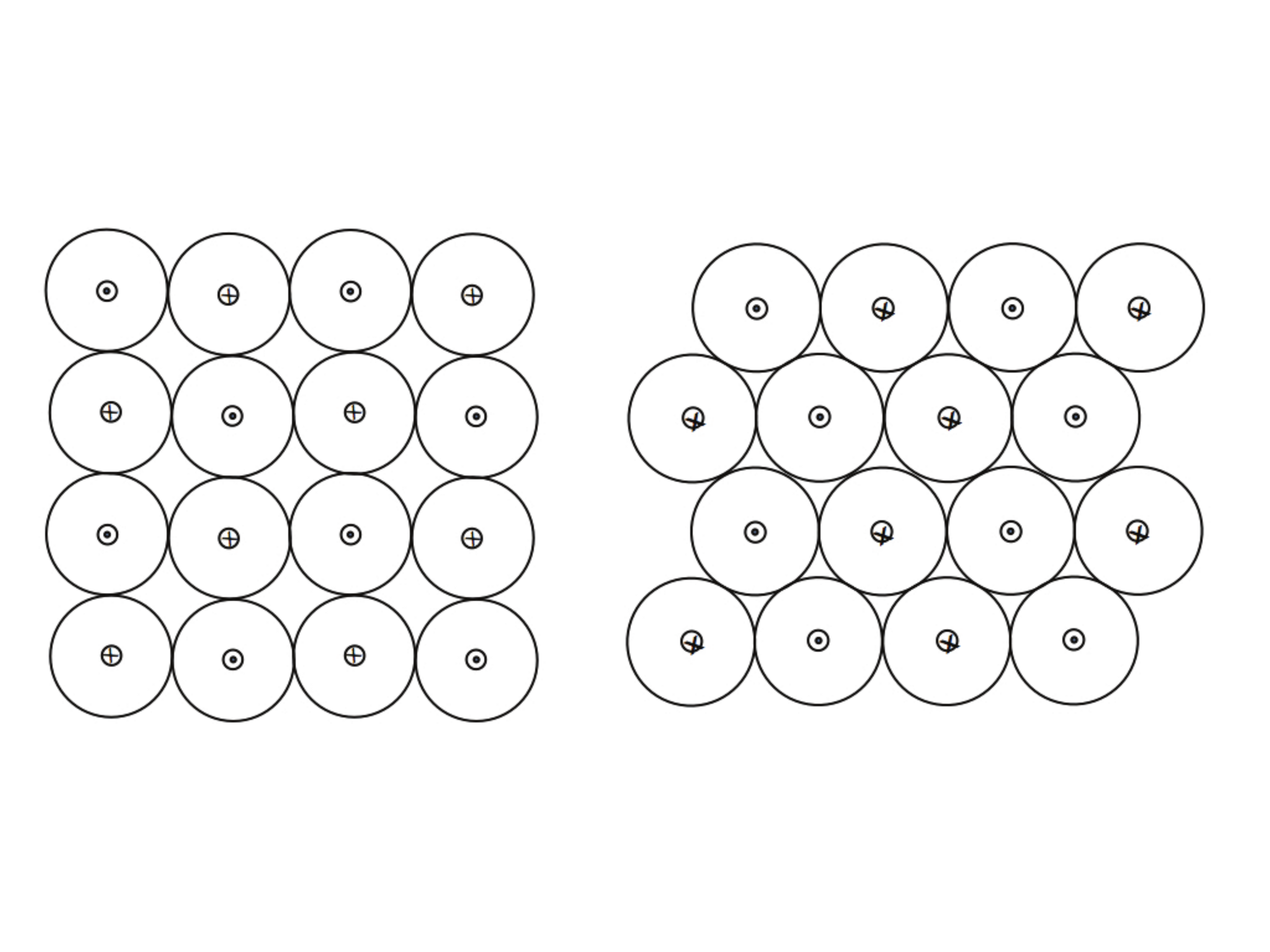}
\includegraphics[width=0.39\linewidth]{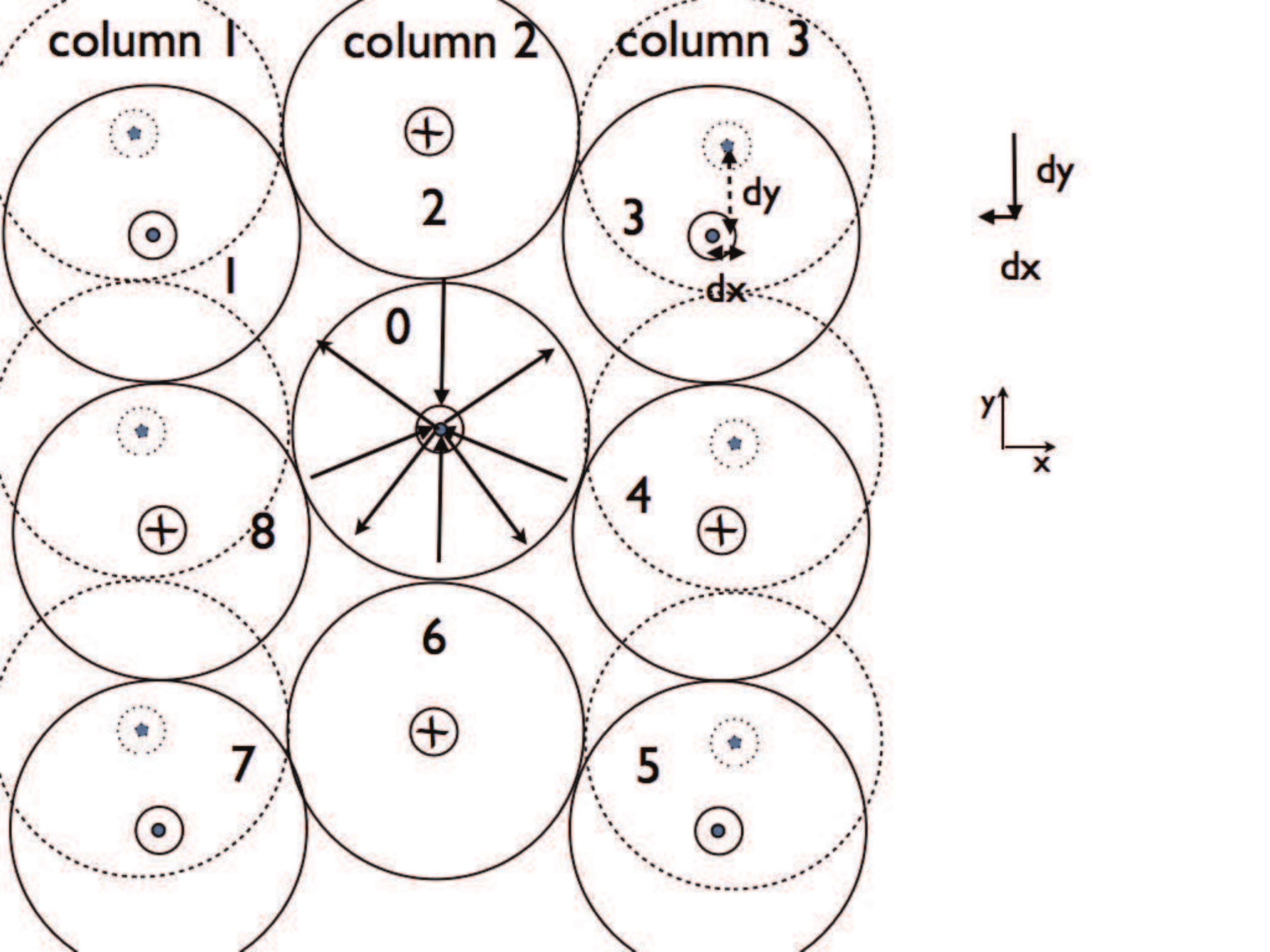}
\caption{Qualitative picture of the development of  shear-type (parallel) instability. The initial magnetic 2D ABC structure ({\it Left Panel}) is unstable to long wavelength instability whereby each consecutive row is shifted by a half wave-length ({\it Right Panel}). A current sheet  forms in-between the islands with co-aligned currents leading to catastrophic merger. Right Panel:  The nature of the instability. Shifting alternating rows (or  columns) of magnetic islands leads to a force dis-balance that amplified the perturbation. In the particular case, shifting first and third column leads to a force along the positive direction on the islands in the second column, amplifying the perturbations.}
\label{inst}
\end{figure}

We identify two basic instability modes which we call the shear mode and the compression mode. Let us first consider a global shear mode, Fig. \ref{inst}.
Let us separate the 2D ABC equilibrium into a set of columns, labeled 1-2-3, Fig. \ref{inst}.
In the initial state the flux tubes in the columns 1 and 3 (dashed lines)  are horizontally aligned with those in column 2. Let us shift  columns 1 and 3 by a value $-dy$ (or, equivalently, shift the column 2 by $+dy$; corresponding horizontal displacement is   $dx= \pm  dy^2/2$. Let us consider forces on the central island (labeled by $0$ in Fig. \ref{inst}). The interaction of the central island with island 4 and  8 produces a force in the positive $y$ direction $F_{y,4} + F_{y,8} \propto dy/2$. The sum of forces $F_{y,1} + F_{y,3}+ F_{y,5} + F_{y,7} \propto dy/2 $ also produces a force in the positive vertical direction. From the symmetry of the problem it is clear that the sum of all the forces from all the islands produce a net force in the positive vertical direction - this amplifies the perturbations, leading to instability. 

We can also compare  directly the corresponding interaction  energies of the two states pictured in Fig. \ref{inst}.
 Let us approximate  each flux tube as a line  current $I$. Up to insignificant overall factor the interaction energy of two currents is $\propto I_1 I_2 \log r_{12}$ where $r_{12}$ is the distance between the two currents. Consider an arbitrary flux tube. It's interaction with flux tubes located in the even row from a given one are the same in two cases. In the second case the interaction with flux tubes located in the odd rows is zero by symmetry, while in the initial state it is
 \be
 E_{odd} = \sum_{n=0} (-1)^n \ln \sqrt{n^2+(2m+1)^2},
 \label{ee}
  \ee
 (here $2m+1, \, m=0,1,2...$  is the number of a row from a given island, $n=0,1,2...$ is the number of the islands in a row from  a given islands). 
 This sum (\ref{ee}) is positive: the  interaction energy with the $n=0$ island is always positive, $\propto \ln ((2m+1) r_0)$; the sum over $n$th and $n+1$ islands is also positive,  $\propto \ln ( (n+1)^2 + (2m +1)^2)/ (n^2 + (2m +1)^2)> 0$. 
 Thus, the energy of the second state in Fig. \ref{inst} is lower - this is the driver of the instability.

\begin{figure}
\includegraphics[width=0.49\linewidth]{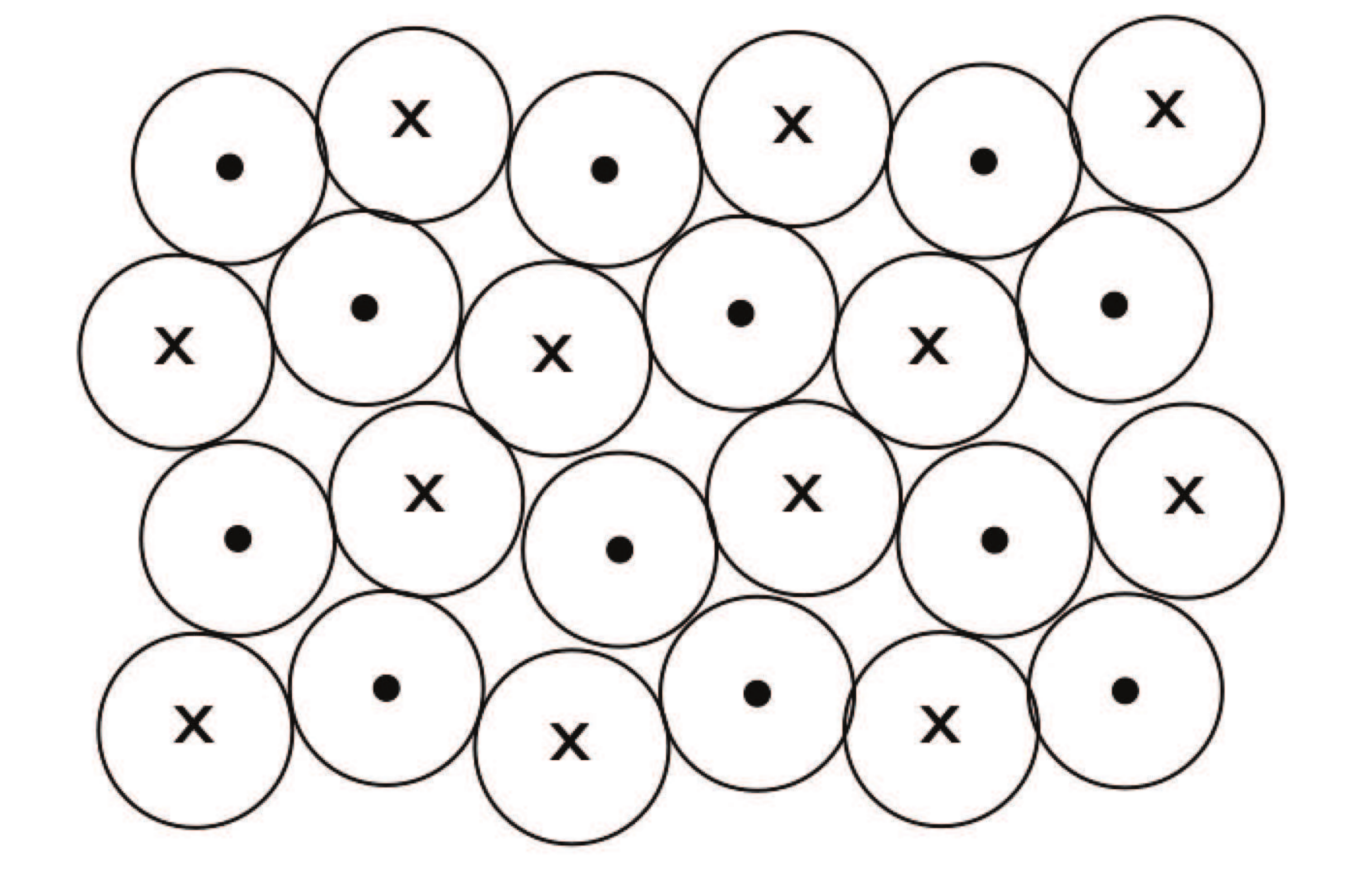}
\includegraphics[width=0.49\linewidth]{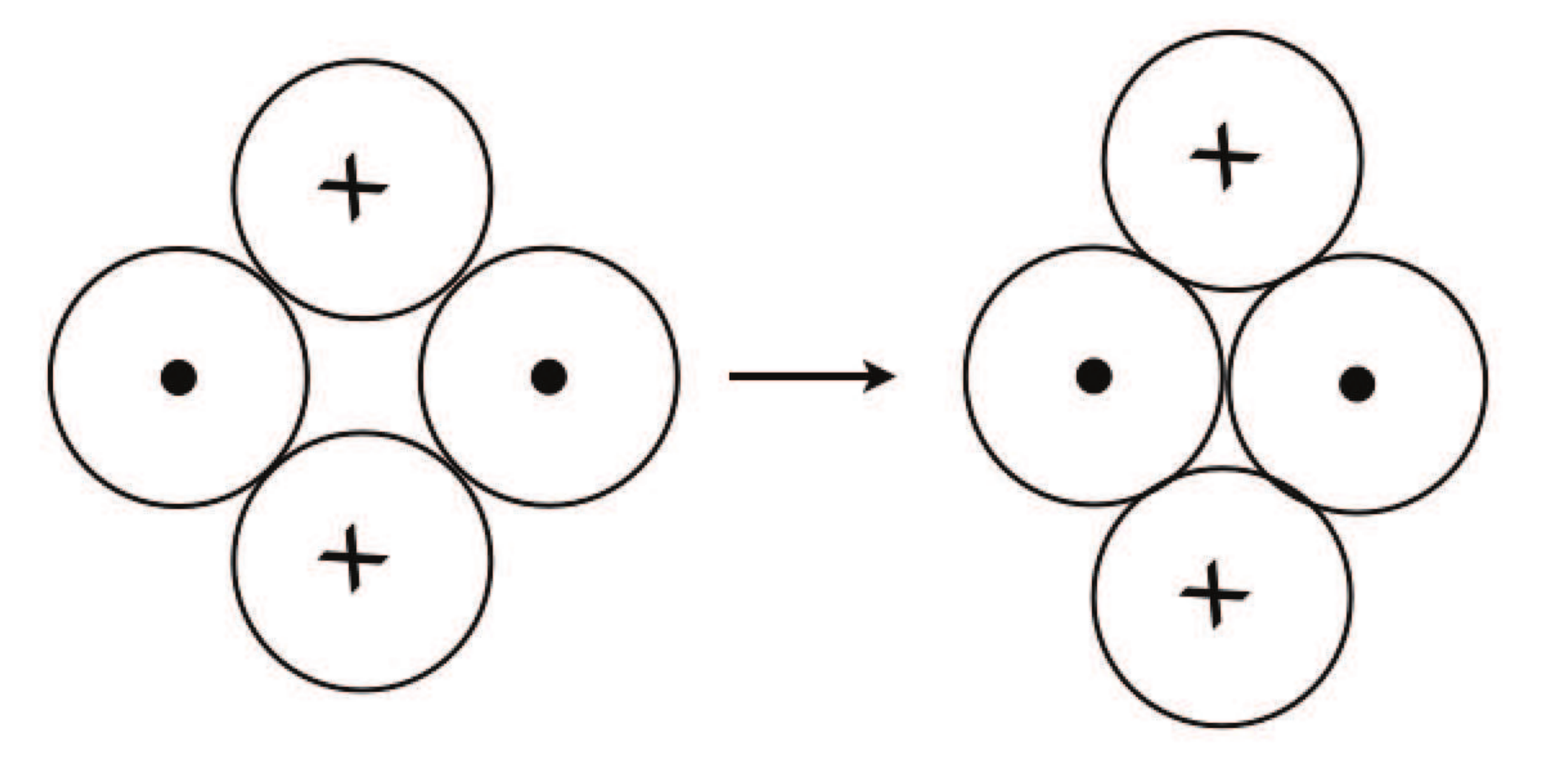}
\caption{Second  (oblique) mode of instability.}
\label{inst-2mode}
\end{figure}

There is another instability mode that we identified, Fig. \ref{inst-2mode}. Instead of coherent shift of the rows/columns it involve local re-arrangement of two pairs of  flux tubes, Fig. \ref{inst-2mode}. This configuration has a lower energy than the initial state: the forces of interaction of each touching pair of currents with all other touching pairs of currents is zero by symmetry. Thus, what is important  is the  change in the energy for each two pairs of currents, left panel. It is negative and is $ \propto - \ln (2/\sqrt{3})$.

This simple model problem demonstrates semi-qualitatively the exponentially growing  instability of a system of  magnetic islands.  The instability initially is ideal and proceeds on the dynamical time scale of an island, so that perturbations grow to non-linear stage in few dynamical time scales. 

The above  analytical  considerations are clearly confirmed by numerical experiments as  we discuss next.

% \clearpage
%%%%%%%

%%%%%%Sergey%%%%
%ssssssssssssssssssssssssssssssssssss
\subsection{2D ABC instability: evolution in the force-free regime}
\label{unstr-latt-ff-1}
%ssssssssssssssssssssssssssssssssssss

\subsubsection{Force-free simulations}

Force-free simulations were carried out in a square domain of size $[-L,L]\times[-L,L]$ with 400
grid cells in each direction. Periodic boundary conditions were imposed on all boundaries. 
Four models with different magnetic Reynolds numbers $Re_m=4\pi\kappa_\parallel L/c$ were investigated.
Fig.~\ref{ff-bme} shows the evolution of the parameter $1-E^2/B^2$, which is a Lorentz-invariant 
measure of the relative electric field strength, for the model with $Re_m=10^3$. Until $t=7$ this parameter 
remains very close to zero throughout the whole domain, indicating absence of current sheets and rapid 
motion in the bulk. Around $t=7$, thin current sheets become visible in between magnetic 
islands (see the top-left panel of Fig.~\ref{ff-bme}).  After this time the evolution proceeds rapidly - 
small islands merge to form larger ones until only two islands remain as one can also see in 
 Fig.~\ref{ff-bz}, which shows the evolution of $B_z$.  Models with higher $Re_m$ 
show very similar evolution, with almost the same time for the onset of mergers and the same rate 
of magnetic dissipation after that.   Fig.~\ref{ff-energy} shows the evolution of the total electromagnetic
energy in the box. One can see that for $t<7$, the energy significantly decreases in models with 
$Re_m=10^3$ and $Re_m=2\times10^3$ due to the Ohmic dissipation, whereas it remains more or less 
unchanged in the models with $Re_m > 5\times10^3$. The first ``knee'' on each curve corresponds to 
the onset of the first wave of merges. At $t>7$ the magnetic dissipation rate no longer depends on $Re_m$. 
The characteristic time scale of the process is very short -- only few light-crossings of the box.

\begin{figure}
\includegraphics[width=0.49\linewidth]{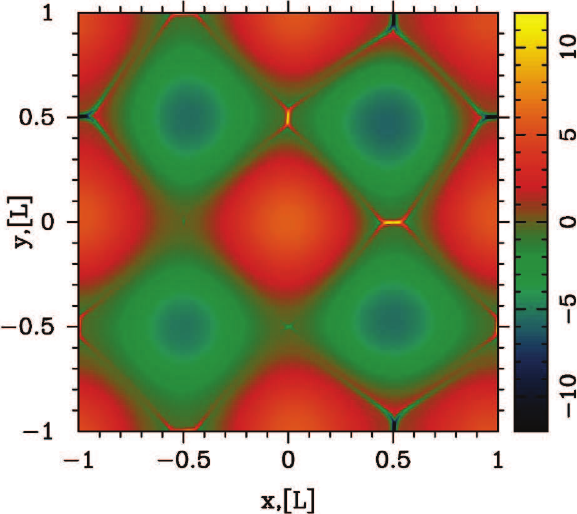}
\includegraphics[width=0.3\linewidth]{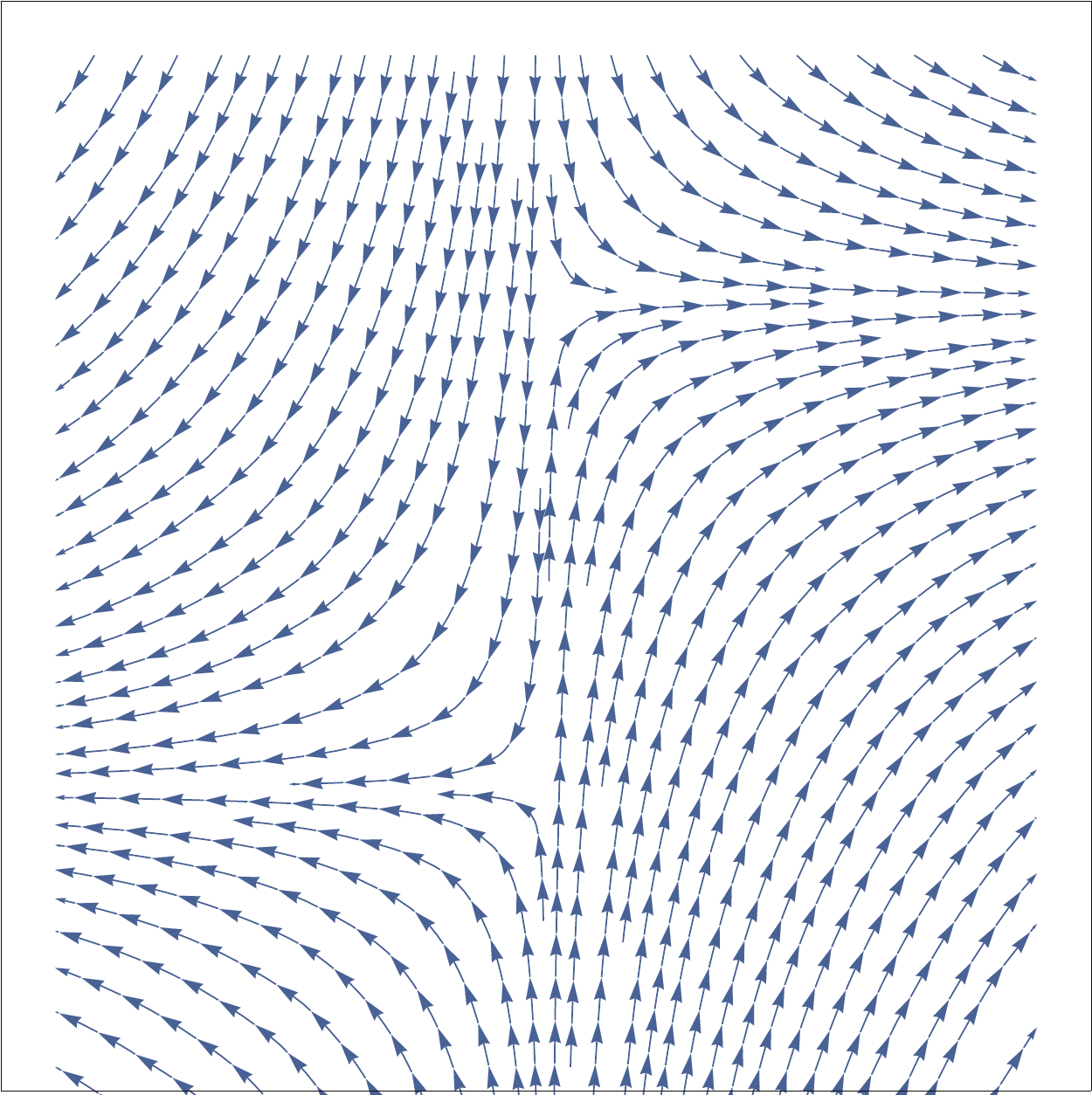}
\caption{ {\it Left Panel}: Initial stage of the development of instability in force-free simulations.
The colour image shows the distribution of the current density $j_z$ at $t=7.4$ in the model 
with $Re_m=10^3$. The newly formed current sheets are clearly visible. 
{\it Right Panel}: The developments of current sheet due to shift of magnetic islands in 
the theoretical model (in the initial configuration 2D ABC magnetic structure the two nearby rows of magnetic islands are shifted by some amount; this produces highly stressed configuration that would lead to X-point collapse). 
}
\label{11}
\end{figure}

\begin{figure}
\centering
\includegraphics[width=.8\textwidth]{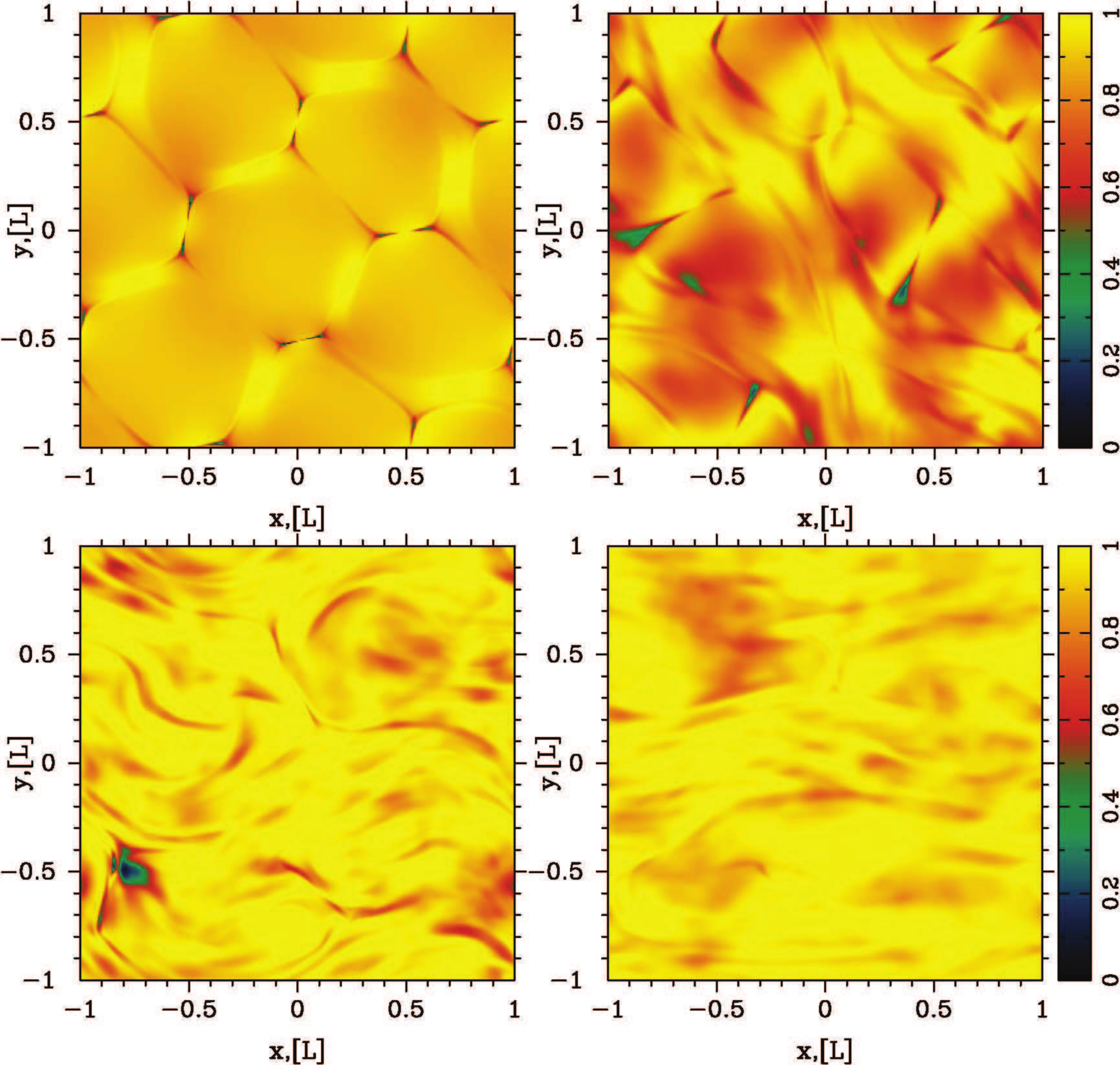}
\caption{X-point collapse and island merging for a set of
 unstressed magnetic islands in force-free simulations. 
We plot $1-E^2/B^2$ at times $t=8.0, 10.0, 15.0$ and 20.  
Compare with results of PIC simulations, Fig.  \protect\ref{fig:abcaccfluid}.
}
\label{ff-bme}
\end{figure}

\begin{figure}
\centering
\includegraphics[width=.8\textwidth]{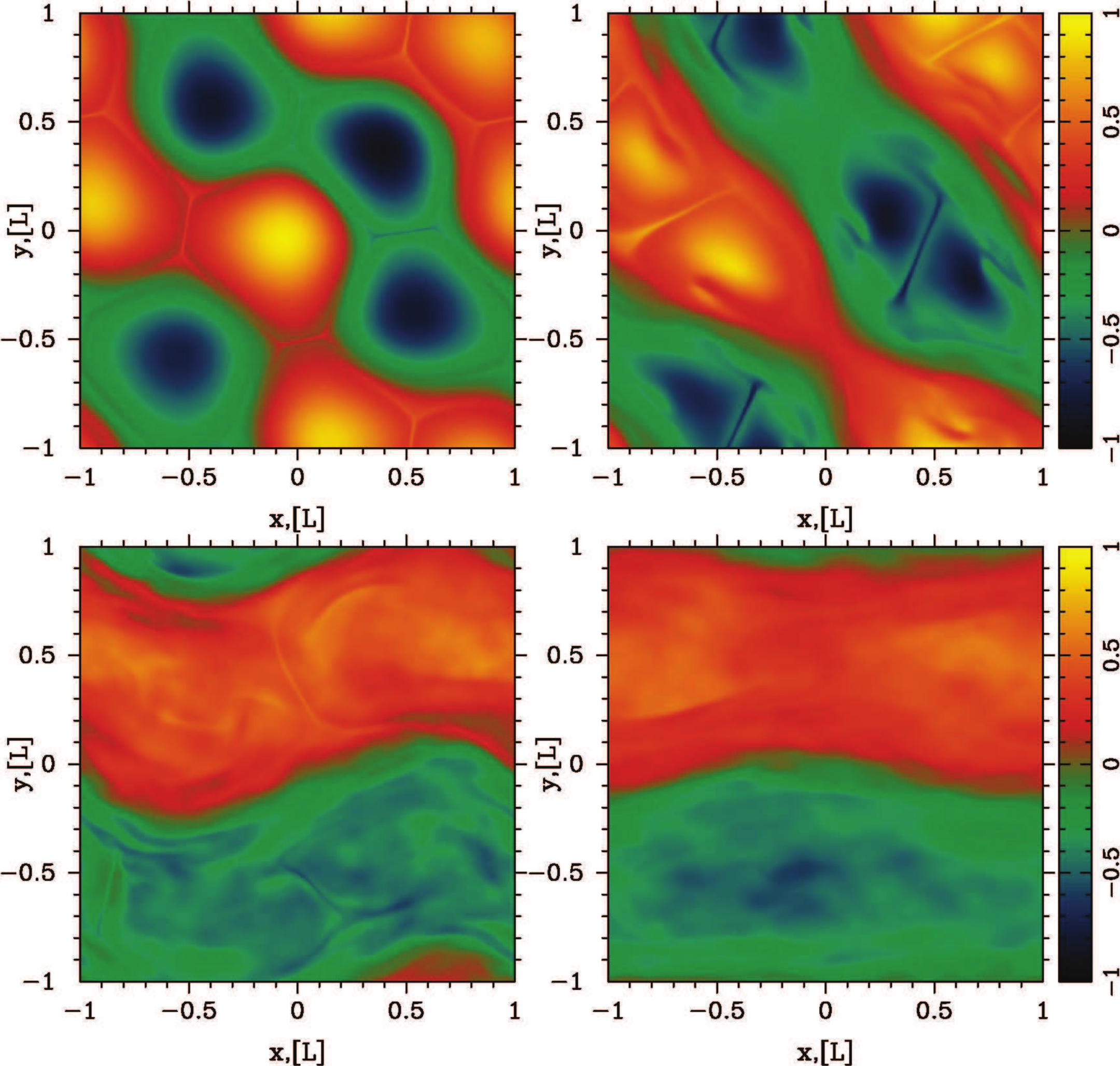}
\caption{X-point collapse and island merging for a set of
 unstressed magnetic islands in force-free simulations.
We plot $B_z/2B_0$ at times $t=8.0, 10.0, 15.0$ and 20.
}
\label{ff-bz}
\end{figure}

\begin{figure}
\centering
\includegraphics[width=.28\textwidth]{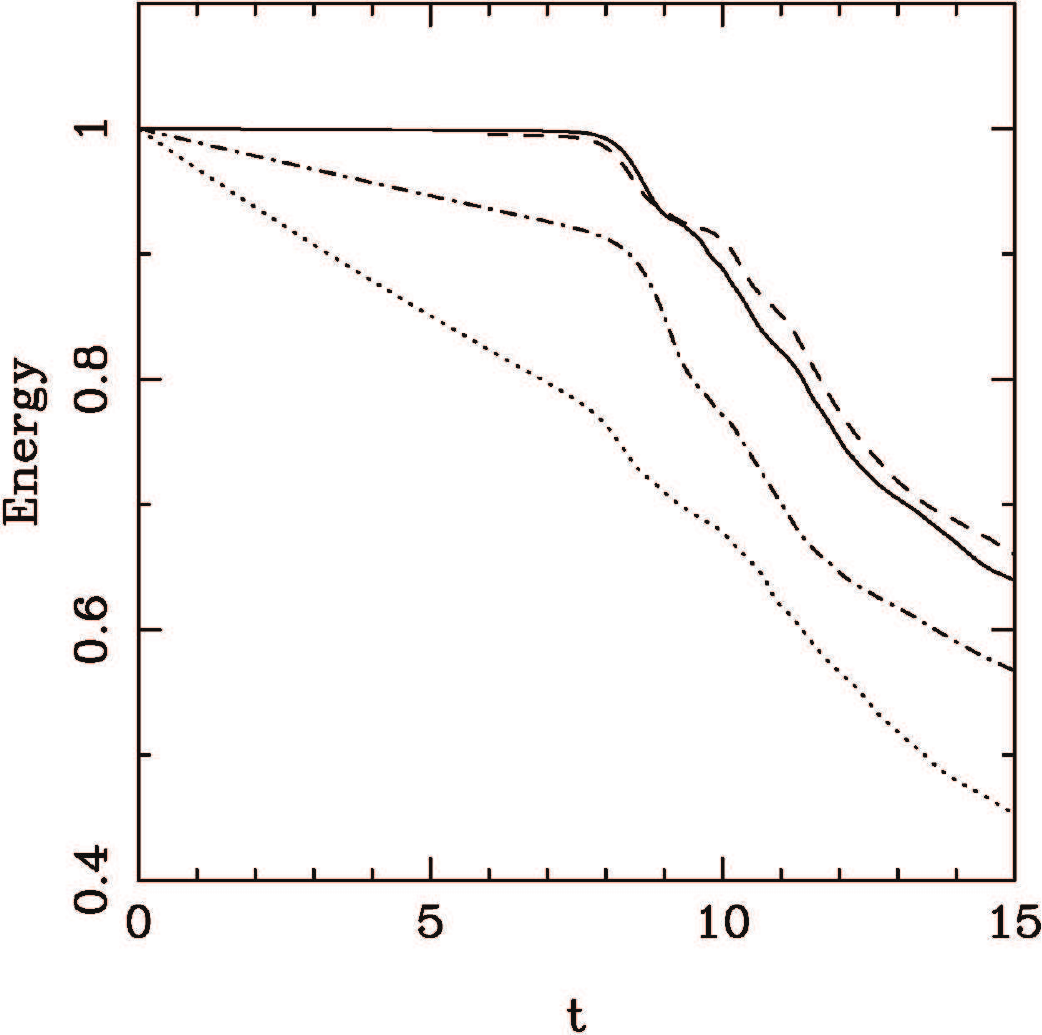}
\includegraphics[width=.6\textwidth]{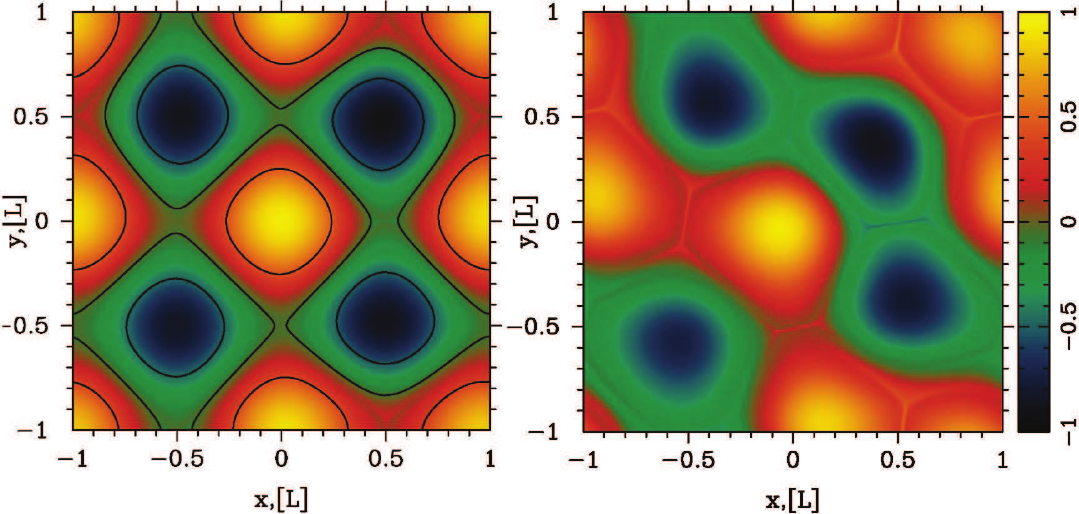}
\caption{Left panel: Evolution of the total electromagnetic energy in the computational domain 
for the force-free models with $Re_m=10^3$ (dotted), $2\times10^3$ (dash-dotted), $5\times10^3$
(dashed) and $2\times10^4$ (solid).  In all models, the merger phase starts around $t=8$, which
corresponds to the first ``knee'' of these curves.  Compare the temporal evolution of the \EM\ in
force-free simulations with those in PICs, Fig. \protect\ref{fig:abctime}.  Middle and Right panels:
Solutions at $t=7.2$ (left) and $t=8$ (right) for the model with $Re_m=10^3$.  The color-coded image
shows $B_z/2 B_0$.  The contours are the magnetic field lines. One can see that some lines have
become common for several islands.
}
\label{ff-energy}
\end{figure}

Overall, the numerical results agree with our theoretical analysis 
({\S} \ref{unstr-latt}-\ref{merger1}).  The fact that the onset of mergers does not 
depend on $Re_m$ supports the conclusion that the evolution starts as {\it an ideal instability, 
driven by the large-scale magnetic stresses.}  This instability leads to the 
relative motion of flux tubes and development of stressed X-points, with highly unbalanced magnetic 
tension. They collapse and form current sheets (see Fig.~\ref{11}).  

Until $t \leq 7$ the instability develops in the linear regime and the initial periodic structure 
is still well preserved. At the non-linear stage, at $t\geq 7$, two new important effects come 
into play.  Firstly, the magnetic reconnection leads to the emergence of closed magnetic field 
lines around two or more magnetic islands of similar polarity (see Fig. \ref{ff-energy}). 
Once formed, the common magnetic shroud pushes the islands towards the reconnection 
region in between via magnetic tension. Thus, this regime can be called a forced reconnection. 
As the reconnection proceeds, more common magnetic field lines develop, increasing the 
driving force.
Secondly, the current of the current sheet separating the merging islands becomes 
sufficiently large to slow down the merger via providing a repulsive force (\S \ref{merger1}). 
Hence, the reconnection rate is determined both by the resistivity of the current sheet and 
the overall magnetic tension of common field lines. 

We have studied the same problem using PIC simulations of highly-magnetized plasma 
and their results are strikingly similar (compare Figs. \ref{11}-\ref{ff-bme}-\ref{ff-bz} with  
Figs. \ref{2}-\ref{fig:abcaccfluid}). Not only they exhibit the same phases of evolution, but 
also very similar timescales.  This shows that the details of microphysics are not 
important and the key role is played by the large-scale magnetic stresses.

%fffffffffffffffffffffffffffffffffffffffffffffffffffffff
%fffffffffffffffffffffffffffffffffffffffffffffffffffffff

%fffffffffffffffffffffffffffffffffffffffffffffffffffffff
%ffffffffffffffffffffffffffffffffffffffffffffffffffffff

%fffffffffffffffffffffffffffffffffffffffffffffffffffffff

% \clearpage
%%%%%%%%

%%%%Lorenzo%%%%
%ABC
%\input{abc_LS1.tex}
%% \clearpage
%%%%Lorenzo%%%%

%%%%Lorenzo%%%%

%sssssssssssssssssssssssssssssssssssssssssssss
\subsection{2D ABC instability: PIC simulations}
\label{unstr-latt-pic}
We study the evolution of 2D ABC structures with PIC simulations, employing the electromagnetic PIC code  TRISTAN-MP \citep{buneman_93,spitkovsky_05} 
 \citep[Independently, a closely related study has been recently published by][]{2016ApJ...826..115N}.

To the best of our knowledge, our investigation is the first to address with PIC simulations the dynamics and particle acceleration of high-magnetization ABC structures.

Our computational domain is a square in the $x-y$ plane with periodic boundary conditions in both directions. The simulation box is initialized with a uniform density of electron-positron plasma at rest, a vanishing electric field and with the magnetic field appropriate for the 2D ABC configuration, as described in Eq.~\ref{Bb}. In addition to the unstressed geometry in Eq.~\ref{Bb}, we also investigate the case of 2D magnetic structures with an initial stress, or with an initial velocity shear, as we specify below.

For our fiducial runs, the spatial resolution is such that the plasma skin depth $\comp$ is resolved with 2.5 cells, but we have verified that our results are the same up to a resolution of $\comp=10$ cells. Since we investigate the case of both cold and hot background plasmas, our definition of the skin depth is $\comp=\sqrt{mc^2[1+(\hat{\gamma}-1)^{-1} kT/m c^2]/4 \pi n e^2}$, where $\hat{\gamma}$ is the adiabatic index. Each cell is initialized with two positrons and two electrons, but we have checked that our results are the same when using up to 16 particles per cell.  In order to reduce noise in the simulation, we filter
the electric current deposited onto the grid by the particles, mimicking the effect of a larger number of particles per cell \citep{spitkovsky_05,belyaev_15}.

Our unit of length is the diameter $L$ of the ABC structures, and time is measured in units of $L/c$. Typically, our domain is a square of side $2L$, but we have investigated also rectangular domains with size $2L\times L$, and large square domains with dimensions $4L\times 4L$. In addition to the shape of the domain, we also vary the flow parameters, such as the temperature of the background plasma ($kT/m c^2=10^{-4}$, $10$ and $10^2$) and the flow magnetization. The general definition of the magnetization is $\sigma=B^2/4\pi w$, where $w=n m c^2+\hat{\gamma}p/(\hat{\gamma}-1)$; here, $w$ is the enthalpy, $p$ is the pressure and $\hat{\gamma}$ is the adiabatic index. In the following, we identify our runs via the mean value $\sigmain$ of the magnetization measured  with the in-plane fields (so, excluding the $B_z$ field). As we argue below, it is the dissipation of the in-plane fields that primarily drives efficient heating and particle acceleration. The mean in-plane field corresponding to $\sigmain$ shall be called $B_{0,\rm in}$, and it will be our unit of measure of the electromagnetic fields. For the 2D ABC geometry, $B_{0,\rm in}$ is equal to the  field $B_0$ in Eq.~\ref{Bb}. We will explore a wide range of magnetizations, from $\sigmain=3$ up to $\sigmain=680$. The mean magnetization of the system, including all the magnetic field components, is $2\sigmain$.

It will be convenient to compare the diameter $L$ of ABC structures to the characteristic Larmor radius $\rhot=\sqrt{\sigmain}\comp$ of the high-energy particles heated/accelerated by reconnection (rather than to the skin depth $\comp$). We will explore a wide range of $L/\rhot$, from $L/\rhot=31$ up to $502$. We will demonstrate that the two most fundamental parameters that characterize a system of 2D ABC structures are the magnetization $\sigmain$ and the ratio $L/\rhot$.

In the following, we will quantify the upper cutoff of the particle energy spectrum as $\gamma_{\rm max}$, defined as
\be\label{eq:ggmax}
\gamma_{\rm max}=\frac{\int \gamma^{n_{\rm cut}} dN/d\gamma\, d\gamma}{\int \gamma^{n_{\rm cut}-1} dN/d\gamma\, d\gamma}
\ee
where $n_{\rm cut}$ is empirically chosen to be $n_{\rm cut}=6$. If the particle energy spectrum takes the form $dN/d\gamma\propto \gamma^{-s} \exp(-\gamma/\gamma_{\rm cut})$ with power-law slope $s$ and exponential cutoff at $\gamma_{\rm cut}$, then our definition yields $\gammamax\sim (n_{\rm cut}-s)\,\gamma_{\rm cut}$.

%%%%%%%%%%%%%%%%%%%%%%%
 \begin{figure}
 \centering
\includegraphics[width=.7\textwidth]{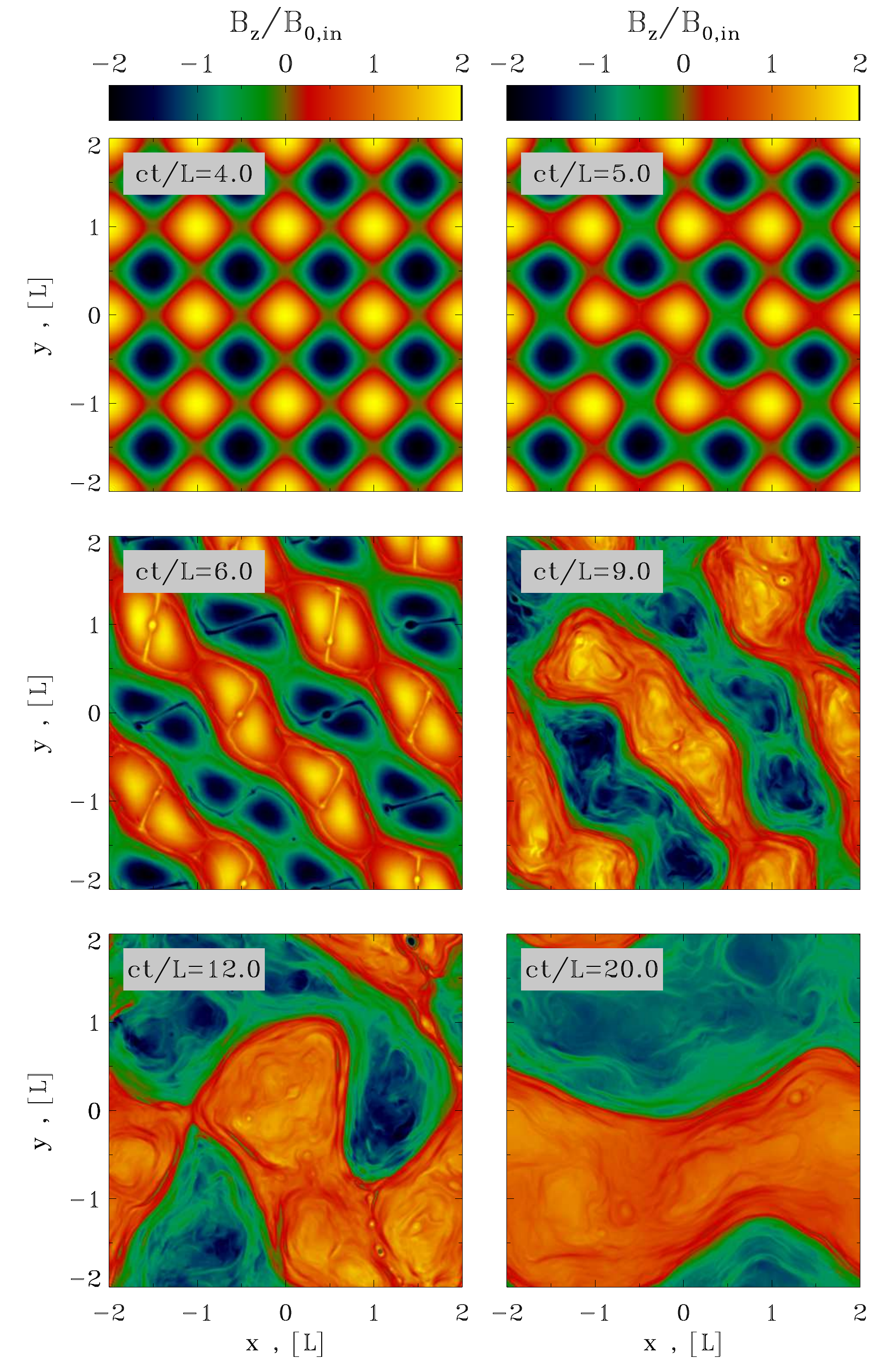} 
\caption{Temporal evolution of the instability of a typical 2D ABC structure (time is indicated in the grey box of each panel). The plot presents the 2D pattern of the out-of-plane field $B_z$ (in units of $B_{0,\rm in}$) from a PIC simulation with $kT/mc^2=10^2$, $\sigmain=42$ and $L=126\,\rhot$, performed within a large square domain of size $4L\times 4L$. The system stays unperturbed until $ct/L\simeq 4$, then it goes unstable via the oblique mode presented in Fig.~\ref{inst}. The instability leads to the formation of current sheets of length $\sim L$ (at $ct/L=6$), and to the merger of islands with $B_z$ fields of the same polarity. At the final time, the box is divided into two regions with $B_z$ fields of opposite polarity. This should be compared with the force-free results in Fig.~\ref{ff-bz}.}
\label{fig:abcfluid_bz} 
\end{figure}
 \begin{figure}
 \centering
\includegraphics[width=.49\textwidth]{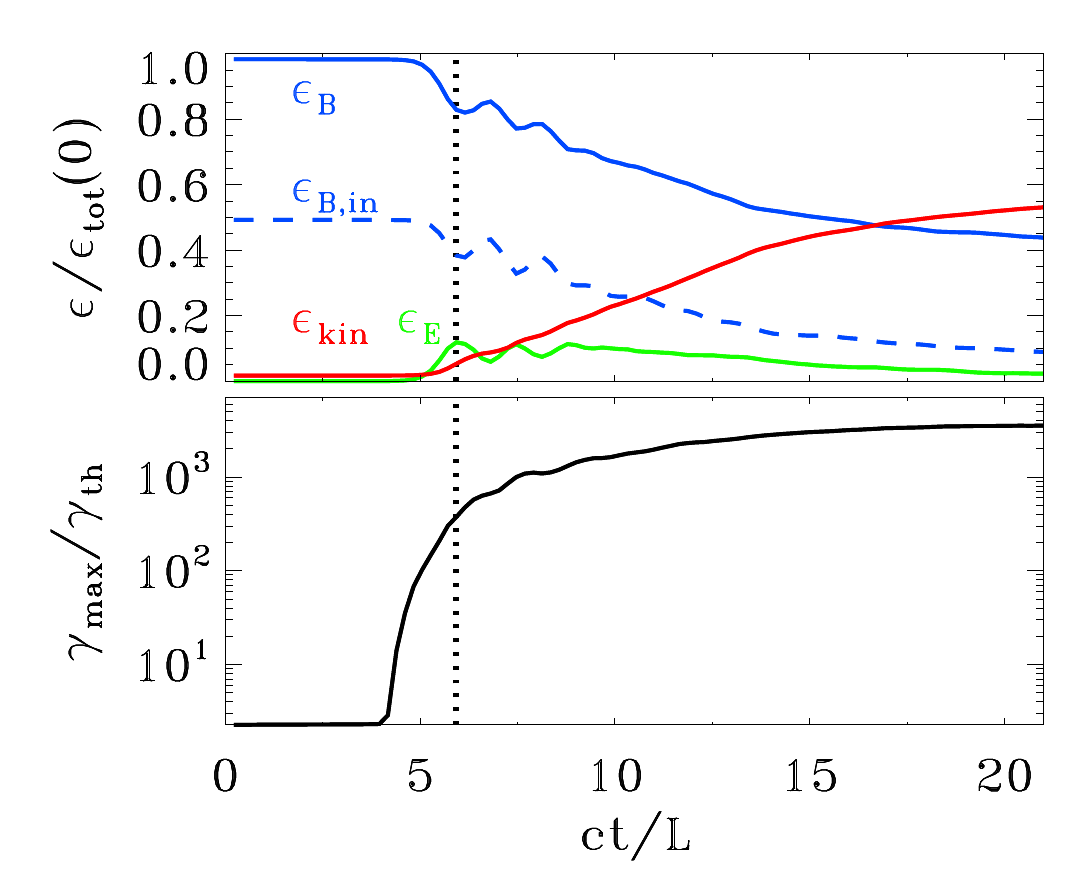} 
\caption{Temporal evolution of various quantities, from a 2D PIC simulation of ABC instability with $kT/mc^2=10^2$, $\sigmain=42$ and $L=126\,\rhot$, performed within a large square domain of size $4L\times 4L$ (the same run as in \fig{abcfluid_bz}). Top panel: fraction of energy in magnetic fields (solid blue), in-plane magnetic fields (dashed blue, with $\epsilon_{B,\rm in}=\epsilon_B/2$ in the initial configuration), electric fields (green) and particles (red; excluding the rest mass energy), in units of the total initial energy.  Compare the temporal evolution of the \EM\ in PIC simulations with those in   force-free, Fig. \protect\ref{ff-energy}.
 Bottom panel: evolution of the maximum Lorentz factor $\gammamax$, as defined in \eq{ggmax}, relative to the thermal Lorentz factor $\gamma_{\rm th}\simeq 1+(\hat{\gamma}-1)^{-1} kT/m c^2$, which for our case is $\gamma_{\rm th}\simeq 300$. The development of the instability at $ct/L\simeq 5$ is accompanied by little field dissipation ($\epsilon_{\rm kin}/\epsilon_{\rm tot}(0)\sim 0.1$) but dramatic particle acceleration. In both panels, the vertical dotted black line indicates the time when the electric energy peaks, which happens shortly before the end of the most violent phase of instability.}
\label{fig:abctime} 
\end{figure}
\subsubsection{The instability of 2D ABC structures}
Figure \ref{fig:abcfluid_bz} illustrates the instability of a typical 2D ABC structure. The plot presents the 2D pattern of the out-of-plane field $B_z$ (in units of $B_{0,\rm in}$) from a PIC simulation with $kT/mc^2=10^2$, $\sigmain=42$ and $L=126\,\rhot$, performed within a large square domain of size $4L\times 4L$. Time is measured in units of $L/c$ and indicated in the grey boxes within each panel. The system does not show any evidence of evolution until $ct/L\sim 4$. This is also confirmed by the temporal tracks shown in \fig{abctime}, where the top panel presents the energy partition among different components. Until $ct/L\sim4$, all the energy is still in the magnetic field (solid blue line), and the state of the system is identical to the initial setup.

Quite abruptly, at $ct/L\sim 5$ (top right panel in \fig{abcfluid_bz}), the system becomes unstable. The pattern of magnetic islands shifts along the oblique direction, in analogy to the mode illustrated in Fig.~\ref{inst}. In agreement with the fluid simulations presented by \cite{2015PhRvL.115i5002E}, our PIC results confirm that this is an ideal instability. The onset time is comparable to what is observed in force-free simulations (see Fig.~\ref{ff-energy}) and it does not appreciably depend on numerical parameters (so, on the level of numerical noise). In particular, we find no evidence for an earlier onset time when degrading our numerical resolution (spatial resolution or number of particles per cell). As we show below, the onset time at $ct/L\sim 5$ is also remarkably independent of physical parameters, with only a moderate trend for later onset times at larger values of $L/\rhot$. We have also investigated the dependence of the onset time on the number of ABC islands in the computational domain (or equivalently, on the box size in units of $L$). We find that square domains with $2L\times 2L$ or $4L\times 4L$ yield similar onset times, whereas the instability is systematically delayed (by a few $L/c$) in rectangular boxes with $2L\times L$, probably because this reduces the number of modes that can go unstable (in fact, we have verified that ABC structures in a  periodic square box of size $L\times L$ are stable).
We conclude that the abrupt evolution observed at times  $ct/L\geq 5$ is physically motivated -  this the stage where the instabilities reaches a non-linear stage.

Following the instability onset, the evolution of the system proceeds on the dynamical time. Within a few $L/c$, as they drift along the oblique direction, neighboring islands with the same $B_z$ polarity come into contact (middle left panel in \fig{abcfluid_bz} at $ct/L=6$), the X-points in between each pair of islands collapse under the effect of large-scale stresses (see Paper I), and thin current sheets are formed. For the parameters of \fig{abcfluid_bz}, the resulting current sheets are so long that they are unstable to the secondary tearing mode \citep{uzdensky_10}, and secondary plasmoids are formed (e.g., see the plasmoid at $(x,y)=(-1.5L,L)$ at $ct/L=6$). Below, we demonstrate that the formation of secondary plasmoids is primarily controlled by the ratio $L/\rhot$. At the X-points, the magnetic energy is converted into particle energy by a reconnection electric field of order $\sim 0.3 B_{\rm in}$, where $B_{\rm in}$ is the in-plane field (so, the reconnection rate is $v_{\rm rec}/c\sim 0.3$).\footnote{This value of the reconnection rate is roughly comparable to the results of solitary X-point collapse presented in Paper I. However, a direct comparison cannot be established, since in that case we either assumed a uniform nonzero guide field or a vanishing guide field. Here, the X-point collapse starts in regions with zero guide field. Yet, as reconnection proceeds, an out-of-plane field of increasing strength is advected into the current sheet.} As shown in the top panel of \fig{abctime}, it is primarily the in-plane field that gets dissipated (compare the dashed and solid blue lines), driving an increase in the electric energy (green) and in the particle kinetic energy (red).\footnote{The out-of-plane field does not dissipate because island mergers always happen between pairs of islands having the same $B_z$ polarity.} In this phase of evolution, the fraction of initial energy released to the particles is still small ($\epsilon_{\rm kin}/\epsilon_{\rm tot}(0)\sim 0.1$), but the particles advected into the X-points experience a dramatic episode of acceleration. As shown in the bottom panel of \fig{abctime}, the cutoff Lorentz factor $\gammamax$ of the particle spectrum presents a dramatic evolution, increasing from the thermal value $\gamma_{\rm th}\simeq 3 kT/m c^2$ (here, $kT/m c^2=10^2$) by a factor of $10^3$ within a couple of dynamical times. It is this phase of extremely fast particle acceleration that we associate with the generation of the Crab flares.

Over one dynamical time, the current sheets in between neighboring islands stretch to their maximum length of $\sim L$. This corresponds to the time when the electric energy peaks, which is indicated by the dotted black line in \fig{abctime}, and it shortly precedes the end of the most violent phase of instability. The peak time of the electric energy will be a useful reference point when comparing runs with different physical parameters. The first island merger event ends at $ct/L\sim 7$ with the coalescence of island cores. At later times, the system of magnetic islands will undergo additional merger episodes (i.e., at $ct/L\sim9$ and 12), with the formation of current sheets and secondary plasmoids (see, e.g., at $(x,y)=(L,-1.5L)$ in the bottom left panel of \fig{abcfluid_bz}), but the upper cutoff of the particle spectrum will not change by more than a factor of three (see the bottom panel in \fig{abctime} at $ct/L\gtrsim 6$). As a consequence, the spectral cutoff at the final time is not expected to be significantly dependent on the number of lengths $L$ in the computational domain, as we have indeed verified. Rather than dramatic acceleration of a small number of ``lucky'' particles, the late phases result in substantial heating of the bulk of the particles, as illustrated by the red line in the top panel of \fig{abctime}. As a consequence of particle heating, the effective magnetization of the plasma is only $\sigmain\sim 1$ after the initial merger episode (compare the dashed blue and solid red lines at $ct/L\sim9$ and 12 in the top panel of \fig{abctime}), which explains why additional island mergers do not lead to dramatic particle acceleration.
The rate of dissipation of magnetic energy into particle energy at late times is roughly consistent with the results of force-free simulations (see Fig.~\ref{ff-energy}). At late times, the system saturates when only two regions are left in the box, having $B_z$ fields of opposite polarity (bottom right panel in \fig{abcfluid_bz}). In the final state, most of the in-plane magnetic energy has been trasferred to the particles (compare the red and dashed blue lines in the top panel of \fig{abctime}).

%%%%%%%%%%%%%%%%%%%%%%%
 \begin{figure}
 \centering
\includegraphics[width=.49\textwidth]{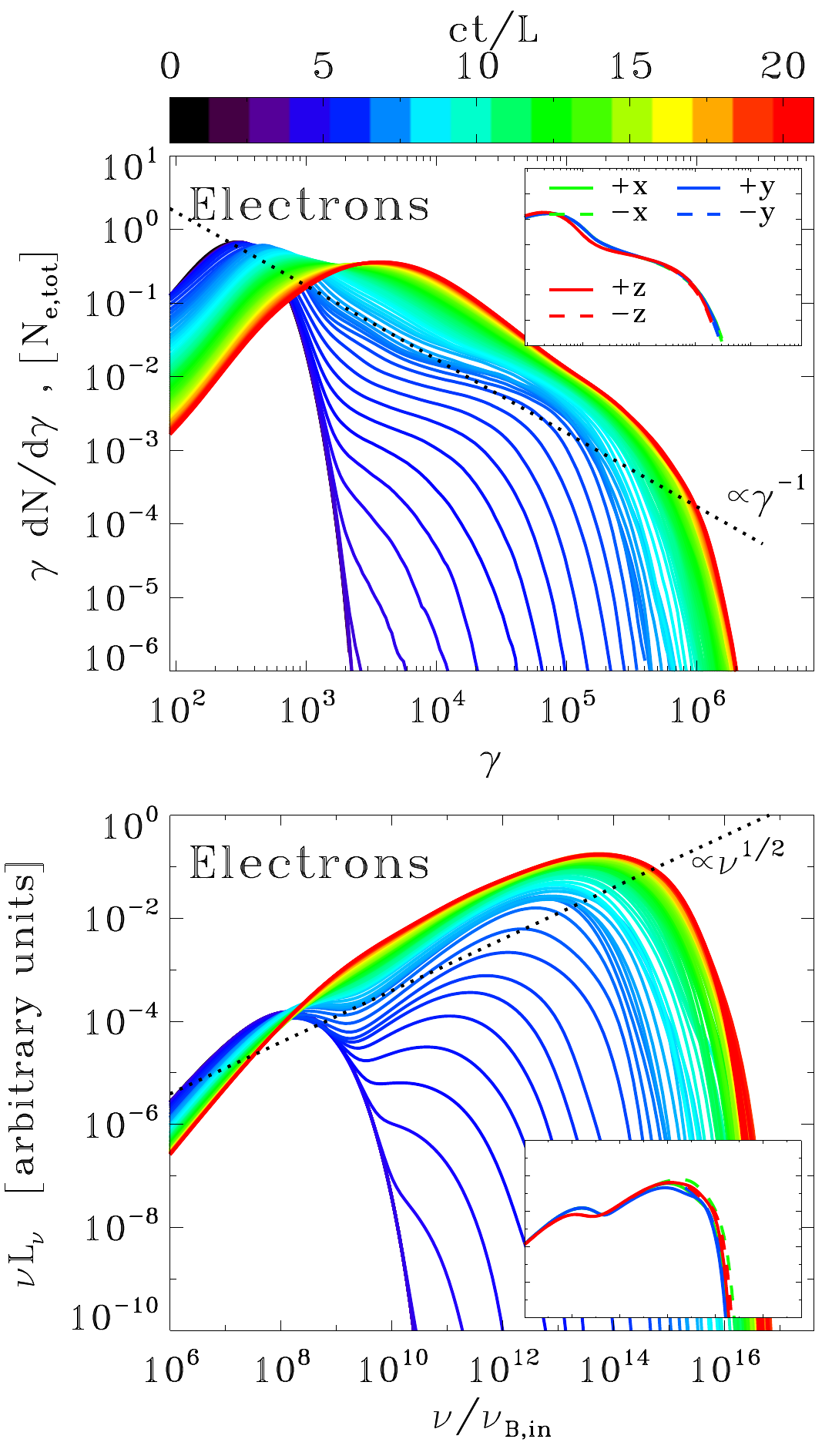} 
\caption{Particle spectrum and synchrotron spectrum from a 2D PIC simulation of ABC instability with $kT/mc^2=10^2$, $\sigmain=42$ and $L=126\,\rhot$, performed within a large square domain of size $4L\times 4L$ (the same run as in \fig{abcfluid_bz} and \fig{abctime}). Time is measured in units of $L/c$, see the colorbar at the top. Top panel: evolution of the electron energy spectrum normalized to the total number of electrons. At late times, the spectrum approaches a distribution of the form $\gamma dN/d\gamma\propto \gamma^{-1}$, corresponding to equal energy content in each decade of $\gamma$ (compare with the dotted black line). The inset in the top panel shows the electron momentum spectrum along different directions (as indicated in the legend), at the time when the electric energy peaks (as indicated by the dotted black line in \fig{abctime}). Bottom panel: evolution of the angle-averaged synchrotron spectrum emitted by electrons. The frequency on the horizontal axis is normalized to $\nu_{B,\rm in}=\sqrt{\sigmain}\omega_{\rm p}/2\pi$. At late times, the synchrotron spectrum approaches $\nu L_\nu\propto \nu^{1/2}$ (compare with the dotted black line), which just follows from the electron spectrum $\gamma dN/d\gamma\propto \gamma^{-1}$. The inset in the bottom panel shows the synchrotron spectrum at the time indicated in \fig{abctime} (dotted black line) along different directions (within a solid angle of $\Delta \Omega/4\pi\sim 3\times10^{-3}$), as indicated in the legend in the inset of the top panel. The resulting  isotropy of the synchrotron emission is consistent with the particle distribution illustrated in the inset of the top panel.}
\label{fig:abcspec} 
\end{figure}
 \begin{figure}
 \centering
\includegraphics[width=.49\textwidth]{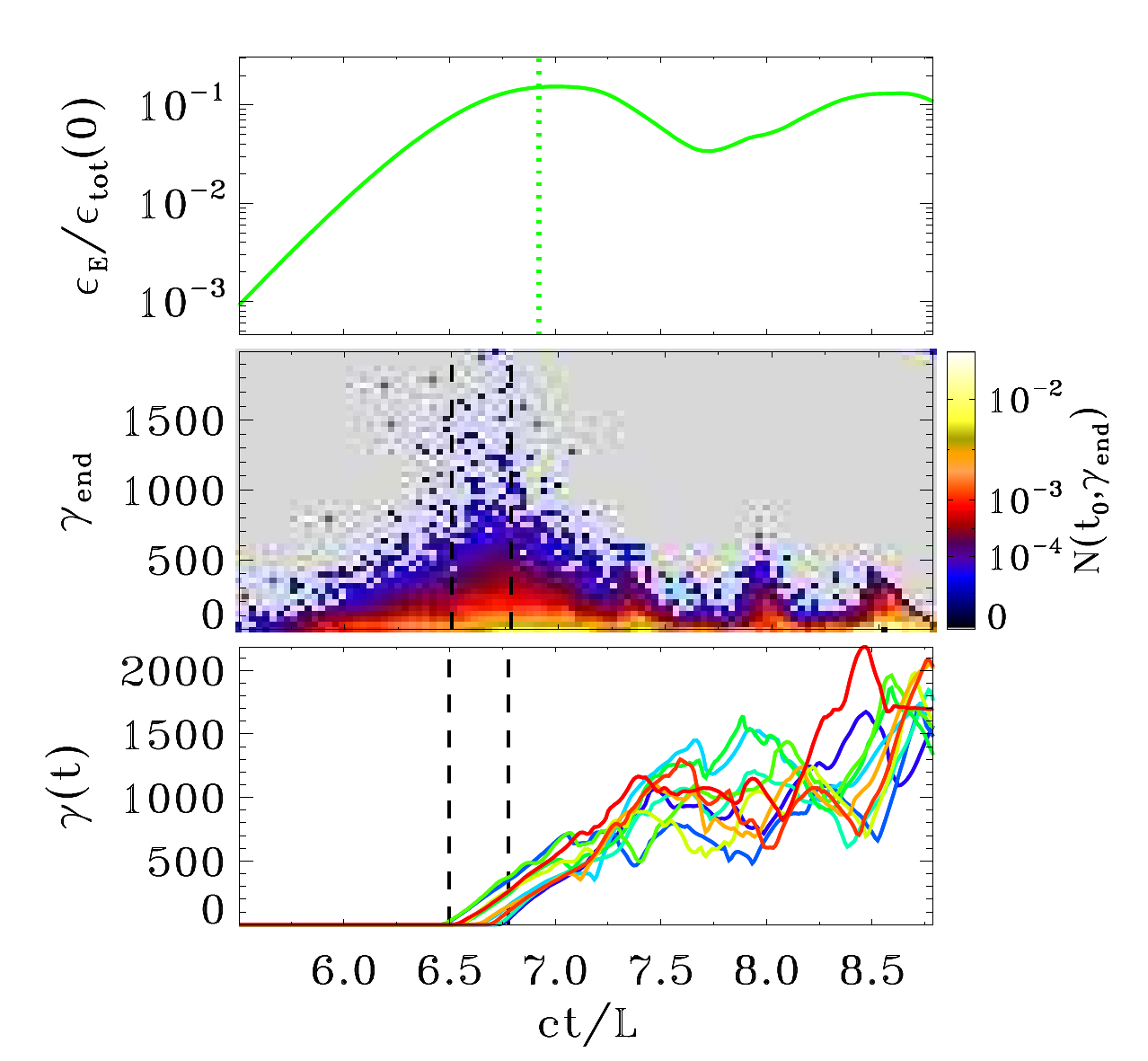} 
\caption{Physics of particle acceleration, from  a 2D PIC simulation of ABC instability with $kT/mc^2=10^{-4}$, $\sigmain=42$ and $L=251\,\rhot$, performed within a square domain of size $2L\times 2L$. Top panel: temporal evolution of the electric energy, in units of the total energy at the initial time. The dotted green vertical line marks the time when the electric energy peaks. Middle panel: 2D histogram of accelerated particles, normalized to the total number of particles. On the vertical axis we plot the final particle Lorentz factor $\gamma_{\rm end}$ (measured at the final time $ct/L=8.8$), while the horizontal axis shows the particle injection time $ct_0/L$, when the particle Lorentz factor first exceeded  the threshold $\gamma_{\rm 0}=30$. Bottom panel: we select all the particles that exceed the threshold $\gamma_{\rm 0}=30$ within a given time interval (chosen to be $6.5\leq ct_0/L\leq6.8$, as indicated by the vertical dashed black lines in the middle and bottom panels), and we plot the temporal evolution of the Lorentz factor of the ten particles that at the final time reach the highest energies.}
\label{fig:abcacctime} 
\end{figure}
 \begin{figure}
 \centering
\includegraphics[width=.49\textwidth]{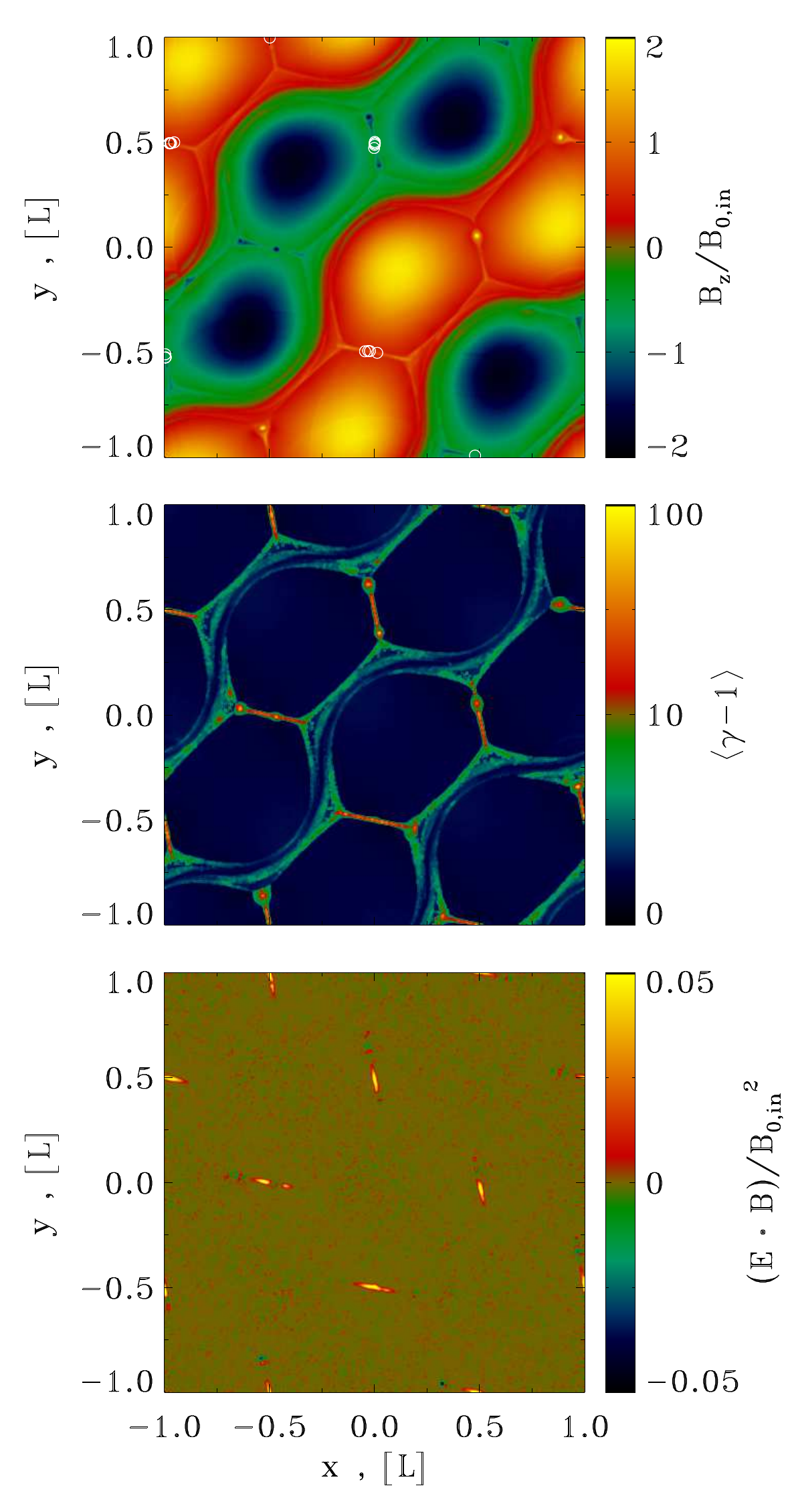} 
\caption{Physics of particle injection into the acceleration process, from  a 2D PIC simulation of ABC instability with $kT/mc^2=10^{-4}$, $\sigmain=42$ and $L=251\,\rhot$, performed within a square domain of size $2L\times 2L$ (same run as in \fig{abcacctime}). We plot the 2D ABC structure at $ct/L=6.65$. Top panel: 2D plot of the out-of-plane field $B_z$, in units of $B_{0,\rm in}$. Among the particles that exceed the threshold $\gamma_{\rm 0}=30$ within the interval $6.5\leq ct_0/L\leq6.8$ (as indicated by the vertical dashed black lines in the middle and bottom panels of \fig{abcacctime}), we select the 20 particles that at the final time reach the highest energies, and
 with open white circles we plot their locations at the injection time $t_0$. Middle panel: 2D plot of the mean kinetic energy per particle $\langle\gamma-1\rangle$. Bottom panel: 2D plot of $\bmath{E}\cdot\bmath{B}/B_{0,\rm in}^2$, showing in red and yellow the regions of charge starvation. Comparison of the top panel with the bottom panel shows that particle injection is localized in the charge-starved regions.}
\label{fig:abcaccfluid} 
\end{figure}

\begin{figure}
\includegraphics[width=0.9\linewidth]{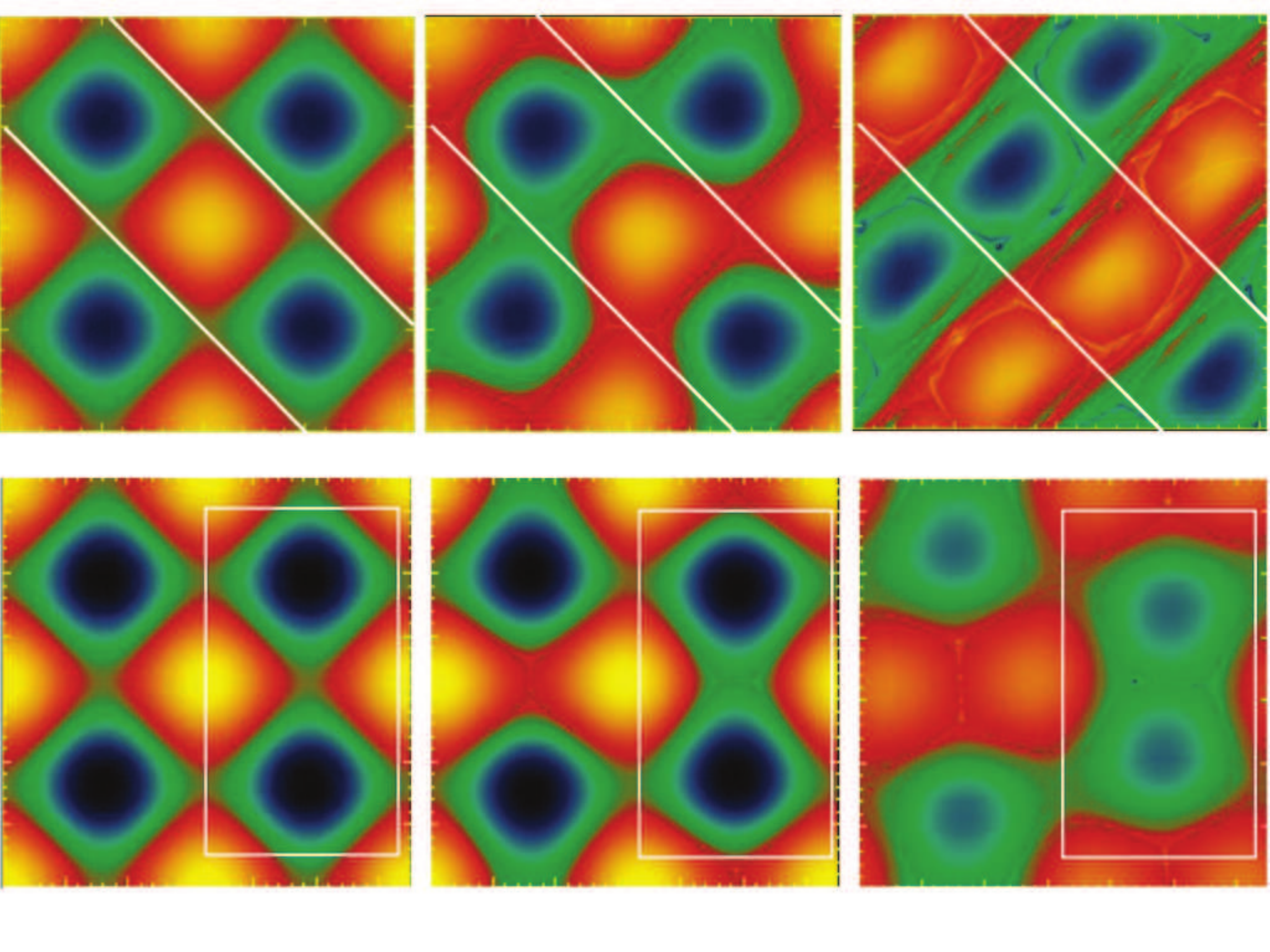}
\caption{Snapshots of the numerical experiment demonstrating  development of different modes of  instability of 2D ABC structures. {\it Upper Panel}: The parallel mode. White line, separating different layers are added as a guide. These pictures clearly demonstrate that during the development of the instability two layers of magnetic islands are shifting with respect to each other. Right Panel:  the compression mode. The white box demonstrates that a pair of aligned currents periodically  approach each other. }
\label{2}
\end{figure}

In  Fig. \ref{2} we compare the development of the two modes of instability, the parallel and the oblique one.
% In addition in \S \ref{idealABC} we consider analytically  the growth of the instability by considering large-scale coherent distortions of  the islands' shape.

\subsubsection{Particle acceleration and emission signatures}
\label{Particleaccelerationandemissionsignatures}
The evolution of the particle spectrum in the top panel of \fig{abcspec} illustrates the acceleration capabilities of the instability of ABC structures. Until $ct/L\sim 4$, the particle spectrum does not show any sign of evolution, and it does not deviate from the initial Maxwellian distribution peaking at $\gamma_{\rm th}\simeq 300$. From $ct/L\sim 4$ up to $ct/L\sim 8$ (from purple to light blue in the plot), the high-energy end of the particle spectrum undergoes a dramatic evolution, whereas the thermal peak is still unaffected (so, only a small fraction of the particles are accelerated). This is the most dramatic phase of particle acceleration, which we associate with the Crab flares. At the end of this explosive phase, the high-energy tail of the particle spectrum resembles a power law $dN/d\gamma\propto\gamma^{-s}$ with a hard slope $s\sim 1.5$.  The angle-averaged synchrotron spectrum (bottom panel in \fig{abcspec}) parallels the extreme evolution of the particle spectrum, with the peak frequency of $\nu L_\nu$ moving from the ``thermal'' value $\sim 10^8\nu_{B,\rm in}$  (dominated by the emission of thermal particles with $\gamma\sim \gamma_{\rm th}$) up to $\sim 10^{13}\nu_{B,\rm in}$ within just a few dynamical times. Here, we have defined $\nu_{B,\rm in}=\sqrt{\sigmain}\omega_{\rm p}/2 \pi$.

The evolution at later times, during subsequent merger episodes, is at most moderate. From $ct/L\sim 8$ up to the final time $ct/L\sim 20$ (from light blue to red in the plot), the upper energy cutoff of the particle spectrum only increases by a factor of three, and the peak synchrotron frequency only by a factor of ten.  The most significant evolution of the electron spectrum at such late times involves the thermal peak of the distribution, which shifts up in energy by a factor of $\sim 10$. This confirms what we have anticipated above, i.e., that subsequent episodes of island mergers primarily result in particle heating, rather than non-thermal acceleration. At the final time, the particle high-energy spectrum resembles a power law $\gamma dN/d\gamma\propto \gamma^{-1}$ (compare with the dotted black line in the top panel) and, consequently, the angle-averaged synchrotron spectrum approaches $\nu L_\nu\propto \nu^{1/2}$ (compare with the dotted black line in the bottom panel).

The inset in the top panel of \fig{abcspec} shows the electron momentum distribution along different directions (as indicated in the legend), at the time when the electric energy peaks (as indicated by the dotted black line in \fig{abctime}). The electron distribution is roughly isotropic. This might appear surprising, since at this time the particles are being dramatically accelerated at the collapsing X-points in between neighboring islands, and in Paper I we have demonstrated that the  particle distribution during X-point collapse is highly anisotropic. The accelerated electrons were beamed along the reconnection outflow and in the direction opposite to  the accelerating electric field (while positrons were parallel to the electric field). Indeed, the particle distribution at each X-point in the collapsing ABC structure shows the same anisotropy as for a solitary X-point. 

The apparent isotropy in the inset of the top panel of \fig{abcspec} is a peculiar result of the ABC geometry. The middle left panel in \fig{abcfluid_bz} shows that in the $x-y$ plane the number of current sheets (and so, of reconnection outflows) oriented along $x$ is roughly comparable to those along $y$. So, no preferred direction for the electron momentum spectrum is expected in the $x-y$ plane. In addition, in 2D ABC structures the accelerating electric field $E_z$ has always the same polarity as the local out-of-plane magnetic field $B_z$ (or equivalently, $\bmath{E}\cdot\bmath{B}>0$ at X-points, as we explicitly show in the bottom panel in \fig{abcaccfluid}). In other words, current sheets that are colored in yellow in the middle left panel of \fig{abcfluid_bz} have $E_z>0$, whereas $E_z<0$ in current sheets colored in blue. Since the two options occur in equal numbers, no difference between the electron momentum spectrum in the $+z$ versus $-z$ direction is expected. Both the spatially-integrated electron momentum distribution (inset in the top panel) and the resulting synchrotron emission (inset in the bottom panel) will then be nearly isotropic, for the case of ABC collapse.

 Figs. \fign{abcacctime} and \fign{abcaccfluid} describe the physics of particle acceleration during the instability of ABC structures. We consider a representative simulation with $kT/mc^2=10^{-4}$, $\sigmain=42$ and $L=251\,\rhot$, performed within a square domain of size $2L\times 2L$. In the middle panel of \fig{abcacctime}, we present a 2D histogram of accelerated particles. On the vertical axis we plot the final particle Lorentz factor $\gamma_{\rm end}$ (measured at $ct/L=8.8$), while the horizontal axis shows the particle injection time $ct_0/L$, when the particle Lorentz factor first exceeded the threshold $\gamma_{\rm 0}=30$ of our choice. The histogram shows that the particles that eventually reach the highest energies are injected into the acceleration process around $6.5\leq ct_0/L\leq 6.8$ (the two boundaries are indicated with vertical dashed black lines in the middle and bottom panels). For our simulation, this interval corresponds to the most violent stage of ABC instability, and it shortly precedes the time when the electric energy peaks (see the dotted green line in the top panel of \fig{abcacctime}) and the current sheets reach their maximum length $\sim L$.\footnote{Even when we extend our time window up to $\sim 15 L/c$, including additional merger episodes, we still find that the highest energy particles are injected during the initial ABC collapse.} In order to reach the highest energies, the accelerated particles should sample the full available potential of the current sheet, by exiting the reconnection layer when it reaches its maximal extent. Since they move at the speed of light along the sheet half-length $\sim 0.5 L$, one can estimate that they should have been injected $\sim 0.5 L/c$ earlier than the peak time of the electric field. This is indeed in agreement with \fig{abcacctime} (compare top and middle panels).

Among the particles injected within $6.5\leq ct_0/L\leq 6.8$, the bottom panel in \fig{abcacctime} shows the energy evolution of the ten positrons that reach the highest energies. We observe two distinct phases: an initial stage of direct acceleration by the reconnection electric field (at $6.5\lesssim ct/L \lesssim7.3$), followed by a phase of stochastic energy gains and losses (from $ct/L \sim7.3$ up to the end). During the second stage, the pre-accelerated particles bounce off the wobbling/merging magnetic islands, and they further increase their energy via a second-order Fermi process \citep[similar to what was observed by][]{2012PhRvL.108m5003H}. At the time of injection, the selected positrons were all localized in  the vicinity of the collapsing X-points. This is shown in the top panel of \fig{abcaccfluid}, where we plot with open white circles the location of the selected positrons at the time of injection, superimposed over the 2D pattern of $B_z$ at $t\sim 6.65 L/c$. In the current sheets resulting from X-point collapse the force-free condition $\bmath{E}\cdot \bmath{B}=0$ is violated (see the yellow regions in the bottom panel of \fig{abcaccfluid}), so that they are sites of efficient particle acceleration (see also the middle panel in \fig{abcaccfluid}, showing the mean particle kinetic energy $\langle\gamma-1\rangle$).

%%%%%%%%%%%%%%%%%%%%%%%
 \begin{figure}
 \centering
\includegraphics[width=.49\textwidth]{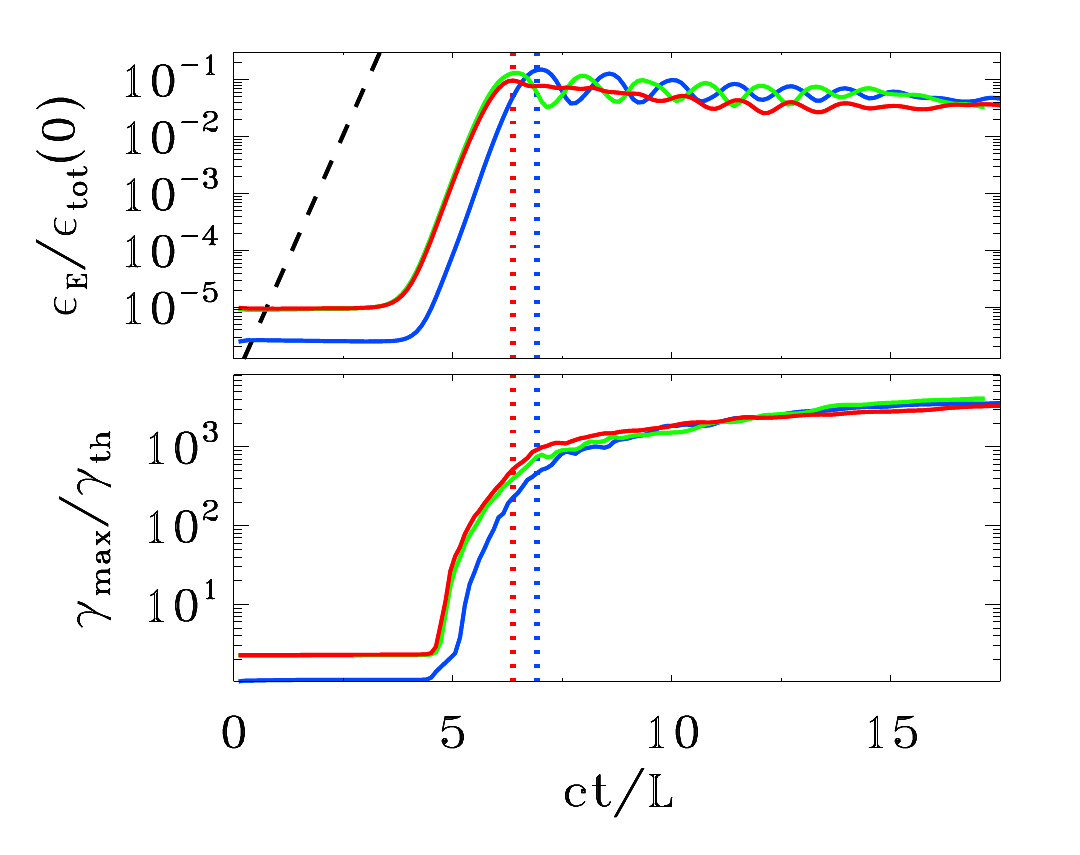} 
\caption{Temporal evolution of the electric energy (top panel, in units of the total initial energy) and of the maximum particle Lorentz factor (bottom panel; $\gammamax$ is defined in \eq{ggmax}, and it is normalized to the thermal Lorentz factor $\gamma_{\rm th}\simeq 1+(\hat{\gamma}-1)^{-1} kT/m c^2$), for a suite of three PIC simulations of ABC collapse with fixed $\sigmain=42$ and fixed $L/\rhot=251$, but different plasma temperatures: $kT/m c^2=10^{-4}$ (blue), $kT/m c^2=10$ (green) and $kT/m c^2=10^{2}$ (red). The dashed black line in the top panel shows that the electric energy grows exponentially as $\propto \exp{(4ct/L)}$. The vertical dotted lines mark the time when the electric energy peaks (colors correspond to the three values of $kT/m c^2$, as described above).}
\label{fig:abcdgamtime} 
\end{figure}
 \begin{figure}
 \centering
\includegraphics[width=.49\textwidth]{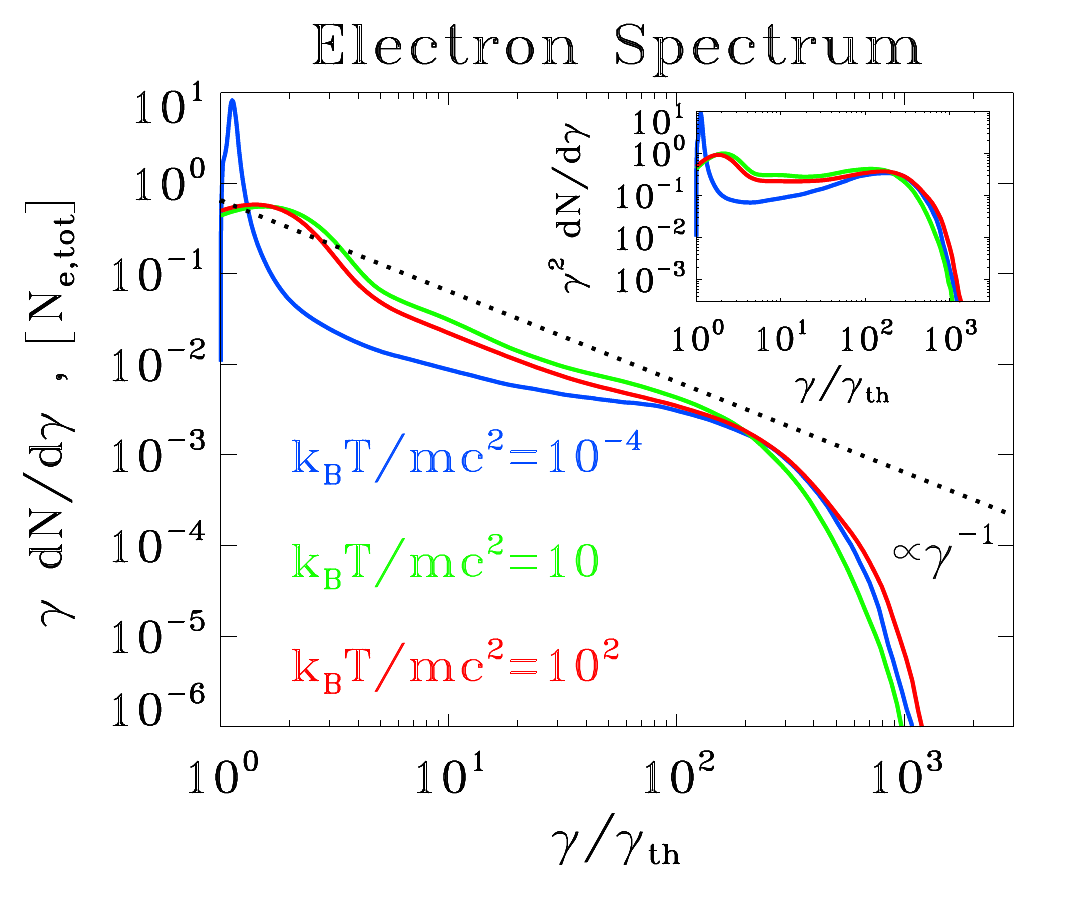} 
\caption{Particle spectrum at the time when the electric energy peaks, for a suite of three PIC simulations of ABC collapse with fixed $\sigmain=42$ and fixed $L/\rhot=251$, but different plasma temperatures: $kT/m c^2=10^{-4}$ (blue), $kT/m c^2=10$ (green) and $kT/m c^2=10^{2}$ (red). The main plot shows $\gamma dN/d\gamma$ to emphasize the particle content, whereas the inset presents $\gamma^2 dN/d\gamma$ to highlight the energy census. The dotted black line is a power law $\gamma dN/d\gamma\propto \gamma^{-1}$, corresponding to equal energy content per decade (which would result in a flat distribution in the inset). The particle Lorentz factor on the horizontal axis is normalized to the thermal value $\gamma_{\rm th}$, to facilitate comparison among the three cases.}
\label{fig:abcdgamspec} 
\end{figure}
\subsubsection{Dependence on the flow parameters}
In this subsection, we explore the dependence of our results on the initial plasma temperature, the in-plane magnetization $\sigmain$ and the ratio $L/\rhot$.

 Figs. \fign{abcdgamtime} and \fign{abcdgamspec} describes the dependence of our results on the flow temperature, for a suite of three simulations of ABC collapse with fixed $\sigmain=42$ and fixed $L/\rhot=251$, but different plasma temperatures: $kT/m c^2=10^{-4}$ (blue), $kT/m c^2=10$ (green) and $kT/m c^2=10^{2}$ (red). The definitions of both the in-plane magnetization $\sigmain$ and the plasma skin depth $\comp$ (and so, also the Larmor radius $\rhot$) account for the flow temperature, as described above. With this choice, Figs. \fign{abcdgamtime} and \fign{abcdgamspec} demonstrate that our results do not appreciably depend on the plasma temperature, apart from an overall shift in the energy scale. In particular, in the relativistic regime $kT/mc^2\gg1$, our results for different temperatures are virtually undistinguishable. The onset of the ABC instability is nearly the same for the three values of temperature we investigate (top panel in \fig{abcdgamtime}), the exponential growth is the same (with a rate equal to $\sim 4 c/L$ for the electric energy, compare with the dashed black line) and the peak time is quite similar (with a minor delay for the non-relativistic case $kT/mc^2=10^{-4}$, in blue). At the onset of the ABC instability, the upper cutoff $\gammamax$ of the particle energy spectrum grows explosively. Once normalized to the thermal value $\gamma_{\rm th}\simeq 1+(\hat{\gamma}-1)^{-1} kT/m c^2$ (which equals $\gamma_{\rm th}\simeq 1$ for $kT/mc^2\ll1$ and $\gamma_{\rm th}\simeq 3 kT/mc^2$ for $kT/mc^2\gg1$), the temporal evolution of $\gammamax$ does not appreciably depend on the initial temperature, especially in the relativistic regime $kT/mc^2\gg1$ (green and red lines).

Similar conclusions hold for the particle spectrum $\gamma dN/d\gamma$ at the time when the electric energy peaks, presented in the main panel of \fig{abcdgamspec}. The spectra of  $kT/mc^2=10$ (green) and $kT/mc^2=10^2$ (red) overlap, once the horizontal axis is normalized to the thermal Lorentz factor $\gamma_{\rm th}$. Their high-energy end approaches the slope $\gamma dN/d\gamma\propto \gamma^{-1}$ indicated with the dotted black line. This corresponds to a distribution with equal energy content per decade (which would result in a flat distribution in the inset, where we plot $\gamma^2 dN/d\gamma$ to emphasize the energy census). The spectrum for $kT/mc^2=10^{-4}$ (blue line) is marginally harder, but its high-energy part is remarkably similar.

%%%%%%%%%%%%%%%%%%%%%%%
 \begin{figure}
 \centering
\includegraphics[width=.49\textwidth]{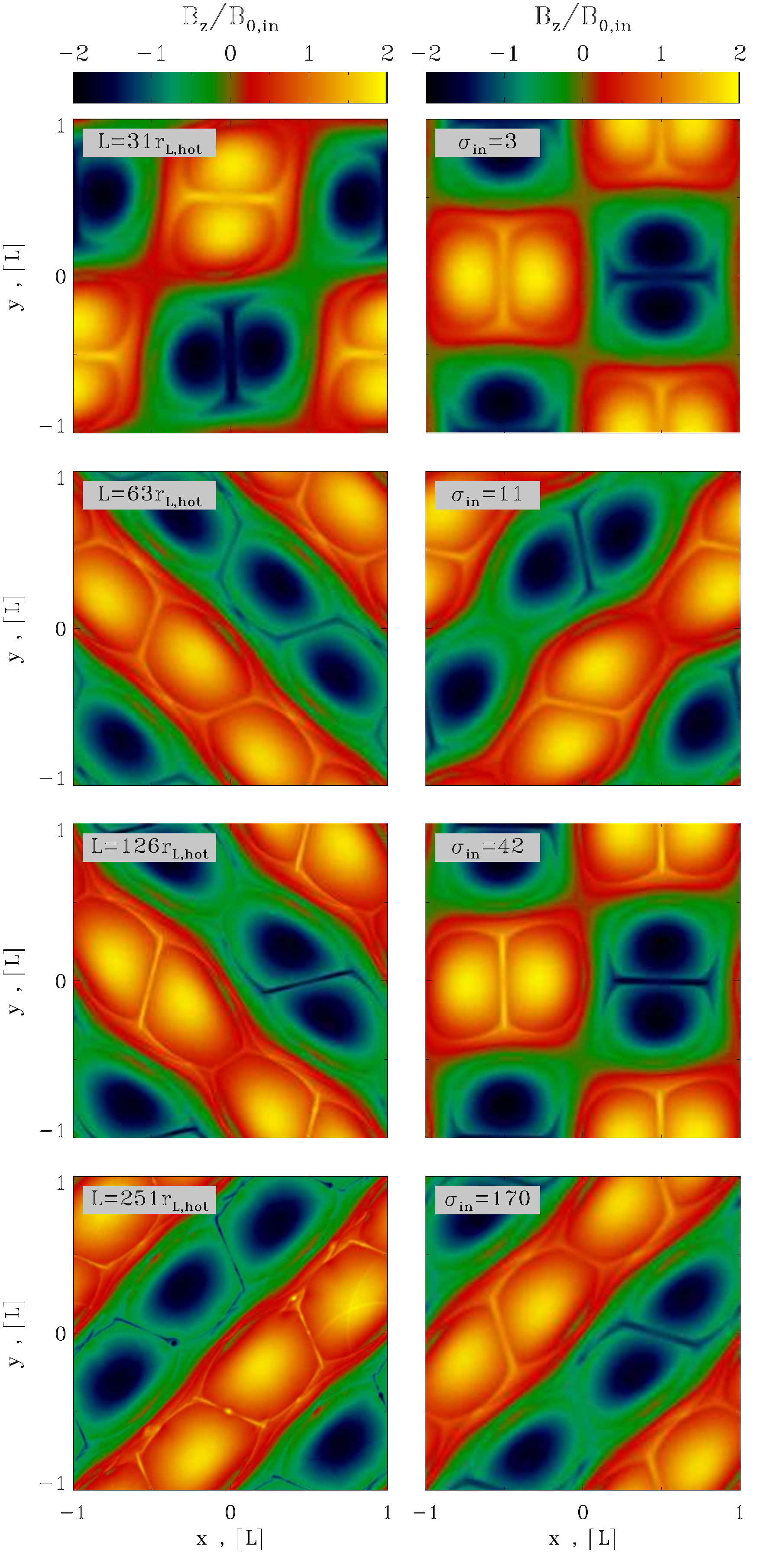} 
\caption{2D pattern of the out-of-plane field $B_z$ (in units of $B_{0,\rm in}$) at the most violent time of ABC instability (i.e., when the electric energy peaks, as indicated by the vertical dotted lines in \fig{abctimecomp}) from a suite of PIC simulations. In the left column, we fix $kT/mc^2=10^{-4}$ and $\sigmain=42$ and we vary the ratio $L/\rhot$, from 31 to 251 (from top to bottom). In the right column, we fix $kT/mc^2=10^2$ and $L/\rhot=63$ and we vary the magnetization $\sigmain$, from 3 to 170 (from top to bottom). In all cases, the domain is a square of size $2L\times 2L$. The 2D structure of $B_z$ in all cases is quite similar, apart from the fact that larger $L/\rhot$ tend to lead to a more pronounced fragmentation of the current sheet. Some cases go unstable via the ``parallel'' mode depicted in Fig.~\ref{inst-2mode}, others via the ``oblique'' mode described in Fig.~\ref{inst}.}
\label{fig:abcfluidcomp} 
\end{figure}
 \begin{figure}
 \centering
\includegraphics[width=.79\textwidth]{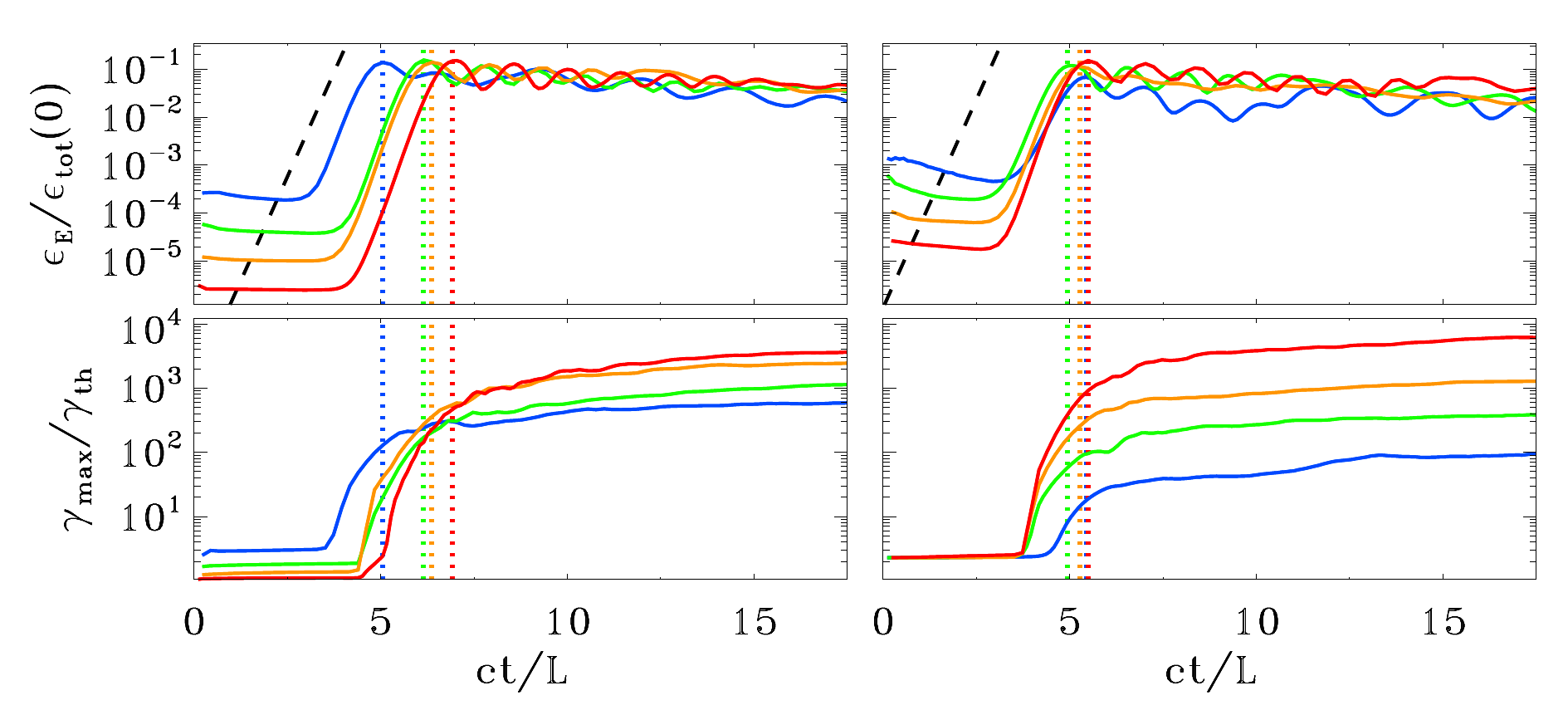} 
\caption{Temporal evolution of the electric energy (top panel, in units of the total initial energy) and of the maximum particle Lorentz factor (bottom panel; $\gammamax$ is defined in \eq{ggmax}, and it is normalized to the thermal Lorentz factor $\gamma_{\rm th}\simeq 1+(\hat{\gamma}-1)^{-1} kT/m c^2$), for a suite of PIC simulations of ABC collapse (same runs as in \fig{abcfluidcomp}). In the left column, we fix $kT/mc^2=10^{-4}$ and $\sigmain=42$ and we vary the ratio $L/\rhot$ from 31 to 251 (from blue to red). In the right column, we fix $kT/mc^2=10^2$ and $L/\rhot=63$ and we vary the magnetization $\sigmain$ from 3 to 170 (from blue to red). The maximum particle energy resulting from ABC collapse increases for increasing $L/\rhot$ at fixed $\sigmain$ (left column) and for increasing $\sigmain$ at fixed $L/\rhot$.
The dashed black line in the top panel shows that the electric energy grows exponentially as $\propto \exp{(4ct/L)}$. The vertical dotted lines mark the time when the electric energy peaks (colors as described above).}
\label{fig:abctimecomp} 
\end{figure}
 \begin{figure}
 \centering
\includegraphics[width=.79\textwidth]{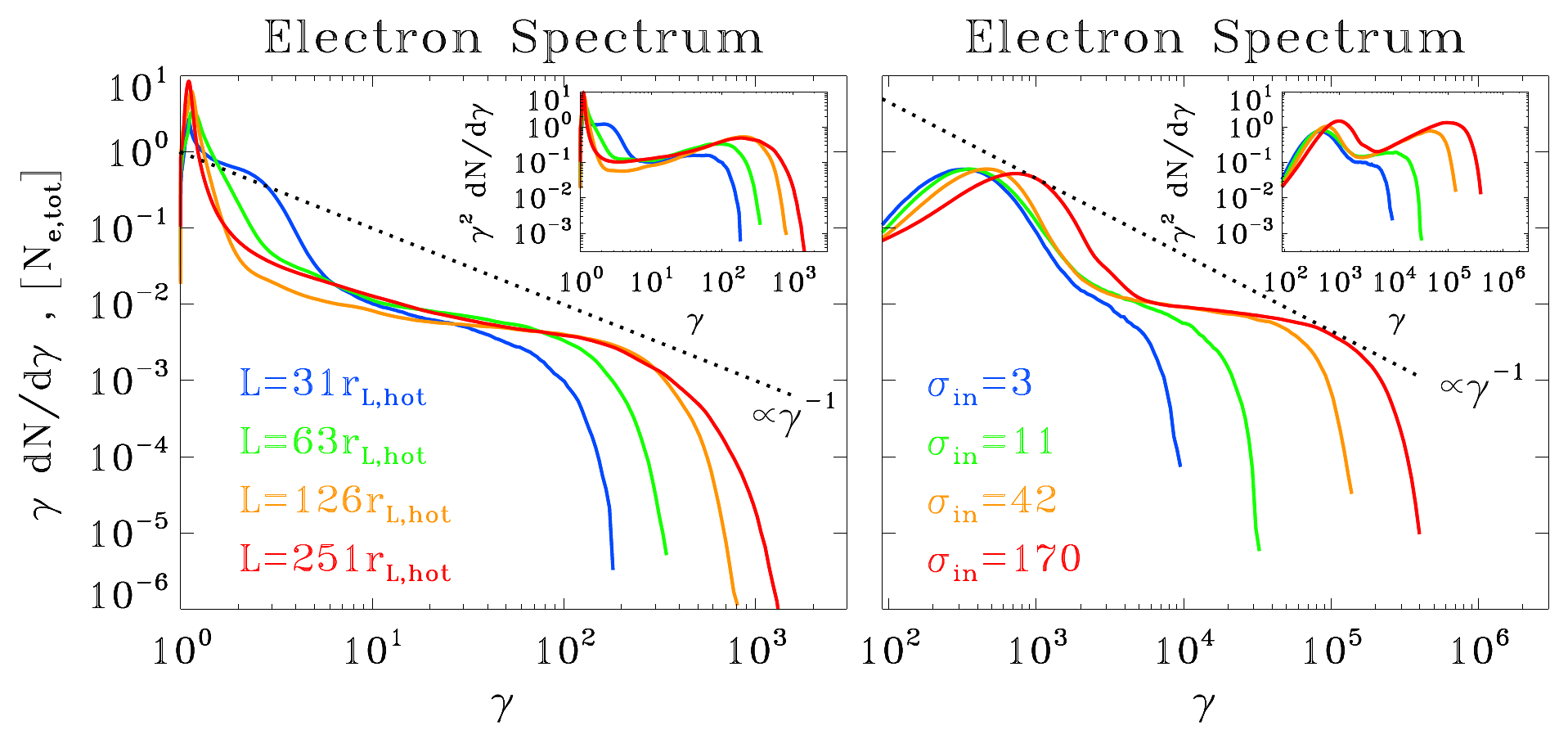} 
\caption{Particle spectrum at the time when the electric energy peaks, for a suite of PIC simulations of ABC collapse (same runs as in \fig{abcfluidcomp} and \fig{abctimecomp}). In the left column, we fix $kT/mc^2=10^{-4}$ and $\sigmain=42$ and we vary the ratio $L/\rhot$ from 31 to 251 (from blue to red, as indicated by the legend). In the right column, we fix $kT/mc^2=10^2$ and $L/\rhot=63$ and we vary the magnetization $\sigmain$ from 3 to 170 (from blue to red, as indicated by the legend). The main plot shows $\gamma dN/d\gamma$ to emphasize the particle content, whereas the inset presents $\gamma^2 dN/d\gamma$ to highlight the energy census. The dotted black line is a power law $\gamma dN/d\gamma\propto \gamma^{-1}$, corresponding to equal energy content per decade (which would result in a flat distribution in the insets). The spectral hardness is not a sensitive function of the ratio $L/\rhot$, but it is strongly dependent on $\sigmain$, with higher magnetizations giving harder spectra, up to the saturation slope of $-1$.}
\label{fig:abcspeccomp} 
\end{figure}

We now investigate the dependence of our results on the magnetization $\sigmain$ and the ratio $L/\rhot$, where $\rhot=\sqrt{\sigmain}\comp$. In \fig{abcfluidcomp}, we present the 2D pattern of the out-of-plane field $B_z$ (in units of $B_{0,\rm in}$) at the most violent time of ABC instability (i.e., when the electric energy peaks, as indicated by the vertical dotted lines in \fig{abctimecomp}) from a suite of PIC simulations in a square domain $2L\times 2L$. In the left column, we fix $kT/mc^2=10^{-4}$ and $\sigmain=42$ and we vary the ratio $L/\rhot$, from 31 to 251 (from top to bottom). In the right column, we fix $kT/mc^2=10^2$ and $L/\rhot=63$ and we vary the magnetization $\sigmain$, from 3 to 170 (from top to bottom). The figure shows that the instability proceeds in a similar way in all the runs, even though some cases go unstable via the ``parallel'' mode depicted in Fig.~\ref{inst-2mode} (e.g., see the top right panel) , others via the ``oblique'' mode described in Fig.~\ref{inst} (e.g., see the bottom left panel). Here, ``parallel'' and ``oblique'' refer to the orientation of the wavevector with respect to the axes of the box.

The evolution is remarkably similar, whether the instability proceeds via the parallel or the oblique mode. Yet, some differences can be found: ({\it i}) In the oblique mode, all the null points are activated, i.e., they all evolve into a long thin current sheet suitable for particle acceleration; in contrast, in the parallel mode only half of the null points are activated (e.g., the null point at $x=0$ and $y=0.5L$ in the top right panel does not form a current sheet, and we have verified that it does not lead to significant particle acceleration). Everything else being equal, this results in a difference by a factor of two in the normalization of the high-energy tail of accelerated particles (i.e., the acceleration efficiency differs by a factor of two), as we have indeed verified. ({\it ii}) The current sheets of the parallel mode tend to stretch longer than for the oblique mode (by roughly a factor of $\sqrt{2}$), resulting in a difference of a factor of $\sqrt{2}$ in the maximum particle energy, everything else being the same. ({\it iii}) in the explosive phase of the instability, the oblique mode tends to dissipate some fraction of the out-of-plane field energy (still, a minor fraction as compared to the dissipated in-plane energy), whereas the parallel mode does not. Regardless of such differences, both modes result in efficient particle acceleration and heating, and in a similar temporal evolution of the rate of magnetic energy dissipation (with the kinetic energy fraction reaching $\epsilon_{\rm kin}/\epsilon_{\rm tot}(0)\sim 0.1$ during the explosive phase, and eventually saturating at $\epsilon_{\rm kin}/\epsilon_{\rm tot}(0)\sim 0.5$).

%\fig{fig1} demonstrates, by plotting the 2D pattern of the mean energy per particle during the explosive phase, that indeed particle acceleration still happens in the regions that stretch the most (so, not at every magnetic null; and in fact, in the plot the regions that are less magnetized at this particular time, shown in blue, are not very capable of efficient particle acceleration). The regions that are efficient in particle acceleration, in turn, are more magnetized, because by effect of reconnection the flow has been advecting some guide field component into the reconnection region, and this is giving the main contribution to the magnetization.

The 2D pattern of $B_z$ shown in \fig{abcfluidcomp} shows a tendency for thinner current sheets at larger $L/\rhot$, when fixing $\sigmain$ (left column in \fig{abcfluidcomp}). Roughly, the thickness of the current sheet is set by the Larmor radius $\rhot$ of the high-energy particles heated/accelerated by reconnection. In the right column, with $L/\rhot$ fixed, the thickness of the current sheet is then a fixed fraction of the box size. In contrast, in the left column, the ratio of current sheet thickness to box size will scale as $\rhot/L$, as indeed it is observed. A long thin current sheet is expected to fragment into a chain of plasmoids/magnetic islands \citep[e.g.,][]{uzdensky_10,2016ApJ...816L...8W}, when the length-to-thickness ratio is much larger than unity.\footnote{The fact that the ratio of lengths most important for regulating the reconnection physics is $L/\rhot$ was already anticipated by \citet{petri_lyubarsky_07} and \citet{2011ApJ...741...39S}, in the context of striped pulsar winds.} It follows that the cases in the right column will display a similar tendency for fragmentation (and in particular, they do not appreciably fragment), whereas the likelihood of fragmentation is expected to increase from top to bottom in the left column. In fact, for the case with $L/\rhot=251$ (left bottom panel), a number of small-scale plasmoids appear in the current sheets (e.g., see the plasmoid at $x=0$ and $y=-0.5L$ in the left bottom panel). We find that as long as $\sigmain\gg1$, the secondary tearing mode discussed by \citet{uzdensky_10} --- that leads to current sheet fragmentation --- appears at $L/\rhot\gtrsim 200$, in the case of ABC collapse.\footnote{This holds for both $kT/mc^2\ll1$ and $kT/mc^2\gg1$, with only a minor evidence for the transition happening at slightly smaller $L/\rhot$ in ultra-relativistic plasmas.} In addition to the runs presented in \fig{abcfluidcomp}, we have confirmed this result by covering the whole $\sigmain-L/\rhot$ parameter space, with $\sigmain$ from 3 to 680 and with $L/\rhot$ from 31 to 502.

We have performed an additional test to check that the current sheet thickness scales as $\rhot$. When the reconnection layers reach their maximal extent (i.e., at the peak of the electric energy), the current sheet length should scale as $L$, whereas its thickness should be $\sim \rhot$. Since the current sheets are characterized by a field-aligned electric field (see the bottom panel in \fig{abcaccfluid}), the fraction of box area occupied by regions with $\bmath{E}\cdot\bmath{B}\ne 0$ should scale as $\rhot/L$. We have explicitly verified that this is indeed the case, both when comparing runs that have the same $L/\rhot$ and for simulations that keep $\sigmain$ fixed.

The reconnection rate in all the cases we have explored is around $v_{\rm rec}/c\sim 0.3-0.5$. It shows a marginal tendency for decreasing with increasing $L/\rhot$ (but we have verified that it saturates at $v_{\rm rec}/c\sim 0.3$ in the limit $L/\rhot\gg1$), and it moderately increases with $\sigmain$ (especially as we transition from the non-relativistic regime to the relativistic regime, but it  saturates at $v_{\rm rec}/c\sim 0.5$ in the limit $\sigmain\gg1$). Our measurements of the inflow speed (which we take as a proxy for the reconnection rate) are easier when the parallel mode dominates, since the inflow direction lies along one of the Cartesian axes, rather than for the oblique mode. In simulations with an aspect ratio of $2L\times L$, the parallel mode is the only one that gets triggered, and most of the measurements quoted above refer to this setup.

In \fig{abctimecomp} we present the temporal evolution of the runs whose 2D structure is shown in \fig{abcfluidcomp}. In the left column, we fix $kT/mc^2=10^{-4}$ and $\sigmain=42$ and we vary the ratio $L/\rhot$ from 31 to 251 (from blue to red). In the right column, we fix $kT/mc^2=10^2$ and $L/\rhot=63$ and we vary the magnetization $\sigmain$ from 3 to 170 (from blue to red). The top panels show that the evolution of the electric energy (in units of the total initial energy) is remarkably similar for all the values of $L/\rhot$ and $\sigmain$ we explore. In particular, the electric energy grows as $\propto \exp{(4ct/L)}$ in all the cases (see the dashed black lines), and it peaks at $\sim 10\%$ of the total initial  energy. The only exception is the trans-relativistic case $\sigmain=3$ and $L/\rhot=63$ (blue line in the top right panel), whose peak value is slightly smaller, due to the lower \Alfven speed. The onset time of the instability is also nearly independent of $\sigmain$ (top right panel), although the trans-relativistic case $\sigmain=3$ (blue line) seems to be growing slightly later. As regard to the dependence of the onset time on $L/\rhot$ at fixed $\sigmain$, the top left panel in \fig{abctimecomp} shows that larger values of $L/\rhot$ tend to grow later, but the variation is only moderate: in all the cases the instability grows around $ct/L\sim 5$.

In the early stages of ABC instability, the cutoff Lorentz factor $\gammamax$ of the particle energy spectrum (as defined in \eq{ggmax}) grows explosively (see the two bottom panels in \fig{abctimecomp}). If the origin of time is set at the onset time of the instability, the maximum energy is expected to grow as
\be\label{eq:ggmax2b}
\gammamax m c^2\sim e v_{\rm rec} B_{\rm in} t
\ee
Due to the approximate linear increase of the in-plane magnetic field with distance from a null point, we find that $B_{\rm in}\propto \sqrt{\sigmain} v_{\rm rec} t/L$. This leads to $\gammamax\propto v_{\rm rec}^2 \sqrt{\sigmain}  t^2/L$, with the same quadratic temporal scaling that was discussed for the solitary X-point collapse. Since the dynamical phase of ABC instability lasts a few $L/c$, the maximum particle Lorentz factor at the end of this stage should scale as $\gammamax\propto v_{\rm rec}^2\sqrt{\sigmain} L\propto v_{\rm rec}^2\sigmain (L/\rhot)$. If the reconnection rate does not significantly depend on $\sigmain$, this implies that $\gammamax\propto L$ at fixed $\sigmain$. The trend for an increasing $\gammamax$ with $L$ at fixed $\sigmain$ is confirmed in the bottom left panel of \fig{abctimecomp}, both at the final time and at the peak time of the electric energy (which is slightly different among the four different cases, see the vertical dotted colored lines).\footnote{In contrast, a comparison at the same $ct/L$ is not very illuminating, since larger values of $L/\rhot$ tend to systematically have later instability onset times.} Similarly, if the reconnection rate does not significantly depend on $L/\rhot$, this implies that $\gammamax\propto \sigmain$ at fixed $L/\rhot$. This linear dependence of $\gammamax$ on $\sigmain$ is  confirmed in the bottom right panel of \fig{abctimecomp}.

The dependence of the particle spectrum on $L/\rhot$ and $\sigmain$ is illustrated in \fig{abcspeccomp} (left and right panel, respectively), where we present the particle energy distribution at the time when the electric energy peaks (as indicated by the colored vertical dotted lines in \fig{abctimecomp}). The particle spectrum shows a pronounced non-thermal component in all the cases, regardless of whether the secondary plasmoid instability is triggered or not in the current sheets (the results presented in \fig{abcspeccomp} correspond to the cases displayed in \fig{abcfluidcomp}). This suggests that  in our setup any acceleration mechanism that relies on plasmoid mergers is not very important, but rather that the dominant source of energy is direct acceleration by the reconnection electric field (see the previous subsection).

 At the time when the electric energy peaks, most of the particles are still in the thermal component (at $\gamma\sim 1$ in the left panel and $\gamma\sim 3 kT/mc^2$ in the right panel of \fig{abcspeccomp}), i.e., bulk heating is still negligible. Yet, a dramatic event of particle acceleration is taking place, with a few lucky particles accelerated much beyond the mean energy per particle $\sim\gamma_{\rm th}\sigmain/2$ (for comparison, we point out that $\gamma_{\rm th}\sim 1$  and $\sigmain=42$ for the cases in the left panel). Since most of the particles are at the thermal peak, this is not violating energy conservation. As described by \citet{2014ApJ...783L..21S} and \citet{2016ApJ...816L...8W}, for a power law spectrum $dN/d\gamma\propto \gamma^{-p}$ of index $1<p<2$ extending from $\gamma\sim \gamma_{\rm th}$ up to $\gamma_{\rm max}$, the fact that the mean energy per particle is $\gamma_{\rm th}\sigmain/2$ yields a constraint on the upper cutoff of the form
\be\label{eq:ggmax }
\gammamax\lesssim \gamma_{\rm th}\left[\frac{\sigmain}{2}\frac{2-p}{p-1}\right]^{1/(2-p)}
\ee
This constraint does not apply during the explosive phase, since most of the particles lie at the thermal peak (so, the distribution does not resemble a single power law), but it does apply at late times (e.g., see the final spectrum in \fig{abcspec}, similar to a power law). However, as the bulk of the particles shift up to higher energies (see the evolution of the thermal peak in \fig{abcspec}), the spectrum at late times tends to be softer than in the early explosive phase (compare light blue and red curves in \fig{abcspec}), which helps relaxing the constraint in \eq{ggmax}.

The spectra in  \fig{abcspeccomp} (main panels for $\gamma dN/d\gamma$, to emphasize the particle content; insets for $\gamma^2 dN/d\gamma$, to highlight the energy census) confirm the trend of $\gammamax$ anticipated above. At fixed $\sigmain=42$ (left panel), we see that $\gammamax\propto L$ ($L$ changes by a factor of two in between each pair of curves);\footnote{The fact that the dependence is a little shallower than linear is due to the fact that the reconnection rate, which enters in the full expression $\gammamax\propto v_{\rm rec}^2\sigmain(L/\rhot)$, is slightly decreasing with increasing $L/\rhot$, as detailed above.} on the other hand, at fixed $L/\rhot=63$ (right panel), we find that $\gammamax\propto \sigmain$ ($\sigmain$ changes by a factor of four in between each pair of curves). 

The spectral hardness is primarily controlled by the average in-plane magnetization $\sigmain$. The right panel in \fig{abcspeccomp} shows that at fixed  $L/\rhot$ the spectrum becomes systematically harder with increasing $\sigmain$, approaching the asymptotic shape $\gamma dN/d\gamma\propto \rm const$ found for plane-parallel steady-state reconnection in the limit of high magnetizations \citep[][]{2014ApJ...783L..21S,2015ApJ...806..167G,2016ApJ...816L...8W}. At large $L/\rhot$, the hard spectrum of the high-$\sigmain$ cases will necessarily run into constraints of energy conservation (see \eq{ggmax }), unless the pressure feedback of the accelerated particles onto the flow structure ultimately leads to a spectral softening (in analogy to the case of cosmic ray modified shocks, see \citealt{amato_06}). This argument seems to be supported by the left panel in \fig{abcspeccomp}. At fixed $\sigmain$ (left panel), the spectral slope is nearly insensitive to $L/\rhot$. The only (marginal) evidence for a direct dependence on $L/\rhot$ emerges at large $L/\rhot$, with larger systems leading to steeper slopes (compare the yellow and red lines in the left panel), which possibly reconciles the increase in $\gammamax$ with the argument of energy conservation illustrated in \eq{ggmax }.

In application to the GeV flares of the Crab Nebula, which we attribute to the explosive phase of ABC instability, we envision an optimal value of $\sigmain$ between $\sim 10$ and $\sim 100$. Based on our results, smaller $\sigmain\lesssim 10$ would correspond to smaller reconnection speeds (in units of the speed of light), and so weaker accelerating electric fields. On the other hand, $\sigmain\gtrsim 100$ would give hard spectra with slopes $p<2$, which would prohibit particle acceleration up to $\gammamax\gg \gamma_{\rm th}$ without violating energy conservation (for the sake of simplicity, here we ignore the potential spectral softening at high $\sigmain$ and large $L/\rhot$ discussed above). 

As we have mentioned above, we invoke the early phases of ABC collapse as an explanation for the Crab GeV flares. As shown in \fig{abcspec}, after this initial stage the maximum particle energy is only increasing by a factor of three (on long timescales, of  order $\sim 10 L/c$). When properly accounting for the rapid synchrotron losses of the highest energy particles (which our simulations do not take into account), it is even questionable whether the spectrum will ever evolve to higher energies, after the initial collapse. This is the reason why we have focused most of our attention on how the spectrum at the most violent time of ABC instability depends on $L/\rhot$ or $\sigmain$. For the sake of completeness, we now describe how the spectrum at late times (corresponding to the bottom right panel in \fig{abcfluid_bz}) depends on the flow conditions. In analogy to what we have described above, the effect of different $kT/mc^2$ is only to change the overall energy scale, and the results are nearly insensitive to the flow temperature, as long as $\comp$ and $\sigmain$ properly account for temperature effects. The role of $\sigmain$ and $L/\rhot$ can be understood from the following simple argument \citep[see also][]{2011ApJ...741...39S}. In the $\sigmain\lesssim 10-100$ cases in which the spectral slope is always $p>2$ and the high-energy spectral cutoff is not constrained by energy conservation, we have argued that $\gammamax/\gamma_{\rm th}\propto \sigmain (L/\rhot)$ (here, we have neglected the dependence on the reconnection speed, for the sake of simplicity). As a result of bulk heating, the thermal peak of the particle distribution shifts at late times --- during the island mergers that follow the initial explosive phase --- from $\gamma_{\rm th}$ up to $\sim \gamma_{\rm th} \sigmain/2$. By comparing the final location of the thermal peak with $\gammamax$, one concludes that there must be a critical value of $L/\rhot$ such that for small $L/\rhot$ the final spectrum is close to a Maxwellian, whereas for large values of $L/\rhot$ the non-thermal component established during the explosive phase is still visible at late times. We have validated this argument with our results, finding that this critical threshold for the shape of the final spectrum is around $L/\rhot\sim 100$.

% \clearpage
%%%%Lorenzo%%%%

%%%%Lorenzo%%%%
 
\section{Driven  evolution of  a system of magnetic islands}
\label{driven}

As we have seen in \S \ref{unstr-latt}, a periodic  2D ABC equilibrium configuration of magnetic islands is unstable to 
merger. Such equilibrium configuration is naturally an idealized simplification. As we demonstrate below,   in the case of a modified initial configuration 
which is compressed along one direction 
the rapid island merger sets up straight away. What could create such a stressed configuration 
and promote rapid reconnection which may give rise to Crab's gamma-ray flares? 
One possibility which comes to mind is a strong shock. Shocks in strongly magnetized relativistic
plasma are often described as weakly compressive and indeed in the shock frame the plasma 
density and the magnetic field strength are almost the same. However, when measured in the fluid
frame, these quantities may experience huge jumps, of the order of the shock Mach number 
\citep{2012MNRAS.422..326K}. This is accompanied by similarly large variation of the flow 
Lorentz factor, which may also accelerate the rate of physical processes via the relativistic 
time dilation effect. PIC simulations of shocked striped wind by \citet{2011ApJ...741...39S} 
spectacularly illustrate this possibility. In these simulations, the shock is triggered
via collision of the wind with a stationary wall. At some distance downstream of the shock 
the flow decelerates from high Lorentz factor to full stop, which shortens the time scale 
of all processes via the relativistic time-dilation effect. Reconnection in the current 
sheets accelerates and the magnetic field of stripes dissipates, with its energy used 
to heat the wind plasma. The conversion of magnetic  energy to heat is accompanied by a 
decrease of the specific volume and hence by additional plasma compression. As the result,
relative to the shocked plasma the shock propagates at speed well below the speed of light 
but close to that of a shock in ultra-relativistic unmagnetized plasma, $v_s\simeq c/3$. 
The role of magnetic dissipation of the shock speed has been studied theoretically by 
\citet{2003MNRAS.345..153L}   

In the following, we discuss the results of driven evolution of  ABC structure using both  fluid and PIC simulations. We find that  using both fluid and PIC codes, that imposing finite amplitude perturbations from the beginning triggers the instability much faster.
% In \S \ref{unstr-latt} we studied spontaneous evolution of the 2D ABC equilibrium. The evolution is characterized by the initial linear stage,  that  becomes highly non-linear after 5-7 dynamical times. In more realistic situations large scale perturbation of the equilibrium can trigger much faster evolution. We consider these next.

% \clearpage
%%%%Lorenzo%%%%

%%%%%%%Sergey%%%%%%

\subsection{Evolution of  compressed 2D ABC equilibrium:force-free simulations}
\label{Force-freeABCdriven} 

\begin{figure}[ht]
\includegraphics[width=0.49\linewidth]{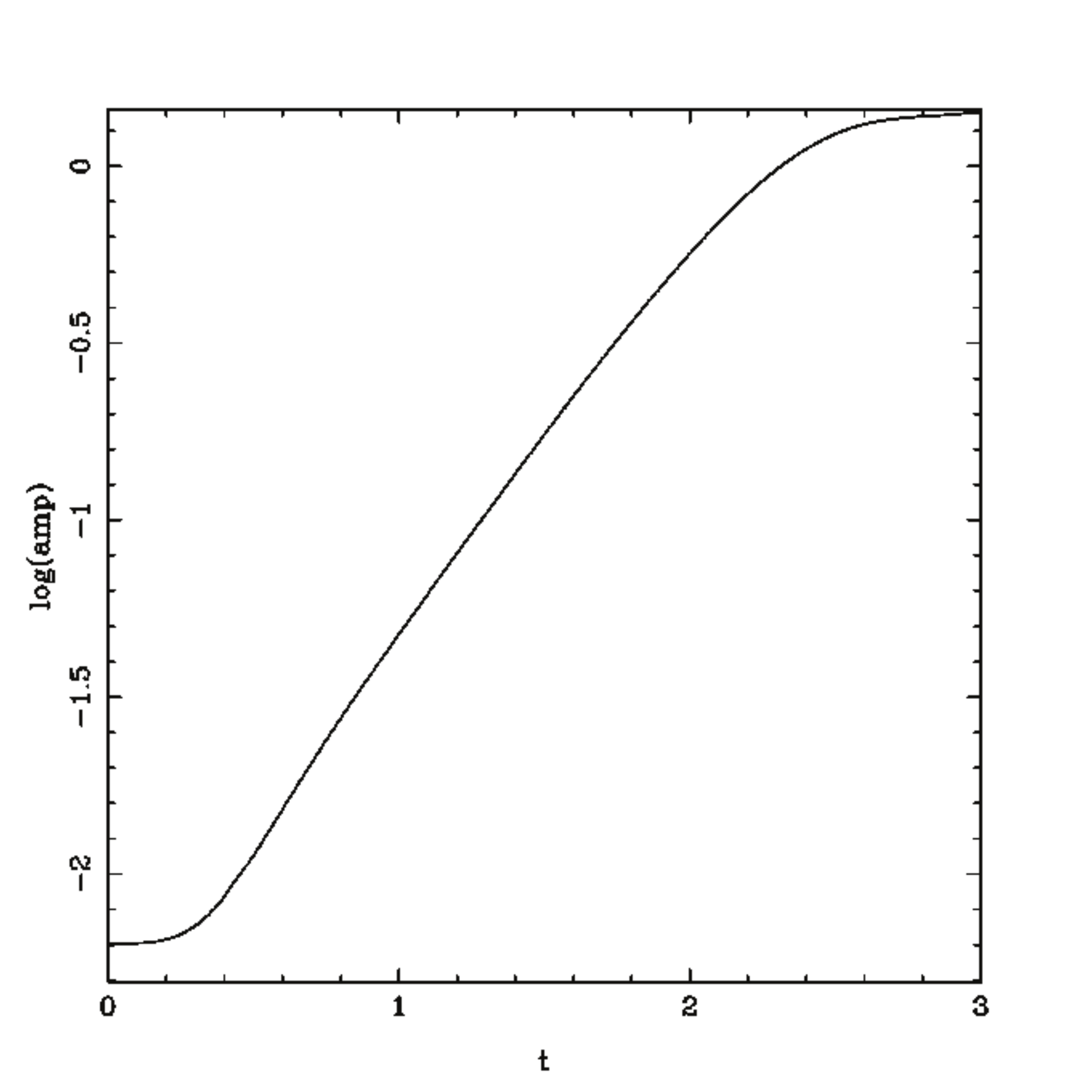}
\caption{{\it Left Panel} Electromagnetic energy and perturbation amplitude of force-free simulations as functions of time for a case of initial  state with finite amplitude perturbations, chosen to have a large contribution from unstable modes. In comparison with the unstressed system of magnetic islands  ({\it Left Panel} Fig. \protect  \ref{ff-energy}), the instability starts to grow immediately. }
\label{10}
\end{figure}

In force-free simulations we  added an initial perturbation $\delta A_z= \delta a *(\sin(kx)+\sin(ky)) * \cos(k(y-x)/2)$, where $ \delta a$ is an amplitude of a perturbations. We find that this leads to the earlier onset of the non-linear stage of instability depending on the amplitude of perturbations, Fig. \ref{10}.

To find configurations that immediately display a large reconnecting electric field, we turn to the stressed single island case.  
This is just a cut-out of the muliple-island ABC fields, adopting an aspect ratio $R\ne1$.  We choose $R=0.9$ as initial perturbation.  We choose three different values for the parallel conductivity $\kappa_{||}\in[500,1000,2000]$ and set the perpendicular component to zero $\kappa_\perp=0$.

\begin{figure}
\begin{center}
  \includegraphics[height=3cm]{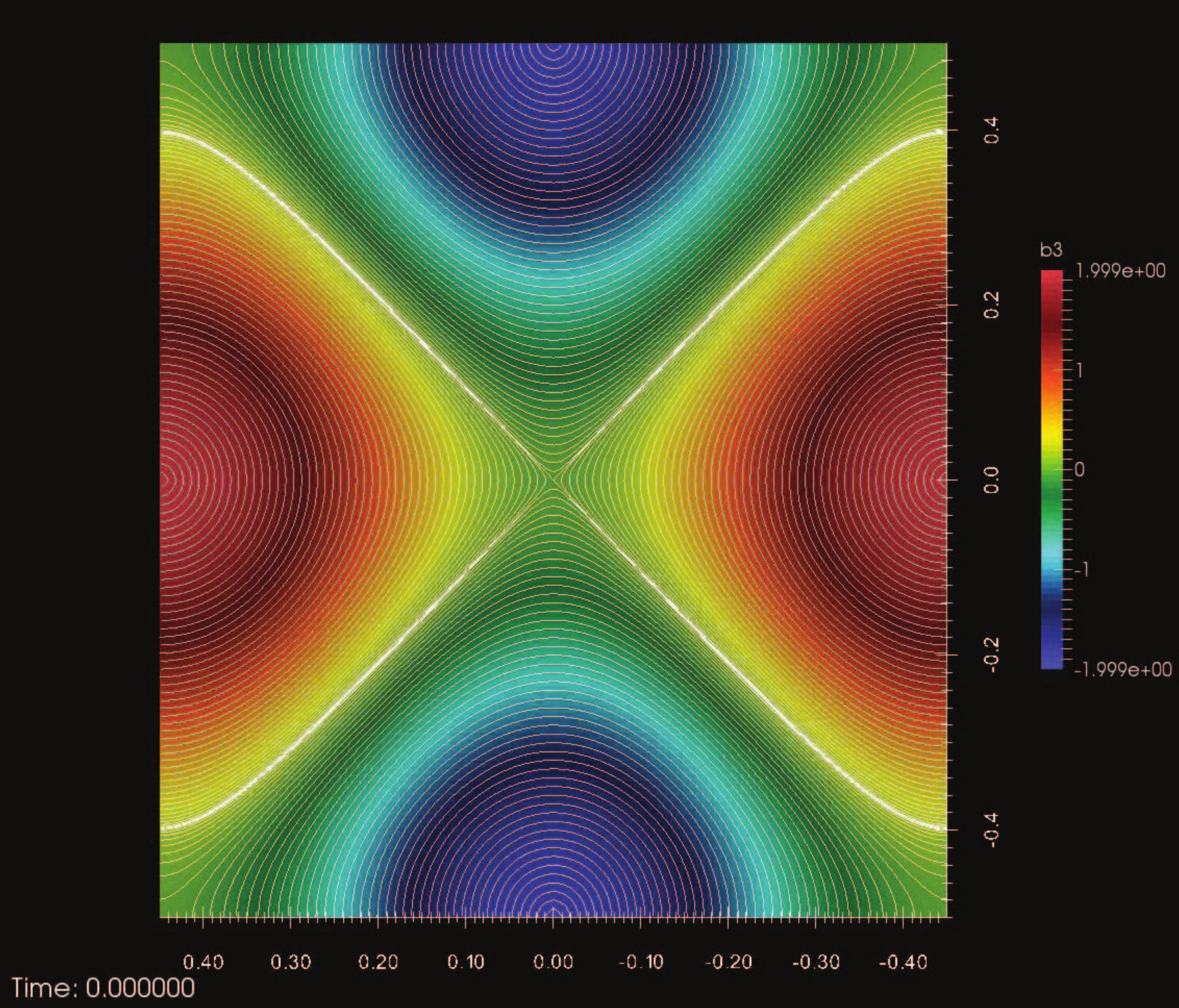}
  \includegraphics[height=3cm]{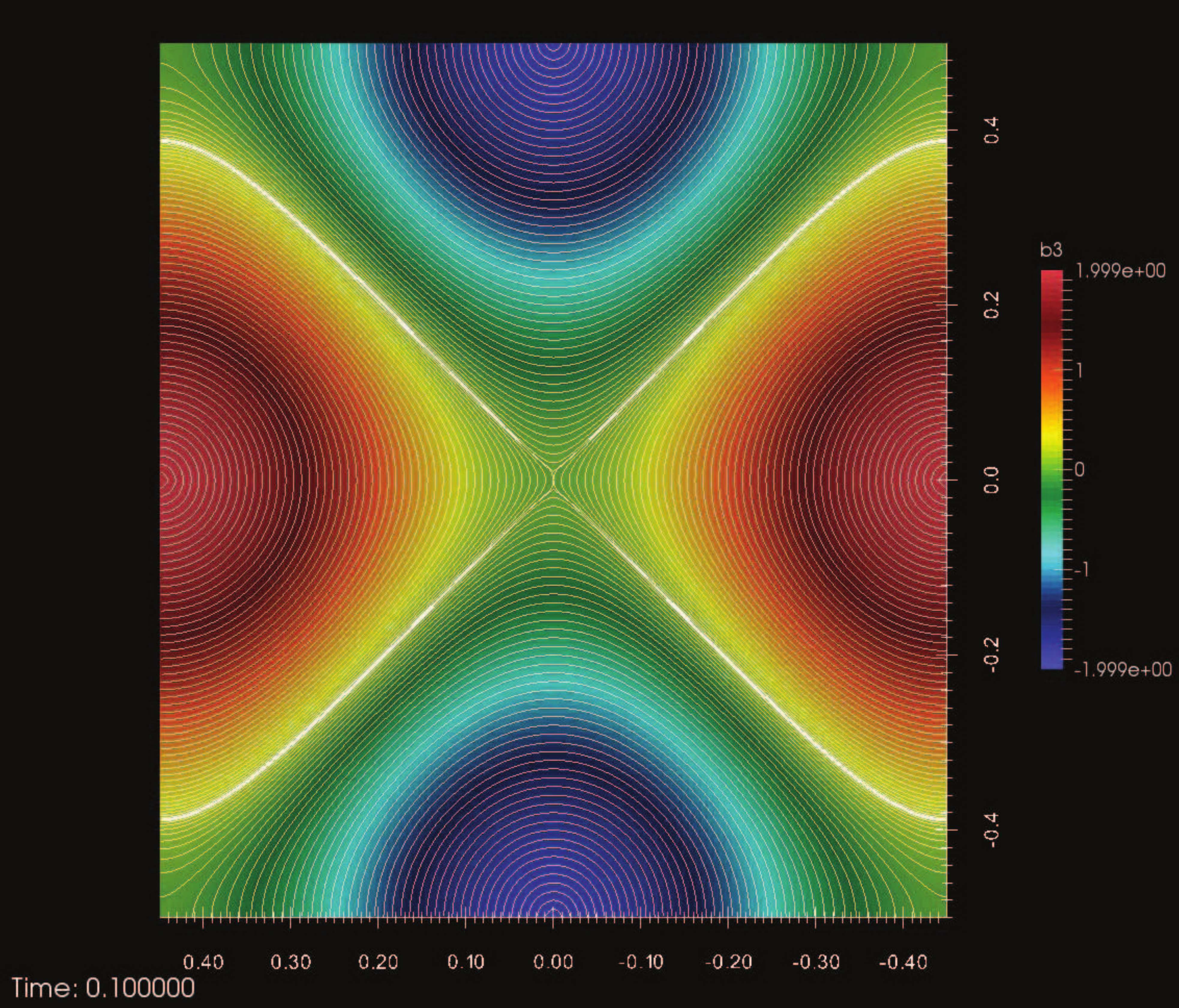}
  \includegraphics[height=3cm]{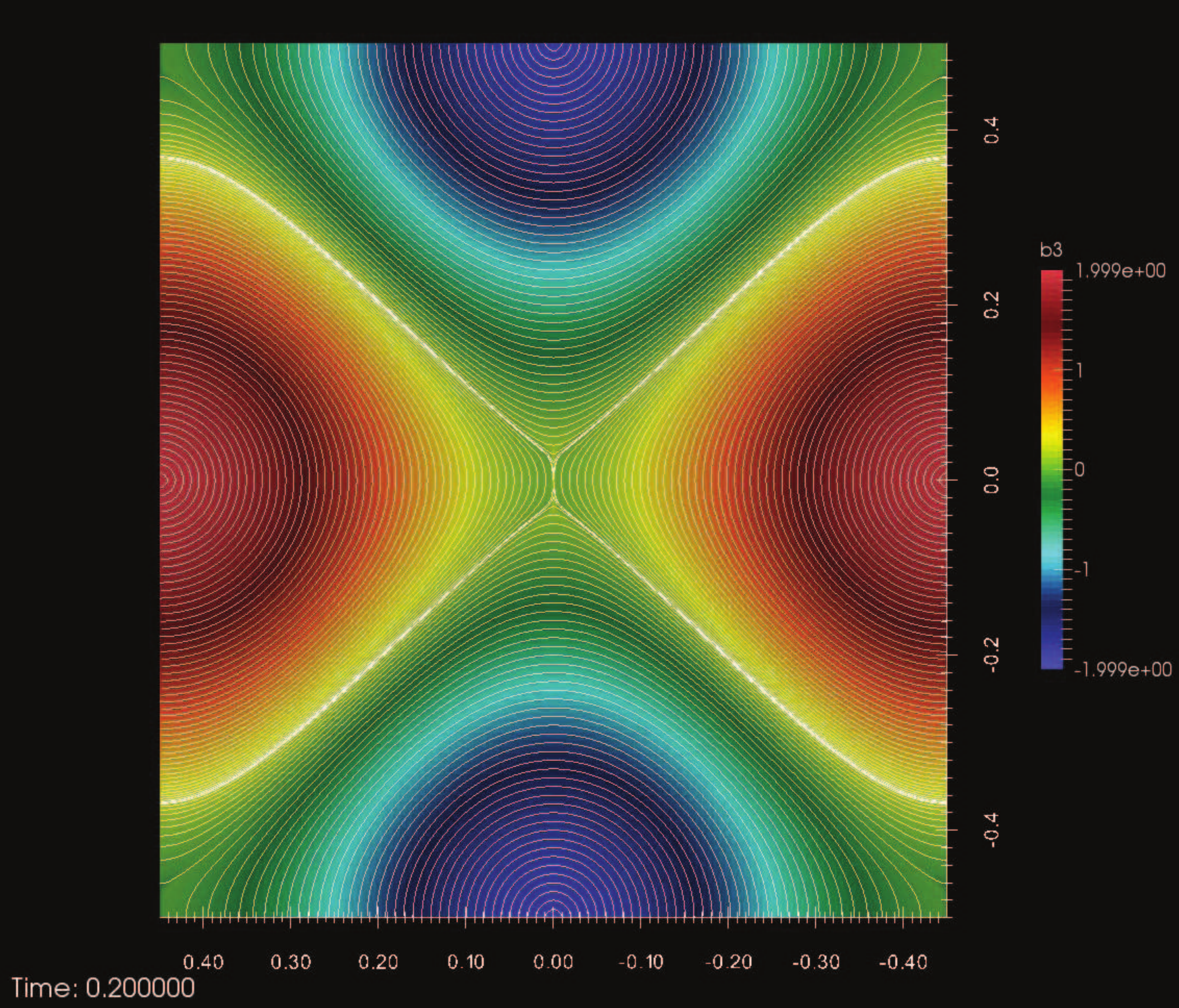}
  \includegraphics[height=3cm]{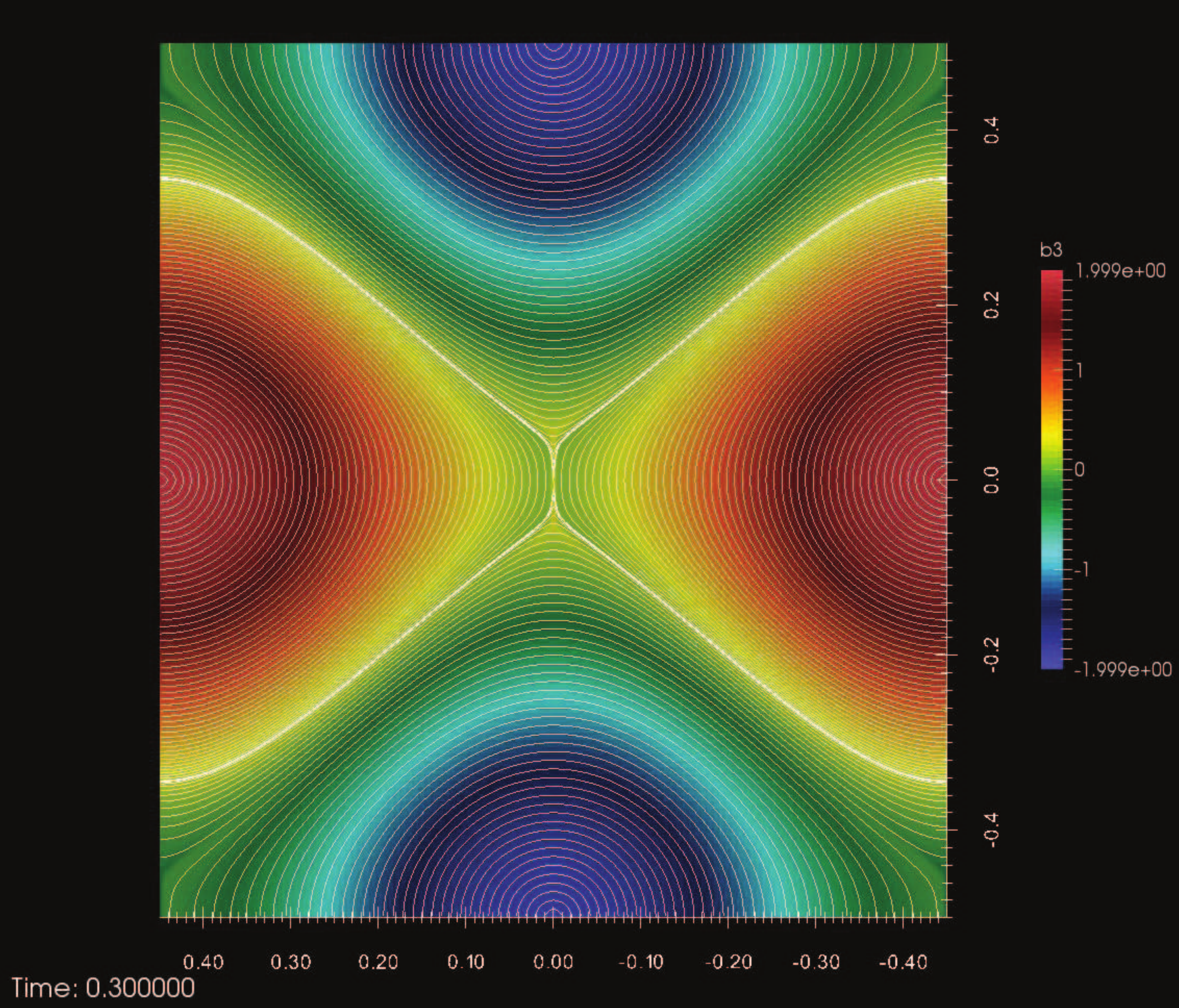}
  \includegraphics[height=3cm]{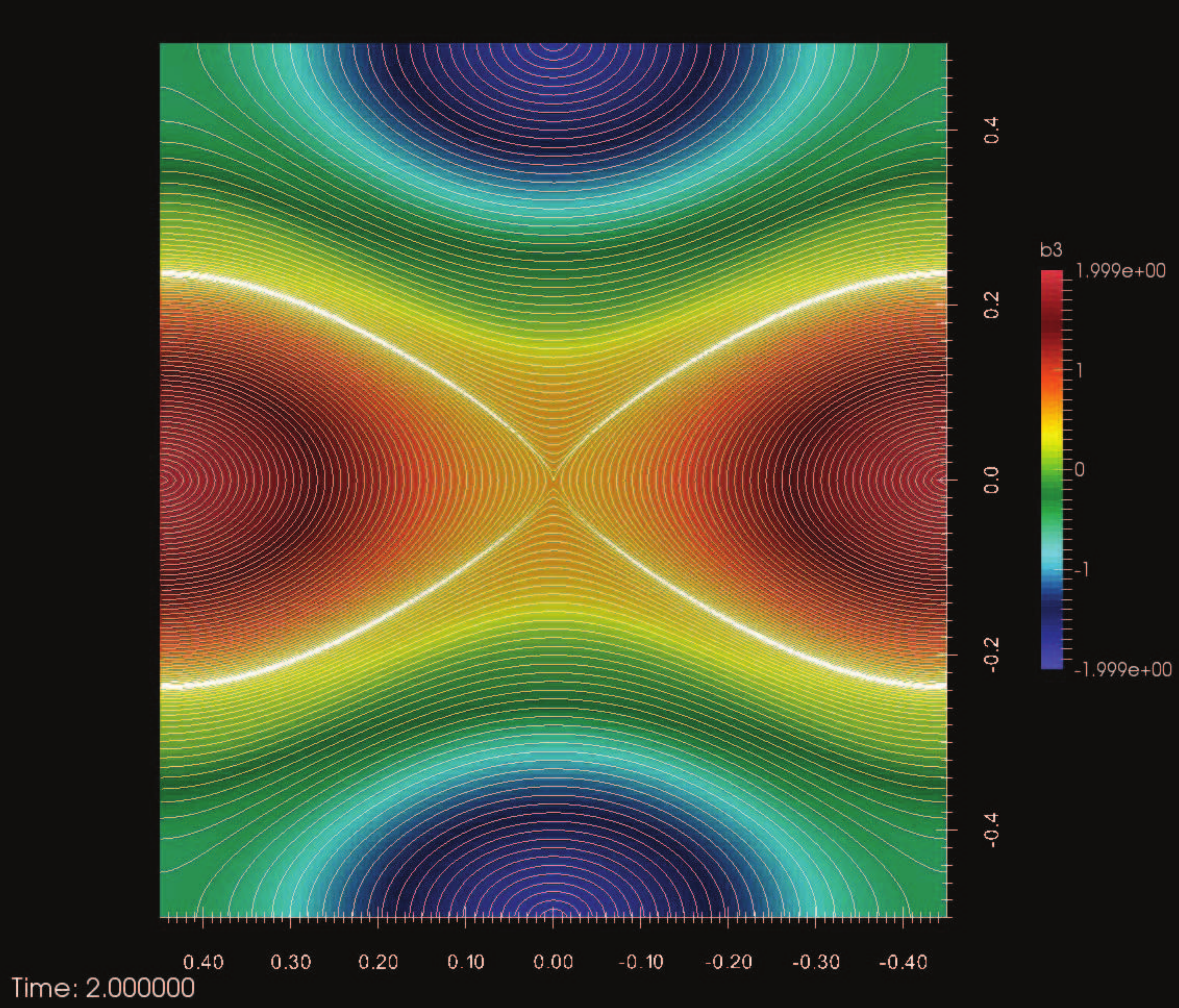}
  \includegraphics[height=3cm]{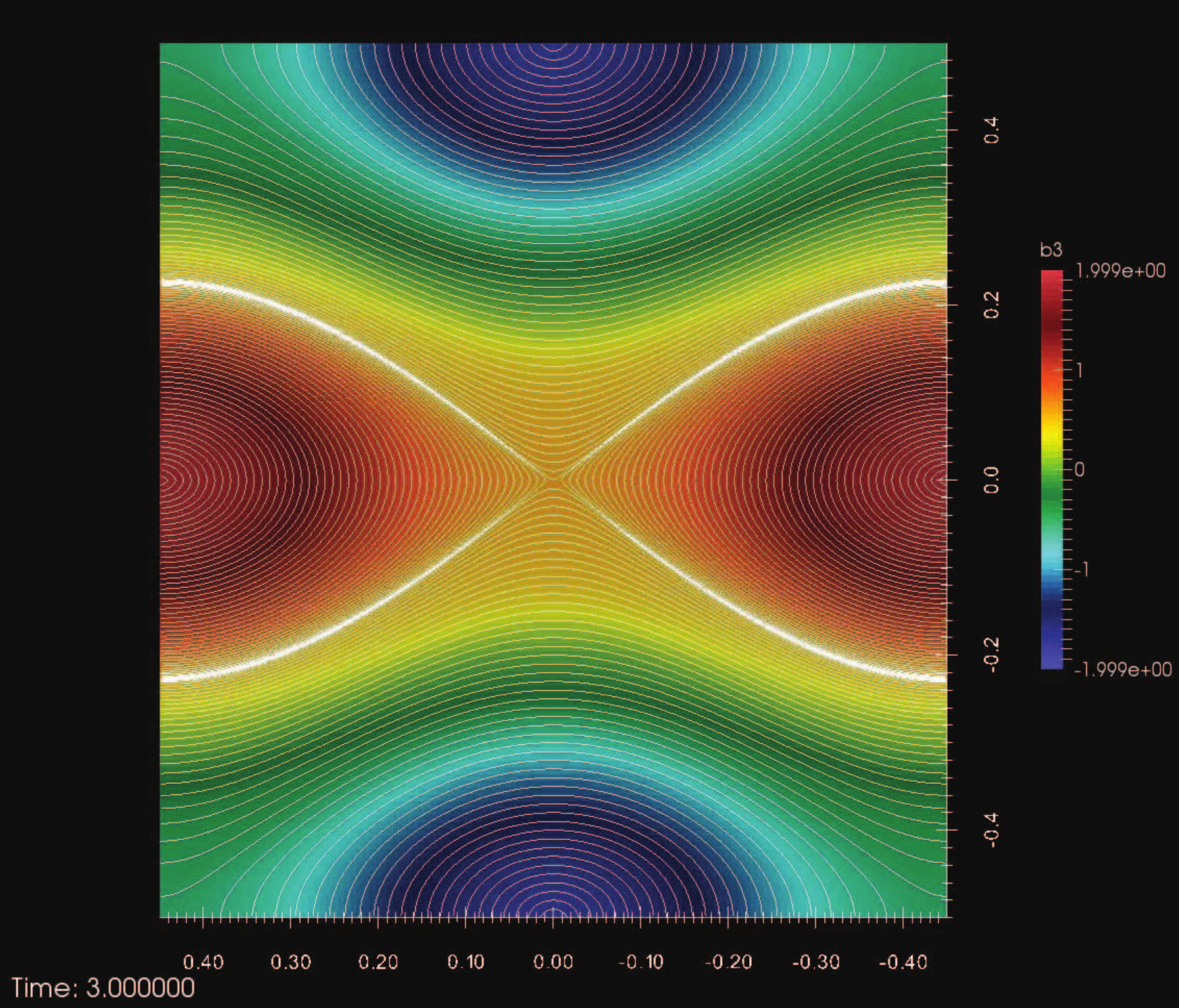}
\caption{Time evolution of the out-of-plane field, case $\kappa_{||}=2000$, triggered collapse of magnetic islands.  The initial X-point collapses and forms a rapidly expanding current-sheet.  After $t=1$, the fast evolution is over and the islands oscillate.  Times are $t\in[0,0.1,0.2,0.3,2,3]$.  
}
\label{fig:island-kpar-bz}
\end{center}
\end{figure}

\begin{figure}
\begin{center}
  \includegraphics[height=3cm]{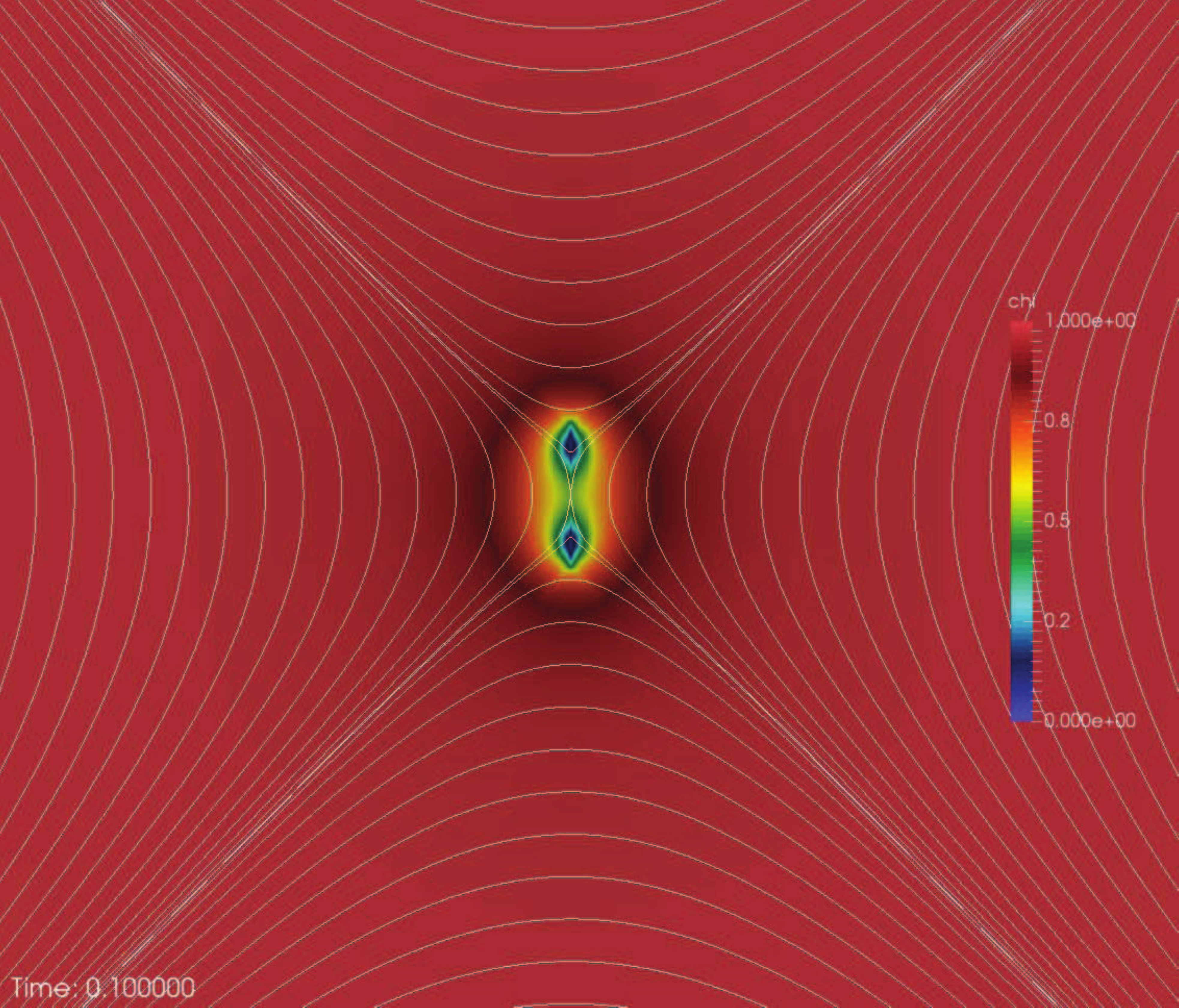}
  \includegraphics[height=3cm]{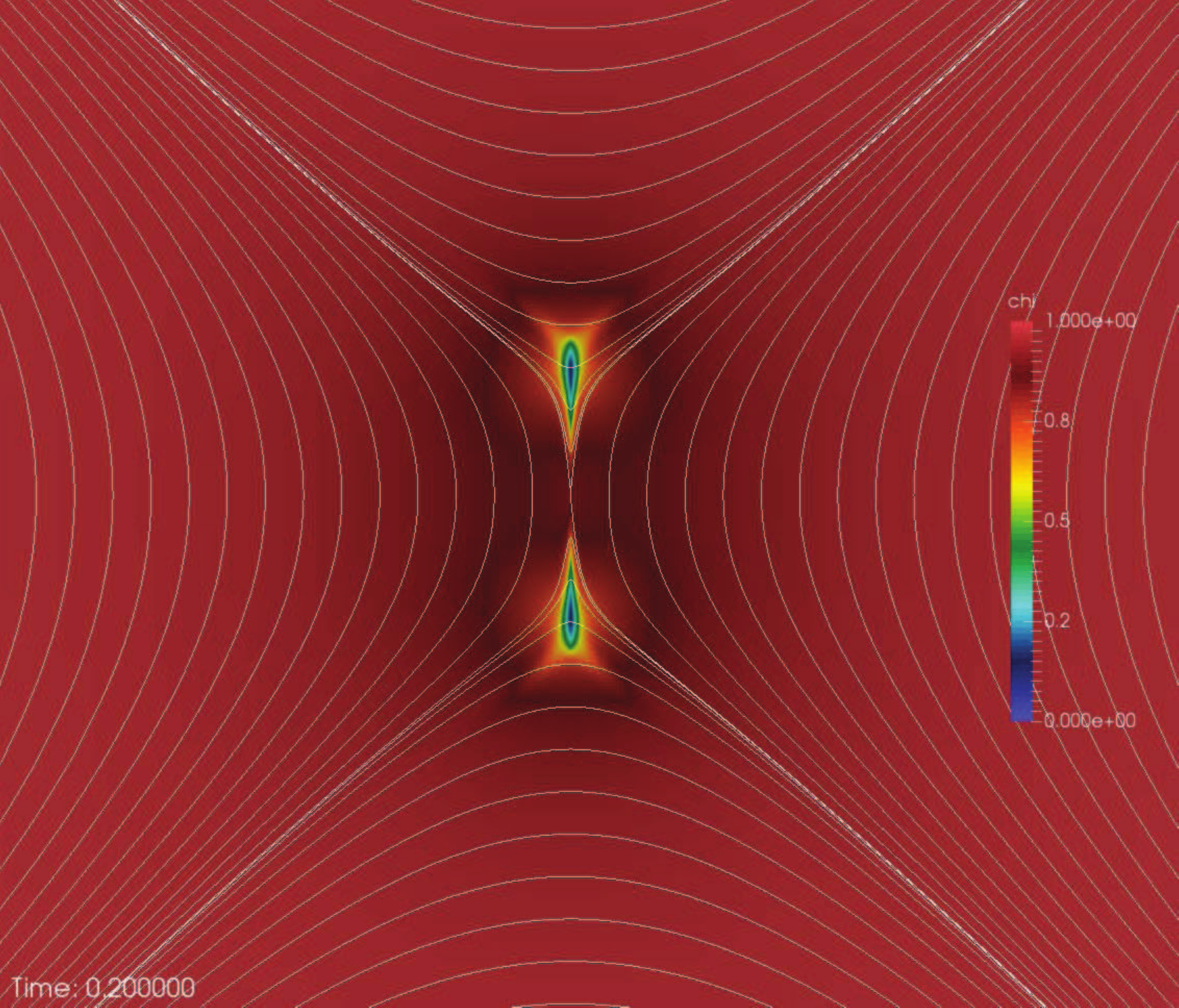}
  \includegraphics[height=3cm]{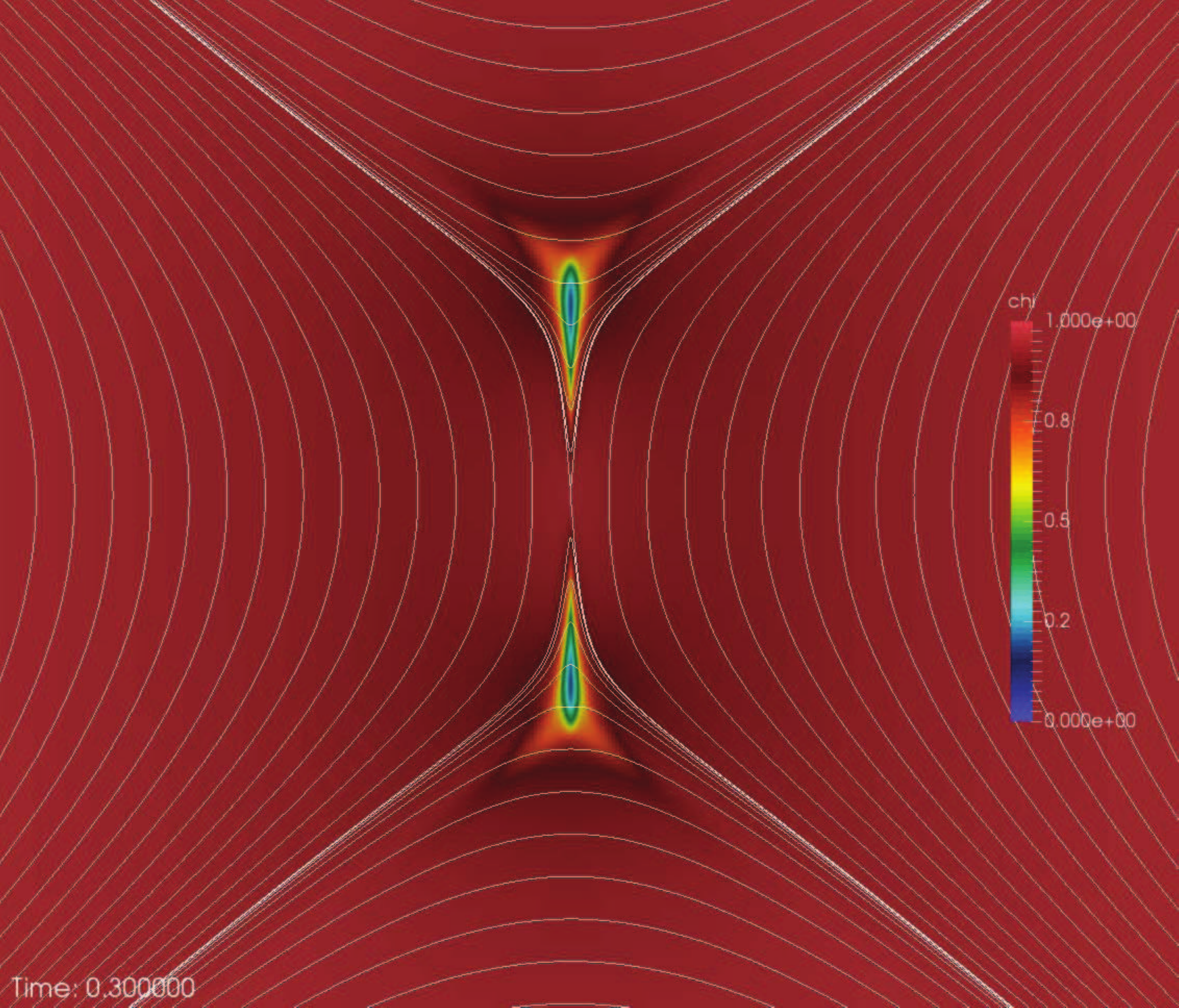}
  \includegraphics[height=3cm]{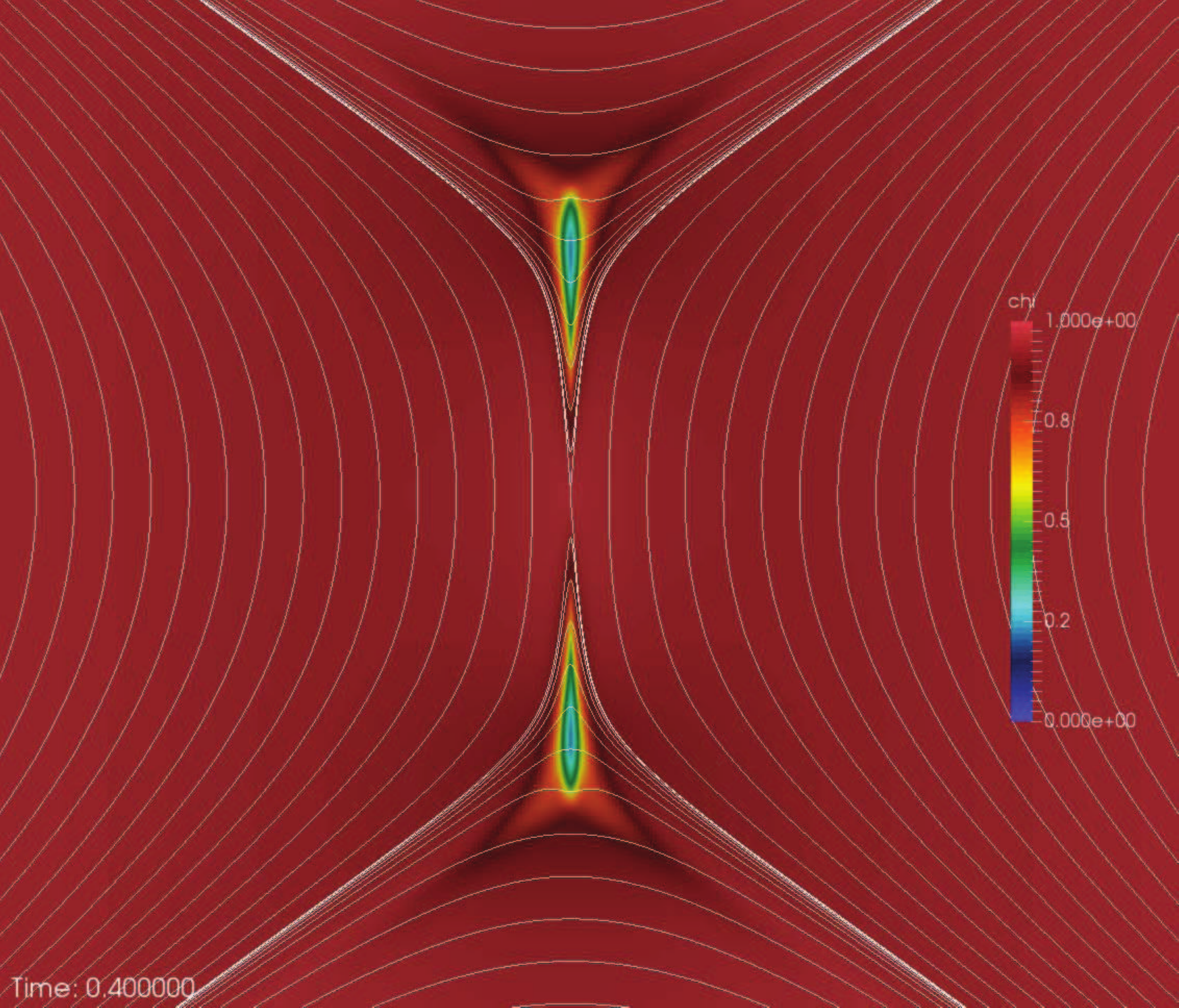}
  \includegraphics[height=3cm]{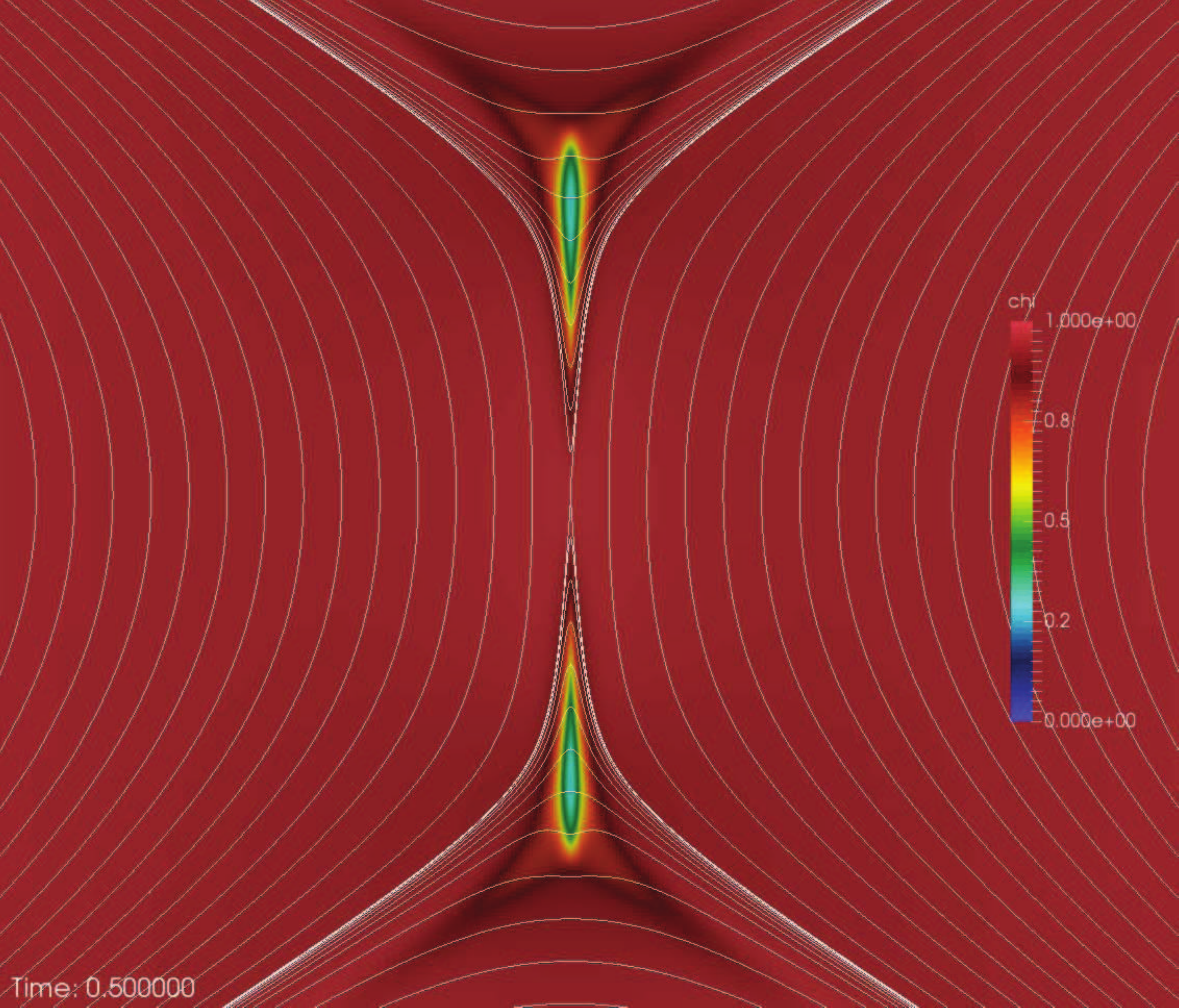}
  \includegraphics[height=3cm]{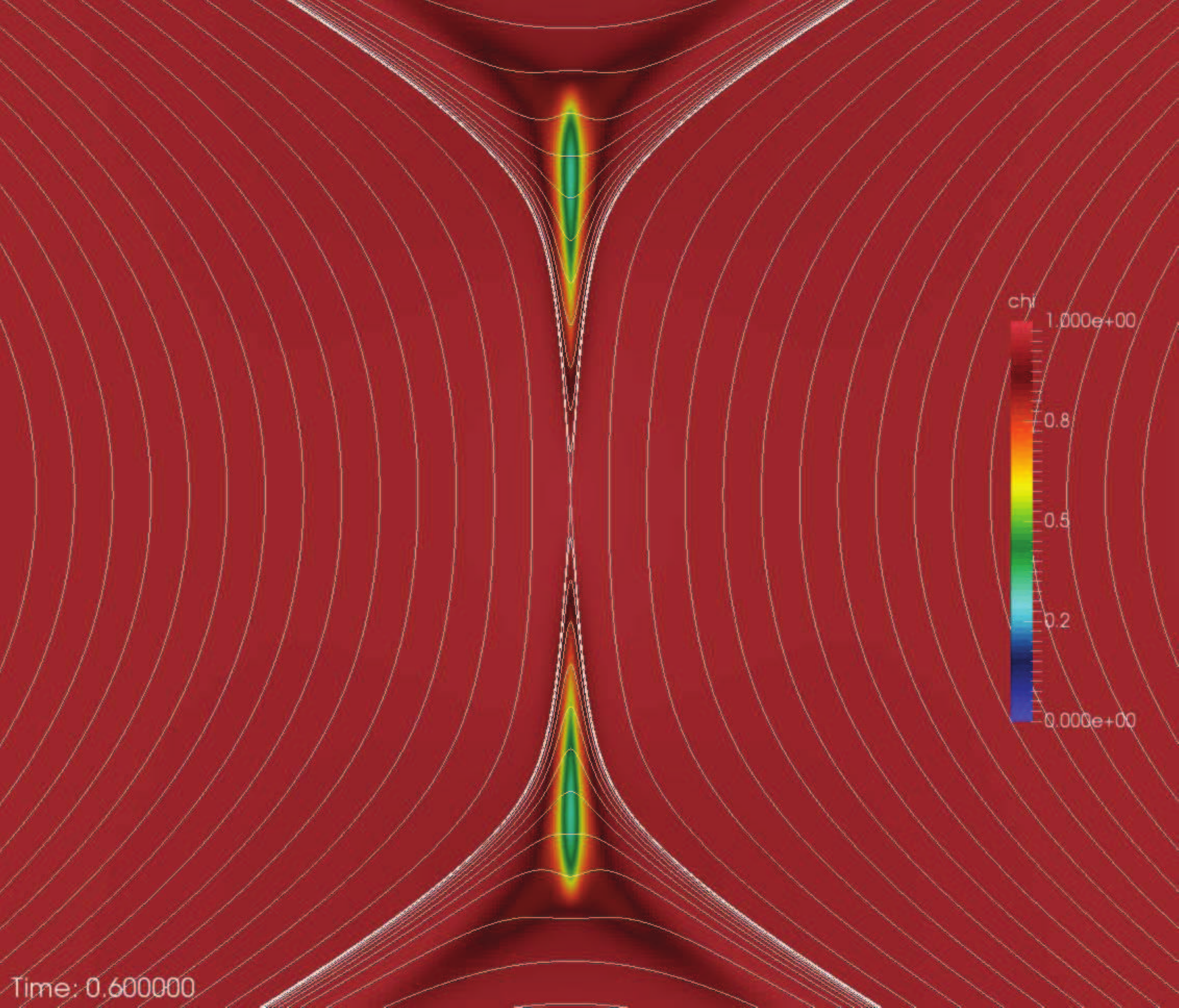}
\caption{Zoom into the central region showing the quantity $1-E^2/B^2$, triggered collapse of magnetic islands.  At $t=0.1$, two disconnected regions of $E\sim B$ exist which rapidly expand.  $E/B$ in these regions gets smaller as time progresses.  Actually, before $t=0.1$, $E/B$ reaches values as high as $3$ for this case with $\kappa_{||}=2000$ (see Fig. \ref{fig:island-kpar}).  
}
\label{fig:island-kpar-chi}
\end{center}
\end{figure}

\begin{figure}
\begin{center}
\includegraphics[height=5cm,angle=-90]{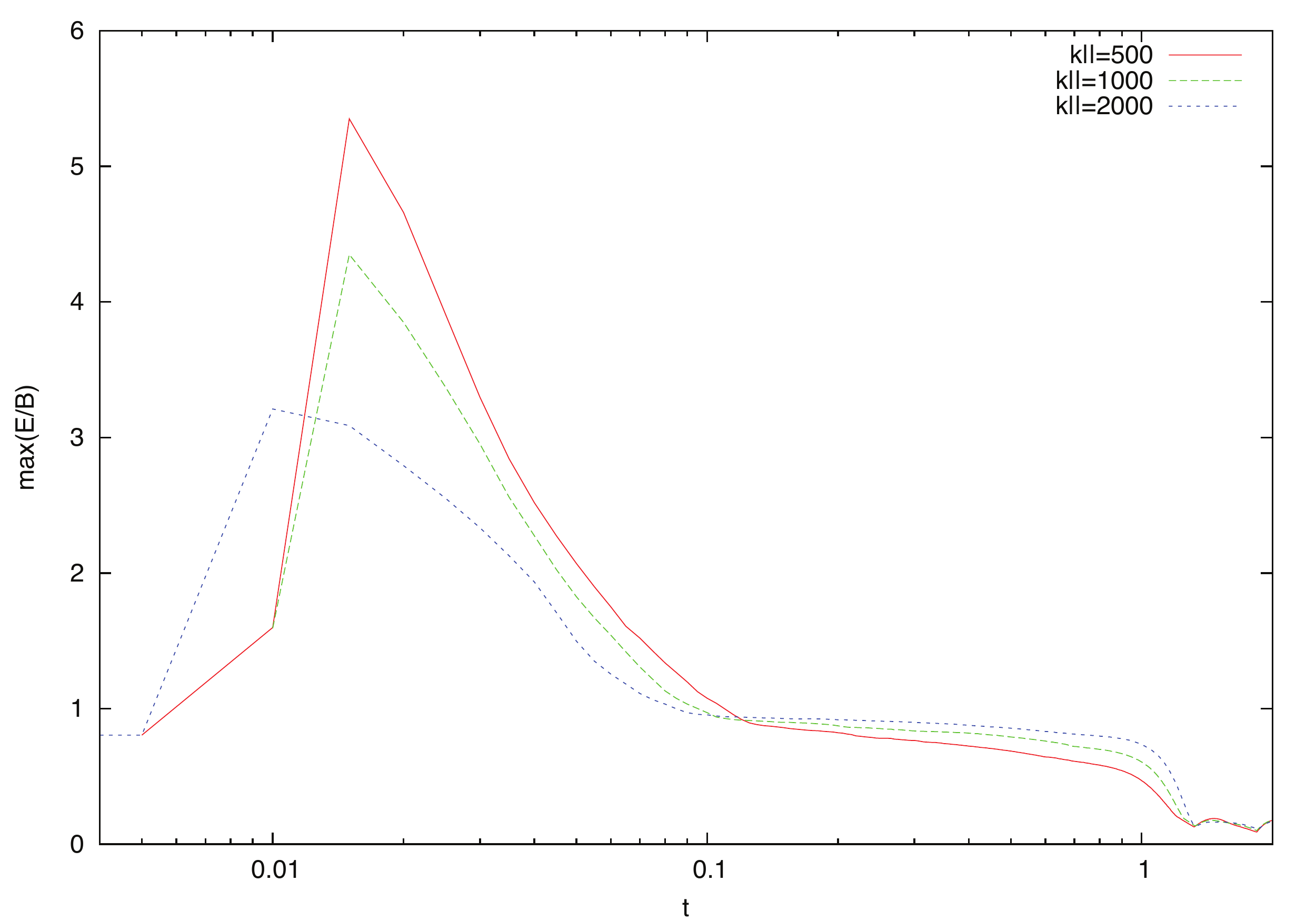}
\includegraphics[height=5cm,angle=-90]{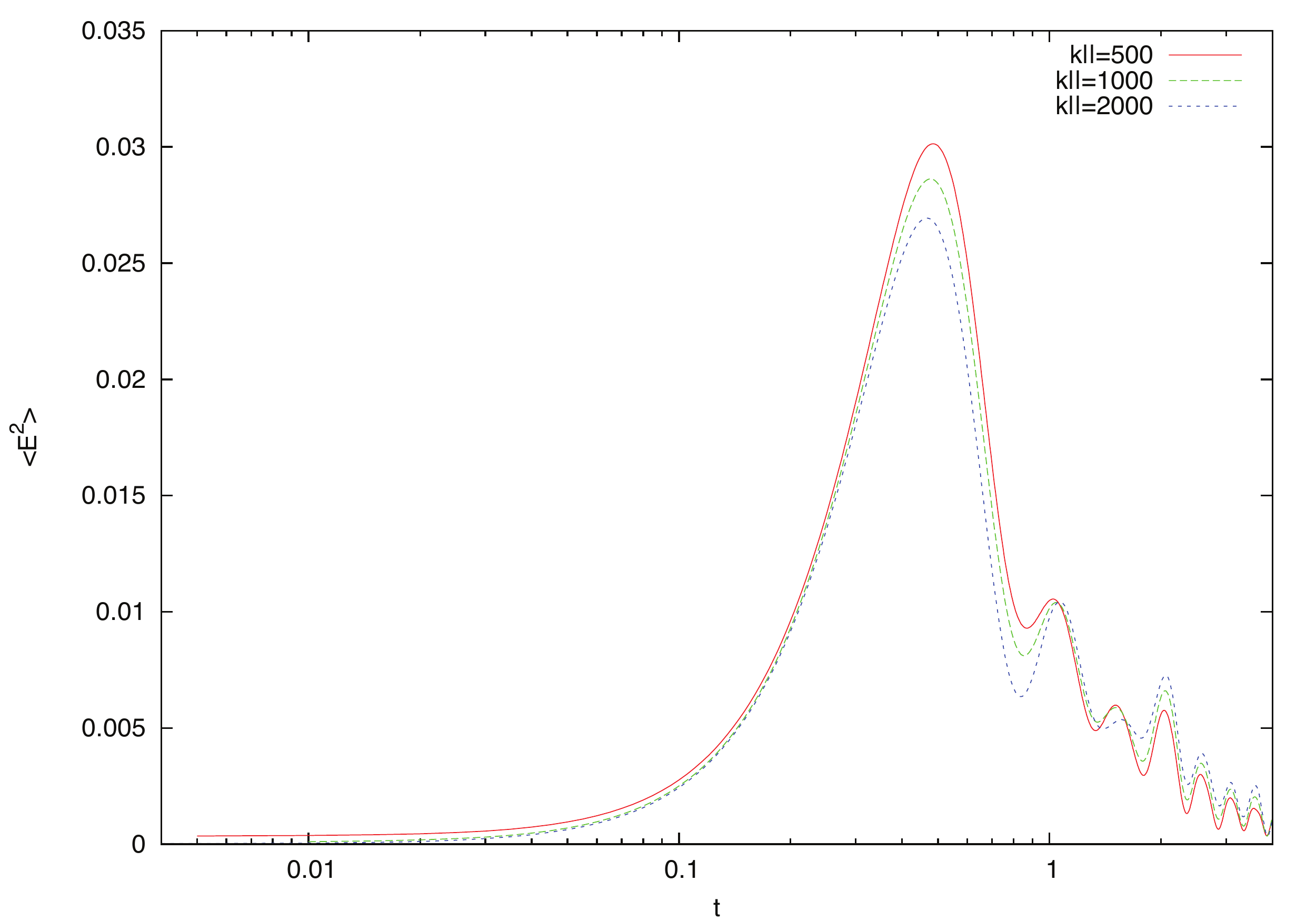}
\includegraphics[height=5cm,angle=-90]{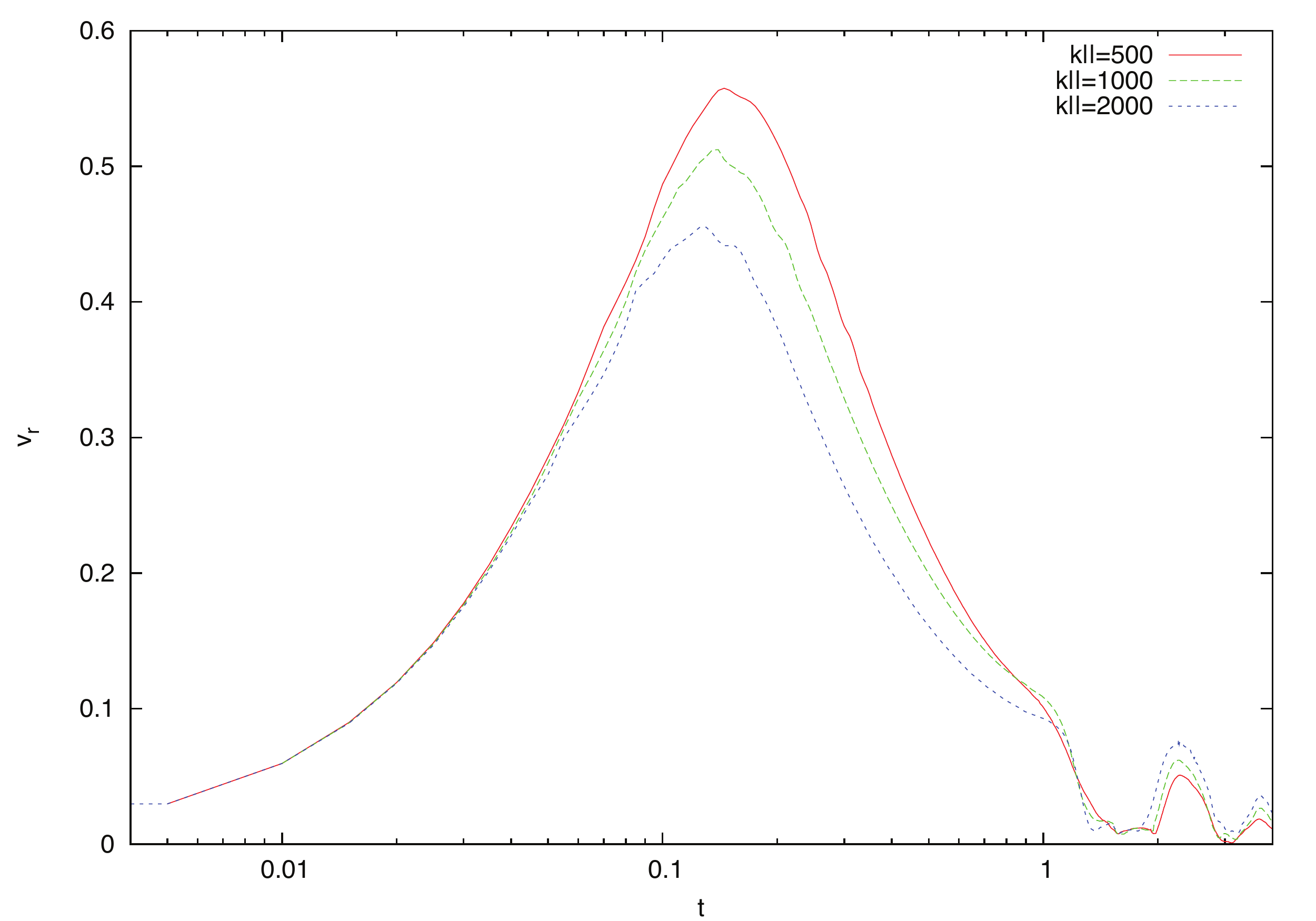}
\caption{Evolution of critical quantities in the stressed island case for various conductivities. 
Immediately a region of $E>B$ is established.  
All cases agree in the initial growth rates of $v_r$ and $\langle E^2\rangle$ and in the over-all time of the evolution.  
This result is consistent with the  initial \emph{ideal} evolution of the instability.  
}
\label{fig:island-kpar}
\end{center}
\end{figure}

The over-all evolution is displayed in Figs. \ref{fig:island-kpar-bz}, \ref{fig:island-kpar-chi} and \ref{fig:island-kpar}.
%showing the out-of-plane magnetic field component together with in-plane field-lines. 
% An animation of this configuration can be found in \emph{animations/stressed-island}.
%
%

%sssssssssssssssssssssssssssssssssssssssssssssssssssssssssssssssssss
\subsection{Collision-triggered merger of magnetic islands: MHD simulations}
\label{Shock-triggered}
%sssssssssssssssssssssssssssssssssssssssssssssssssssssssssssssssssss

Next, we study the role of collisions in triggering magnetic reconnection and dissipation
using the framework of resistive force-free electrodynamics (magnetodynamics).   
In the fluid frame, the structure of colliding flows is that of the unstressed 2D ABC configuration.
In order obtain its structure in the lab-frame one can simply apply the Lorentz transformation for the 
electromagnetic field. Hence for the flow moving along the $x$ axis to the left with speed $v$, 
we have 
   
\ba &&
B_x = -\sin(2\pi y/L)B_0\,,
\nn &&
B_y = \Gamma \sin(2\pi \Gamma x/L) B_0 \,,
\nn &&
B_z =\Gamma\left[\left(\cos(2\pi \Gamma x/L)+\cos(2\pi  y/L)\right)\right] B_0  \,,
\nn &&
E_x=0 \,,
\nn &&
E_y =-v \Gamma\left[\left(\cos(2\pi \Gamma x/L)+\cos(2\pi  y/L)\right)\right] B_0  \,,
\nn &&
E_z = v \Gamma \sin(2\pi \Gamma x/L) B_0 \,,
\label{coll-flow}
\ea
where $\Gamma$ is the Lorentz factor.  Via replacing $v$ with $-v$, we obtain the solution for 
the flow moving to right with the same speed. To initiate the collision, one can introduce at $t=0$ 
a discontinuity at $x=0$ so that the flow is moving to the left for $x>0$ and to the right for $x<0$. 
The corresponding jumps at $x=0$ are 
\be 
  [B_x]=[B_z]=[E_x]=[E_z=0],\qquad [B_y]=2B_y^{(r)}, \qquad [E_y]=2E_y^{(r)}\,,
\label{eq-jumps}
\ee         
where index $(r)$ refers to the value to the right of the discontinuity. 
Instead of studying the domain which includes both flows, one can treat $x=0$ as a 
boundary with the boundary conditions described by Eqs. (\ref{eq-jumps}) and deal only with 
the left (or the right) half of the domain. This is exactly what we did in our simulations. 

In the simulations, we set $B_0=1$, $\Gamma=3$ and used the computational 
domain $(0,3L)\times(-L,L)$, with a uniform grid of $600\times400$ computational cells. 
The boundary at $x=3L$ is treated as an inlet of the flow described by Eq. (\ref{coll-flow}). 
Hence the solution at this boundary is given by these equation with  
$x$ replaced by $x+vt$. The magnetic Reynolds number is $Re_m=10^3$.   

\begin{figure}
\centering
\includegraphics[width=0.4\textwidth]{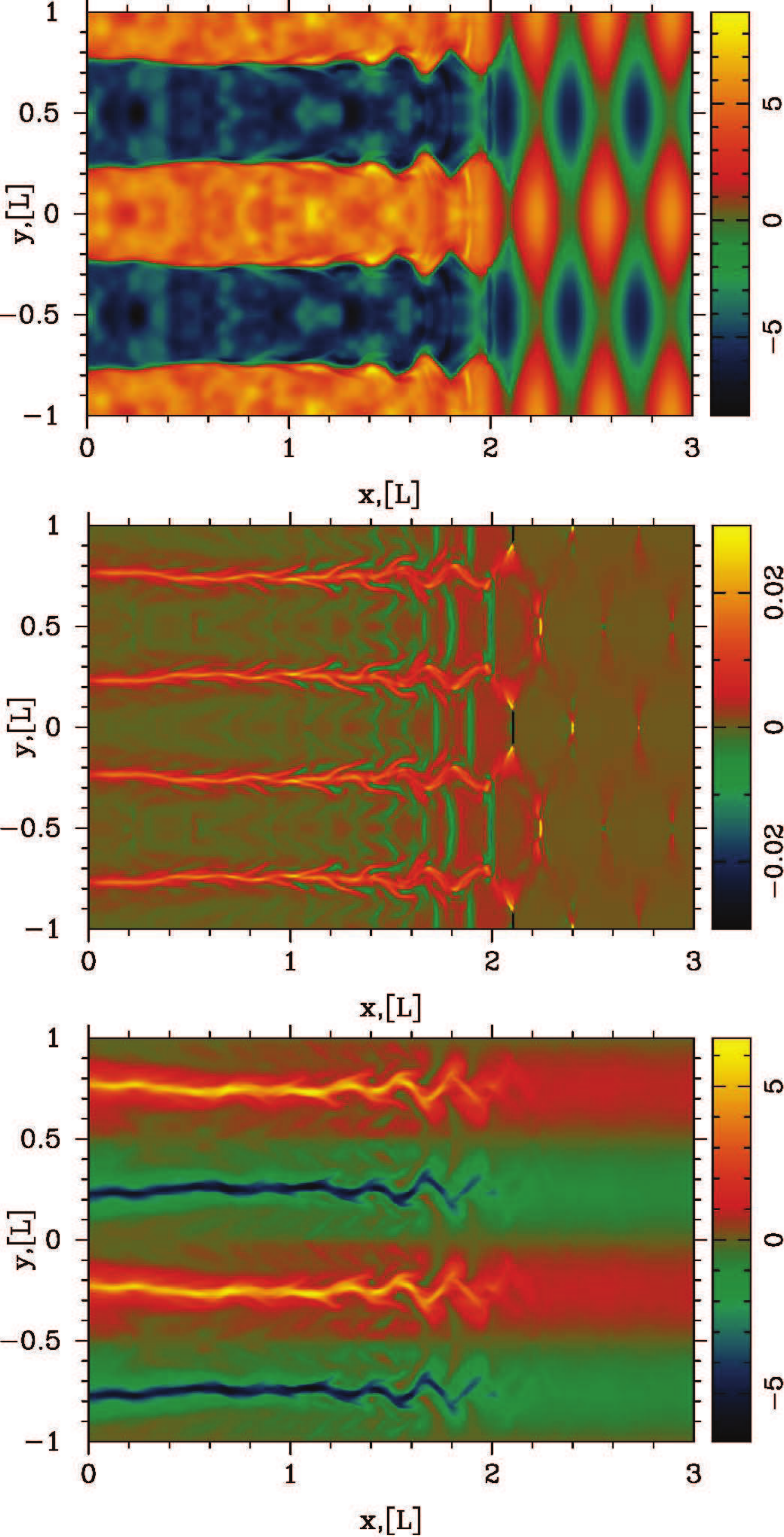}
\caption{Collision-induced merger of magnetic islands. The plots show the distribution of
$B_z$ (top), $B\cdot E$ (middle) and $B_x$ at time $t=4$. 
}
\label{sic-snap}
\end{figure}

Figure~\ref{sic-snap} illustrates the solution at $t=4$. To the right of $x=2$ the flow 
structure is the same as in the initial solution. In the plot of $B_z$ (top panel), one 
can clearly see the lattice of magnetic islands. It appears compressed along the x axis, 
which is a signature of the relativistic length contraction effect. To the left of $x=2$, 
the islands rapidly merge and form horizontal stripes of oppositely directed $B_z$. 
The in-plane magnetic field is concentrated in the current sheets separating these stripes.

\begin{figure}
\centering
\includegraphics[width=0.45\textwidth]{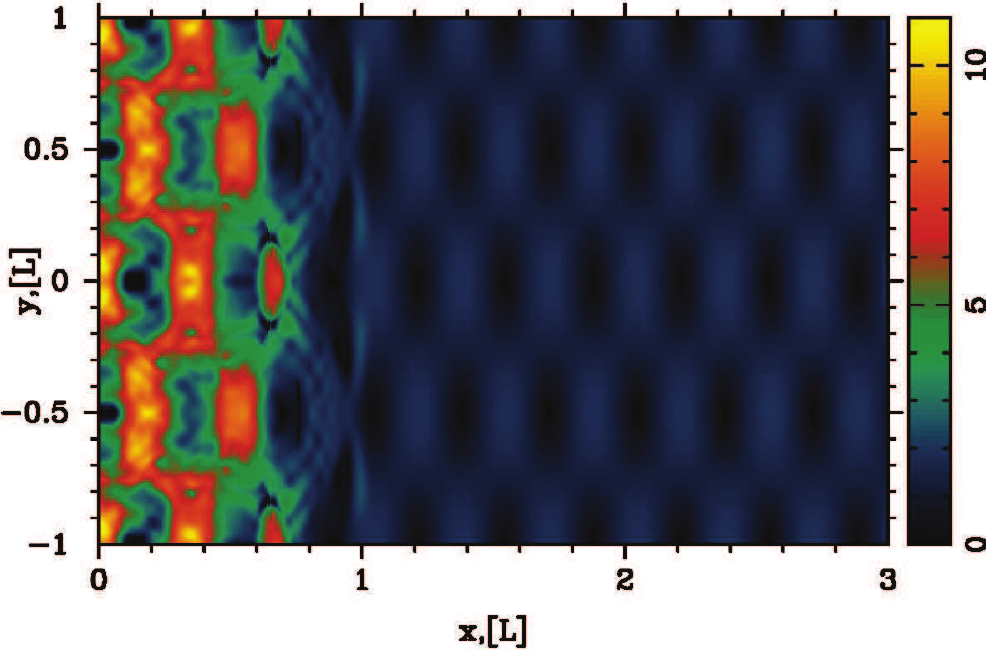}
\includegraphics[width=0.45\textwidth]{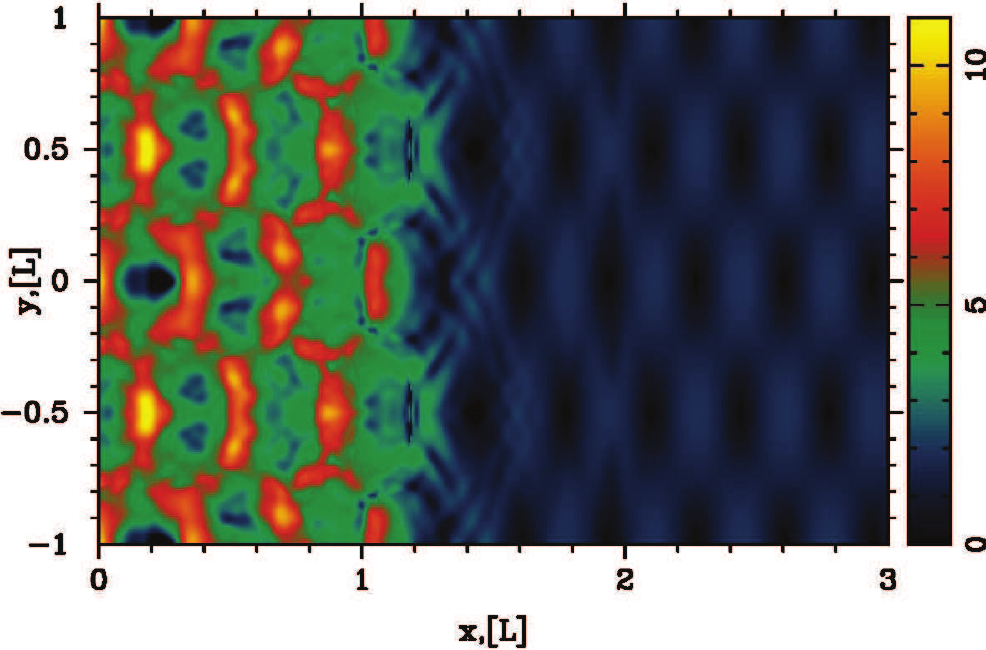}
\includegraphics[width=0.45\textwidth]{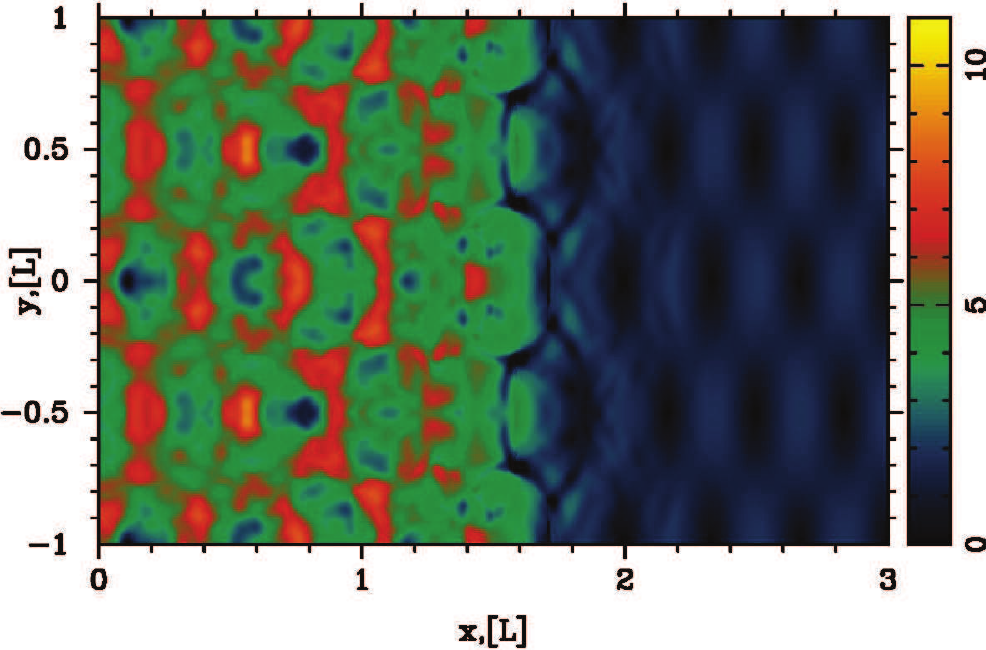}
\includegraphics[width=0.45\textwidth]{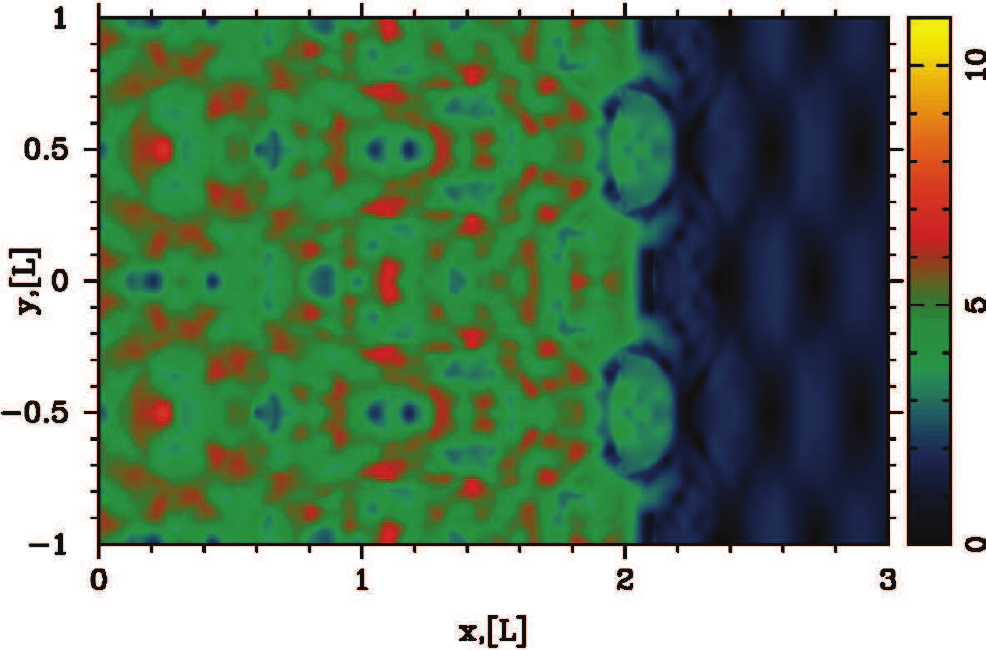}
\caption{Shock-induced merger of magnetic islands. The plots show the distribution of 
the magnetic pressure $p_m=(B^2-E^2)/2$ at time $t=1,2,3$ and 4 (increasing from left to 
right and from top to bottom).
}
\label{sic-pm}
\end{figure}

Figure~\ref{sic-pm} shows the time evolution of the flow using the Lorentz-invariant  
parameter $p_m=(B^2-E^2)/2$, which plays the role of magnetic pressure. One can see that 
the evolution can be described as a wave propagation in the direction away from the plane 
of collision. Behind this wave the magnetic pressure increases and the wave speed is significantly 
lower than the speed of light. 

\begin{figure}
\centering
\includegraphics[width=0.25\textwidth]{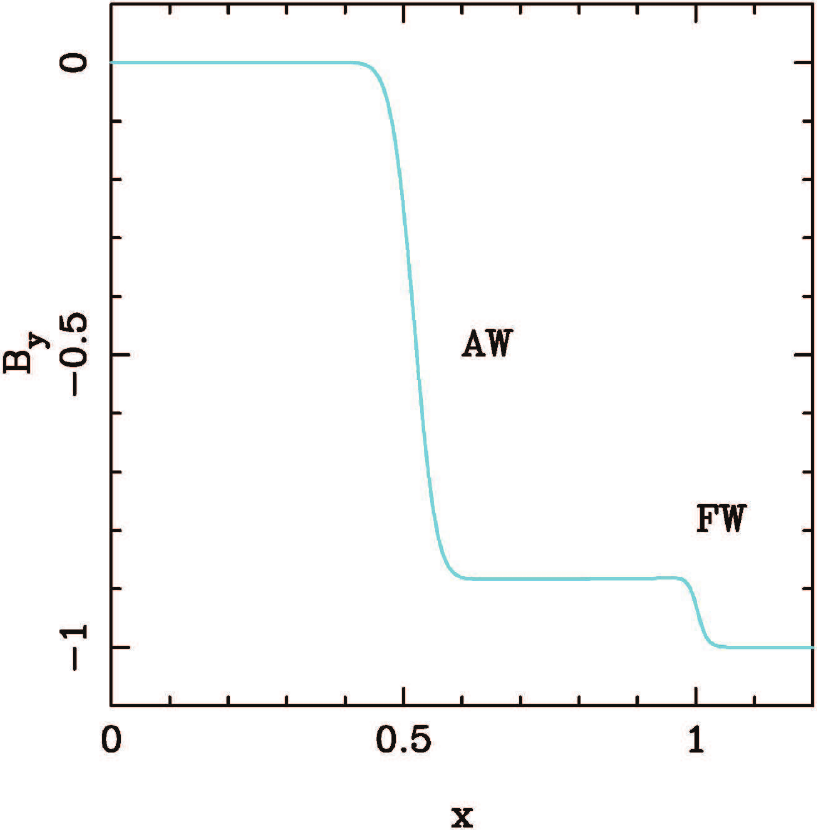}
\includegraphics[width=0.25\textwidth]{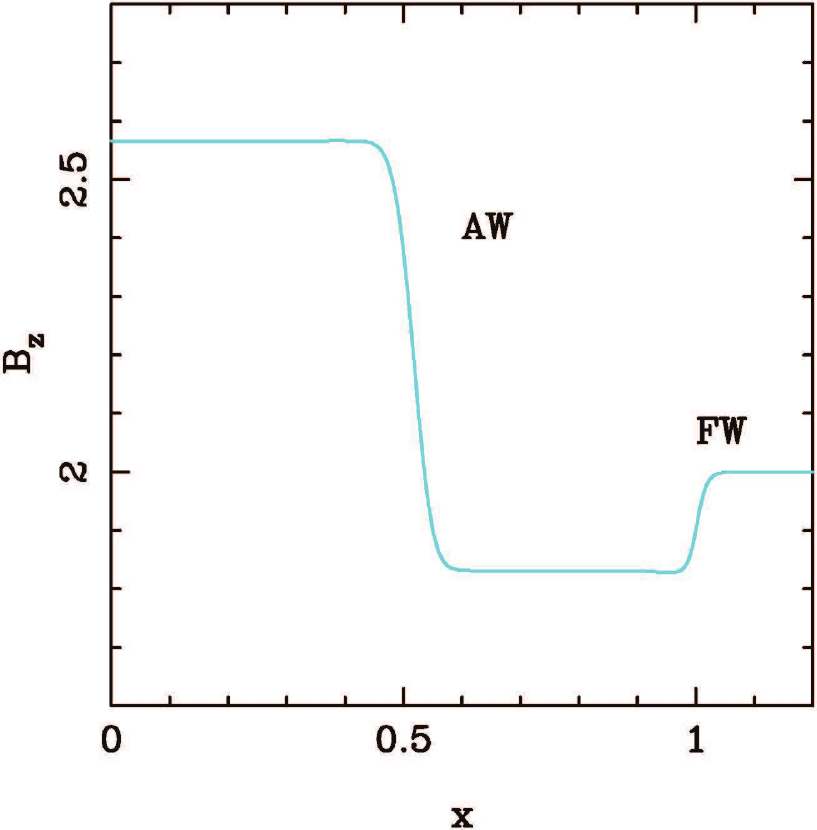}
\includegraphics[width=0.25\textwidth]{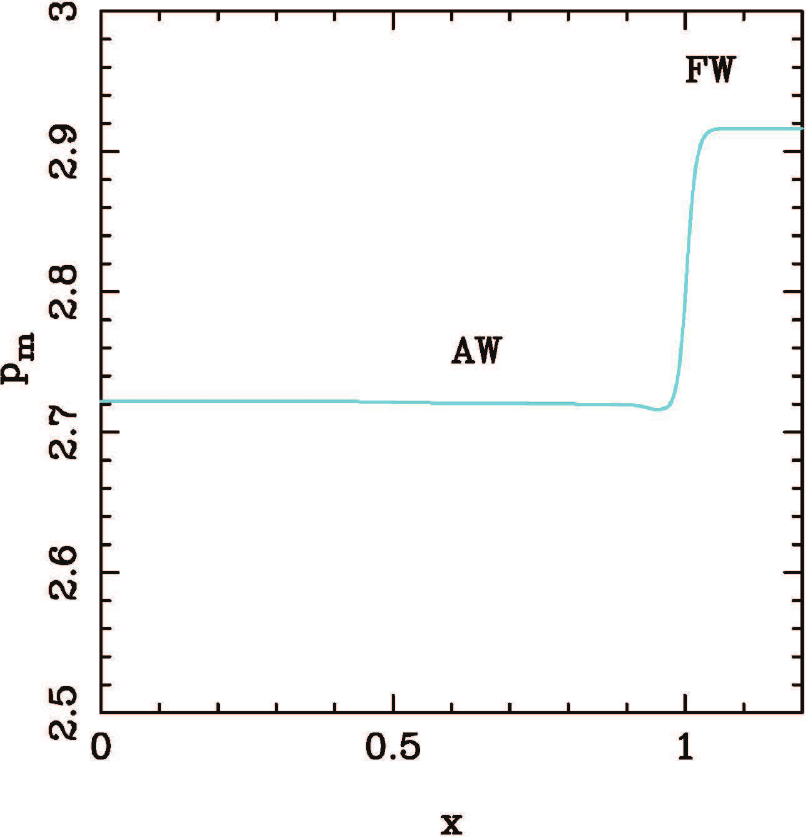}
\caption{Collision of uniform flows.  
{\it Left panel:} The $B_y$ component of the magnetic field at $t=1$. 
Both the fast wave and Alvf\'en waves are easily identified on the plot. 
{\it Middle panel:} The $B_z$ component of the magnetic field.
{\it Right panel:} The magnetic pressure $p_m=(B^2-E^2)/2$ at the same time. 
Only the fast wave can be seen on this plot. }
\label{sic-rp}
\end{figure}

The wave properties are very different from those of the basic hyperbolic waves of magnetodynamics. 
To illustrate this fact, Figure~\ref{sic-rp} shows the solution of the Riemann problem 
$$
\mbox{Left state:}\quad B=(2,1,2), \quad E=(0,0,-v)\,;\qquad 
\mbox{Right state:}\quad B=(2,-1,2), \quad E=(0,0,v)\,, 
$$  
which describes the collision of two uniform flows. 
The collision speed $v$ is the same as before ( $\Gamma=3$). 
The collision creates both the fast wave (FW) and the Alfv\'en wave (AW). 
One can see that, the fast wave propagates with the speed of light and the magnetic 
pressure is invariant across AW (Komissarov, 2002).  

Comparing the results of the test Rienmann problem with the main simulations, we conclude 
that the main wave is not AW because the magnetic pressure is not invariant and not FW as 
its speed is significantly less than the speed of light. In fact, the top-left panel ($t=1$) 
of Figure~\ref{sic-pm} shows a weak linear feature located at $x=1$ which can be identified with 
the FW produced by the collision.  A similar, but weaker feature can also be seen in the 
top-right panel ($t=2$). Presumable, this wave gets scattered by the inhomogeneities of the incoming 
flow and gradually looses coherence. Other fine arc-like features seen in front of the main 
wave (e.g. in the region  $1.2<x<1.7$ ) are likely to be fast waves emitted by the turbulence 
downstream of the main wave.         

The main wave, which we will refer to as the dissipation front, seems to start as an Alfv\'en wave. 
Immediately downstream of the Alfv\'en wave, the $B_z$ component of the magnetic field increases
(see the middle panel of Fig.\ref{sic-rp}), and hence the ABC configuration remains highly 
squashed along the x-axis. However, the flow comes to halt in the x direction and the X-points 
are now highly stressed in the fluid frame\footnote{In magnetodynamics, there no unique way to 
define the fluid frame. If the electric field vanishes in one frame then it also vanishes in 
any other frame moving relative to this one along the magnetic field.}. This leads to their 
immediate collapse and rapid merger of the ABC islands, as described in \S \ref{Force-freeABCdriven}.  The increase of $p_m$ seems to be 
a product of magnetic dissipation that follows the merger.  

Since  the dissipation front is not a well defined surface of fixed shape, it is rather 
difficult to measure its speed. However even naked eye inspection of the plots suggest that it 
monotonically decreases with time with saturation towards the end of the run. 
Crude measurements based on the position of first features showing strong deviation from 
the properties of incoming flow give 
$v_{df}\simeq 0.73\,,0.47\,,0.45\,,0.40\,,0.38$ and 0.37. 

To see how the magnetic dissipation can influence the front speed, 
we analyze the energy and momentum conservation in our problem. (This approach is similar 
to that used by \citep{2003MNRAS.345..153L}.) To simplify the analysis we 
ignore the complicated electromagnetic structure of flow and assume that the magnetic field is 
parallel to the front. Denote the front position as $x_{df}$ and use indexes 1 and 2  to 
indicate the flow parameters upstream of the front and downstream of its dissipation 
zone the front respectively. Since our equations do not include particles pressure and energy,  
and the dissipated electromagnetic energy simply vanishes this analysis is only relevant 
for the case of rapid radiative cooling.  

Since the flow grounds to halt downstream of the dissipation front, the momentum 
conservation in the computational  box $[0,x_r]$ 
\be
   \frac{d}{dt}\left(-v_1B_1^2(x_b-x_{df}) \right) = \frac{B_2^2}{2}-\frac{1}{2}(B_1^2+v_1^2 B_1^2) \,, 
\ee
where $v_1=E_1/B_1$. Hence
\be
   B_1^2 v_{df} v_1=\frac{B_2^2}{2}-\frac{1}{2}B_1^2(1+v_1^2) \,. 
\ee
Similarly, we obtain the energy conservation law  
\be
   \frac{B_2^2}{2}v_{df} - \frac{1}{2}v_{df}B_1^2 (1+v_1^2)= v_1B_1^2-Q_{d} \,, 
\ee
where $Q_{d}$ is the dissipation rate of the front. Solving the last two equation for $B_2$ and $v_{df}$,  
we obtain the simple result for the velocity of the shock
\be 
   v_s^2=1-\alpha \,, 
\label{shock-speed}
\ee 
where $\alpha=Q_d/v_1B_1^2$ is the fraction of the incoming energy flux which is dissipated and 
lost in the dissipation front, the dissipation efficiency. 
From this we find that in the absence of dissipation the shock speed equals to the 
speed of light and thus recover the result for ideal magnetodynamics. 
In the opposite case of total dissipation, the front speed vanishes, as expected.  
The final value of $v_s=0.37$ found in our simulations corresponds to the dissipation  
efficiency $\alpha=86$\%.

% \clearpage
%%%%%%%%%%

%%%%%%%Lorenzo%%%%%%
%%%%%%%%%%%%%%%%%%%%%%%%%%%%%%%%
 \begin{figure}
 \centering
\includegraphics[width=.49\textwidth]{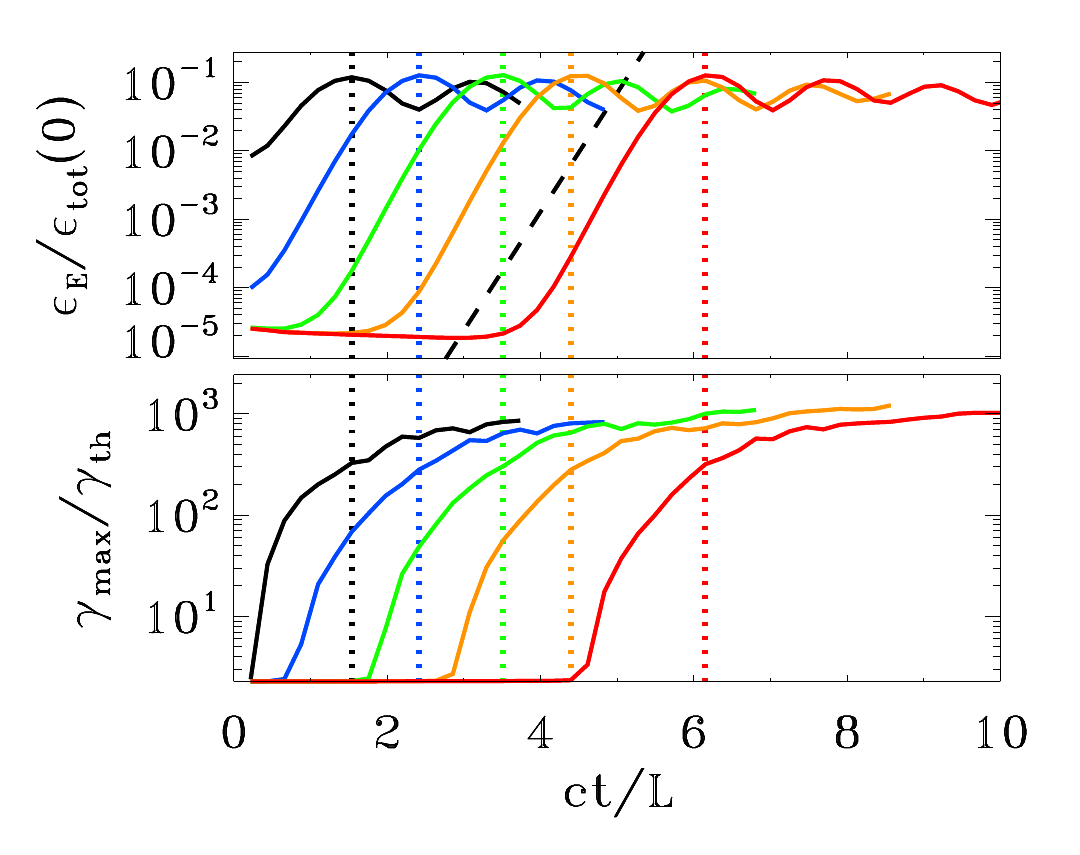} 
\caption{Temporal evolution of the electric energy (top panel, in units of the total initial energy) and of the maximum particle Lorentz factor (bottom panel; $\gammamax$ is defined in \eq{ggmax}, and it is normalized to the thermal Lorentz factor $\gamma_{\rm th}\simeq 1+(\hat{\gamma}-1)^{-1} kT/m c^2$), for a suite of five PIC simulations of ABC collapse with fixed $kT/mc^2=10^2$, $\sigmain=42$ and $L/\rhot=126$, but different magnitudes of the initial velocity shear (see \eq{vel1}): $v_{\rm push}/c=10^{-1}$ (black), $v_{\rm push}/c=10^{-2}$ (blue), $v_{\rm push}/c=10^{-3}$ (green), $v_{\rm push}/c=10^{-4}$ (yellow) and $v_{\rm push}/c=0$ (red). The dashed black line in the top panel shows that the electric energy grows exponentially as $\propto \exp{(4ct/L)}$. The vertical dotted lines mark the time when the electric energy peaks (colors correspond to the five values of $v_{\rm push}/c$, as described above).}
\label{fig:abcshrtime} 
\end{figure}
 \begin{figure}
 \centering
\includegraphics[width=.49\textwidth]{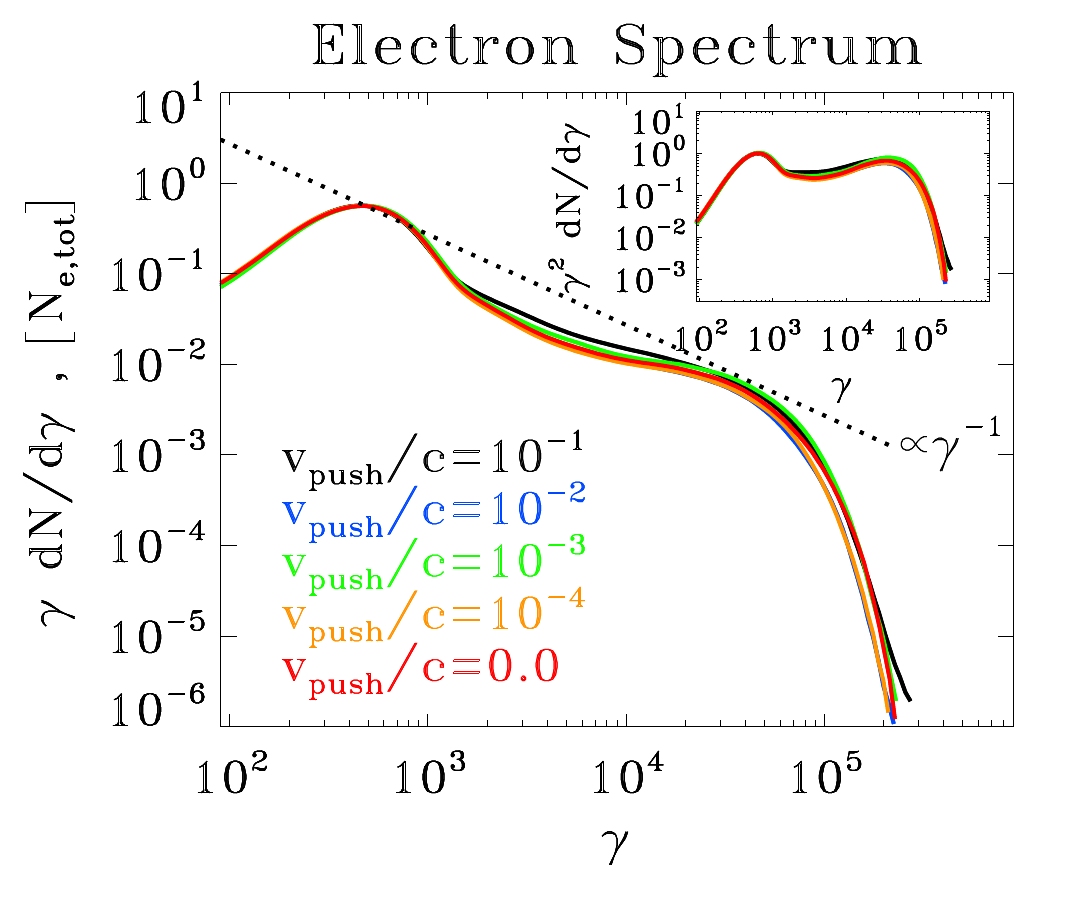} 
\caption{Particle spectrum at the time when the electric energy peaks, for a suite of five PIC simulations of ABC collapse with fixed $kT/mc^2=10^2$, $\sigmain=42$ and $L/\rhot=126$, but different magnitudes of the initial velocity shear (same runs as in \fig{abcshrtime}): $v_{\rm push}/c=10^{-1}$ (black), $v_{\rm push}/c=10^{-2}$ (blue), $v_{\rm push}/c=10^{-3}$ (green), $v_{\rm push}/c=10^{-4}$ (yellow) and $v_{\rm push}/c=0$ (red). The main plot shows $\gamma dN/d\gamma$ to emphasize the particle content, whereas the inset presents $\gamma^2 dN/d\gamma$ to highlight the energy census. The dotted black line is a power law $\gamma dN/d\gamma\propto \gamma^{-1}$, corresponding to equal energy content per decade (which would result in a flat distribution in the inset).}
\label{fig:abcshrspec} 
\end{figure}

\subsection{Driven ABC evolution: PIC simulation}
\label{drivenABCPIC}

\subsubsection{2D ABC structures with initial shear}
Next  we investigate a driven evolution of the ABC structures using PIC simulations. Inspired by the development of the oblique mode described in Fig.~\ref{inst}, we set up our system with an initial velocity in the oblique direction, so that two neighboring chains of ABC islands will shear with respect to each other. More precisely, we set up an initial velocity profile of the form
\begin{eqnarray}\label{eq:vel1}
v_x&=&-v_{\rm push}\cos[\pi (x+y)/L]/\sqrt{2} \\
v_y&=&-v_x\nonumber\\
v_z&=&0\nonumber
\end{eqnarray}
and we explore the effect of different values of $\vpush$.\footnote{In addition to initializing a particle distribution with a net bulk velocity, we also set up the resulting $-\bmath{v}\times \bmath{B}/c$ electric field.}

\fig{abcshrtime} shows that the evolution of the electric energy (top panel, in units of the total initial energy) is remarkably independent of $\vpush$, apart from an overall  shift in the onset time. In all the cases, the electric energy grows exponentially as $\propto \exp{(4ct/L)}$ (compare with the black dashed line) until it peaks at $\sim 10\%$ of the total initial energy (the time when the electric energy peaks is indicated with the vertical colored dotted lines). In parallel, the maximum particle energy $\gammamax$ grows explosively (bottom panel in \fig{abcshrtime}), with a temporal profile that is nearly indentical in all the cases (once again, apart from an overall time shift). The initial value of the electric energy  scales as $\vpush^2$ for large $\vpush$  (black for $\vpush/c=10^{-1}$ and blue for $\vpush/c=10^{-2}$), just as a consequence of the electric field $-\bmath{v_{\rm push}}\times \bmath{B}/c$ that we initialize. At smaller values of $\vpush$, the initial value of the electric energy is independent of $\vpush$ (green to red curves in the top panel). Here, we are sensitive to the electric field required to build up the particle currents implied by the steady ABC setup. Overall, the similarity of the different curves in \fig{abcshrtime} (all the way to the undriven case of $\vpush=0$, in red) confirms that the oblique ``shearing'' mode is a natural instability avenue for 2D ABC structures.

As a consequence, it is not surprising that the particle spectrum measured at the time when the electric energy peaks (as indicated by the vertical colored lines in \fig{abcshrtime}) bears no memory of the driving speed $\vpush$. In fact, the five curves in \fig{abcshrspec} nearly overlap.

%%%%%%%%%%%%%%%%%%%%%%%%%%%%%%%
 \begin{figure}
 \centering
\includegraphics[width=.7\textwidth]{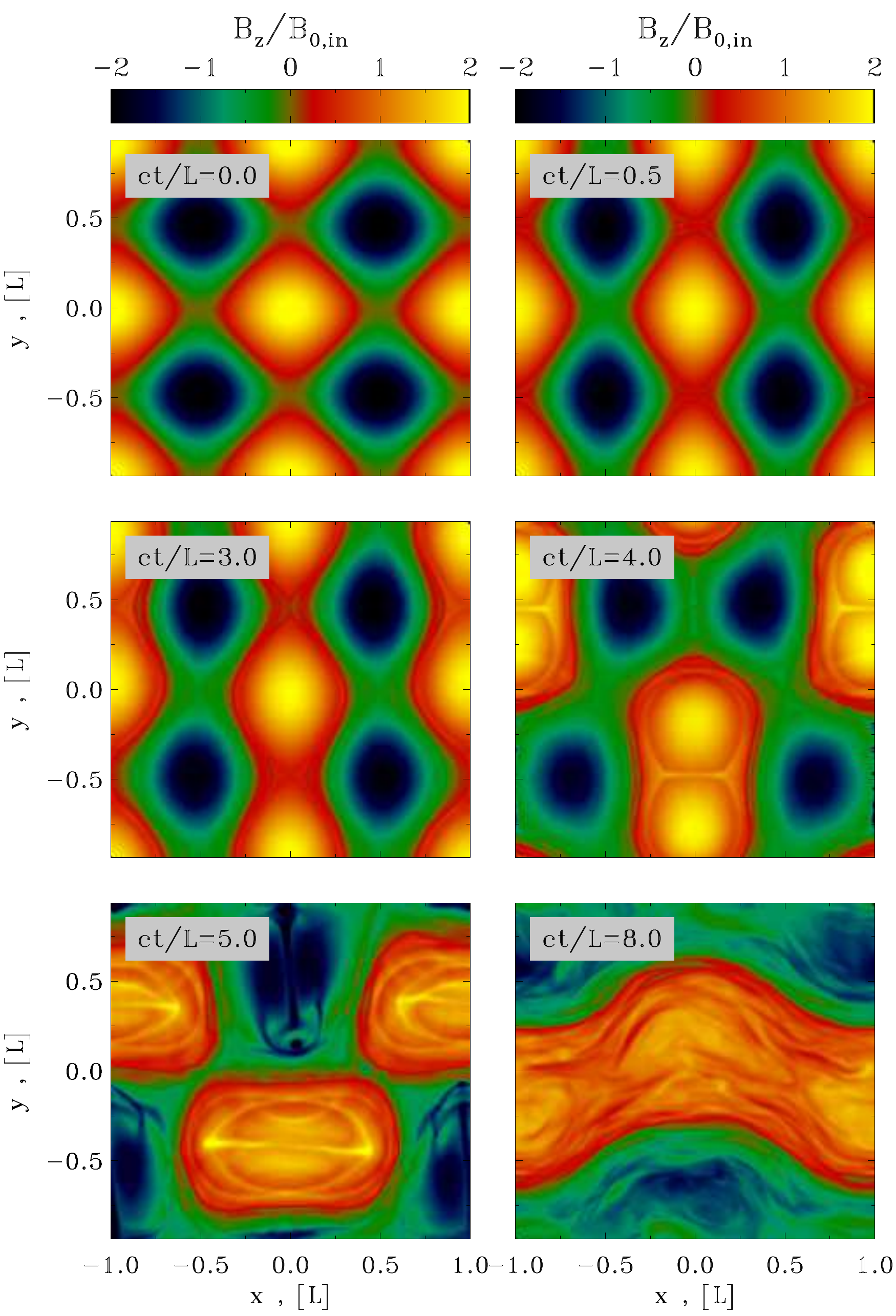} 
\caption{Temporal evolution of the instability of a typical 2D ABC structure (time is indicated in the grey box of each panel) with an initial stress of $\lambda=0.94$ (see Paper I for the definition of $\lambda$ in the context of solitary X-point collapse). The plot presents the 2D pattern of the out-of-plane field $B_z$ (in units of $B_{0,\rm in}$) from a PIC simulation with $kT/mc^2=10^{-4}$, $\sigmain=42$ and $L=126\,\rhot$, performed within a domain of size $2L\times 2\lambda L$. The system is initially squeezed along the $y$ direction. This leads to X-point collapse and formation of current sheets (see the current sheet at $x=0$ and $y=0.5L$ in the top right panel). The resulting reconnection outflows push away neighboring magnetic structures (the current sheet at $x=0$ and $y=0.5L$ pushes away the two blue islands at $y=0.5L$ and $x=\pm0.5L$). The system is now squeezed along the $x$ direction (middle left panel), and it goes unstable on a dynamical time  (middle right panel), similarly to the case of spontaneous (i.e., unstressed) ABC instability. The subsequent evolution closely resembles the case of spontaneous ABC instability shown in \fig{abcfluid_bz}.}
\label{fig:abcstrfluid} 
\end{figure}
 \begin{figure}
 \centering
\includegraphics[width=.49\textwidth]{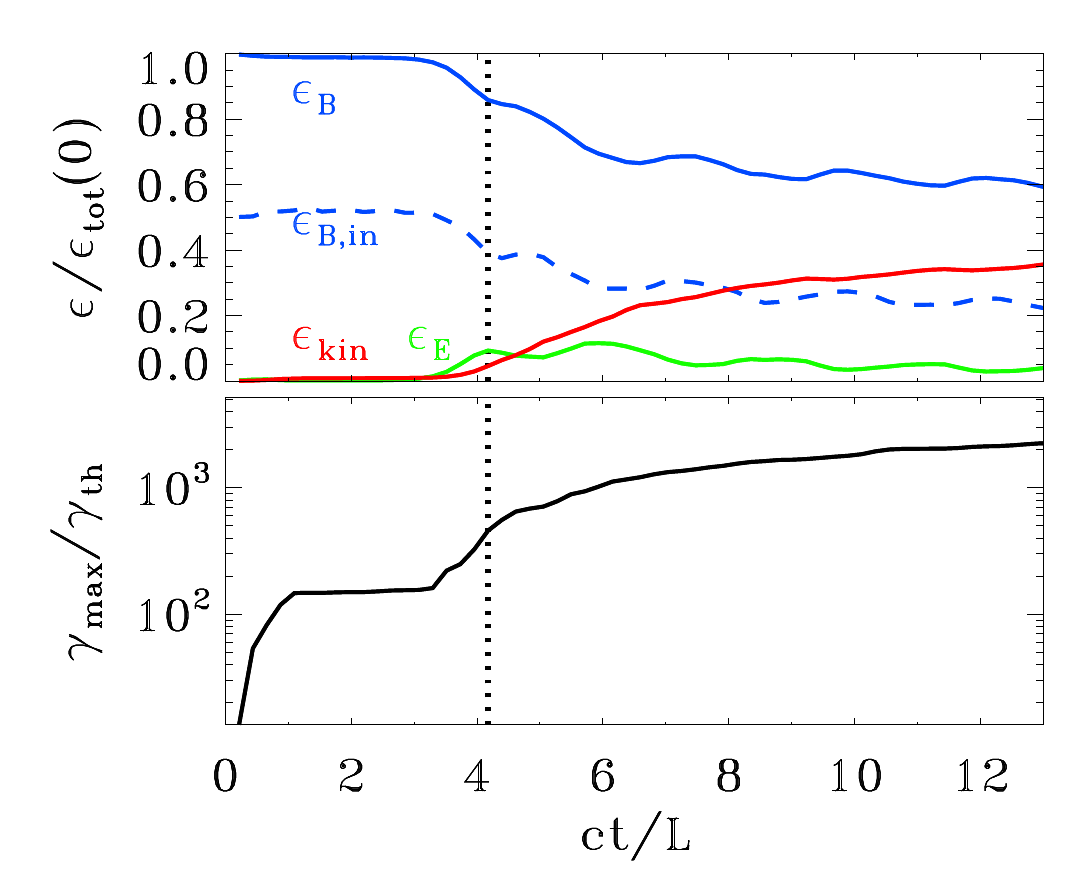} 
\caption{Temporal evolution of various quantities, from a 2D PIC simulation of ABC instability with $kT/mc^2=10^{-4}$, $\sigmain=42$ and $L=126\,\rhot$ and stress parameter $\lambda=0.94$, performed within a domain of size $2L\times 2\lambda L$ (the same run as in \fig{abcstrfluid}). Top panel: fraction of energy in magnetic fields (solid blue), in-plane magnetic fields (dashed blue, with $\epsilon_{B,\rm in}=\epsilon_B/2$ in the initial configuration), electric fields (green) and particles (red; excluding the rest mass energy), in units of the total initial energy. Bottom panel: evolution of the maximum Lorentz factor $\gammamax$, as defined in \eq{ggmax}, relative to the thermal Lorentz factor $\gamma_{\rm th}\simeq 1+(\hat{\gamma}-1)^{-1} kT/m c^2$, which for our case is $\gamma_{\rm th}\simeq 1$. The early growth of $\gammamax$ up to $\gammamax/\gamma_{\rm th}\sim 1.5\times 10^2$ is due to particle acceleration at the current sheets induced by the initial stress. The subsequent development of the ABC instability at $ct/L\simeq 4$ is accompanied by little field dissipation ($\epsilon_{\rm kin}/\epsilon_{\rm tot}(0)\sim 0.1$) but dramatic particle acceleration, up to $\gammamax/\gamma_{\rm th}\sim 10^3$. In both panels, the vertical dotted black line indicates the time when the electric energy peaks.}
\label{fig:abcstrtime} 
\end{figure}
 \begin{figure}
 \centering
\includegraphics[width=.49\textwidth]{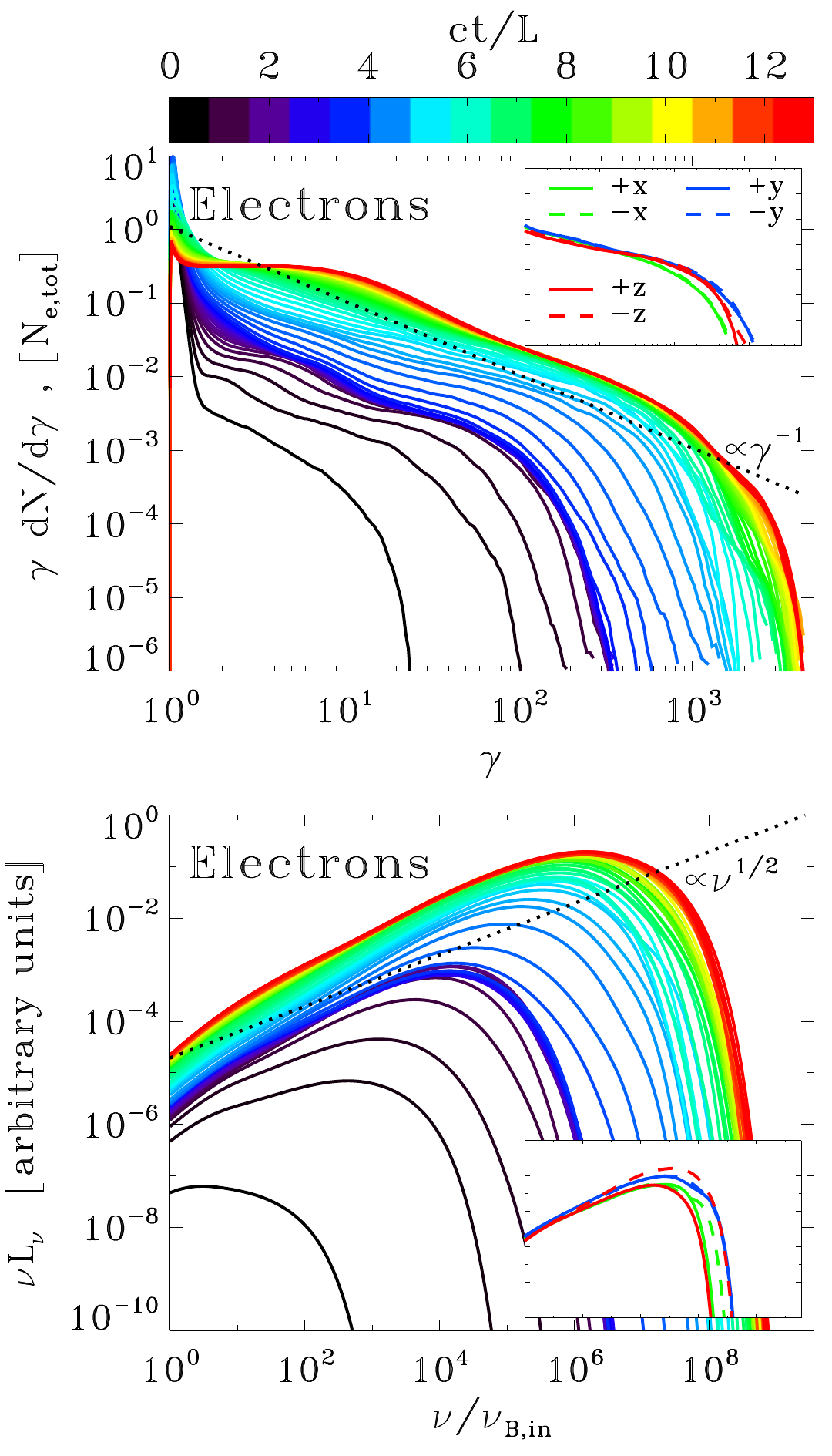} 
\caption{Particle spectrum and synchrotron spectrum from a 2D PIC simulation of ABC instability with $kT/mc^2=10^{-4}$, $\sigmain=42$ and $L=126\,\rhot$ and stress parameter $\lambda=0.94$, performed within a  domain of size $2L\times 2 \lambda L$ (the same run as in \fig{abcstrfluid} and \fig{abcstrtime}). Time is measured in units of $L/c$, see the colorbar at the top. Top panel: evolution of the electron energy spectrum normalized to the total number of electrons. At late times, the spectrum approaches a distribution of the form $\gamma dN/d\gamma\propto \gamma^{-1}$, corresponding to equal energy content in each decade of $\gamma$ (compare with the dotted black line). The inset in the top panel shows the electron momentum spectrum along different directions (as indicated in the legend), at the time when the electric energy peaks (as indicated by the dotted black line in \fig{abcstrtime}). Bottom panel: evolution of the angle-averaged synchrotron spectrum emitted by electrons. The frequency on the horizontal axis is normalized to $\nu_{B,\rm in}=\sqrt{\sigmain}\omega_{\rm p}/2\pi$. At late times, the synchrotron spectrum approaches $\nu L_\nu\propto \nu^{1/2}$ (compare with the dotted black line), which just follows from the electron spectrum $\gamma dN/d\gamma\propto \gamma^{-1}$. The inset in the bottom panel shows the synchrotron spectrum at the time indicated in \fig{abcstrtime} (dotted black line) along different directions (within a solid angle of $\Delta \Omega/4\pi\sim 3\times10^{-3}$), as indicated in the legend in the inset of the top panel. }
\label{fig:abcstrspec} 
\end{figure}

\subsubsection{2D ABC structures with initial stress}
We now investigate the evolution of ABC structures in the presence of an initial stress, quantified by the stress parameter $\lambda$ (see Paper I for the definition of $\lambda$ in the context of solitary X-point collapse). The unstressed cases discussed so far would correspond to $\lambda=1$. The simulation box will be a rectangle with size $2L$ along the $x$ direction and $2\lambda L$ along the $y$ direction, with periodic boundary conditions.

\fig{abcstrfluid} shows the temporal evolution of the 2D pattern of the out-of-plane field $B_z$ (in units of $B_{0,\rm in}$) from a PIC simulation with $kT/mc^2=10^{-4}$, $\sigmain=42$ and $L=126\,\rhot$. The system is initially squeezed along the $y$ direction, with a stress parameter of $\lambda=0.94$ (top left panel). The initial stress leads to X-point collapse and formation of current sheets (see the current sheet at $x=0$ and $y=0.5L$ in the top right panel). This early phase results in a minimal amount of magnetic energy dissipation (see the top panel in \fig{abcstrtime}, with the solid blue line indicating the magnetic energy and the red line indicating the particle kinetic energy), but significant particle acceleration. As indicated in the bottom panel of \fig{abcstrtime}, the high-energy cutoff $\gammamax$ of the particle distribution grows quickly (within one dynamical time) up to $\gammamax/\gamma_{\rm th}\sim 1.5\times10^2$. At this point ($ct/L\gtrsim 1$), the increase in the maximum particle energy stalls. In \fig{abcstrfluid}, we find that this corresponds to a phase in which the system tends to counteract the initial stress. In particular, the reconnection outflows emanating from the current sheets in the top right panel of \fig{abcstrfluid} push away neighboring magnetic structures (the current sheet at $x=0$ and $y=0.5L$ pushes away the two blue islands at $y=0.5L$ and $x=\pm0.5L$). The system gets squeezed along the $x$ direction (middle left panel), i.e., the stress is now opposite to the initial stress. Shortly thereafter, the ABC structure goes unstable on a dynamical timescale  (middle right panel), with a pattern similar to the case of spontaneous (i.e., unstressed) ABC instability. In particular, the middle right panel in \fig{abcstrfluid} shows that in this case the instability proceeds via the ``parallel'' mode depicted in Fig.~\ref{inst-2mode} (but other simulations are dominated by the ``oblique'' mode sketched in Fig.~\ref{inst}). The subsequent evolution closely resembles the case of spontaneous ABC instability shown in \fig{abcfluid_bz}, with the current sheets stretching up to a length $\sim L$ and the merger of islands having the same $B_z$ polarity (bottom left panel in \fig{abcstrfluid}), until only two regions remain in the box, with $B_z$ fields of opposite polarity (bottom right panel in \fig{abcstrfluid}).

The second stage of evolution --- resembling the spontaneous instability of unstressed ABC structures --- leads to a dramatic episode of particle acceleration (see the growth in $\gammamax$ in the bottom panel of \fig{abcstrtime} at $ct/L\sim 4$), with the energy spectral cutoff extending beyond $\gammamax/\gamma_{\rm th}\sim 10^3$ within one dynamical time. This fast increase in $\gammamax$ is indeed reminiscent of what we had observed in the case of unstressed ABC instability (compare with the bottom panel in \fig{abctime} at $ct/L\sim 5$). In analogy to the case of unstressed ABC structures, the instability leads to dramatic particle acceleration, but only minor energy dissipation (the mean kinetic energy reaches a fraction $\sim 0.1$ of the overall energy budget). Additional dissipation of magnetic energy into particle heat (but without much non-thermal particle acceleration) occurs at later times ($ct/L\gtrsim 6$) during subsequent island mergers, once again imitating the evolution of unstressed ABC instability. 

The two distinct evolutionary phases --- the early stage driven by the initial stress, and the subsequent dynamical ABC collapse resembling the unstressed case --- are clearly apparent in the evolution of the particle energy spectrum (top panel in \fig{abcstrspec}) and of the angle-averaged synchrotron emission (bottom panel in \fig{abcstrspec}). The initial stress drives fast particle acceleration at the resulting currrent sheets (from black to dark blue in the top panel). Once the stress reverses, as part of the self-consistent  evolution of the system (top right panel in \fig{abcstrfluid}), the particle energy spectrum freezes (see the clustering of the dark blue lines). Correspondingly, the angle-averaged synchrotron spectrum stops evolving (see the clustering of the dark blue lines in the bottom panel). A second dramatic increase in the particle and emission spectral cutoff (even more dramatic than the initial growth) occurs between $ct/L\sim 3$ and $ct/L\sim 6$ (dark blue to cyan curves in \fig{abcstrspec}), and it directly corresponds to the phase of ABC instability resembling the unstressed case. The particle spectrum quickly extends up to $\gammamax\sim 10^3$ (cyan lines in the top panel), and correspondigly the peak of the $\nu L_\nu$ emission spectrum shifts up to $\sim \gammamax^2\nu_{B,\rm in}\sim10^6 \nu_{B,\rm in}$ (cyan lines in the bottom panel). At times later than $c/L\sim 6$, the evolution proceeds slower, similarly to the case of unstressed ABC instability: the high-energy cutoff in the particle spectrum shifts up by only a factor of three before saturating (green to red curves in the top panel), and the peak frequency of the synchtron spectrum increases by less than a factor of ten (green to red curves in the bottom panel). Rather than non-thermal particle acceleration, the late evolution is accompanied by substantial heating of the bulk of the particles, with the peak of the particle spectrum shifting from $\gamma\sim1$ up to $\gamma\sim 10$ after $ct/L\sim 6$ (see also the red line in the top panel of \fig{abcstrtime} at $ct/L\gtrsim 6$, indicating efficient particle heating). Once again, this closely parallels the late time evolution of unstressed ABC instability. 

Finally, we point out that the particle distribution at the time when the electric energy peaks (indicated by the vertical dotted line in \fig{abcstrtime}) is nearly isotropic, as indicated by the inset in the top panel of \fig{abcstrspec}. In turn, this results in quasi-isotropic synchrotron emission (inset in the bottom panel). We refer to the unstressed case in \fig{abcspec} for an explanation of this result, which is peculiar to the ABC geometry employed here.

%%%%%%%%%%%%%%%%%%%%%
 \begin{figure}
 \centering
\includegraphics[width=.49\textwidth]{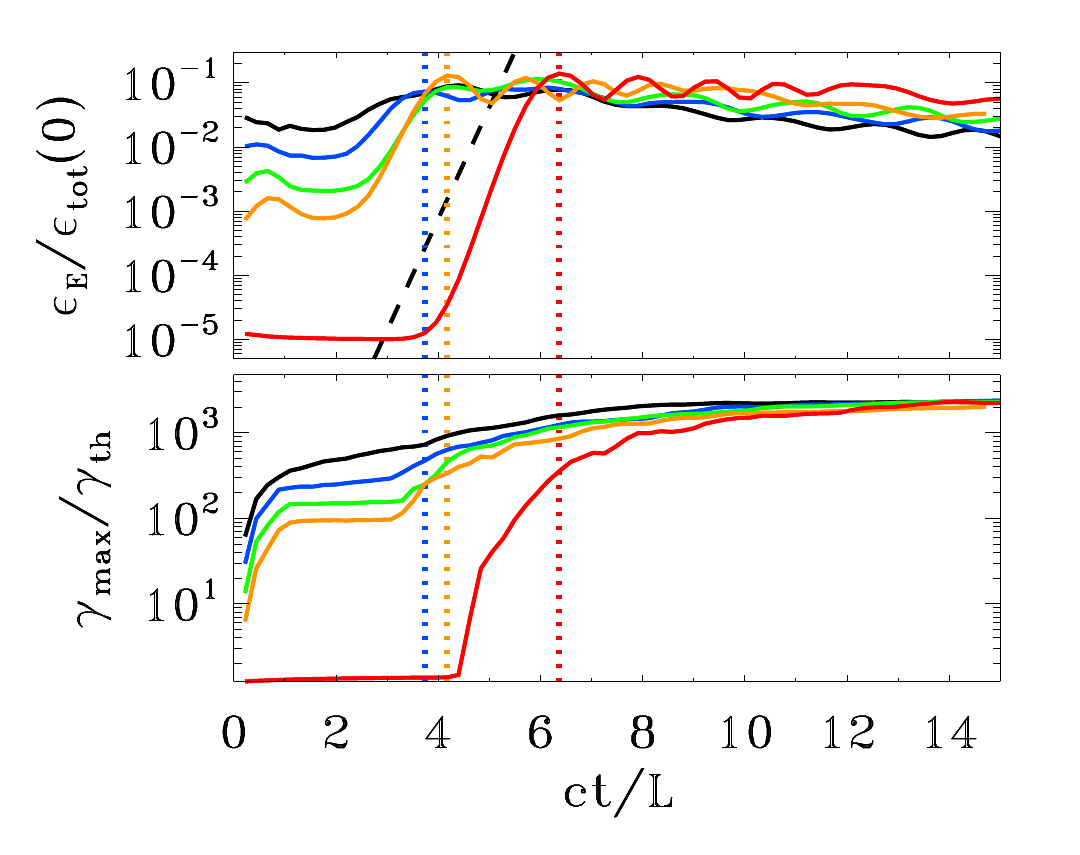} 
\caption{Temporal evolution of the electric energy (top panel, in units of the total initial energy) and of the maximum particle Lorentz factor (bottom panel; $\gammamax$ is defined in \eq{ggmax}, and it is normalized to the thermal Lorentz factor $\gamma_{\rm th}\simeq 1+(\hat{\gamma}-1)^{-1} kT/m c^2$), for a suite of five PIC simulations of ABC collapse with fixed $kT/mc^2=10^{-4}$, $\sigmain=42$ and $L/\rhot=126$, but different magnitudes of the initial stress: $\lambda=0.78$ (black), 0.87 (blue), 0.94 (green), 0.97 (yellow) and 1 (red; i.e., unstressed). The early phases (until $ct/L\sim 3$) bear memory of the prescribed stress parameter $\lambda$, whereas the evolution subsequent to the ABC collapse at $ct/L\sim 4$ is remarkably similar for different values of $\lambda$. The dashed black line in the top panel shows that the electric energy grows exponentially as $\propto \exp{(4ct/L)}$, during the dynamical phase of the ABC instability. The vertical dotted lines mark the time when the electric energy peaks (colors correspond to the five values of $\lambda$, as described above; the black, green and yellow vertical lines overlap).}
\label{fig:abcstrtimecomp} 
\end{figure}
 \begin{figure}
 \centering
\includegraphics[width=.49\textwidth]{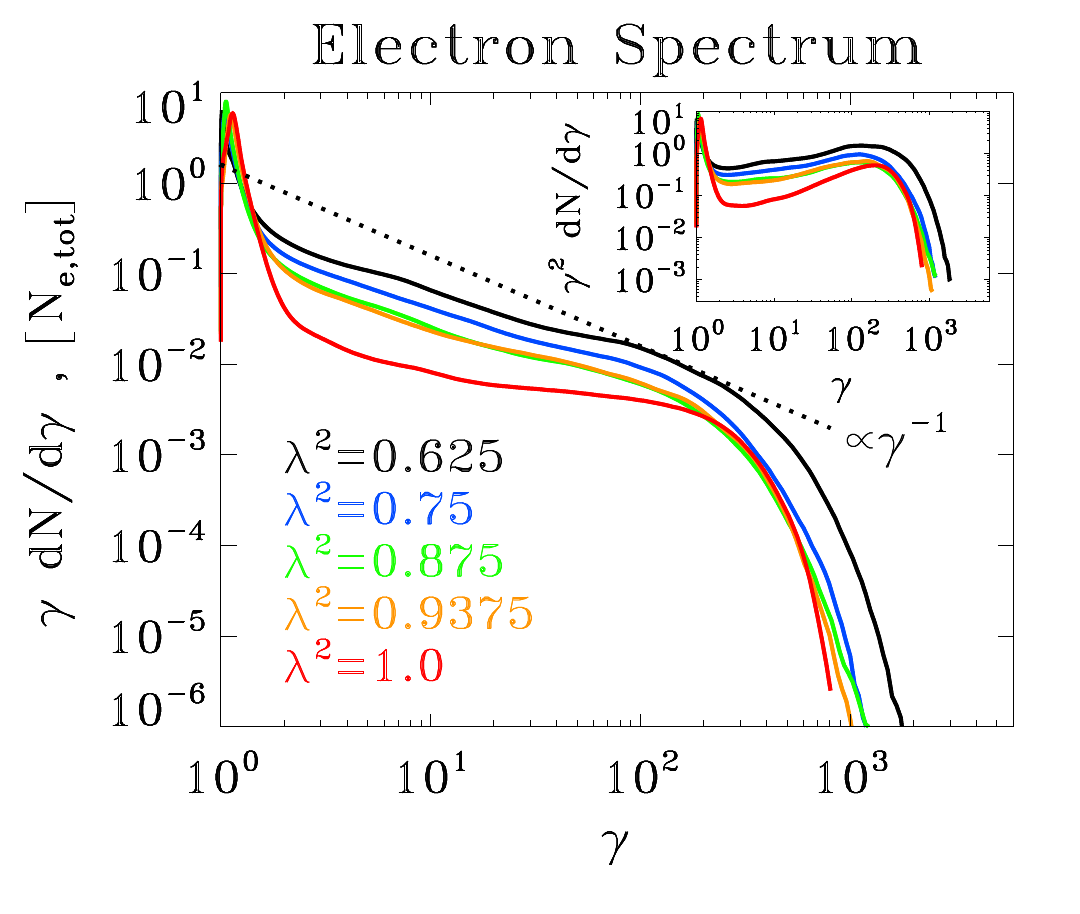} 
\caption{Particle spectrum at the time when the electric energy peaks, for a suite of five simulations of ABC collapse with fixed $kT/mc^2=10^{-4}$, $\sigmain=42$ and $L/\rhot=126$, but different magnitudes of the initial stress (same runs as in \fig{abcstrtimecomp}): $\lambda=0.78$ (black), 0.87 (blue), 0.94 (green), 0.97 (yellow) and 1 (red; i.e., unstressed). The main plot shows $\gamma dN/d\gamma$ to emphasize the particle content, whereas the inset presents $\gamma^2 dN/d\gamma$ to highlight the energy census. The dotted black line is a power law $\gamma dN/d\gamma\propto \gamma^{-1}$, corresponding to equal energy content per decade (which would result in a flat distribution in the inset). The particle spectrum at the time when the electric energy peaks has little dependence on the initial stress parameter $\lambda$, and it resembles the result of the spontaneous ABC instability (red curve).}
\label{fig:abcstrspeccomp} 
\end{figure}

We conclude this subsection by investigating the effect of the initial stress, as quantified by the squeezing parameter $\lambda$. In Figs.~\fign{abcstrtimecomp} and \fign{abcstrspeccomp}, we present the results of a suite of five PIC simulations of ABC collapse with fixed $kT/mc^2=10^{-4}$, $\sigmain=42$ and $L/\rhot=126$, but different magnitudes of the initial stress: $\lambda=0.78$ (black), 0.87 (blue), 0.94 (green), 0.97 (yellow) and 1 (red; i.e., unstressed). In all the cases, the early phase (until $ct/L\sim 3$) bears memory of the prescribed stress parameter $\lambda$. In particular, the value of the electric energy at early times increases for decreasing $\lambda$ (top panel in \fig{abcstrtimecomp} at $ct/L\lesssim 3$), as the initial stress becomes stronger. In parallel, the process of particle acceleration initiated at the stressed X-points leads to a maximum particle energy $\gammamax$ that reaches higher values for stronger stresses (i.e., smaller $\lambda$, see the bottom panel of \fig{abcstrtimecomp} at $ct/L\lesssim 3$). While the early stage is sensitive to the value of the initial stress, the dramatic evolution happening at $ct/L\sim 4$ is remarkably similar for all the values of $\lambda\neq1$ explored in \fig{abcstrtimecomp} (black to yellow curves). In all the cases, the electric energy grows exponentially as $\propto \exp{(4ct/L)}$. Both the growth rate and the peak level of the electric energy ($\sim 0.1$ of the total energy) are remarkably insensitive to $\lambda$, and they resemble the unstressed case $\lambda=1$ (red curve in \fig{abcstrtimecomp}), aside from a temporal offset of $\sim 2L/c$. The fast evolution occurring at $ct/L\sim 4$ leads to dramatic particle acceleration (bottom panel in \fig{abcstrtimecomp}), with the high-energy spectral cutoff reaching $\gammamax\sim 10^3$ within a dynamical time, once again imitating the results of the unstressed case, aside from a temporal shift of $\sim 2L/c$. 

As we have just described, the electric energy peaks during the second phase, when the stressed systems evolve in close similarity with the unstressed setup. In \fig{abcstrspeccomp}, we present a comparison of the particle spectra for different $\lambda$ (as indicated in the legend), measured at the peak time of the electric energy (as indicated by the vertical dotted lines in \fig{abcstrtimecomp}). The spectral shape for all the stressed cases is nearly identical, especially if the initial stress is not too strong (i.e., with the only exception of the black line, corresponding to $\lambda=0.78$). In addition, the spectral cutoff is in remarkable agreement with the result of the unstressed setup (red curve in \fig{abcstrspeccomp}), confirming that the evolution of the stressed cases closely parallels the spontaneous ABC instability. 

We conclude that this setup --- with an initially imposed stress --- does not introduce additional advantages in driving the instability of the ABC structure (unlike the shearing setup described in the previous subsection, which directly leads to ABC collapse). Still, the perturbation imposed onto the system eventually leads to ABC instability, which proceeds in close analogy to the unstressed case. Particle acceleration to the highest energies is not achieved in the initial phase, which still bears memory of the imposed stress, but rather in the quasi-spontaneous ABC collapse at later times. In order to further test this claim, we have performed the following experiment. After the initial evolution (driven by the imposed stress), we artificially reduce the energies of the highest energy particles (but still keeping them ultra-relativistic, to approximately preserve the electric currents). The subsequent evolution of the high-energy particle spectrum is similar to what we observe when we do not artificially ``cool'' the particles, which is another confirmation of the fact that the long-term physics is independent from the initial stress.

% \clearpage
%%%%%%%%%%

%%%%%%%%%%%

%sssssssssssssssssssssssssssssssssssssssssssss
\section{Merging flux tubes carrying  zero total current}
 \label{fluxtubes}
%sssssssssssssssssssssssssssssssssssssssssssss

In Sections \ref{unstr-latt}-\ref{driven} we considered evolution of the unstable configurations -- those of the stressed  X-point and  2D ABC configurations (both stressed and unstressed). We found that during the development of instability of the   2D ABC configurations the  particles are efficiently accelerated during the initial dynamical phase of the merger of current-carrying flux tubes (when the evolution is mostly due to the X-point collapse).  There are two   key feature of the preceding  model that are specific to the initial set-up:  (i)  each flux tube carries non-zero poloidal current; (ii) the initial 2D ABC configuration is an  unstable equilibrium. It is  not clear how generic  these conditions and how  the details of  particle acceleration are affected by these specific properties.

  In this section we relax these conditions. We investigate a merger of two flux tubes with zero total current. We  will study, using relativistic MHD and PIC simulations,   the evolution of colliding flux tubes with various  internal structure. Importantly, all the cases under investigation have a  common property: they carry zero net poloidal current (either completely distributed electric currents as in the case of configurations (\ref{eq:lundquist}) and  (\ref{sec:modlundquist}), or balanced by the surface current, (\ref{eq:bphi})). Thus, two flux tubes are not attracted to each other - at least initially.  All the configurations considered have a common property: the current in the core of the flux tube is balanced by the reverse current in the outer parts. 
% \clearpage
%%%%%%%%%%

 %%%%Oliver%%%%
 %\subsection{MHD simulation}
%ssssssssssssssssssssssssssssssssssssssssssssssss 
\subsection{Force-free simulations}
%ssssssssssssssssssssssssssssssssssssssssssssssss 

%sssssssssssssssssssssssssssssssssssssss
\subsubsection{Merger of Lundquist's magnetic ropes }\label{sec:lundquist}
First we consider Lundquist's force-free cylinders, surrounded by uniform magnetic field,
\be
\bB_L(r\le r_{\rm j}) = J_1 (r \alpha) {\bf e}_\phi +  J_0 (r \alpha) {\bf e}_z\,,
\label{eq:lundquist}
\ee
Here, $J_{0},\,J_{1}$ are Bessel functions of zeroth and first order and the constant $\alpha\simeq 3.8317$ is the first root of $J_{0}$.  
We chose to terminate this solution at the first zero of $J_1$, which we denote 
as $r_j$ and hence continue with $B_z=B_z(r_j)$ and $B_\phi=0$ for $r>r_j$. Thus the total current of the flux tube is zero.
As the result, the azimuthal field vanishes at the boundary of the rope, whereas the poloidal one changes sign inside the rope. 

We start with the  position of two ropes  just touching each other and set the centre positions to be $\mathbf{x_c}=(-r_{\rm j},0)$ and $\mathbf{x_c}=(r_{\rm j},0)$. The evolution is very slow, given the fact that at the contact the reconnecting field vanishes.
 To speed things up, we ``push'' the ropes towards each other by imposing a drift velocity. That is, we initialize the electric field 
\be
    \bmath{E}=-(1/c)\vpr{v_{\rm kick}}{B}\, , \label{eq:vkick}
\ee    
where $\mathbf{v}_{\rm kick}=(\pm v_{\rm kick},0,0)$ inside the ropes and here we set $v_{\rm kick}=0.1c$.  

The numerical setup is as follows.
We have adopted the force-free algorithm of \cite{2007MNRAS.374..415K} to create a physics module in the publicly available code MPI-AMRVAC\footnote{https://gitlab.com/mpi-amrvac/amrvac} \cite{2014ApJS..214....4P,Keppens2012718}. We simulate a 2D Cartesian domain with $x\in[-6,6]$ and $y\in[-3,3]$ at a base-resolution of $128\times64$ cells.  In the following, the scales are given in units of $r_{\rm j}$.  During the evolution, we ensure that the FWHM of the current density in the current sheet is resolved by at least 16 cells via local adaptive mesh refinement of up to eight levels, each increasing the resolution by a factor of two.  Boundary conditions are set to ``periodic''.  

The results are illustrated in Figure 2, which shows the distribution of $B_z$ and $\chi\equiv1-E^2/B^2$ at $t=0,2,5,6,9$.  
%
%fffffffffffffffffffffffffffffffffffffffffffffffffffffffffffffffffffffffffffffffffffffffff
\begin{figure}
\begin{center}
\includegraphics[height=3.cm]{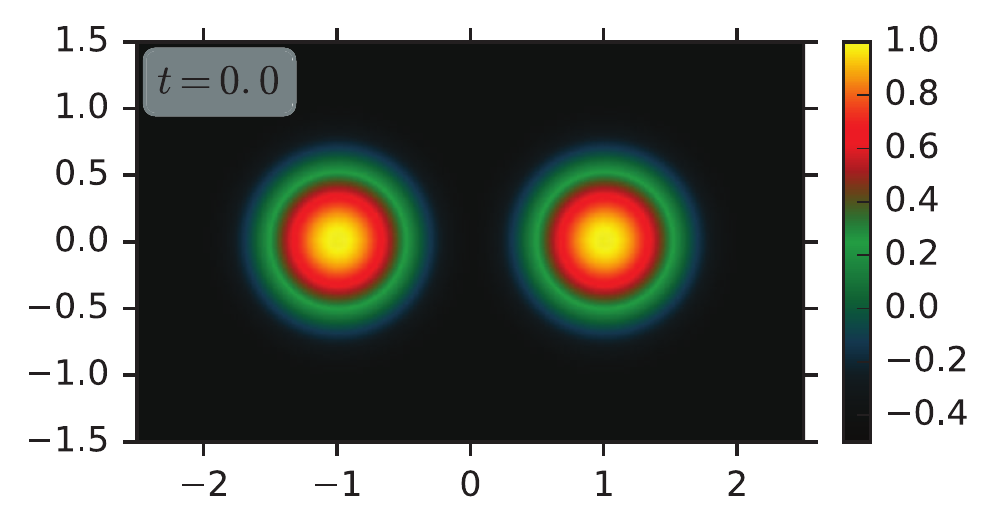}
\includegraphics[height=3.cm]{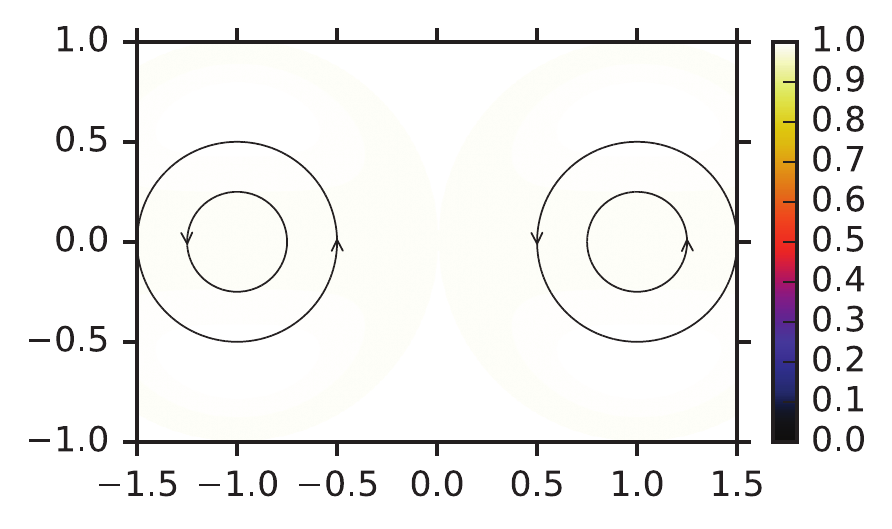}\\
\includegraphics[height=3.cm]{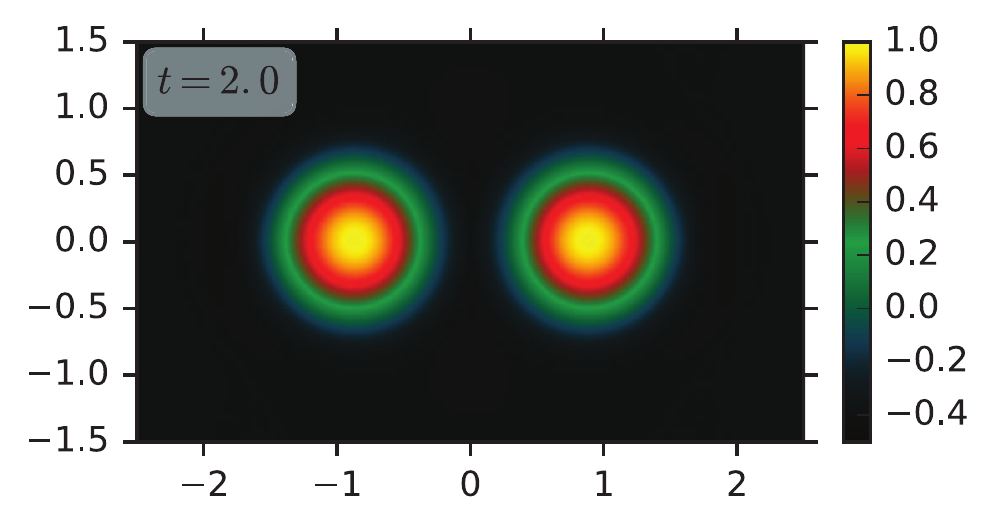}
\includegraphics[height=3.cm]{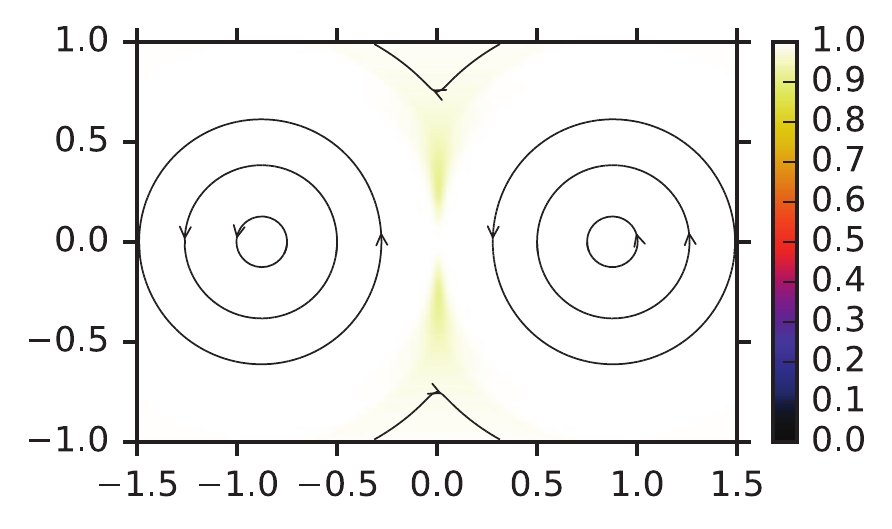}\\
\includegraphics[height=3.cm]{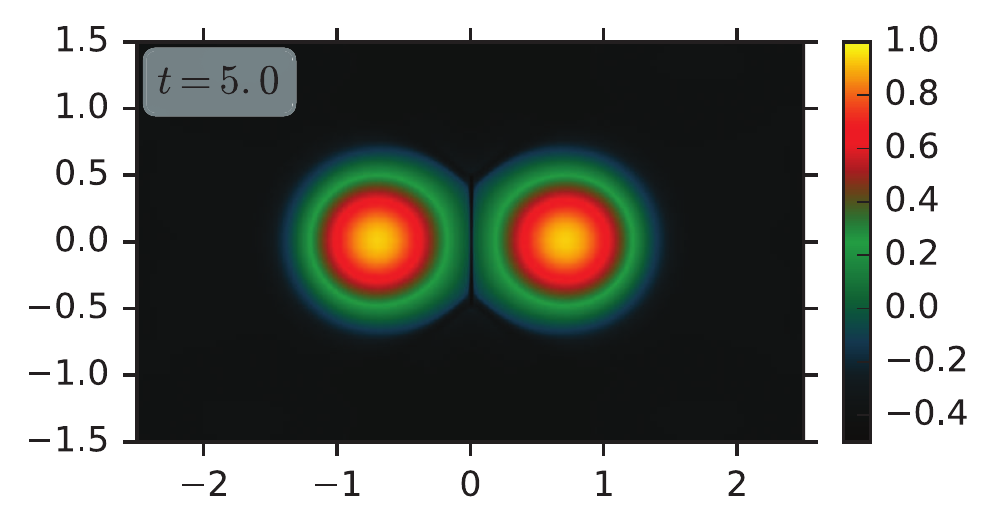}
\includegraphics[height=3.cm]{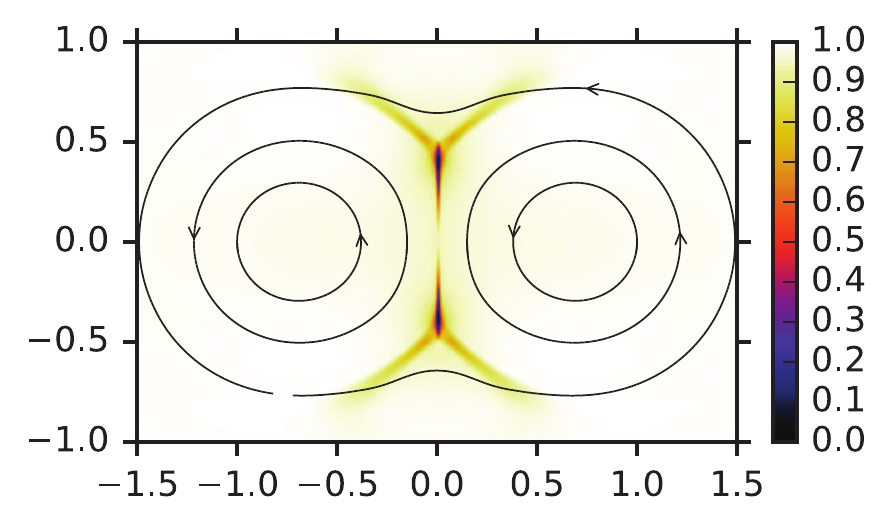}\\
\includegraphics[height=3.cm]{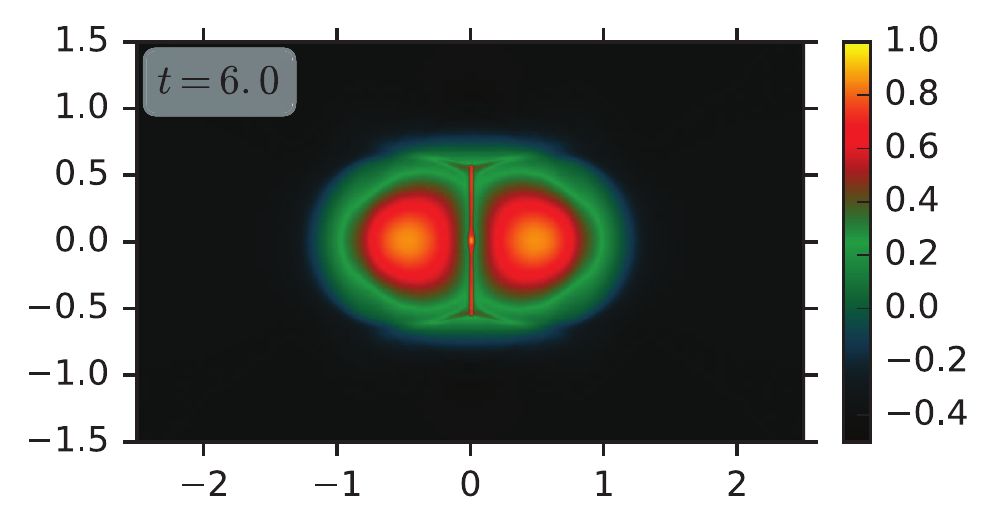}
\includegraphics[height=3.cm]{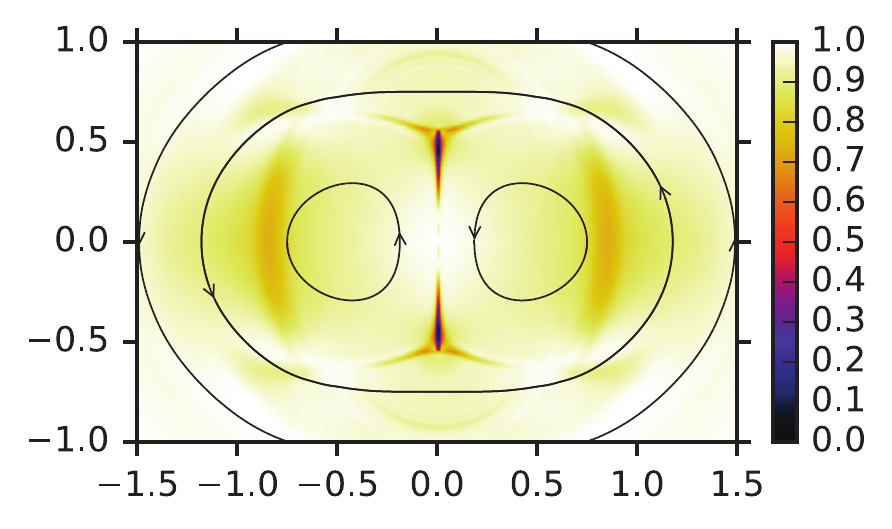}\\
\includegraphics[height=3.cm]{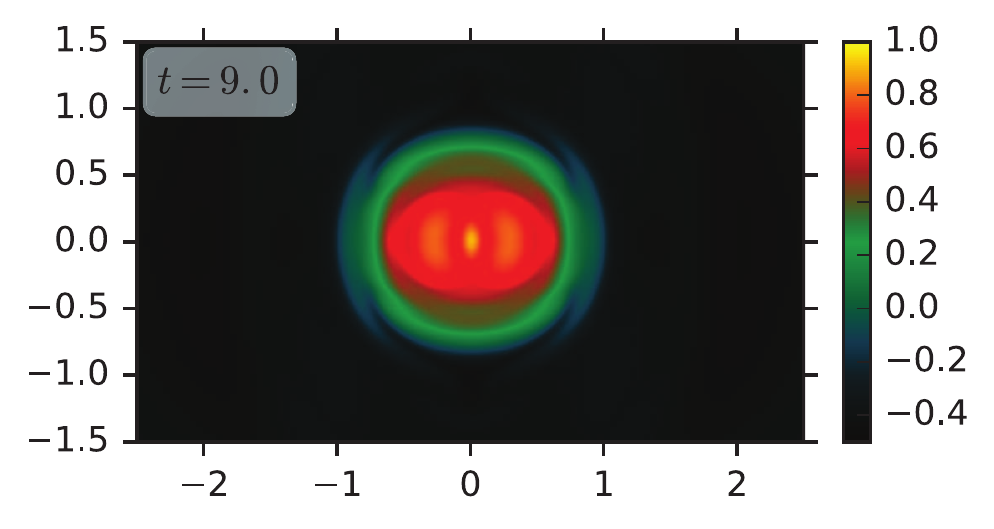}
\includegraphics[height=3.cm]{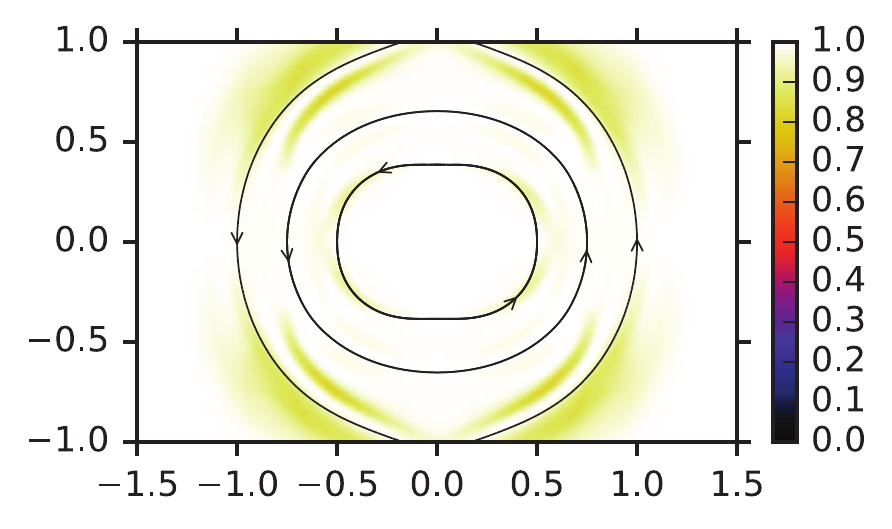}
\caption{Merging Lundquist flux tubes with initial kick velocity $v_{\rm kick}=0.1\rm c$ and $\eta_{||}=1/1000$.  From top to bottom, we show snapshots at $t=[0,2,5,6,9]$ where the coordinates are given in units of initial flux tube radius $r_{\rm j}$ and time is measured in units of $c/r_{\rm j}$.  \textit{Left:} Magnitude of out-of-plane component $B_z$.  
\textit{Right:} Plots of $\chi=1-E^2/B^2$ indicating that regions with $E\sim B$ emerge in the outflow region of the current-sheet.  
Starting from the small kick velocity, the merger rate starts with initial small kick velocity and speeds up until $t\sim5$.  
Exemplary field lines are traced and shown as black lines in the right panels.  }
\label{fig:ff-lundquist}
\end{center}
\end{figure}
%fffffffffffffffffffffffffffffffffffffffffffffffffffffffffffffffffffffffffffffffffffffffff
%
 Interestingly, around the time $t\approx5$, the cores (with $B_z$ of the same sign) of the flux tubes 
begin to merge. This leads to zero guide field in the current sheet and increased reconnection
rate. The sudden increase in reconnection rate leads to a strong wave being emitted from the current sheet which is also seen in the decrease to $\chi\approx0.7$ in the outer part of the flux tubes, i.e. in the fourth panel of figure \ref{fig:ff-lundquist}.  

\begin{figure}
\begin{center}
\includegraphics[height=5cm]{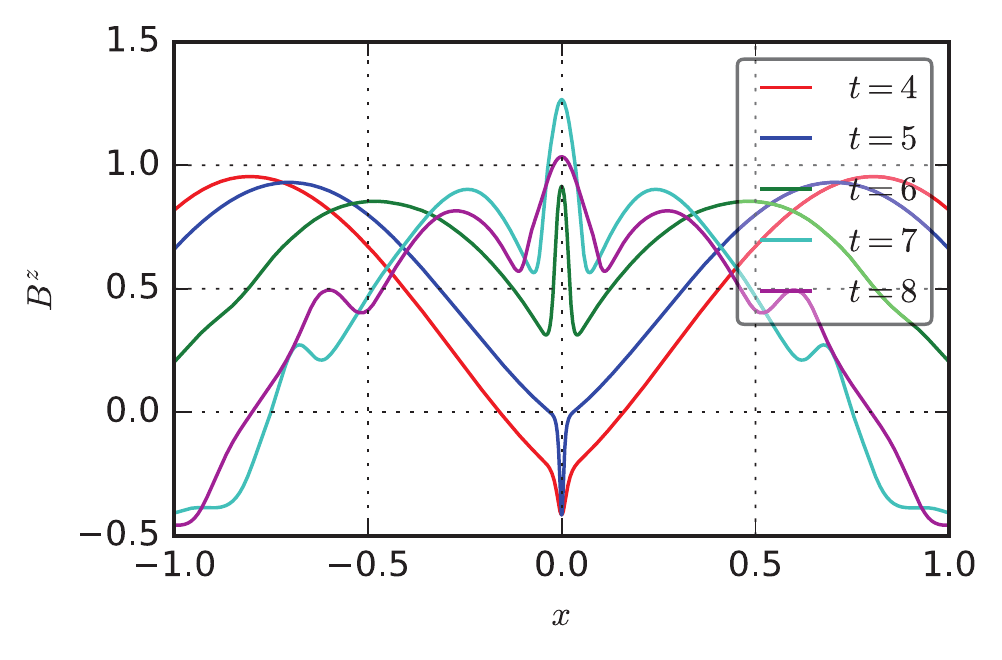}
\includegraphics[height=5cm]{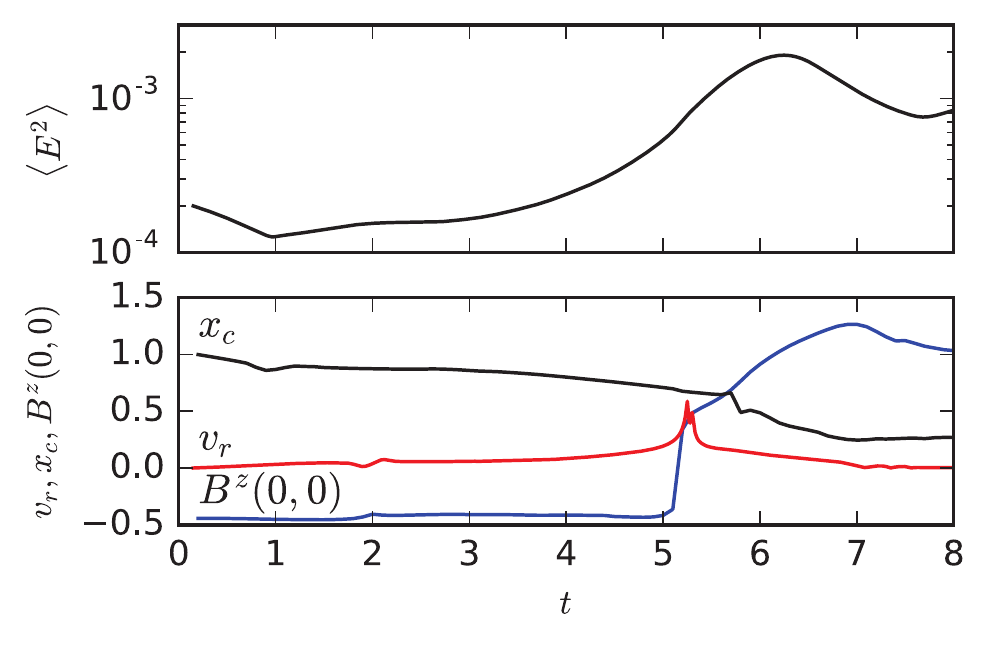}
\caption{Merging Lundquist flux tubes.  
\textit{Left:} cuts along the $y=0$ plane, showing the profile of the guide-field for various times.  Coordinates are given in terms of $r_{\rm j}$ and times in units of $r_j/c$.  
At $t\simeq5.2$, the guide field in the current sheet at $x=0$ changes sign which results in a sharp rise in the reconnection rate and accelerated merger of the cores.  
\textit{Right:} Domain averaged electric field $\langle E^2\rangle$ (top) and other quantities of interest (bottom): reconnection rate, measured as drift velocity through $x=\pm0.1$ (red), $x$-coordinate of the core quantified as peak in $B^z$ (black) and guide field at the origin $B^z(0,0)$ (blue).  When the guide-field changes sign, the reconnection rate spikes at $v_r\simeq0.6c$.  This equivalent PIC result is discussed in Figure \ref{fig:lundveltime}}
\label{fig:ff-lundquist-cuts}
\end{center}
\end{figure}

The influence of the guide field is investigated in Figure \ref{fig:ff-lundquist-cuts}.  In the left panel, we show cuts of $B^z$ along the $y=0$ plane.  As the in-plane magnetic flux reconnects in the current sheet, the out-of-plane component remains and accumulates, leading to a steep profile with magnetic pressure that opposes the inward motion of reconnecting field lines.  
At $t=5.2$, the guide field changes sign and the reconnection rate spikes rapidly at a value of $v_r\approx0.6$.  This is clearly seen in the right panel of figure \ref{fig:ff-lundquist-cuts}.  
Thereafter, a guide field of the opposite sign builds up and the reconnection rate slows down again.  
The position of the flux-tube core as quantified by the peak of $B^z$ evolves for $t<r_{\rm j}/c$ according to the prescribed velocity kick Eq. (\ref{eq:vkick}) but then settles for a slower evolution governed by the resistive timescale.

%fffffffffffffffffffffffffffffffffffffffffffffffffffffffffffffffffffffffffffffffffffffffff
%fffffffffffffffffffffffffffffffffffffffffffffffffffffffffffffffffffffffffffffffffffffffff

\subsection{Modified Lundquist's magnetic ropes}\label{sec:modlundquist}

\subsubsection{Description of setup}

For the future analysis, we consider a modified version of Lundquist's force-free cylinders discussed in the previous section \ref{sec:lundquist}.  
The toroidal field is the same as given by equation \ref{eq:lundquist}, but the vertical field reads 
\begin{align}
B_{z}(r\le r_{\rm j})    &= \sqrt{J_{0}(\alpha r)^{2} + C}
\end{align}
within a flux tube and is set to $B_{z}(r_{\rm j})$ 
in the external medium.  As a result, the sign-change of the guide-field is avoided and only positive values of $B_z$ are present.  The additional constant $C$ sets the minimum (positive) vertical magnetic field component.   In the following, we always set $C=0.01$.  

In this section, we will investigate dependence of reconnection rate and electric field magnitude on the kick velocity and magnetic Reynolds number.

\subsubsection{Overall evolution}

\begin{figure}
\begin{center}
\includegraphics[height=3.cm]{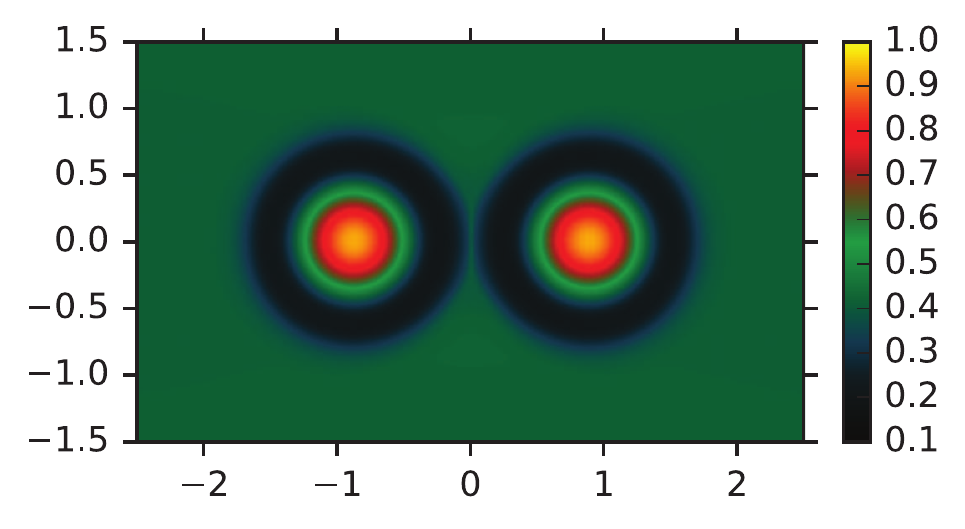}
\includegraphics[height=3.cm]{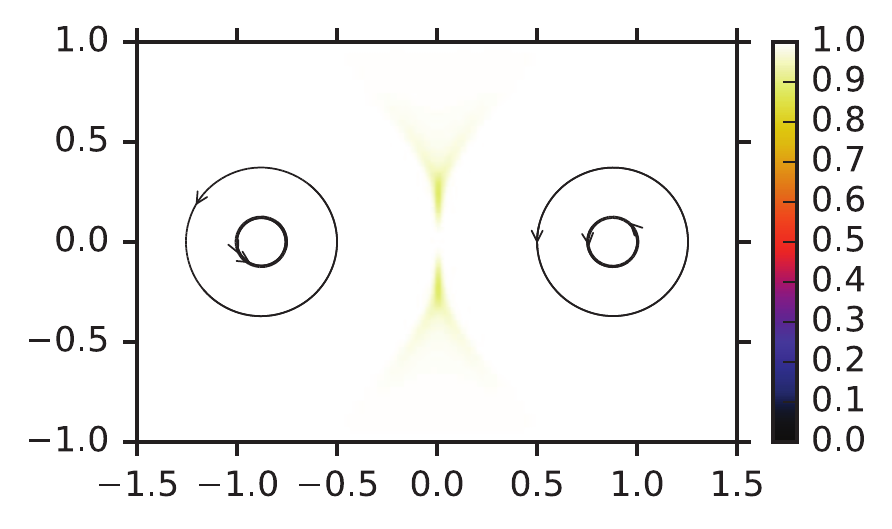}\\
\includegraphics[height=3.cm]{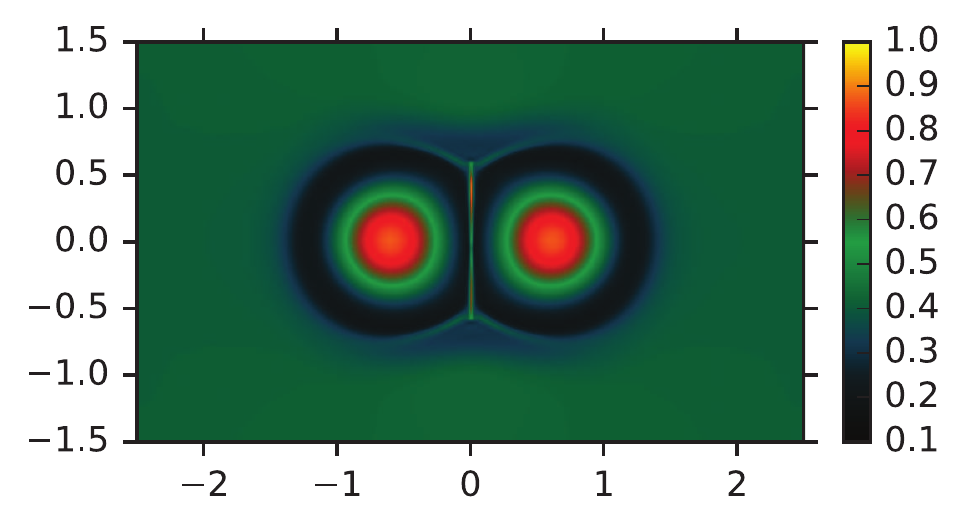}
\includegraphics[height=3.cm]{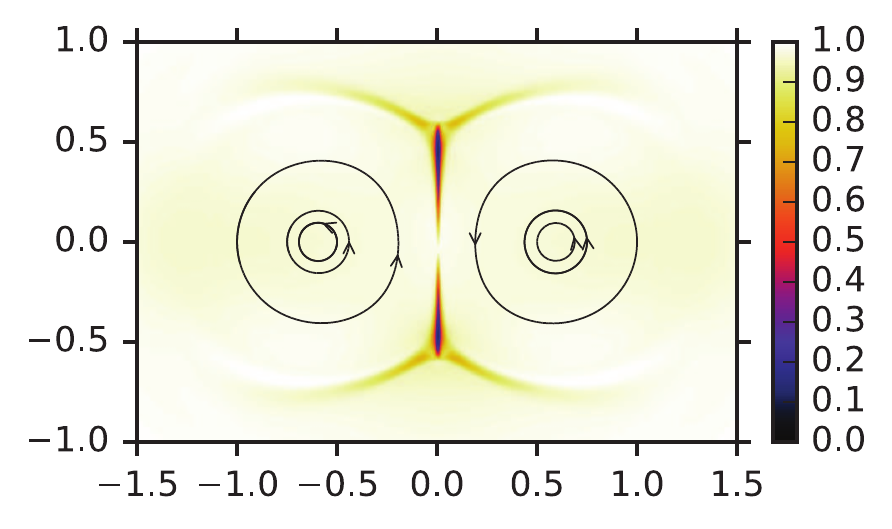}\\
\includegraphics[height=3.cm]{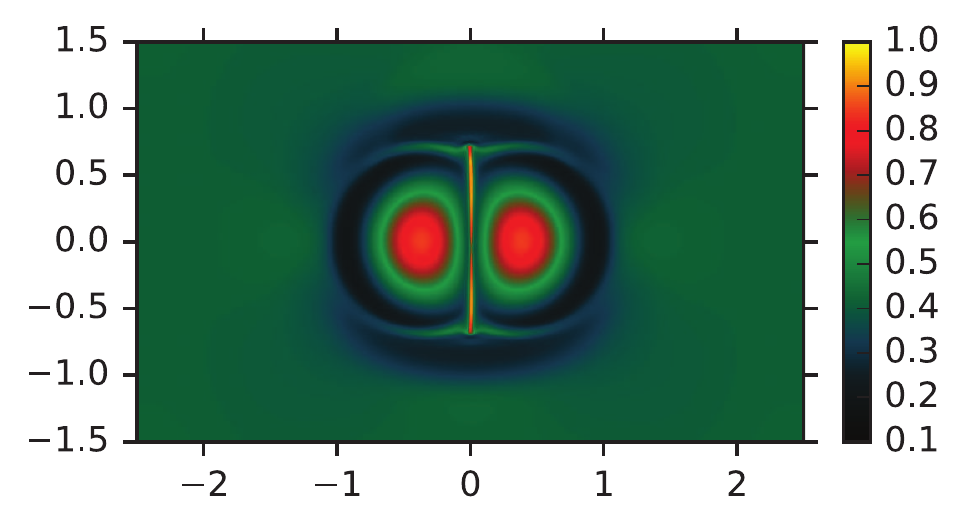}
\includegraphics[height=3.cm]{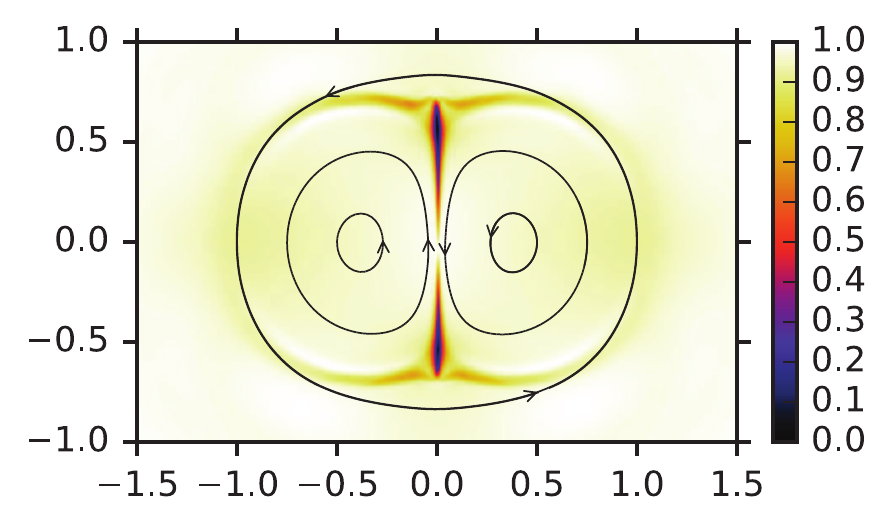}\\
\includegraphics[height=3.cm]{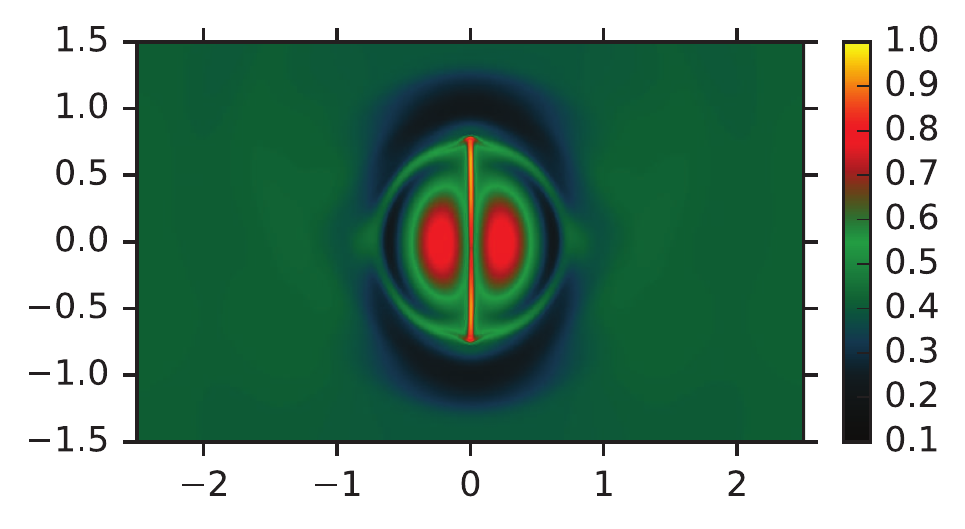}
\includegraphics[height=3.cm]{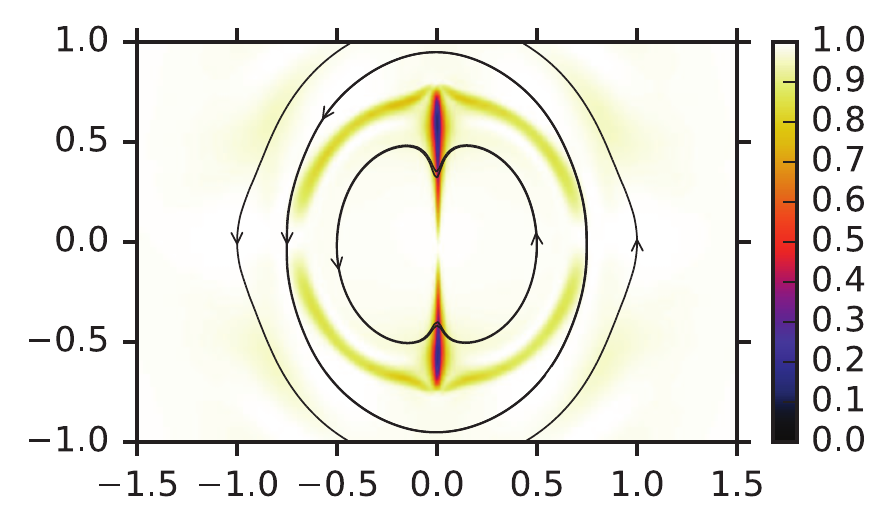}\\
\includegraphics[height=3.cm]{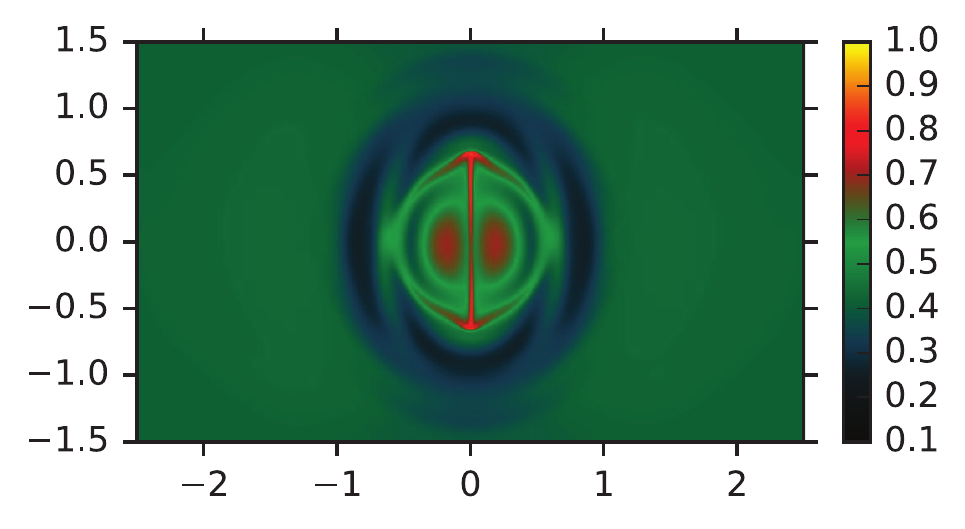}
\includegraphics[height=3.cm]{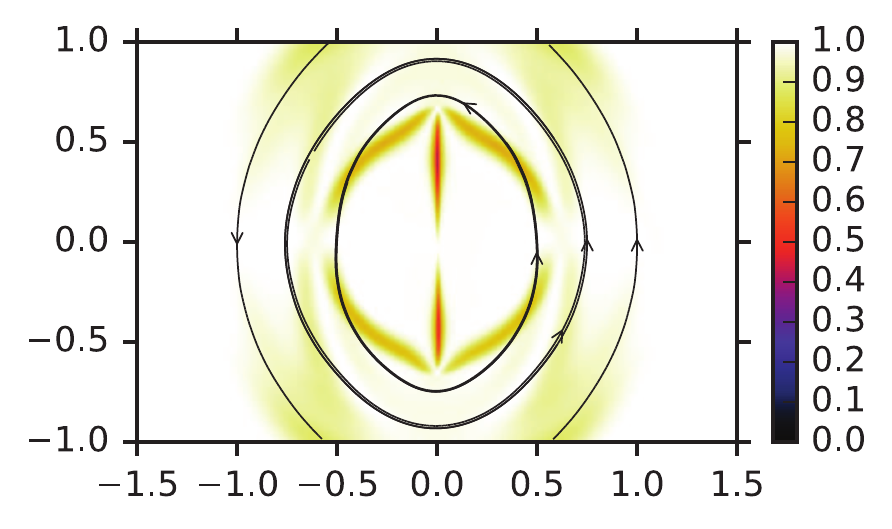}\\
\includegraphics[height=3.cm]{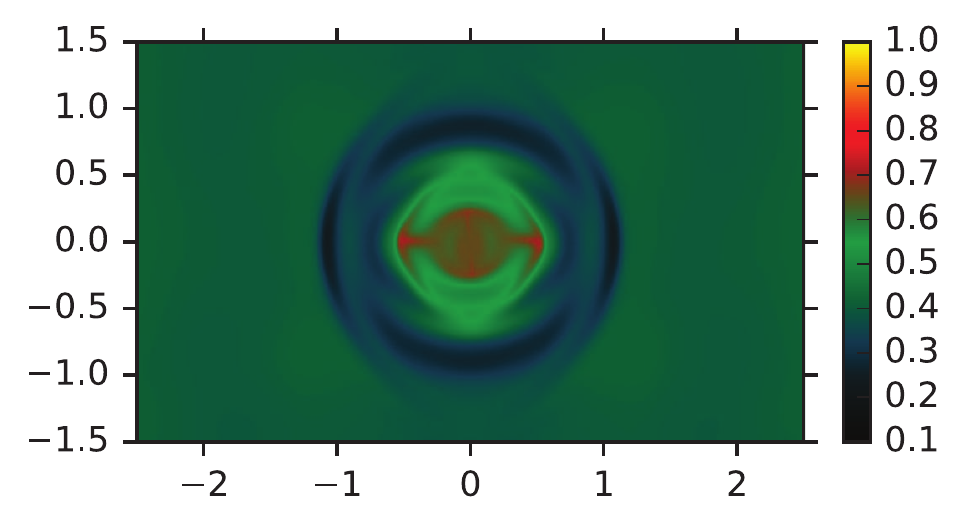}
\includegraphics[height=3.cm]{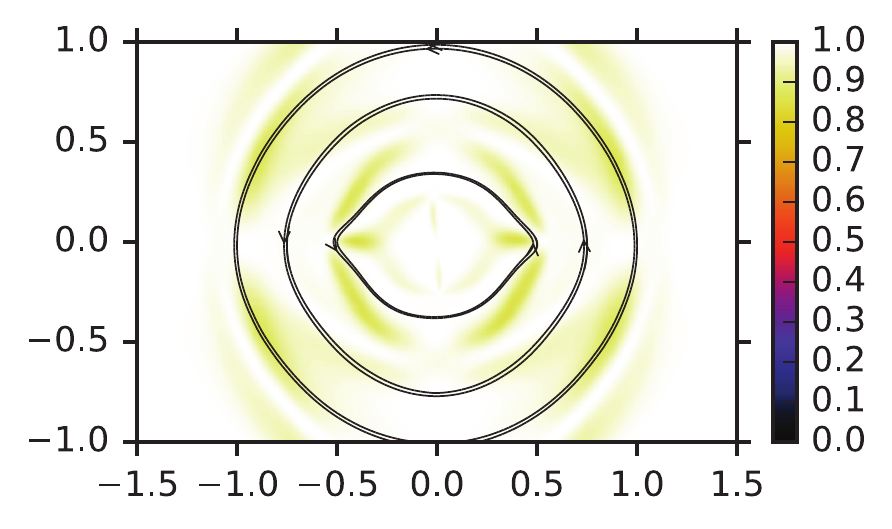}
\caption{Merger of 2D modified Lundquist flux tubes with initial kick velocity $v_{\rm kick}=0.03\rm c$ and $\eta_{||}=1000$.  From to to bottom, we show snapshots at $t=[6,9,10,11,1214]$ where the coordinates are given in units of initial flux tube radius $r_{\rm j}$ and time is measured in units of $c/r_{\rm j}$.  \textit{Left:} Magnitude of out-of-plane component $B_z$.  
\textit{Right:} Plots of $\chi=1-E^2/B^2$ indicating that regions with $E\sim B$ emerge in the outflow region of the current-sheet.  
Starting from the small kick velocity, the merger rate starts with initial small kick velocity and speeds up until $t\sim9$.  
Exemplary field lines are traced and shown as black lines in the right panels.  }
\label{fig:modLundquistOverview}
\end{center}
\end{figure}

As the current vanishes on the surface of the flux tubes, the initial Lorentz force also vanishes and the flux tubes approach each other on the time scale given by the kick velocity.  Then a current sheet is formed at the intersection which reconnects in-plane magnetic flux resulting in an engulfing field.  This evolution is illustrated in figure \ref{fig:modLundquistOverview} for an exemplary run with $(v_{\rm kick},\eta_{||})=(0.03 c,10^{-3})$, showing out-of plane magnetic field strength $B_z$ and the previously introduced parameter $\chi=1-E^2/B^2$.  
It can be seen that in the outflow region of the current sheet, $\chi$ assumes values as small as $\chi_{\rm min}=0.09$ but the distributed region of $\chi\approx0.7$ that is observed for the unmodified Lundquist tubes is absent.  
After the peak reconnection rate which for the shown parameters is reached at $t\simeq 9$, the system undergoes oscillations between prolate and oblate shape during which the current sheet shrinks.  The oscillation frequency corresponds to a light-crossing time across the structure.  

\subsubsection{Dependence on kick velocity}
\begin{figure}
\begin{center}
\includegraphics[height=4cm]{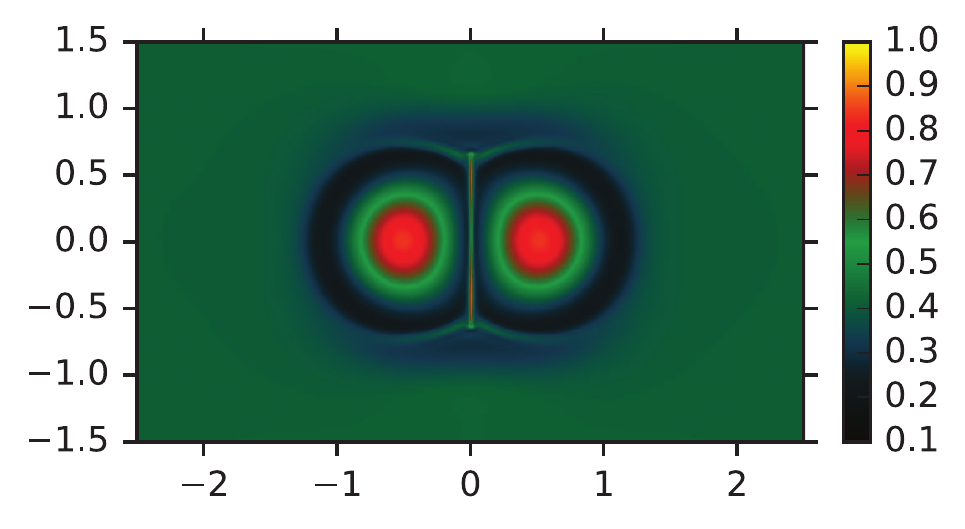}
\includegraphics[height=4cm]{vc0p03bz0009.pdf}
\includegraphics[height=4cm]{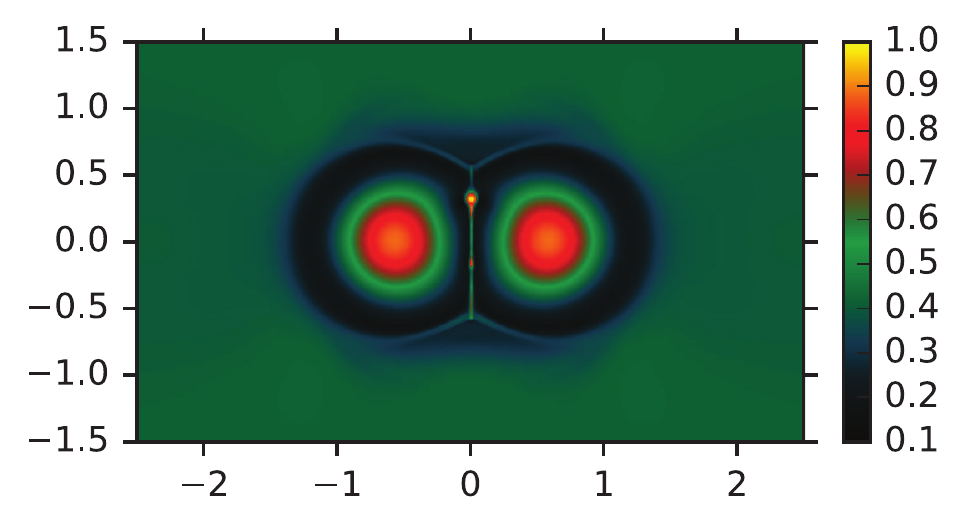}
\includegraphics[height=4cm]{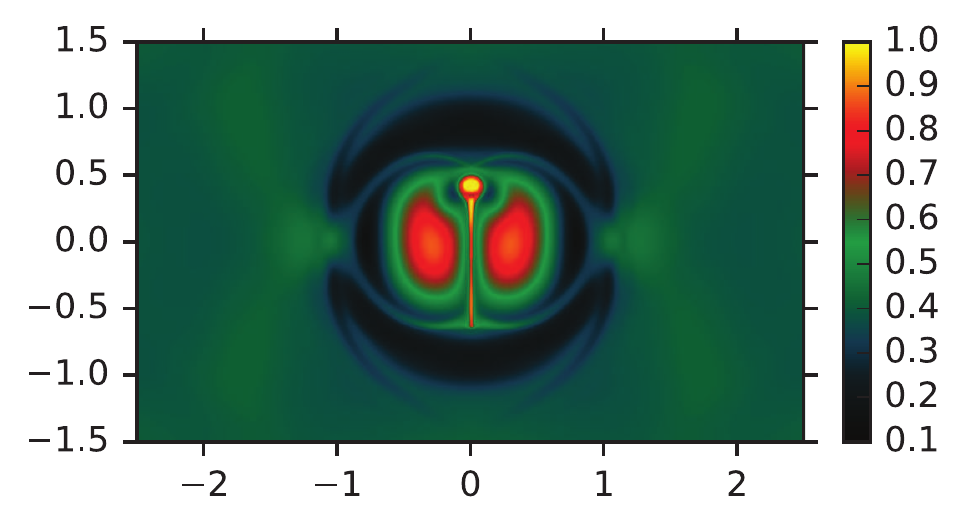}
\caption{Merger of 2D modified Lundquist flux tubes showing $B_z$ with varying kick velocity: $v_{\rm kick}=0.01c$ at $t=12$ (top left), $v_{\rm kick}=0.03c$ at $t=9$ (top right), $v_{\rm kick}=0.1c$ at $t=6$ (bottom left) and $v_{\rm kick}=0.3c$ at $t=5$ (bottom right).  The snapshot times are chosen to reflect the maximum reconnection rate $v_r$, measured as inflow velocity at $x=\pm0.05$.  See text for details.  }
\label{fig:modLundquistKickOverview}
\end{center}
\end{figure}

We now vary the magnitude of the initial perturbation in the interval $v_{\rm kick}\in[0.03c,0.3c]$.  Figure \ref{fig:modLundquistKickOverview} gives an overview of the morphology when the peak reconnection rate is reached.  The reconnection rate is measured as inflowing drift velocity at $x=\pm0.05$ and is averaged over a vertical extent of $\Delta y=0.1$.  For the lower three kick velocities, we obtain a very similar morphology when the peak reconnection rate is reached.  With $v_{\rm kick}=0.3c$, the flux tubes rebound, resulting in emission of waves in the ambient medium.  
As the flux tubes dynamically react to the large perturbation, it is not possible to force a higher reconnection rate by smashing them together with high velocity, they simply bounce off due to the magnetic tension in each flux tube.  
However, high velocity perturbations of $v_{\rm kick}=0.1c$ and $v_{\rm kick}=0.3c$ lead to the formation of plasmoids which are absent in the cases with small perturbation and this in turn can impact on the reconnection rate.  

\begin{figure}
\begin{center}
\includegraphics[height=7cm]{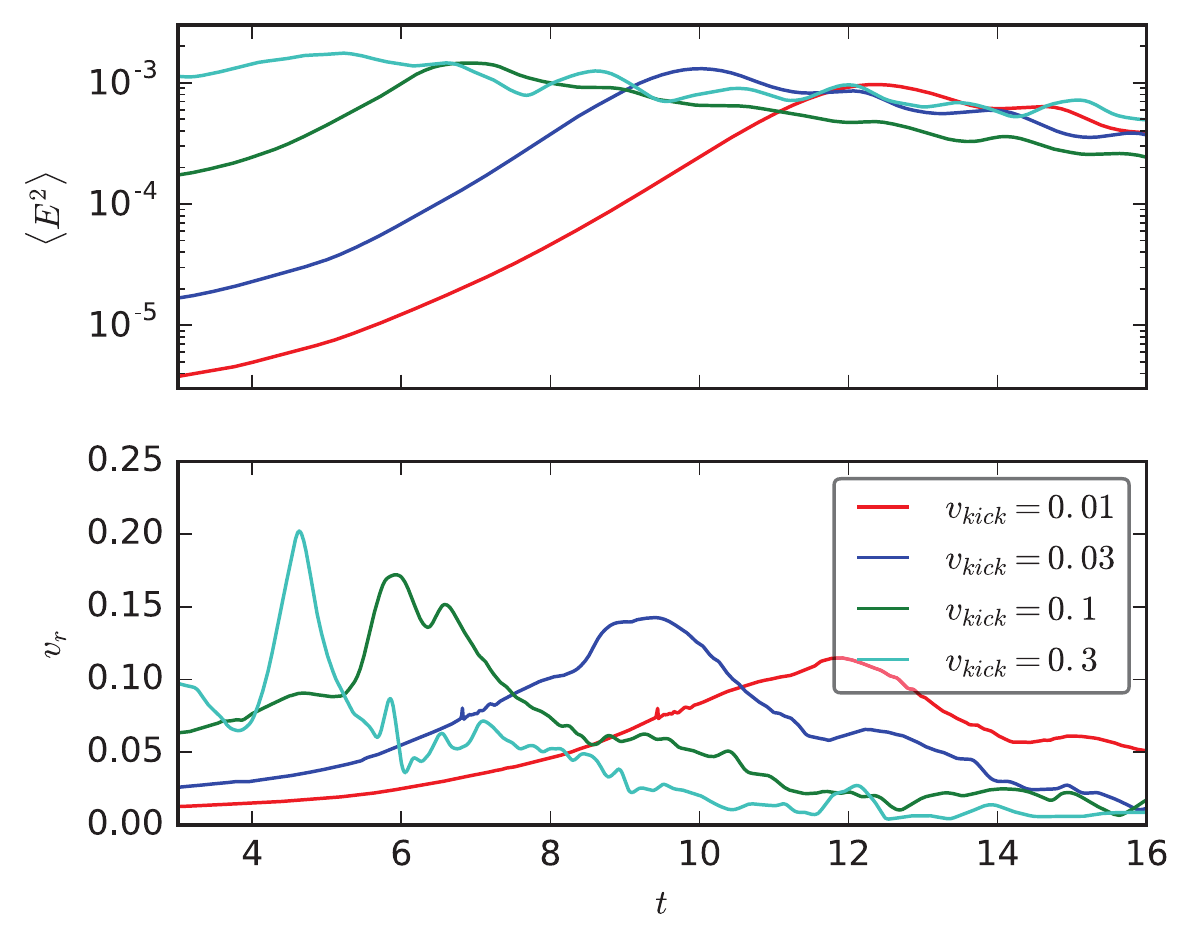}
\caption{Merger of 2D modified Lundquist flux tubes showing the evolution of domain averaged electric field $\langle E^2 \rangle$ (top) and reconnection rate $v_r$ (bottom).  We vary the initial driving velocity from $v_{\rm}=0.01c$ to $v_{\rm}=0.3c$.  See text for discussion.  
}
\label{fig:modLundquistKick-e2}
\end{center}
\end{figure}

A more quantitative view is shown in Figure \ref{fig:modLundquistKick-e2} illustrating the evolution of the mean electric energy in the domain $\langle E^2 \rangle(t)$ and reconnection rate as previously defined.  One can see that all simulations reach a comparable electric field strength in the domain, independent of the initial perturbation.  The electric field grows exponentially with comparable growth rate and also the saturation values are only weakly dependent on the initial perturbation.  
Indeed, comparing highest and lowest kick velocities, we find that $\langle E^2 \rangle_{\rm max}$ is within a factor of two for a range of $30$ in the velocity perturbation.  

In the evolution of $\langle E^2 \rangle(t)$, one can also read off the oscillation period of $P\approx 1 r_{\rm j}/c$, most pronounced in the strongly perturbed case.  
The span of peak reconnection rates is similar to the mean electric field strength with a range from $v_{\rm r}=0.2$ down to $v_r={0.11}$.  The reconnection rate saturates at $\sim0.1c$ for vanishing initial perturbation.  When plasmoids are triggered, as for the case $v_{kick}=0.1$ and $v_{kick}=0.3$, the run of $v_{\rm r}$ shows additional substructure leading to secondary peaks as seen e.g. in the green curve on the lower panel.  

\subsubsection{Scaling with magnetic Reynolds number}\label{sec:mod-lund-res}

We next investigate the scaling of reconnection rate with magnetic Reynolds number while keeping a fixed kick velocity of $0.1c$.  As astrophysical Reynolds numbers are typically much larger than what can be achieved numerically, this is an important check for the applicability of the results.  For the force-free case, we simply define the Lundquist number
\begin{equation}
S = \frac{r_{\rm j} c}{\eta_{||}} \simeq \eta_{||}^{-1}
\end{equation}
and check the scaling via the resistivity parameter $\eta_{||}$.  

\begin{figure}
\begin{center}
\includegraphics[height=7cm]{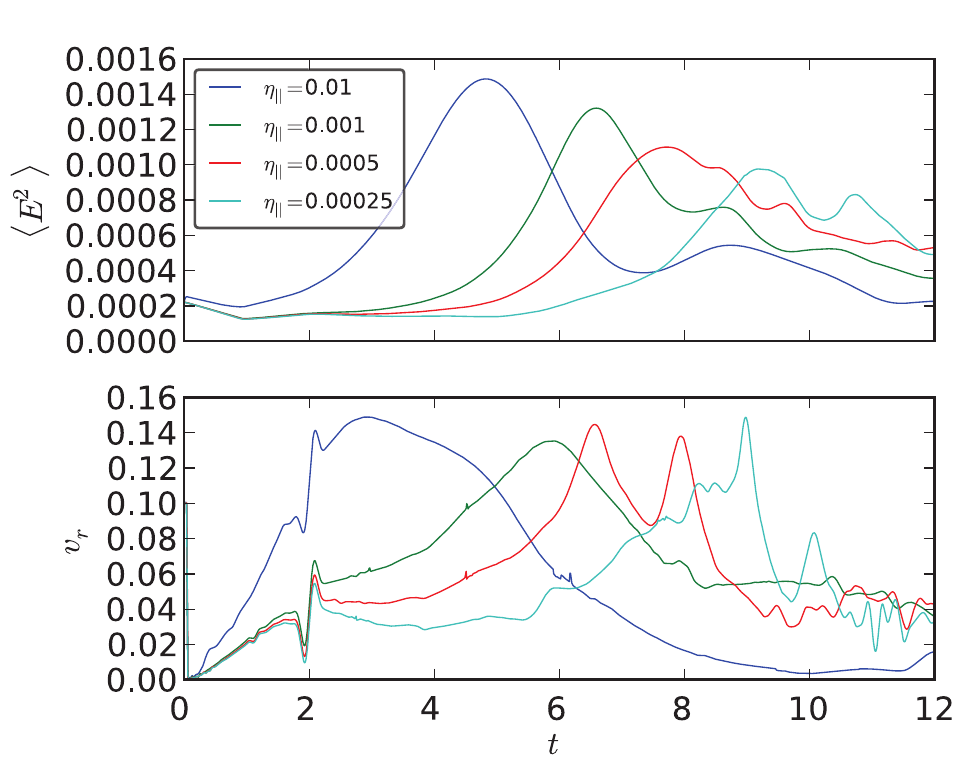}
\includegraphics[height=7cm]{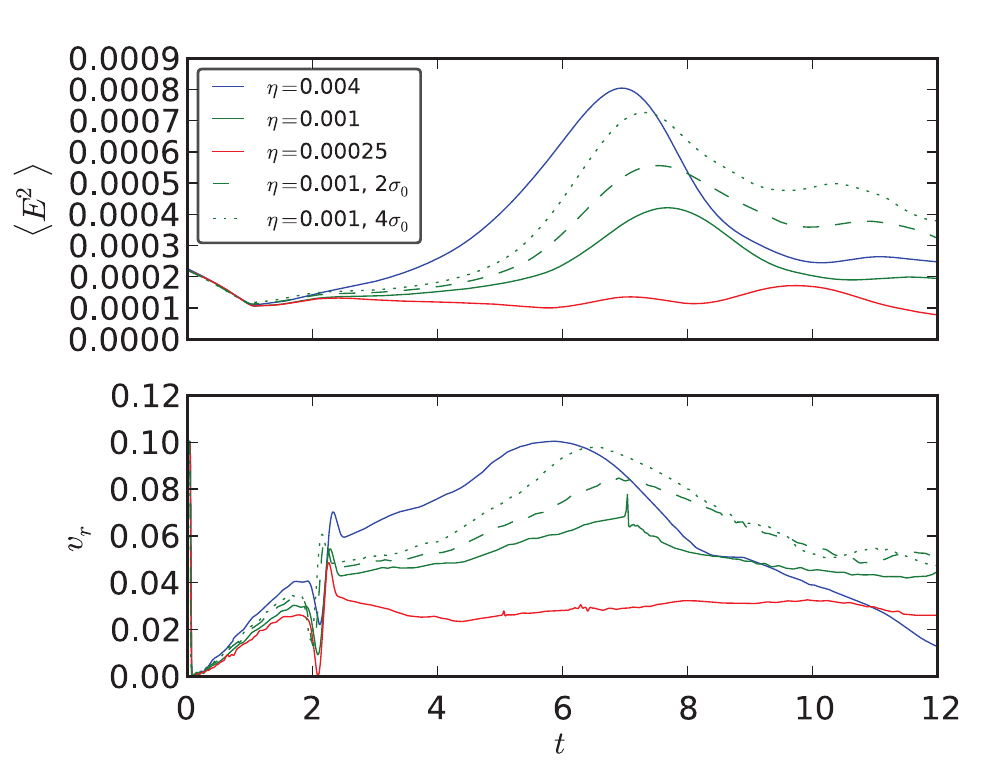}
\caption{Merger of 2D modified Lundquist flux tubes for different resistivities $\eta_{||}$ with initial kick velocity of $0.1c$. 
Mean squared electric field in the domain \textit{(top)} and reconnection speed in units of $c$ \textit{(bottom)}.  
\textit{Left:} Force-free dynamics.  The initial evolution depends on the resistivity but comparable values for the reconnection rate of $v_{\rm r}\approx 0.15 c$ are obtained for all considered Reynolds numbers.  Note that in the case $\eta_{||}=1/4000$ plasmoids appear at $t\simeq 8$ which are not observed in the other runs.  
\textit{Right:}
Corresponding resistive MHD evolution with $\sigma\in[1.7,10]$ and doubled (dashed) and quadrupled (dotted) magnetization.  One can see that the force-free result is approached both in the evolution of $\langle E^2\rangle$ and $v_{\rm r}$.  
}
\label{fig:lmod-res-e2}
\end{center}
\end{figure}

Figure \ref{fig:lmod-res-e2} (left panel), shows the run of the domain averaged electric field and the reconnection rate as in Figure \ref{fig:modLundquistKick-e2}.  
In general, the initial evolution progresses faster when larger resistivities are considered, suggesting a scaling with the resistive time.  However, once a significant amount of magnetic flux has reconnected, the electric energy in the domain peaks at values within a factor of $1.5$ for a significant range of Reynolds numbers $\eta_{||}\in[1/100,1/4000]$.  Very good agreement is found for the peak  reconnection rate of $v_{\rm r}\simeq0.15c$.  The $\eta_{||}=1/4000$ run developed plasmoids at $t\simeq8$ which were not observed in the other runs.  

We have also conducted a range of experiments in resistive relativistic magnetohydrodynamics (RMHD).  The setup is as the force-free configuration described above, however we now choose constant values for density and pressure such that the magnetisation in the flux rope has the range $\sigma\in[1.7,10]$ and the Alfv\'en velocity ranges in $v_A\in[0.8,0.95]$.  The adiabatic index was chosen as $\gamma=4/3$.  Figure \ref{fig:lmod-res-e2} (right panel) shows the resulting evolution of electric field and reconnection rate for a variety of resistivity parameter $\eta\in[1/250,1/4000]$.  Note that the green curves in the left and right panel correspond to the same resistivity parameter of $\eta_{||},\eta=1/1000$ and cyan (left) and red (right) curves both correspond to $\eta_{||},\eta=1/4000$.  
The initial evolution up to the wave-reflection feature at $t=2$ is very similar in both realisations.  Afterwards the evolution differs:  while the reconnection rate in the force-free run with $\eta_{||}=1/4000$ increases to reach a peak of $v_{\rm r}\simeq 0.15c$ at $t\simeq9$, $v_{\rm r}$ remains approximately constant in the corresponding RMHD run.  In contrast to the force-free scenario, the RMHD runs did not feature any plasmoids at the Reynolds numbers considered here.  
Better agreement is found for $\eta_{||},\eta=1/1000$ (green curves in both panels).  Here, the force-free reconnection rate reaches its peak value of $v_{\rm r}\simeq0.14$ at $t\simeq6$ and the RMHD peaks at $t\simeq7$ with $v_{\rm r}\simeq0.07c$.  Increasing the magnetisation improves the match between both cases:  In the green dashed (dotted) curve, we have doubled (quadrupled) the magnetization by lowering plasma density and pressure in the SRMHD runs at $\eta=1/1000$.  One can see that the evolution speeds up and higher reconnection rates are achieved:  For the highest magnetisation run, we reach the peak reconnection rate of $\simeq 0.1c$ and the peak is reached already at $t\simeq6.2$ very similar to the corresponding force-free run which peaked at $t\simeq6$ with $v_{\rm r}\simeq 0.13 c$.  This indicates that the RMHD model indeed approaches the force-free limit for increasing magnetization.  

\subsection{Core-envelope magnetic ropes}

\subsubsection{Description of setup}

We now adopt a setup where the current does no return volumetrically within the flux tube, but returns as a current sheet on the surface.  The solution is based on the ``core-envelope'' solution of \citet{ssk-mj99}, with the gas pressure replaced by the pressure of the poloidal field.  Specifically, the profiles read 

\be
B^\phi(r) = \left\{
\begin{array}{ccl}
  B_m(r/r_m) &;& r<r_m \\
  B_m(r_m/r) &;& r_m<r<r_j
  \\ 0&;& r>r_j
\end{array}
\right. ,
\label{eq:bphi}
\ee
\be
B_z^2(r)/2 = \left\{
\begin{array}{ccl}
  p_0\left[\alpha+\frac{2}{\beta_m}(1-(r/r_m)^{2})\right] &;& r<r_m \\
  \alpha p_0 &;& r_m<r<r_j \\
  p_0 &;& r>r_j
\end{array}
\right. ,
\label{eq:pressure}
\ee
where
\be
\beta_m= \frac{2 p_0}{B_m^2},\qquad \alpha=1-(1/\beta_m)(r_m/r_j)^2\,,
\ee
$r_j$ is the outer radius of the rope and $r_m$ is the radius of its core. In the simulations, $r_m=0.5$, $B_m=1$, 
$\alpha=0.2$ and the coordinates are scaled to $r_{\rm j}=1$. As in the previous cases, two identical current tubes are centred  at $(x,y)=(-1,0)$ and $(x,y)=(1,0)$. They are at rest and barely touch each other at $(x,y)=(0,0)$. 
In this scenario, since the current closes discontinuously at the surface of the flux tubes, the Lorentz-force acting on the outer field lines is non-vanishing.  Thus we can dispense with the initial perturbation and always set $v_{\rm kick}=0$.  

\subsubsection{Overall evolution}

\begin{figure}
\begin{center}
\includegraphics[height=3.cm]{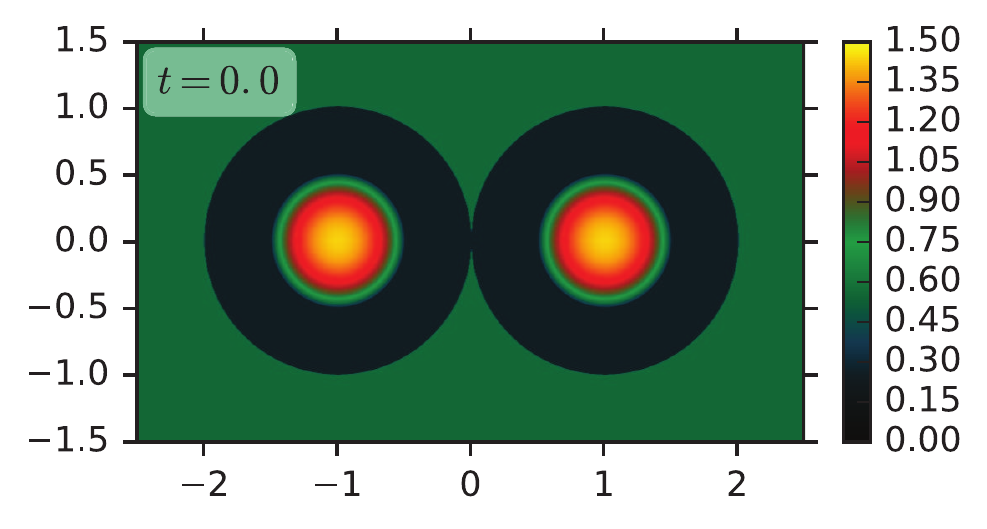}
\includegraphics[height=3.cm]{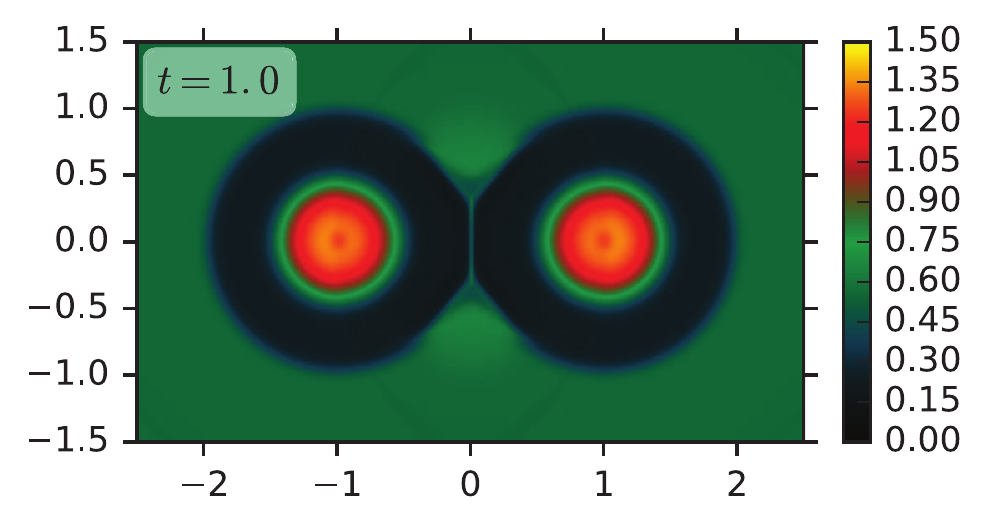}
\includegraphics[height=3.cm]{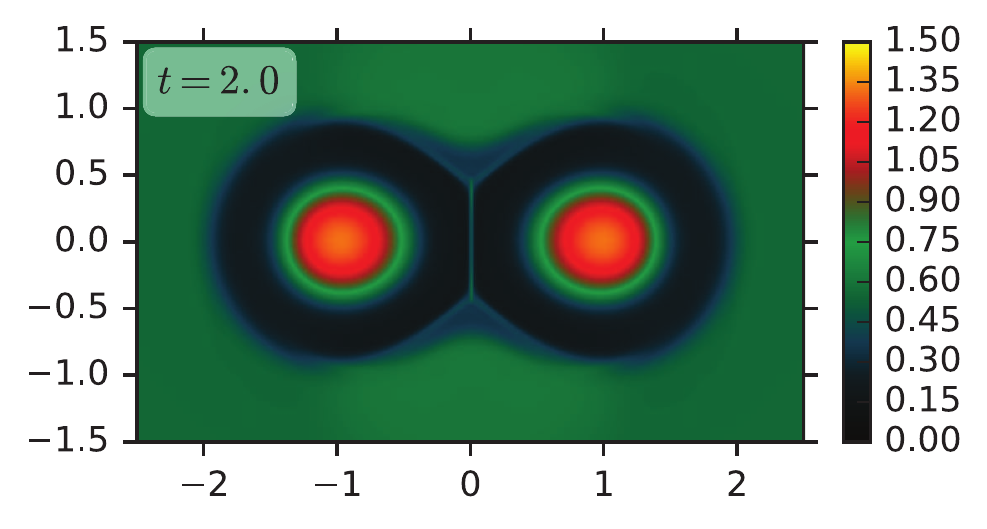}
\includegraphics[height=3.cm]{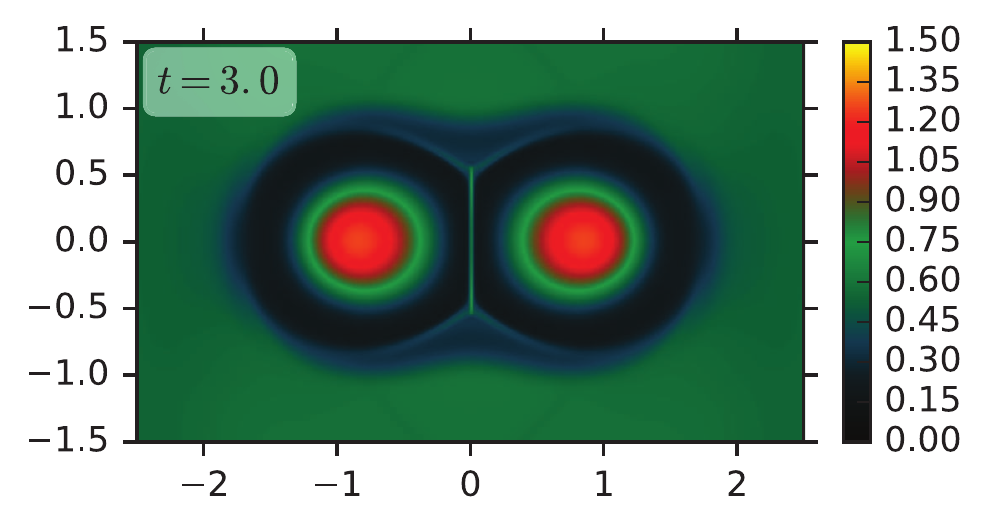}r
\includegraphics[height=3.cm]{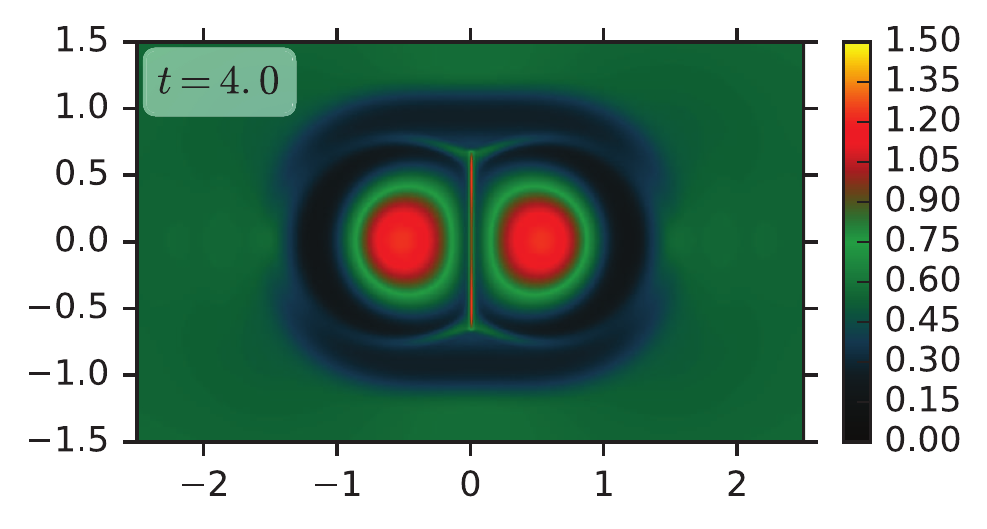}
\includegraphics[height=3.cm]{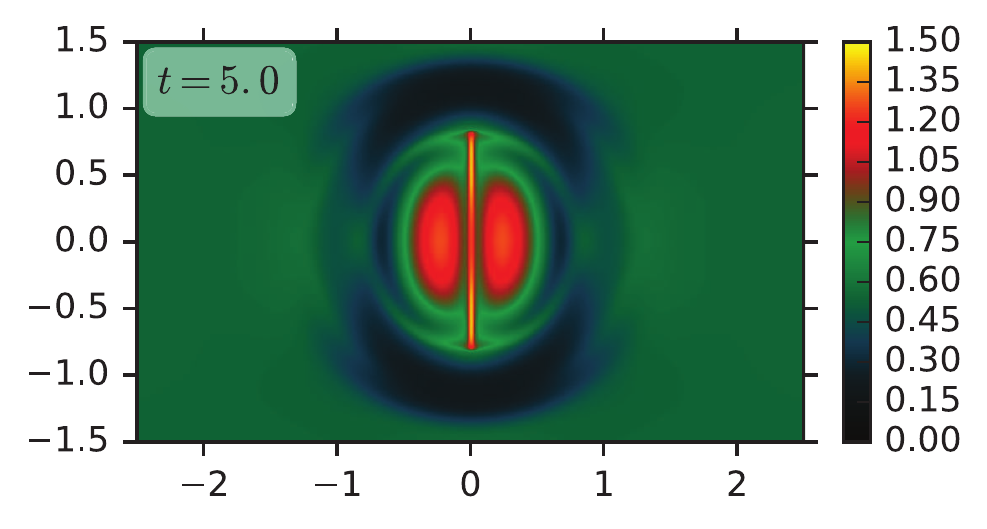}
\caption{Merger of 2D core-envelope flux tubes with $\eta_{||}=500$.  We show snapshots showing $B^z$ at $t=[1,2,3,4,5]$ as indicated in each panel, where the coordinates are given in units of initial flux tube radius $r_{\rm j}$ and time is measured in units of $c/r_{\rm j}$.  This figure is to be compared to the PIC results as shown in Figure \ref{fig:corefluid}.  }
\label{fig:CEOverview}
\end{center}
\end{figure}

The overall evolution is characterised as follows:  
Immediately, a current-sheet forms at the point (0,0) and two Y-points get expelled in vertical direction with initial speed of $c$.  The expansion of the Y-points is thereafter slowed down by the accumulating pressure of $B^z$ in the ambient medium.  
Snapshots of $B^z$ for the case $\eta_{||}=1/500$ are shown in figure \ref{fig:CEOverview}.  The stage of most rapid evolution of the cores is at $t\simeq4$.  At this time, the morphology is comparable to the modified Lundquist ropes, i.e. Figure \ref{fig:modLundquistKickOverview}.  

\begin{figure}
\begin{center}
\includegraphics[height=5cm]{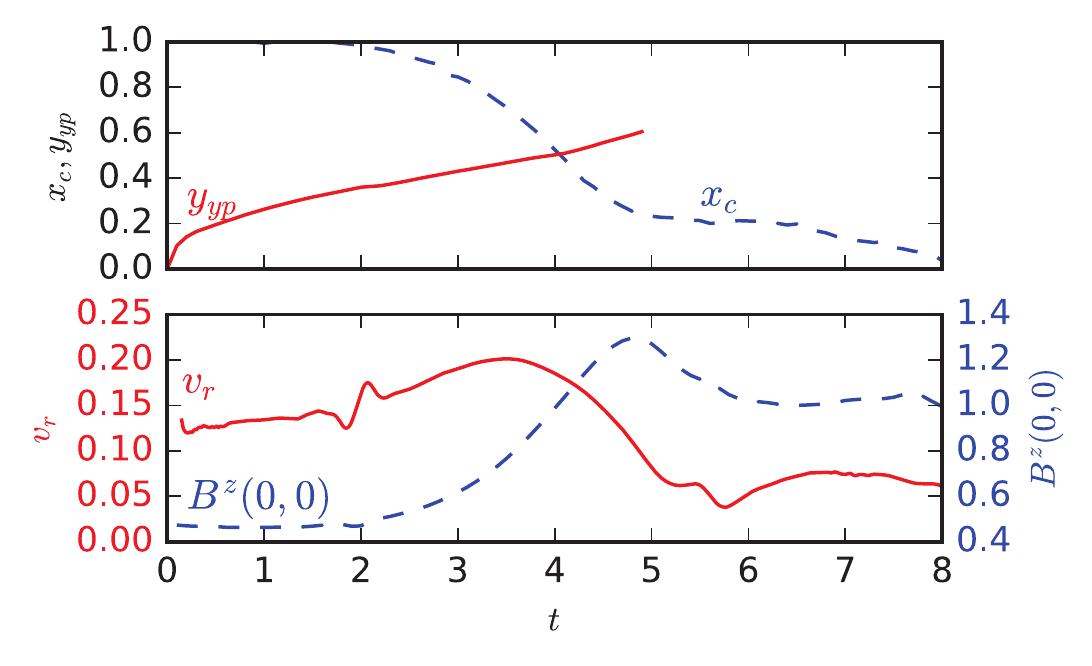}
\caption{Merger of 2D core-envelope flux tubes with $\eta_{||}=500$.  
\textit{Top:}  Position of the upper Y-Point (solid red) quantified as peak of $E^x$ on the $x=0$ line and right core position (blue dashed) obtained from the peak of $B^z$ on the $x=0$ line. 
\textit{Bottom:} Reconnection rate, measured as drift velocity through $x=\pm0.1$ (solid red) and guide field at the origin $B^z(0,0)$ (blue dashed).  This figure can be compared to the PIC results shown in Figure \ref{fig:coretime}.  }
\label{fig:ff-ce-rate}
\end{center}
\end{figure}

We quantify the evolution of Y-point $(0,y_{yp})$ via the peak in $E^x$ along the $x=0$ line and the core location $(x_c,0)$ as peak of $B^z$ along the $y=0$ line.  The data is shown in the top panel of figure \ref{fig:ff-ce-rate}.  
One can see the Y-point slowing down from its initial very fast motion.  
Gradually, the cores accelerate their approach, reaching a maximal velocity at $t\simeq4$.  The evolution stalls as the guide-field at $(0,0)$ reaches its maximum and magnetic pressure in the current sheet balances the tension of the encompassing field.  

%

%fffffffffffffffffffffffffffffffffffffffffffffffffffffffffffffffffffffffffffffffffffffffff
%fffffffffffffffffffffffffffffffffffffffffffffffffffffffffffffffffffffffffffffffffffffffff

\subsubsection{Scaling with magnetic Reynolds number}

\begin{figure}
\begin{center}
\includegraphics[height=8cm]{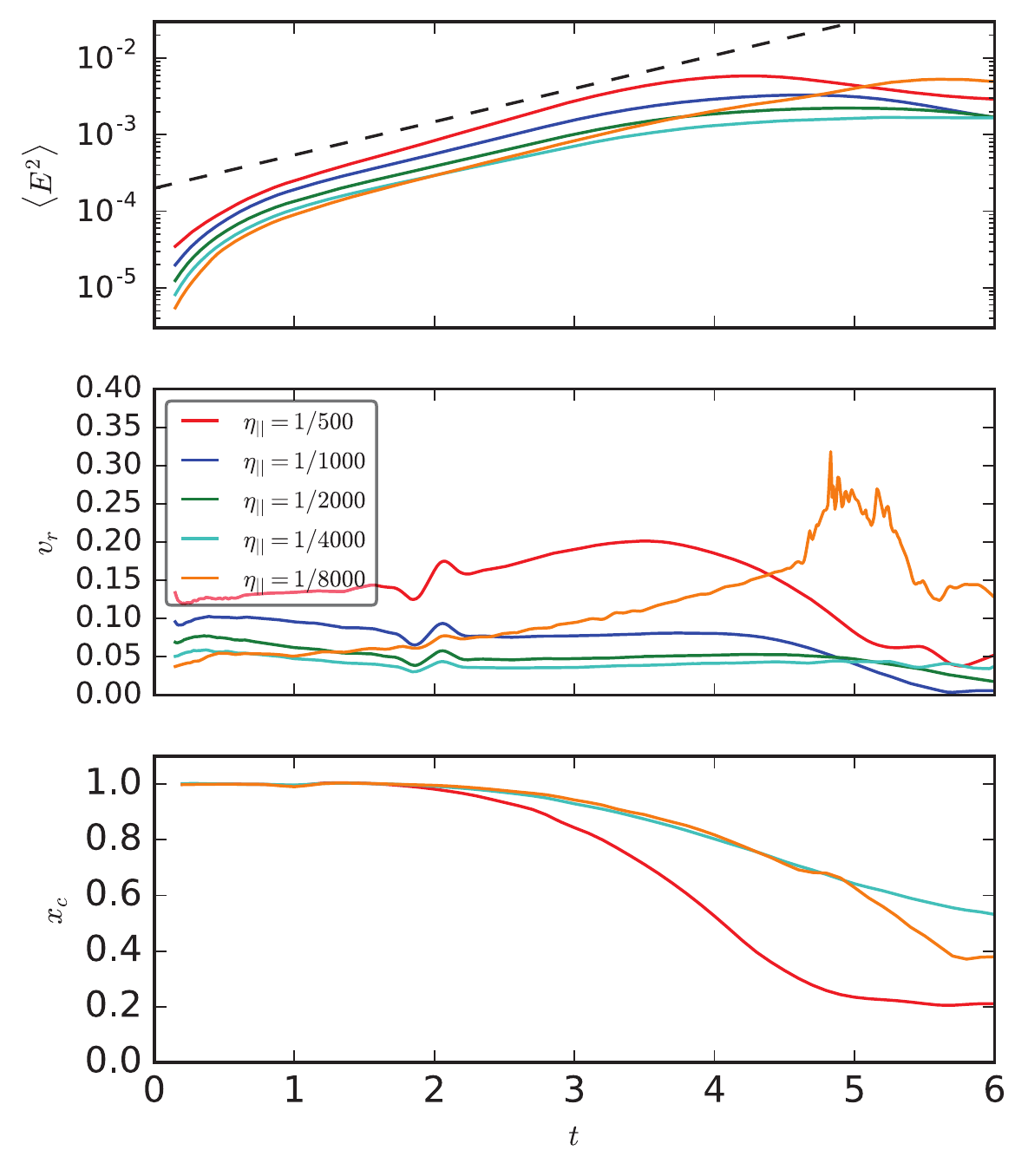}
\caption{Merger of 2D core-envelope flux tubes for different resistivities $\eta_{||}\in[1/500,1/8000]$.  
\textit{Top:} 
Domain averaged electric field strength $\langle E^2 \rangle$.  After one light-crossing time we observe a growth according to $\propto \exp(tc/r_{\rm j})$ indicated as black dashed line.  
\textit{Middle:}
Reconnection rate, measured as drift velocity through $x=\pm0.1$.  The dynamical evolution for $\eta_{||}=1/8000$ is very different from the cases $\eta_{||}=1/1000,1/2000,1/4000$.  In those cases, one large plasmoid would form and remain stationary in the current  sheet until very late times ($t>6$).  In the lowest resistivity case however, many small plasmoids are created which expel flux continuously to both sides.  Thus the magnetic tension surrounding the structure grows more rapidly, leading to the explosive increase of reconnection at $t\simeq4.5$.  
\textit{Bottom:}
Core position for the three selected cases $\eta_{||}=1/500,1/4000,1/8000$.  In the regime $\eta_{||}\in [1/500,1/4000]$, the merging of the cores slows down as electric conductivity increases.  
For $\eta_{||}=1/8000$ however, this trend is halted and reversed as the motion of the cores actually speeds up after the peak in reconnection rate at $t\simeq5$.   This figure can be compared to the PIC results shown in Figure \ref{fig:coretimecomp}.}
\label{fig:ff-ce-compare}
\end{center}
\end{figure}

As for the modified Lundquist ropes, section \ref{sec:mod-lund-res}, we investigate the dependence of the dynamics with magnetic Reynolds number by varying the resistivity parameter.  An overview of relevant quantities is given in Figure \ref{fig:ff-ce-compare} for the range $\eta_{||}\in[1/500,1/8000]$.  
After the onset-time of $\approx r_{\rm j}/c$, all runs enter into a phase where the electric energy grows according to $\propto \exp(tc/r_{\rm j})$.  As expected, the electric energy decreases when the conductivity is increased.  Between $t\simeq4$ and $t\simeq5$, the growth of the electric energy saturates and the evolution slows down.  
In general, we observe two regimes: Up to $\eta_{||}=1/4000$, the reconnection rate decreases with increasing conductivity, as would be expected in a Sweet-Parker scaling.  In this regime, also the motion of the cores slows down when $\eta_{||}$ is decreased.  Whereas the most rapid motion for $\eta_{||}=1/500$ was found at $t\simeq4$, for $\eta_{||}=1/4000$ it is delayed to $t\simeq5$ (c.f. \ref{fig:ff-ce-compare}, lower panel).  

The case with the highest conductivity, $\eta_{||}=1/8000$ breaks the trend of the low conductivity cases.  Here, the reconnection rate increases continuously up to $t\simeq4.5$ at which point it rapidly flares to a large value of $v_{\rm r}\simeq0.25 c$ where it remains for somewhat less than $r_{\rm j}/c$.  
Also the slowing of the core-motion is halted at $\eta_{||}=1/8000$:  Up to $t\simeq5$, the core position lines up well with the case $\eta_{||}=1/4000$.  Just after the flare, the flux tube merger is accelerated as evidenced by the rapid decline of $x_c$.  

%fffffffffffffffffffffffffffffffffffffffffffffffffffffffffffffffffffffffffffffffffffffffff
%fffffffffffffffffffffffffffffffffffffffffffffffffffffffffffffffffffffffffffffffffffffffff

\begin{figure}
\begin{center}
\includegraphics[width=\textwidth]{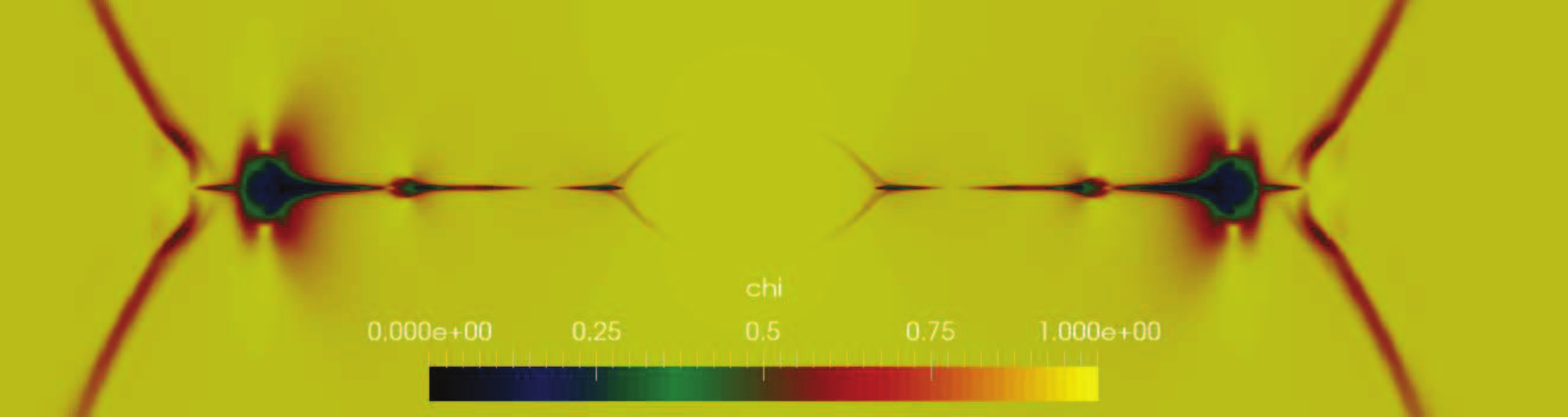}
\includegraphics[width=\textwidth]{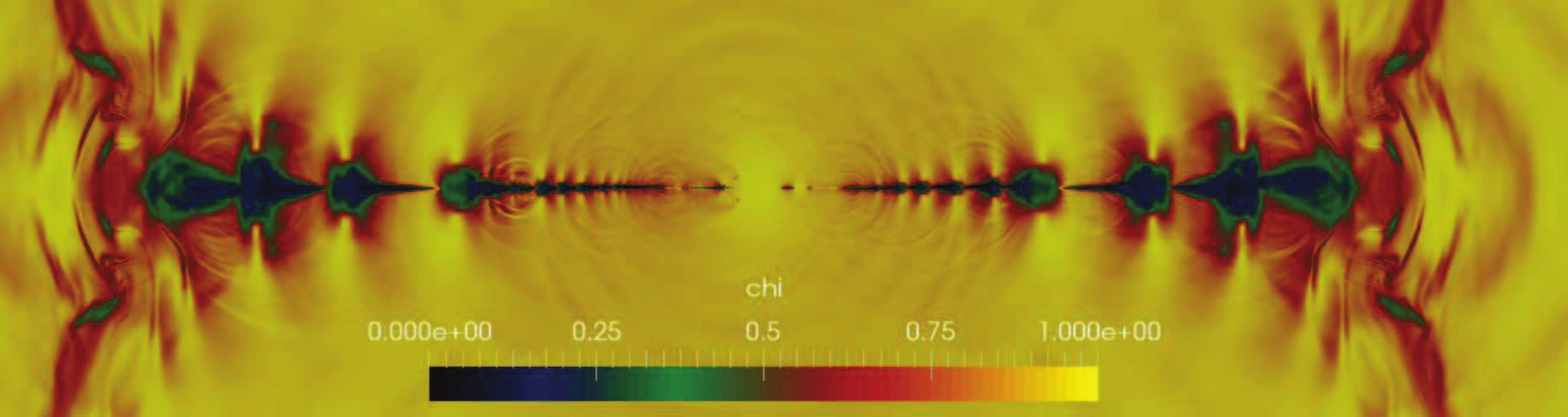}
\includegraphics[width=\textwidth]{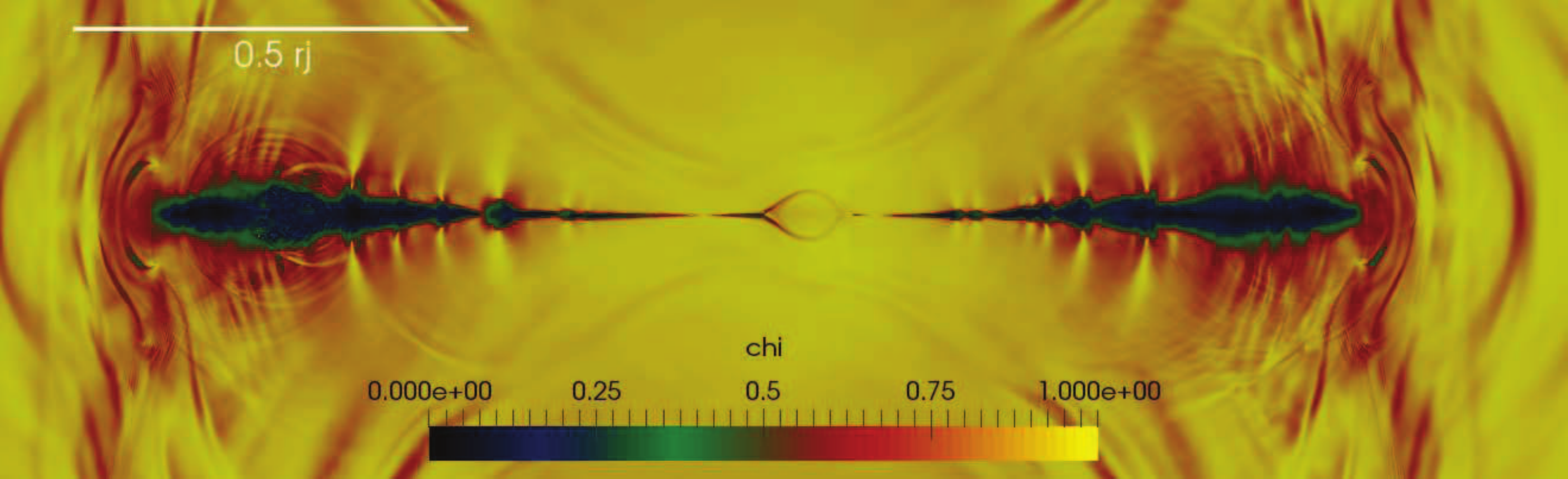}
\caption{Merger of 2D core-envelope flux tubes.
Comparison of $\chi=1-E^2/B^2$ in the central region for the two highest conductivity cases at the time of the reconnection flare at $t=5$: $\eta_{||}=1/4000$ (top) and $\eta_{||}=1/8000$ (middle).  
One can clearly see the different qualitative behaviour:  In the former case, a single large plasmoid grows in the centre of the current sheet, in the latter case, instead small scale plasmoids are expelled in both directions.  At this time, there are also clear differences in the position of the Y-point:  in the high conductivity case, the Y-point is notably further out as the evolution has progressed faster.  As the evolution progresses, a large connected charge starved area develops as seen in the bottom panel for $\eta_{||}=1/8000$ at $t=5.5 r_{\rm j}/c$.  
}
\label{fig:ff-plasmoids}
\end{center}
\end{figure}

It is interesting to ask why the dynamics changed so dramatically for the case $\eta_{||}=1/8000$.  In figure \ref{fig:ff-plasmoids}, we show snapshots of $\chi=1-E^2/B^2$ at the time $t=5$ for the two highest conductivity cases.  
At this point, in the lower conductivity case, a single large central plasmoid has formed that remains in the current sheet for the entire evolution.  This qualitative picture also holds for the cases $\eta_{||}=1/1000,1/2000$.  Flux that accumulates in the central plasmoid does not get expelled from the current sheet and so does not add to the magnetic tension which pushes the ropes together.  Furthermore, the single island opposes the merger via its magnetic pressure of the accumulated $B^z$.  
For the case $\eta_{||}=1/8000$, the large-scale plasmoid does not form, instead smaller plasmoids are expelled symmetrically in both directions.  
The expelled flux continues to contribute to the large scale stresses and the merger of the islands is accelerated further which is reflected also in the continued exponential rise of the electric energy shown in figure \ref{fig:ff-ce-compare}.  
As in the case of the Lundquist and modified Lundquist flux tubes and as required for rapid particle acceleration, macroscopic regions of $E>B$ are found also in this configuration in the outflow regions of the current sheet, illustrated in figure \ref{fig:ff-plasmoids}.  

%fffffffffffffffffffffffffffffffffffffffffffffffffffffffffffffffffffffffffffffffffffffffff
%fffffffffffffffffffffffffffffffffffffffffffffffffffffffffffffffffffffffffffffffffffffffff

In conclusion, our force-free experiments of coalescing flux tubes show that large scale stresses can lead to high reconnection rates in the range of $0.1 c - 0.3 c$, independent of the magnetic Reynolds number.  The dynamics is highly nonlinear and depends on the details of the reconnection in the current sheet:  When a single large plasmoid is formed in the current sheet, rapid instability can be avoided.  This was the case in the three intermediate resistivity cases for the core-envelope configuration considered here.  
Also the original Lundquist tubes showed a rapid reconnection flare of $v_{\rm r}\simeq0.6c$.  The reason is the sign change of the $B^z$ component, leading to vanishing guide field and a missing restoring force in the current sheet.  As guide field with opposite polarity builds up in the current sheet thereafter, the fast reconnection is quickly stalled.

%\begin{thebibliography}{3} \expandafter\ifx\csname natexlab\endcsname\relax\def\natexlab#1{#1}\fi

%\bibitem[{Keppens {et~al.}(2012)Keppens, Meliani, van Marle, Delmont, Vlasis,\& van~der Holst}]{Keppens2012718}Keppens, R., Meliani, Z., van Marle, A., Delmont, P., Vlasis, A., \& van~der  Holst, B. 2012, Journal of Computational Physics, 231, 718

% \bibitem[{{Komissarov} {et~al.}(2007){Komissarov}, {Barkov}, \& {Lyutikov}}]{KomissarovBarkov2007} {Komissarov}, S.~S., {Barkov}, M., \& {Lyutikov}, M. 2007, \mnras, 374, 415

% \bibitem[{{Porth} {et~al.}(2014){Porth}, {Xia}, {Hendrix}, {Moschou}, \&  {Keppens}}]{PorthXia2014} {Porth}, O., {Xia}, C., {Hendrix}, T., {Moschou}, S.~P., \& {Keppens}, R. 2014,  \apjs, 214, 4

%\end{thebibliography}

% \clearpage
%%%%Lorenzo%%%%

%%%%Lorenzo%%%%
%Lundquist and core envelope
%\subsection{}%
%sssssssssssssssssssssssssssssssssssssssssssss
\subsection{PIC simulations of 2D flux tube mergers: simulation setup}
\label{PICfluxtubemergers}
We study the evolution of 2D flux tubes with PIC simulations, employing the electromagnetic PIC code  TRISTAN-MP \citep{buneman_93,spitkovsky_05}. We analyze two possible field configurations: ({\it i}) Lundquist magnetic ropes (see Eq.~\ref{eq:lundquist}), containing zero total current (so, the azimuthal magnetic field vanishes at the boundaries of the rope); ({\it ii}) core-envelope ropes (see Eqs.~\ref{eq:bphi} and \ref{eq:pressure}) this configuration has   nonzero total volumetric  current, which is cancelled by the surface current.

Our computational domain is a rectangle in the $x-y$ plane, with periodic boundary conditions in both directions. The simulation box is initialized with a uniform density of electron-positron plasma at rest and with the magnetic field appropriate for the Lundquist or core-envelope configuration. Since the azimuthal magnetic field vanishes at the boundaries of the Lundquist ropes, the evolution is very slow. Actually, we do not see any sign of evolution up to the final time $\sim 20\, c/\rj$ of our simulation of undriven Lundquist ropes. For this reason, we push by hand the two ropes toward each other, with a prescribed $\vpush$ whose effects will be investigated below. It follows that inside the ropes we start with the magnetic field in Eq.~\ref{eq:lundquist} and with the electric field $\bmath{E}=-\bmath{\vpush}\times \bmath{B}/c$, whereas the electric field is initially zero outside the ropes. 

The azimuthal magnetic field does not vanish at the boundaries of the core-envelope ropes, and the system evolves self-consistently (so, we do not need to drive the system by hand). Here, both the azimuthal field and the poloidal field are discontinuous at the boundary of the ropes. A particle current would be needed to sustain the curl of the field. In our setup, the computational particles are initialized at rest, but such electric current gets self-consistently built up within a few timesteps. To facilitate comparison with the force-free results, our core-envelope configuration has $r_{\rm m}/\rj=0.5$ and $\alpha=0.2$, the same values as in the force-free simulation of Fig.~\ref{fig:CEOverview}.

For our fiducial runs, the spatial resolution is such that the plasma skin depth $\comp$ is resolved with 2.5 cells, but we have verified that our results are the same up to a resolution of $\comp=10$ cells. We only investigate the case of a cold  background plasma, with initial thermal dispersion $kT/mc^2=10^{-4}$.\footnote{For a hot background plasma, the results are expected to be the same (apart from an overall shift in the energy scale), once the skin depth and the magnetization are properly defined. We point to Sect.~\ref{unstr-latt-pic} for a demonstration of this claim in the context of ABC structures.} Each cell is initialized with two positrons and two electrons, but we have checked that our results are the same when using up to 64 particles per cell.  In order to reduce noise in the simulation, we filter
the electric current deposited onto the grid by the particles, mimicking the effect of a larger number of particles per cell \citep{spitkovsky_05,belyaev_15}.

Our unit of length is the radius $\rj$ of the flux ropes, and time is measured in units of $\rj/c$. Our domain is typically a large rectangle of size $10\rj$ along $x$ (i.e., along the direction connecting the centers of the two ropes) and of size $6\rj$ along $y$ (but we have tested that a smaller domain with $5\rj\times 3\rj$ gives the same results). A large domain is needed to minimize the effect of the boundary conditions on the evolution of the system. 

We explore the dependence of our results on the flow magnetization. We identify our runs via the mean value $\sigmain$ of the magnetization measured within the flux ropes using the initial in-plane fields (so, excluding the $B_z$ field). As we argue below, it is the dissipation of the in-plane fields that primarily drives efficient heating and particle acceleration. In addition, this choice allows for a direct comparison of our results between the two configurations (Lundquist and core-envelope) and with the ABC structures of Sect.~\ref{unstr-latt-pic}.
The mean in-plane field corresponding to $\sigmain$ shall be called $B_{0,\rm in}$, and it will be our unit of measure of the electromagnetic fields. We will explore a wide range of magnetizations, from $\sigmain=3$ up to $\sigmain=170$. 

It will be convenient to compare the radius $\rj$ of the ropes to the characteristic Larmor radius $\rhot=\sqrt{\sigmain}\comp$ of the high-energy particles heated/accelerated by reconnection (rather than to the skin depth $\comp$). We will explore a wide range of $\rj/\rhot$, from $\rj/\rhot=31$ up to $245$. We will demonstrate that the two most fundamental parameters that characterize the system  are the magnetization $\sigmain$ and the ratio $\rj/\rhot$.

%%%%%%%%%%%%%%%%%%%%%%%%%%%%%%%
\begin{figure}
\centering
\includegraphics[width=.34\textwidth]{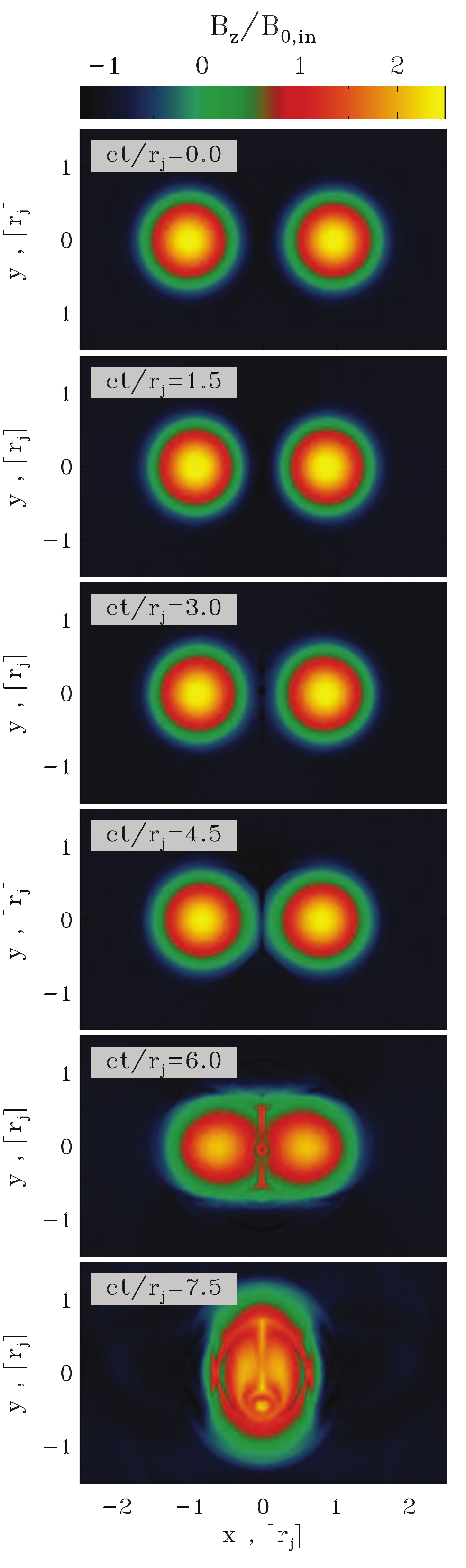} 
\includegraphics[width=.34\textwidth]{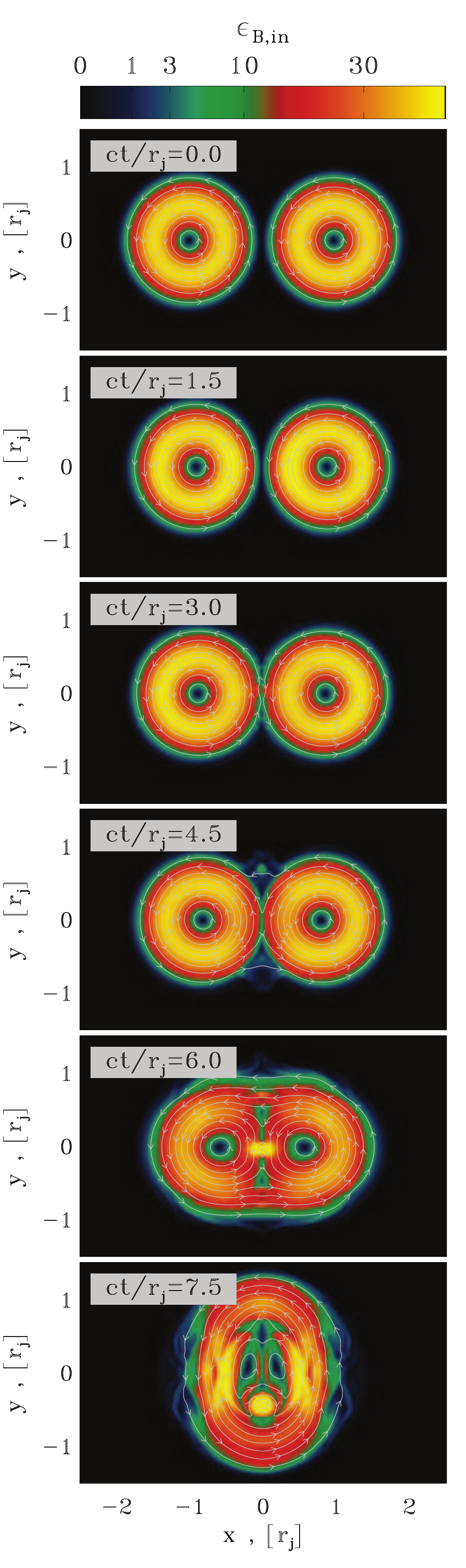} 
\caption{Temporal evolution of 2D Lundquist ropes (time is measured in $c/\rj$ and indicated in the grey box of each panel, increasing from top to bottom). The plot presents the 2D pattern of the out-of-plane field $B_z$ (left column; in units of $B_{0,\rm in}$) and of the in-plane magnetic energy fraction $\epsilon_{B,\rm in}=(B_x^2+B_y^2)/8 \pi n m c^2$ (right column; with superimposed magnetic field lines), from a PIC simulation with $kT/mc^2=10^{-4}$, $\sigmain=43$ and $\rj=61\,\rhot$, performed within a rectangular domain of size $10\rj\times 6\rj$ (but we only show a small region around the center). The evolution is driven with a velocity $\vpush/c=0.1$. Reconnection at the interface between the two flux ropes (i.e., around $x=y=0$) creates an envelope of field lines engulfing the two islands, whose tension force causes them to merge on a dynamical time. This figure should be compared with the force-free result in Fig. \ref{fig:ff-lundquist} (even though $\vpush/c=0.3$ in that case).}
\label{fig:lundfluid} 
\end{figure}
\begin{figure}
\centering
\includegraphics[width=.49\textwidth]{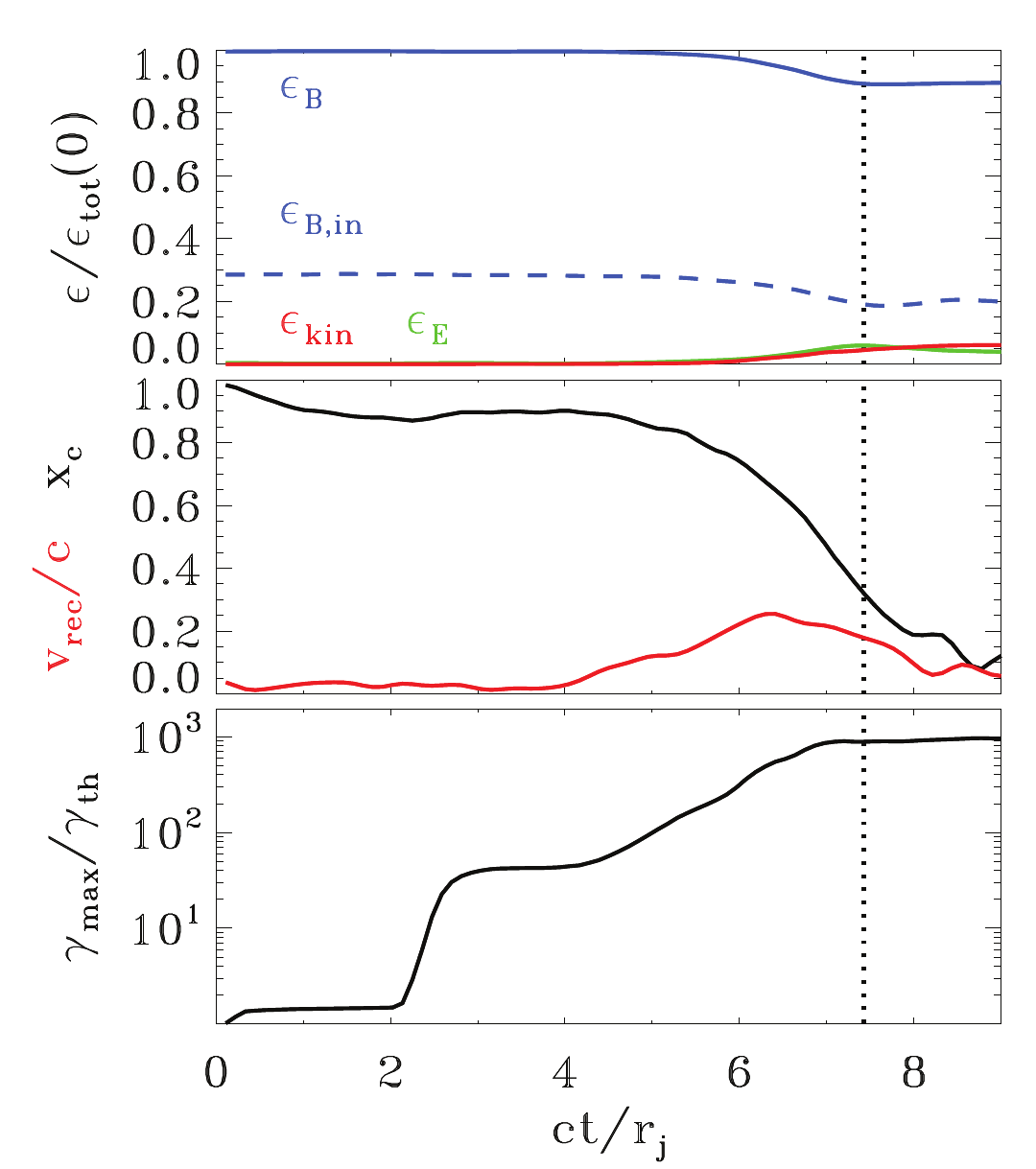} 
\caption{Temporal evolution of various quantities, from a 2D PIC simulation of Lundquist ropes with $kT/mc^2=10^{-4}$, $\sigmain=43$ and $\rj=61\,\rhot$ (the same as in \fig{lundfluid}), performed within a rectangular domain of size $5\rj\times 3\rj$. The evolution is driven with a velocity $\vpush/c=0.1$. Top panel: fraction of energy in magnetic fields (solid blue), in-plane magnetic fields (dashed blue), electric fields (green) and particles (red; excluding the rest mass energy), in units of the total initial energy. Middle panel: reconnection rate $v_{\rm rec}/c$ (red), defined as the  inflow speed along the $x$ direction averaged over a square of side equal to $\rj/2$ centered at $x=y=0$; and location $x_{\rm c}$ of the core of the rightmost flux rope (black), in units of $\rj$.
Bottom panel: evolution of the maximum Lorentz factor $\gammamax$, as defined in \eq{ggmax}, relative to the thermal Lorentz factor $\gamma_{\rm th}\simeq 1+(\hat{\gamma}-1)^{-1} kT/m c^2$, which for our case is $\gamma_{\rm th}\simeq 1$. In response to the initial push, $\gammamax$ increases at $ct/\rj\sim 2$ up to $\gammamax/\gamma_{\rm th}\sim 40$, before stalling. At this stage, the cores of the two islands have not significantly moved (black line in the middle panel). At $ct/\rj\sim 5$, the tension force of the common envelope of field lines starts pushing the two islands toward each other (and the $x_{\rm c}$ location of the rightmost rope decreases, see the middle panel), resulting in a merger event occurring on a dynamical timescale. During the merger (at $ct/\rj\sim 6$), the reconnection rate peaks (red line in the middle panel), a fraction of the in-plane magnetic energy is transferred to the plasma (compare dashed blue and solid red lines in the top panel), and particles are quickly accelerated up to $\gammamax/\gamma_{\rm th}\sim 10^3$ (bottom panel). In all the panels, the vertical dotted black line indicates the time when the electric energy peaks, shortly after the most violent phase of the merger event.}
\label{fig:lundtime} 
\end{figure}
\begin{figure}
\centering
\includegraphics[width=.49\textwidth]{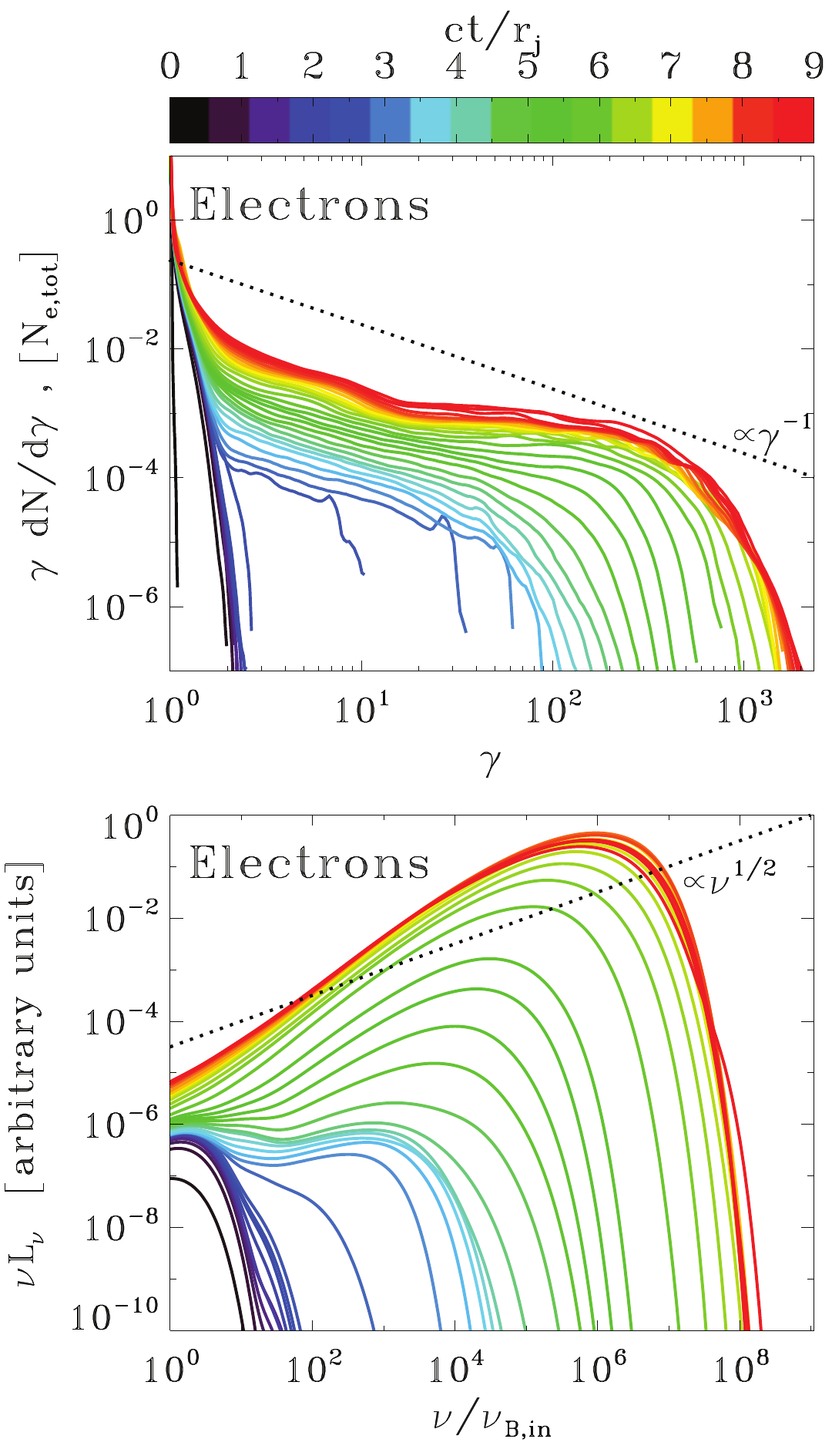} 
\caption{Particle energy spectrum and synchrotron spectrum from a 2D PIC simulation of Lundquist ropes with $kT/mc^2=10^{-4}$, $\sigmain=43$ and $\rj=61\,\rhot$ (the same as in \fig{lundfluid} and \fig{lundtime}), performed within a  domain of size $10\rj\times 6\rj$. The evolution is driven with a velocity $\vpush/c=0.1$.  Time is measured in units of $\rj/c$, see the colorbar at the top. Top panel: evolution of the electron energy spectrum normalized to the total number of electrons. At late times, the high-energy spectrum approaches a hard distribution $\gamma dN/d\gamma\propto {\rm const}$ (for comparison, the dotted black line shows the case $\gamma dN/d\gamma\propto \gamma^{-1}$ corresponding to equal energy content in each decade of $\gamma$). Bottom panel: evolution of the angle-averaged synchrotron spectrum emitted by electrons. The frequency on the horizontal axis is in units of $\nu_{B,\rm in}=\sqrt{\sigmain}\omega_{\rm p}/2\pi$. At late times, the synchrotron spectrum approaches a power law with $\nu L_\nu\propto \nu$, which just follows from the fact that the electron spectrum at high energies is close to $\gamma dN/d\gamma\propto {\rm const}$. This is harder than the dotted line, which indicates the slope $\nu L_\nu\propto \nu^{1/2}$ resulting from an electron spectrum $\gamma dN/d\gamma\propto \gamma^{-1}$.}
\label{fig:lundspec} 
\end{figure}
\begin{figure}
\centering
\includegraphics[width=.49\textwidth]{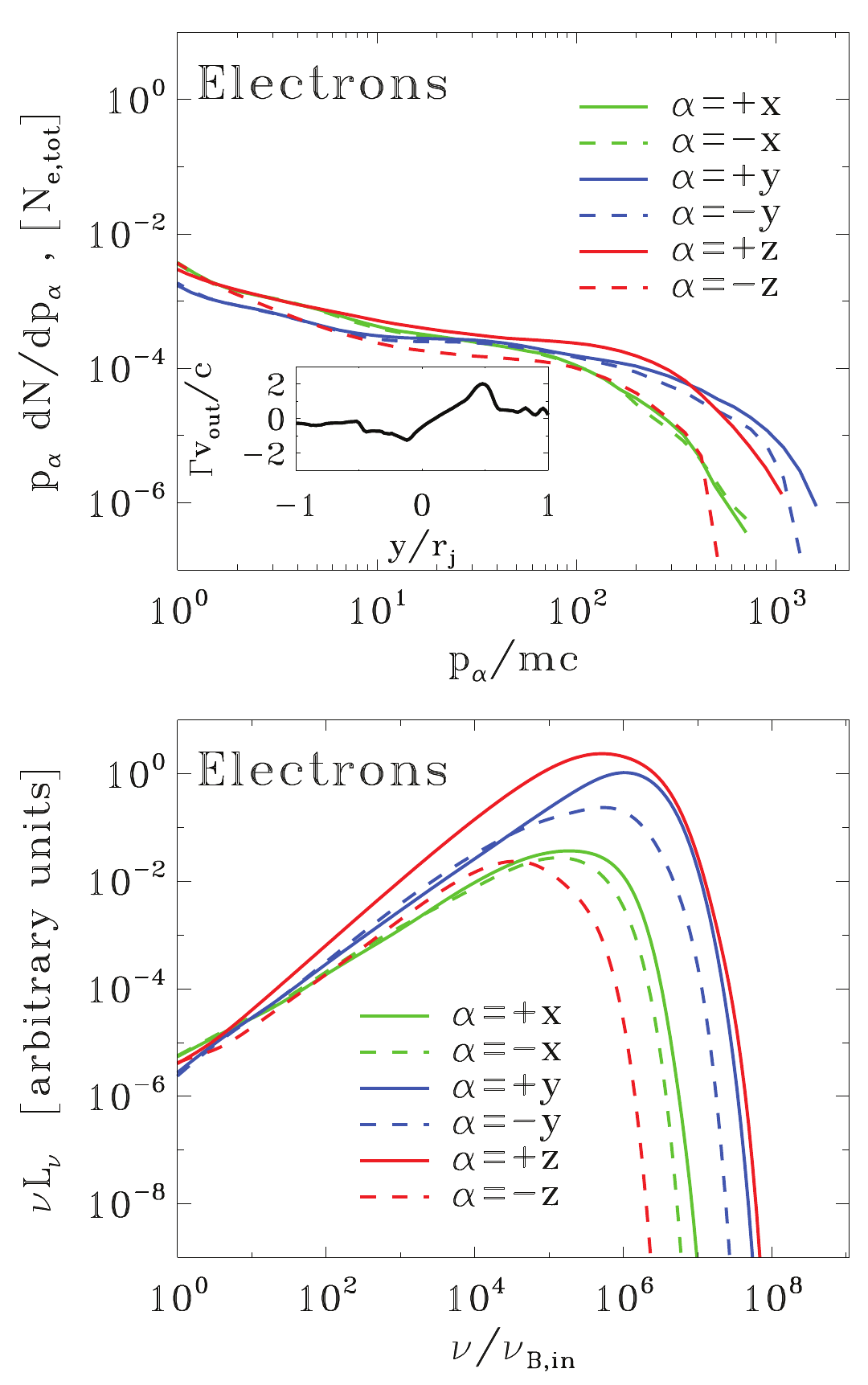} 
\caption{Particle momentum spectrum and anisotropy of the synchrotron emission from a 2D PIC simulation of Lundquist ropes with $kT/mc^2=10^{-4}$, $\sigmain=43$ and $\rj=61\,\rhot$ (the same as in \fig{lundfluid}, \fig{lundtime} and \fig{lundspec}), performed within a  domain of size $10\rj\times 6\rj$. The evolution is driven with a velocity $\vpush/c=0.1$.  Top panel: electron momentum spectrum along different directions (as indicated in the legend), at the time when the electric energy peaks (as indicated by the vertical dotted black line in \fig{lundtime}). The highest energy electrons are beamed along the direction $y$ of the reconnection outflow (blue lines) and along the direction $+z$ of the accelerating electric field (red solid line; positrons will be beamed along $-z$, due to the opposite charge).  The inset shows the 1D profile along $y$ of the bulk four-velocity in the outflow direction (i.e., along $y$), measured at $x=0$.  Bottom panel: synchrotron spectrum at the time indicated in \fig{lundtime} (vertical dotted black line) along different directions (within a solid angle of $\Delta \Omega/4\pi\sim 3\times10^{-3}$), as indicated in the legend. The resulting  anisotropy of the synchrotron emission is consistent with the particle anisotropy illustrated in the top panel.}
\label{fig:lundspecmom} 
\end{figure}
\subsection{PIC simulations of 2D flux tube mergers: Lundquist ropes}
The temporal evolution of the merger of two Lundquist ropes is shown in \fig{lundfluid}.
The plot presents the 2D pattern of the out-of-plane field $B_z$ (left column; in units of $B_{0,\rm in}$) and of the in-plane magnetic energy fraction $\epsilon_{B,\rm in}=(B_x^2+B_y^2)/8 \pi n m c^2$ (right column; with superimposed magnetic field lines), from a PIC simulation with $kT/mc^2=10^{-4}$, $\sigmain=43$ and $\rj=61\,\rhot$. The magnetic ropes are initially driven toward each other with a speed $\vpush/c=0.1$ (below, we study the dependence on the driving speed). Time is measured in units of $\rj/c$ and indicated in the grey boxes within each panel. The figure should be compared with the force-free result in Fig. \ref{fig:ff-lundquist} (even though $vpush/c=0.3$ in that case).

As the two magnetic ropes slowly approach, driven by the initial velocity push (compare the first and second row in \fig{lundfluid}), reconnection is triggered in the plane $x=0$, as indicated by the formation and subsequent ejection of small-scale plasmoids (third and fourth row in \fig{lundfluid}). So far (i.e., until $ct/\rj\sim 4.5$), the cores of the two islands have not significantly moved (black line in the middle panel of \fig{lundtime}, indicating the $x_{\rm c}$ location of the center of the rightmost island), the reconnection speed is quite small (red line in the middle panel of \fig{lundtime}) and no significant energy exchange has occurred from the fields to the particles (compare the in-plane magnetic energy, shown by  the dashed blue line in the top panel of \fig{lundtime}, with the particle kinetic energy, indicated with the red line).\footnote{We remark that the energy balance described in the top panel of \fig{lundtime} is dependent on the overall size of the box, everything else being the same. More specifically, while the in-plane fields (which are the primary source of dissipation) are nonzero only within the flux ropes, the out-of-plane field pervades the whole domain. It follows that for a larger box the relative ratios between $\epsilon_{\rm kin}$ (red line; or $\epsilon_{\rm E}$, green line) and $\epsilon_{B,\rm in}$ (dashed blue line) will not change, whereas they will all decrease as compared to the total magnetic energy $\epsilon_{B}$ (blue solid line).} The only significant evolution before $ct/\rj\sim 4.5$ is the reconnection-driven increase in the maximum particle Lorentz factor up to $\gammamax/\gamma_{\rm th}\sim 40$ occurring at $ct/\rj\sim 2.5$ (see the black line in the bottom panel of \fig{lundtime}).

As a result of reconnection, an increasing number of field lines, that initially closed around one of the ropes, are now engulfing both magnetic islands. Their tension force causes the two ropes to approach and merge on a quick (dynamical) timescale, starting at $ct/\rj\sim 4.5$ and ending at $ct/\rj\sim 7.5$ (see that the distance of the rightmost island from the center rapidly decreases, as indicated by the black line in the middle panel of \fig{lundtime}). The tension force drives the particles in the flux ropes toward the center, with a fast reconnection speed peaking at $v_{\rm rec}/c\sim 0.3$ (red line in the middle panel of \fig{lundtime}).\footnote{This value of the reconnection rate is roughly comparable to the results of solitary X-point collapse presented in Paper I. However, a direct comparison cannot be established, since in that case we either assumed a uniform nonzero guide field or a vanishing guide field, whereas here the guide field strength is not uniform in space.} The reconnection layer at $x=0$  stretches up to a length of $\sim 2\rj$, it becomes unstable to the secondary tearing mode \citep{uzdensky_10}, and secondary plasmoids are formed (e.g., see the plasmoid at $(x,y)\sim(0,0)$ at $ct/\rj=6$). Below, we demonstrate that the formation of secondary plasmoids is primarily controlled by the ratio $\rj/\rhot$. In the central current sheet, it is primarily the in-plane field that gets dissipated (compare the dashed and solid blue lines in the top panel of \fig{lundtime}), driving an increase in the electric energy (green) and in the particle kinetic energy (red).\footnote{The out-of-plane field for the Lundquist configuration in Eq.~\ref{eq:lundquist} switches sign inside the flux ropes. In force free simulations, this leads to the formation of large areas with $E>B$ in the outermost regions of the ropes. We do not observe the formation of such features in PIC simulations.}  In this phase of evolution, the fraction of initial energy released to the particles is small ($\epsilon_{\rm kin}/\epsilon_{\rm tot}(0)\sim 0.1$), but the particles advected into the central X-point experience a dramatic episode of acceleration. As shown in the bottom panel of \fig{lundtime}, the cutoff Lorentz factor $\gammamax$ of the particle spectrum presents a dramatic evolution, increasing up to $\gammamax/\gamma_{\rm th}\sim 10^3$ within a couple of dynamical times. It is this phase of extremely fast particle acceleration that we associate with the generation of the Crab flares.

The two distinct evolutionary phases --- the early stage driven by the initial velocity push, and the subsequent dynamical merger driven by large-scale electromagnetic stresses --- are clearly apparent in the evolution of the particle energy spectrum (top panel in \fig{lundspec}) and of the angle-averaged synchrotron emission (bottom panel in \fig{lundspec}). The initial velocity push drives reconnection in the central current sheet, which leads to fast particle acceleration (from dark blue to cyan in the top panel). This is only a transient event, and the particle energy spectrum then freezes (see the clustering of the cyan lines, and compare with the phase at $2.5\lesssim ct/\rj\lesssim 4$ in the bottom panel of \fig{lundtime}). Correspondingly, the angle-averaged synchrotron spectrum stops evolving (see the clustering of the cyan lines in the bottom panel). A second dramatic increase in the particle and emission spectral cutoff (even more dramatic than the initial growth) occurs between $ct/\rj\sim 4.5$ and $ct/\rj\sim 7$ (cyan to yellow curves in \fig{lundspec}), and it directly corresponds to the dynamical merger of the two magnetic ropes, driven by large-scale stresses. The particle spectrum quickly extends up to $\gammamax\sim 10^3$ (yellow lines in the top panel), and correspondingly the peak of the $\nu L_\nu$ emission spectrum shifts up to $\sim \gammamax^2\nu_{B,\rm in}\sim10^6 \nu_{B,\rm in}$ (yellow lines in the bottom panel). The system does not show any sign of evolution at times later than $c/\rj\sim 7.5$ (see the clustering of the yellow to red lines). At late times, the high-energy spectrum approaches a hard distribution $\gamma dN/d\gamma\propto {\rm const}$ (for comparison, the dotted black line shows the case $\gamma dN/d\gamma\propto \gamma^{-1}$ corresponding to equal energy content in each decade of $\gamma$). The synchrotron spectrum approaches a power law with $\nu L_\nu\propto \nu$, which just follows from the fact that the electron spectrum at high energies is close to $\gamma dN/d\gamma\propto {\rm const}$. This is harder than the dotted line, which indicates the slope $\nu L_\nu\propto \nu^{1/2}$ resulting from an electron spectrum $\gamma dN/d\gamma\propto \gamma^{-1}$.

The particle distribution  is significantly anisotropic. In the top panel of \fig{lundspecmom}, we plot the electron momentum spectrum at the time when the electric energy peaks (see the vertical black dotted line in \fig{lundtime}) along different directions, as indicated in the legend. The highest energy electrons are beamed along the direction $y$ of the reconnection outflow (blue lines) and along the direction $+z$ anti-parallel to the accelerating electric field (red solid line; positrons will be beamed along $-z$, due to the opposite charge). This is consistent with our results for solitary X-point collapse in Paper I. Most of the anisotropy is to be attributed to the ``kinetic beaming'' discussed by \citet{2012ApJ...754L..33C}, rather than beaming associated with the bulk motion (which is only marginally relativistic, see the inset in the top panel of \fig{lundspecmom}). The   anisotropy of the synchrotron emission (bottom panel in \fig{lundspecmom}) is consistent with the particle anisotropy illustrated in the inset of the top panel.

%%%%%%%%%%%%%%%%%%%%%%%%%%%%%
\begin{figure}
\centering
\includegraphics[width=.79\textwidth]{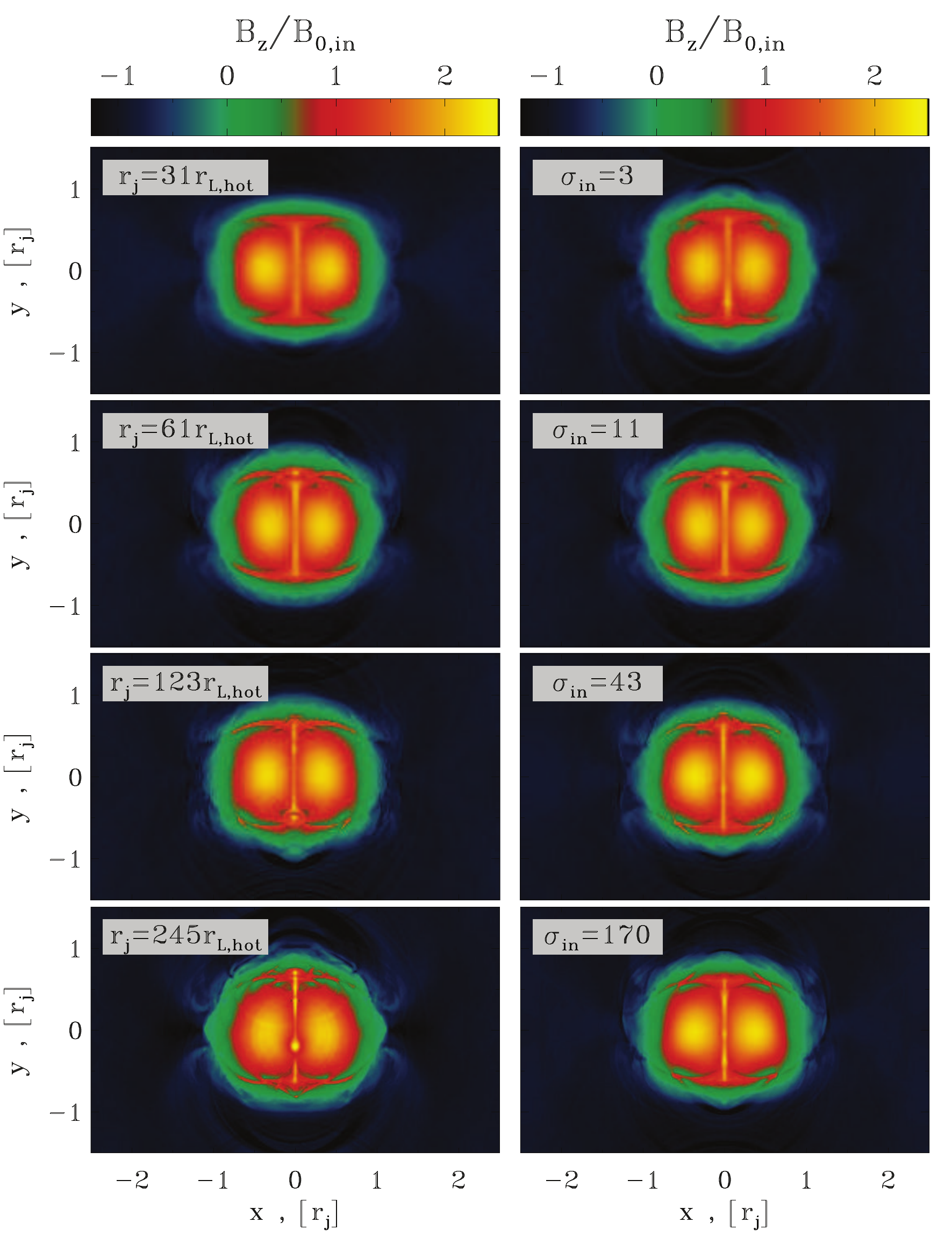} 
\caption{2D pattern of the out-of-plane field $B_z$ (in units of $B_{0,\rm in}$) at the most violent time of the merger of Lundquist ropes (i.e., when the electric energy peaks, as indicated by the vertical dotted lines in \fig{lundtimecomp}) from a suite of PIC simulations. The evolution is driven with a velocity $\vpush/c=0.1$. In the left column, we fix $kT/mc^2=10^{-4}$ and $\sigmain=11$ and we vary the ratio $\rj/\rhot$, from 31 to 245 (from top to bottom). In the right column, we fix $kT/mc^2=10^{-4}$ and $\rj/\rhot=61$ and we vary the magnetization $\sigmain$, from 3 to 170 (from top to bottom). In all cases, the simulation box is a rectangle of size $5\rj\times 3\rj$. The 2D structure of $B_z$ in all cases is quite similar, apart from the fact that larger $\rj/\rhot$ tend to lead to a more pronounced fragmentation of the current sheet.}
\label{fig:lundfluidcomp} 
\end{figure}
\begin{figure}
\centering
\includegraphics[width=.79\textwidth]{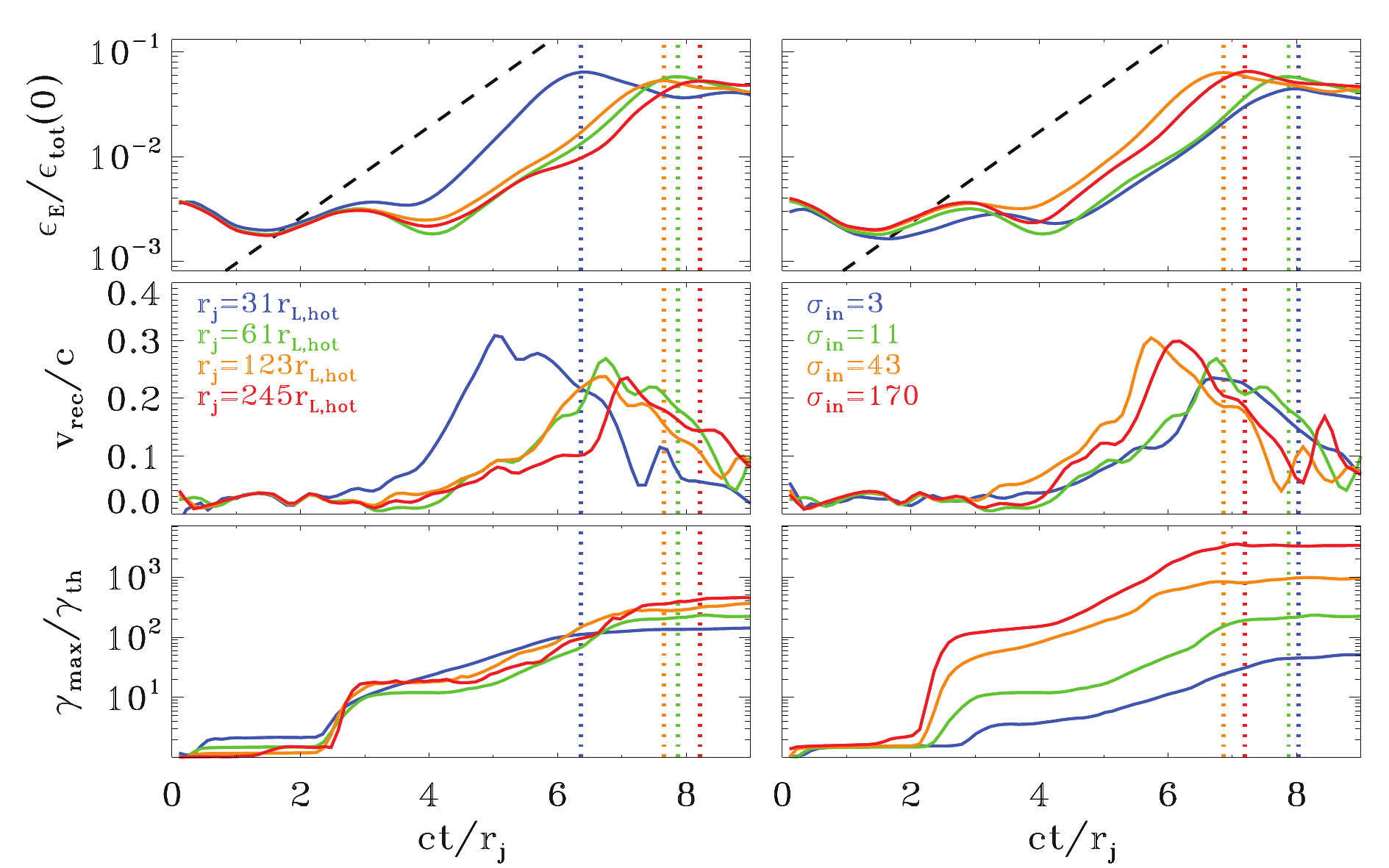} 
\caption{Temporal evolution of the electric energy (top panel; in units of the total initial energy), of the reconnection rate (middle panel; defined as the mean inflow velocity in a square of side $\rj/2$ centered at $x=y=0$) and of the maximum particle Lorentz factor (bottom panel; $\gammamax$ is defined in \eq{ggmax}, and it is normalized to the thermal Lorentz factor $\gamma_{\rm th}\simeq 1+(\hat{\gamma}-1)^{-1} kT/m c^2$), for a suite of PIC simulations of Lundquist ropes (same runs as in \fig{lundfluidcomp}). In all the cases, the evolution is driven with a velocity $\vpush/c=0.1$. In the left column, we fix $kT/mc^2=10^{-4}$ and $\sigmain=11$ and we vary the ratio $\rj/\rhot$ from 31 to 245 (from blue to red, as indicated in the legend). In the right column, we fix $kT/mc^2=10^{-4}$ and $\rj/\rhot=61$ and we vary the magnetization $\sigmain$ from 3 to 170 (from blue to red, as indicated in the legend). The maximum particle energy $\gammamax mc^2$ resulting from the merger increases for increasing $\rj/\rhot$ at fixed $\sigmain$ (left column) and for increasing $\sigmain$ at fixed $\rj/\rhot$.
The dashed black line in the top panel shows that the electric energy grows exponentially as $\propto \exp{(ct/\rj)}$. The vertical dotted lines mark the time when the electric energy peaks (colors as described above).}
\label{fig:lundtimecomp} 
\end{figure}
\begin{figure}
\centering
\includegraphics[width=.79\textwidth]{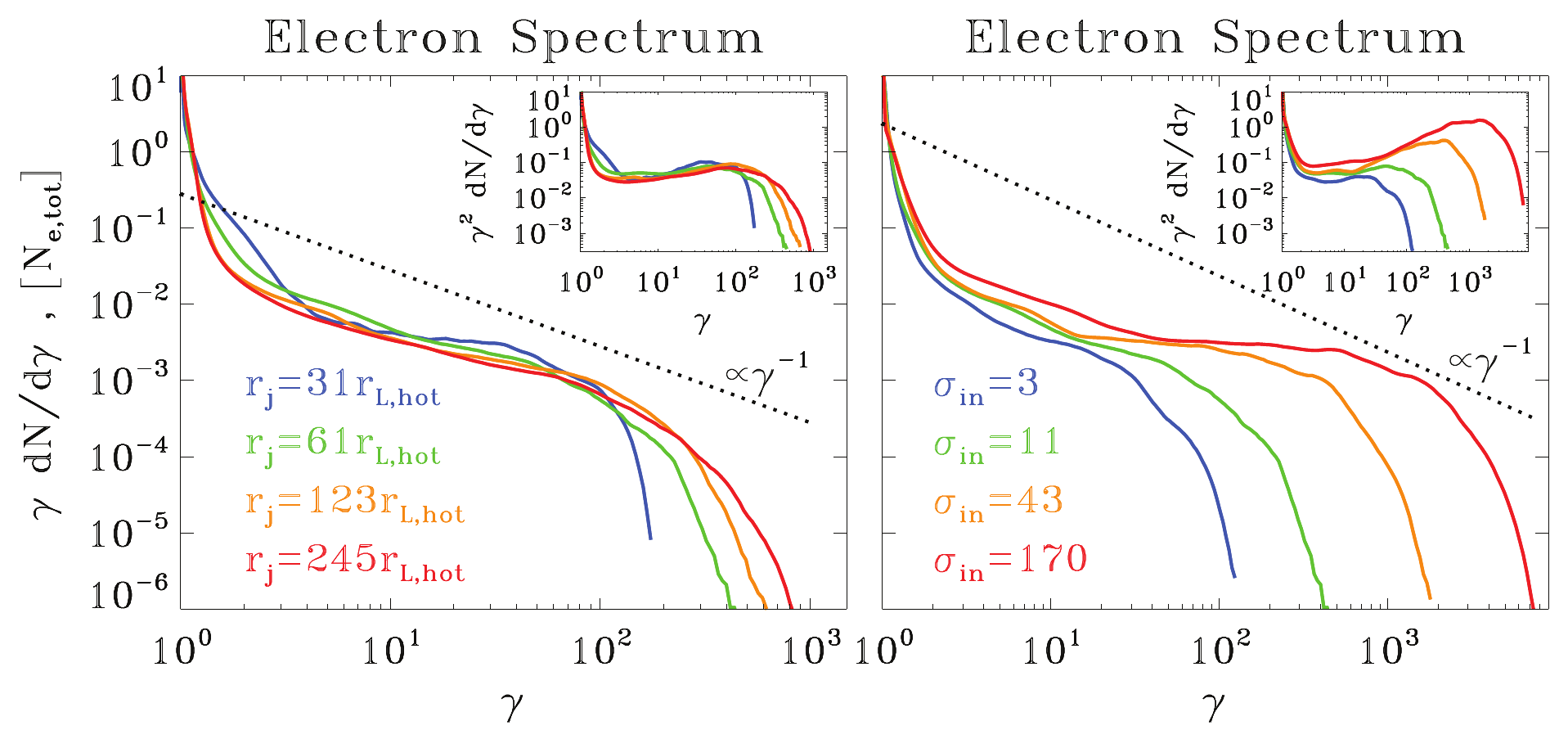} 
\caption{Particle spectrum at the time when the electric energy peaks, for a suite of PIC simulations of Lundquist ropes (same runs as in \fig{lundfluidcomp} and \fig{lundtimecomp}). In all the cases, the evolution is driven with a velocity $\vpush/c=0.1$.  In the left column, we fix $kT/mc^2=10^{-4}$ and $\sigmain=11$ and we vary the ratio $\rj/\rhot$ from 31 to 245 (from blue to red, as indicated by the legend).  In the right column, we fix $kT/mc^2=10^{-4}$ and $\rj/\rhot=61$ and we vary the magnetization $\sigmain$ from 3 to 170 (from blue to red, as indicated by the legend). The main plot shows $\gamma dN/d\gamma$ to emphasize the particle content, whereas the inset presents $\gamma^2 dN/d\gamma$ to highlight the energy census. The dotted black line is a power law $\gamma dN/d\gamma\propto \gamma^{-1}$, corresponding to equal energy content per decade (which would result in a flat distribution in the insets). The spectral hardness is not a sensitive function of the ratio $\rj/\rhot$, but it is strongly dependent on $\sigmain$, with higher magnetizations giving harder spectra, up to the saturation slope of $-1$.}
\label{fig:lundspeccomp} 
\end{figure}
\subsubsection{Dependence on the flow conditions}
We now investigate the dependence of our results on the magnetization $\sigmain$ and the ratio $\rj/\rhot$, where $\rhot=\sqrt{\sigmain}\comp$. In \fig{lundfluidcomp}, we present the 2D pattern of the out-of-plane field $B_z$ (in units of $B_{0,\rm in}$) during the most violent phase of rope merger (i.e., when the electric energy peaks, as indicated by the vertical dotted lines in \fig{lundtimecomp}) from a suite of PIC simulations in a rectangular domain of size $5\rj\times 3\rj$. In the left column, we fix $kT/mc^2=10^{-4}$ and $\sigmain=11$ and we vary the ratio $\rj/\rhot$, from 31 to 245 (from top to bottom). In the right column, we fix $kT/mc^2=10^{-4}$ and $\rj/\rhot=61$ and we vary the magnetization $\sigmain$, from 3 to 170 (from top to bottom). The evolution is driven with a velocity $\vpush/c=0.1$ in all the cases.  

The 2D pattern of $B_z$ presented in \fig{lundfluidcomp} shows that the merger proceeds in a similar way in all the runs. The only difference is that  larger $\rj/\rhot$ lead to thinner current sheets, when fixing $\sigmain$ (left column in \fig{lundfluidcomp}). Roughly, the thickness of the current sheet is set by the Larmor radius $\rhot$ of the high-energy particles heated/accelerated by reconnection. In the right column, with $\rj/\rhot$ fixed, the thickness of the current sheet is then a fixed fraction of the box size. In contrast, in the left column, the ratio of current sheet thickness to box size will scale as $\rhot/\rj$, as indeed it is observed. A long thin current sheet is expected to fragment into a chain of plasmoids/magnetic islands \citep[e.g.,][]{uzdensky_10,2016ApJ...816L...8W}, when the length-to-thickness ratio is much larger than unity. It follows that all the cases in the right column will display a similar tendency for fragmentation (and in particular, they do not appreciably fragment), whereas the likelihood of fragmentation is expected to increase from top to bottom in the left column. In fact, for the case with $\rj/\rhot=245$ (left bottom panel), a number of small-scale plasmoids appear in the current sheets (e.g., see the plasmoid at $x\sim0$ and $y\sim-0.1\rj$ in the left bottom panel). We find that as long as $\sigmain\gg1$, the secondary tearing mode discussed by \citet{uzdensky_10} --- that leads to current sheet fragmentation --- appears at $\rj/\rhot\gtrsim 100$, in the case of Lundquist ropes.\footnote{A similar result had been found for the case of ABC collapse, see Sect.\ref{unstr-latt-pic}.}

In \fig{lundtimecomp} we present the temporal evolution of the runs whose 2D structure is shown in \fig{lundfluidcomp}. In the left column, we fix $kT/mc^2=10^{-4}$ and $\sigmain=11$ and we vary the ratio $\rj/\rhot$, from 31 to 245 (from blue to red, as indicated in the legend of the middle panel). In the right column, we fix $kT/mc^2=10^{-4}$ and $\rj/\rhot=61$ and we vary the magnetization $\sigmain$, from 3 to 170 (from blue to red, as indicated in the legend of the middle panel). 
 The top panels show that the evolution of the electric energy (in units of the total initial energy) is  similar for all the values of $\rj/\rhot$ and $\sigmain$ we explore. In particular, the electric energy grows approximately as $\propto \exp{(ct/\rj)}$ in all the cases (compare with the dashed black lines), and it peaks at $\sim 5\%$ of the total initial  energy. The only marginal exception is the trans-relativistic case $\sigmain=3$ and $\rj/\rhot=61$ (blue line in the top right panel), whose peak value is slightly smaller, due to the lower \Alfven speed. The onset time of the instability is also nearly independent of $\sigmain$ (top right panel), although the low-magnetization cases $\sigmain=3$ and 11 (blue and green lines) seem to be growing slightly later. As regard to the dependence of the onset time on $\rj/\rhot$ at fixed $\sigmain$, the top left panel in \fig{lundtimecomp} shows that larger values of $\rj/\rhot$ tend to grow later, but the variation is only moderate (with the exception of the case with the smallest $\rj/\rhot=31$, blue line in the top left panel, that grows quite early).

The peak reconnection rate in all the cases we have explored is around $v_{\rm rec}/c\sim 0.2-0.3$ (middle row in \fig{lundtimecomp}). It marginally decreases with increasing $\rj/\rhot$ (but we have verified that it saturates at $v_{\rm rec}/c\sim 0.25$ in the limit $\rj/\rhot\gg1$, see the middle left panel in \fig{lundtimecomp}), and it moderately increases with $\sigmain$ (especially as we transition from the non-relativistic regime to the relativistic regime, but it  saturates at $v_{\rm rec}/c\sim 0.3$ in the limit $\sigmain\gg1$, see the middle right panel in \fig{lundtimecomp}). We had found similar values and trends for the peak reconnection rate in the case of ABC collapse, see Sect.\ref{unstr-latt-pic}.

In the evolution of the maximum particle Lorentz factor $\gammamax$ (bottom row in \fig{lundtimecomp}), one can distinguish two phases. At early times ($ct/\rj\sim 2.5$), the increase in $\gammamax$ is moderate, when reconnection is triggered in the central region by the initial velocity push. At later times  ($ct/\rj\sim 6$), as the two magnetic ropes merge on a dynamical timescale, the maximum particle Lorentz factor grows explosively. Following the same argument detailed in Sect.\ref{unstr-latt-pic}, we estimate that the high-energy cutoff of the particle spectrum at the end of the merger  event (which lasts for a few $\rj/c$) should scale as $\gammamax/\gamma_{\rm th}\propto v_{\rm rec}^2\sqrt{\sigmain} \rj\propto v_{\rm rec}^2\sigmain (\rj/\rhot)$. If the reconnection rate does not significantly depend on $\sigmain$, this implies that $\gammamax\propto \rj$ at fixed $\sigmain$. The trend for a steady increase of $\gammamax$ with $\rj$ at fixed $\sigmain$ is confirmed in the bottom left panel of \fig{lundtimecomp}, both at the final time and at the peak time of the electric energy (which is slightly different among the four different cases, see the vertical dotted colored lines).\footnote{The fact that the dependence appears slightly sub-linear is due to the fact that the reconnection rate is slightly larger for smaller $\rj/\rhot$.} Similarly, if the reconnection rate does not significantly depend on $\rj/\rhot$, this implies that $\gammamax\propto \sigmain$ at fixed $\rj/\rhot$. This linear dependence of $\gammamax$ on $\sigmain$ is  confirmed in the bottom right panel of \fig{lundtimecomp}.

The dependence of the particle spectrum on $\rj/\rhot$ and $\sigmain$ is illustrated in \fig{lundspeccomp} (left and right panel, respectively), where we present the particle energy distribution at the time when the electric energy peaks (as indicated by the colored vertical dotted lines in \fig{lundtimecomp}). In the main panels we plot $\gamma dN/d\gamma$, to emphasize the particle content, whereas the insets show $\gamma^2 dN/d\gamma$, to highlight the energy census. The particle spectrum shows a pronounced non-thermal component in all the cases, regardless of whether the secondary plasmoid instability is triggered or not in the current sheets (the results presented in \fig{lundspeccomp} correspond to the cases displayed in \fig{lundfluidcomp}). This suggests that  in our setup any acceleration mechanism that relies on plasmoid mergers is not very important, but rather that the dominant source of energy is direct acceleration by the reconnection electric field, in analogy to what we had demonstrated for the solitary X-point collapse and the ABC instability.

 At the time when the electric energy peaks, most of the particles are still in the thermal component (at $\gamma\sim 1$), i.e., bulk heating is  negligible. Yet, a dramatic event of particle acceleration is taking place, with a few lucky particles accelerated much beyond the mean energy per particle $\sim\gamma_{\rm th}\sigmain/2$ (for comparison, we point out that $\gamma_{\rm th}\sim 1$  and $\sigmain=11$ for the cases in the left panel). At fixed $\sigmain=11$ (left panel), we see that the high-energy spectral cutoff scales as $\gammamax\propto \rj$ ($\rj$ changes by a factor of two in between each pair of curves).\footnote{The fact that the dependence is a little shallower than linear is due to the fact that the reconnection rate, which enters in the full expression $\gammamax\propto v_{\rm rec}^2\sigmain(\rj/\rhot)$, slightly decreases with increasing $\rj/\rhot$, as detailed above.} On the other hand, at fixed $\rj/\rhot=61$ (right panel), we find that $\gammamax\propto \sigmain$ ($\sigmain$ changes by a factor of four in between each pair of curves). 

The spectral hardness is primarily controlled by the average in-plane magnetization $\sigmain$. The right panel in \fig{lundspeccomp} shows that at fixed  $\rj/\rhot$ the spectrum becomes systematically harder with increasing $\sigmain$, approaching the asymptotic shape $\gamma dN/d\gamma\propto \rm const$ found for plane-parallel steady-state reconnection in the limit of high magnetizations \citep[][]{2014ApJ...783L..21S,2015ApJ...806..167G,2016ApJ...816L...8W}. At large $\rj/\rhot$, the hard spectrum of the high-$\sigmain$ cases will necessarily run into constraints of energy conservation (see \eq{ggmax}), unless the pressure feedback of the accelerated particles onto the flow structure ultimately leads to a spectral softening (in analogy to the case of cosmic ray modified shocks, see \citealt{amato_06}). This argument seems to be supported by the left panel in \fig{lundspeccomp}. At fixed $\sigmain$, the left panel shows that the spectral slope is nearly insensitive to $\rj/\rhot$, but larger systems seem to lead to steeper slopes, which possibly reconciles the increase in $\gammamax$ with the argument of energy conservation illustrated in \eq{ggmax}.

%In application to the GeV flares of the Crab Nebula, which we attribute to the dynamical phase of rope merger, we envision an optimal value of $\sigmain$ between $\sim 10$ and $\sim 100$. Based on our results, smaller $\sigmain\lesssim 10$ would correspond to smaller reconnection speeds (in units of the speed of light), and so weaker accelerating electric fields. On the other hand, $\sigmain\gtrsim 100$ would give hard spectra with slopes $s<2$, which would prohibit particle acceleration up to $\gammamax\gg \gamma_{\rm th}$ without violating energy conservation (for the sake of simplicity, here we ignore the potential spectral softening at high $\sigmain$ and large $\rj/\rhot$ discussed above).

%%%%%%%%%%%%%%%%%%%%%%%%%%%%%
\begin{figure}
\centering
\includegraphics[width=.49\textwidth]{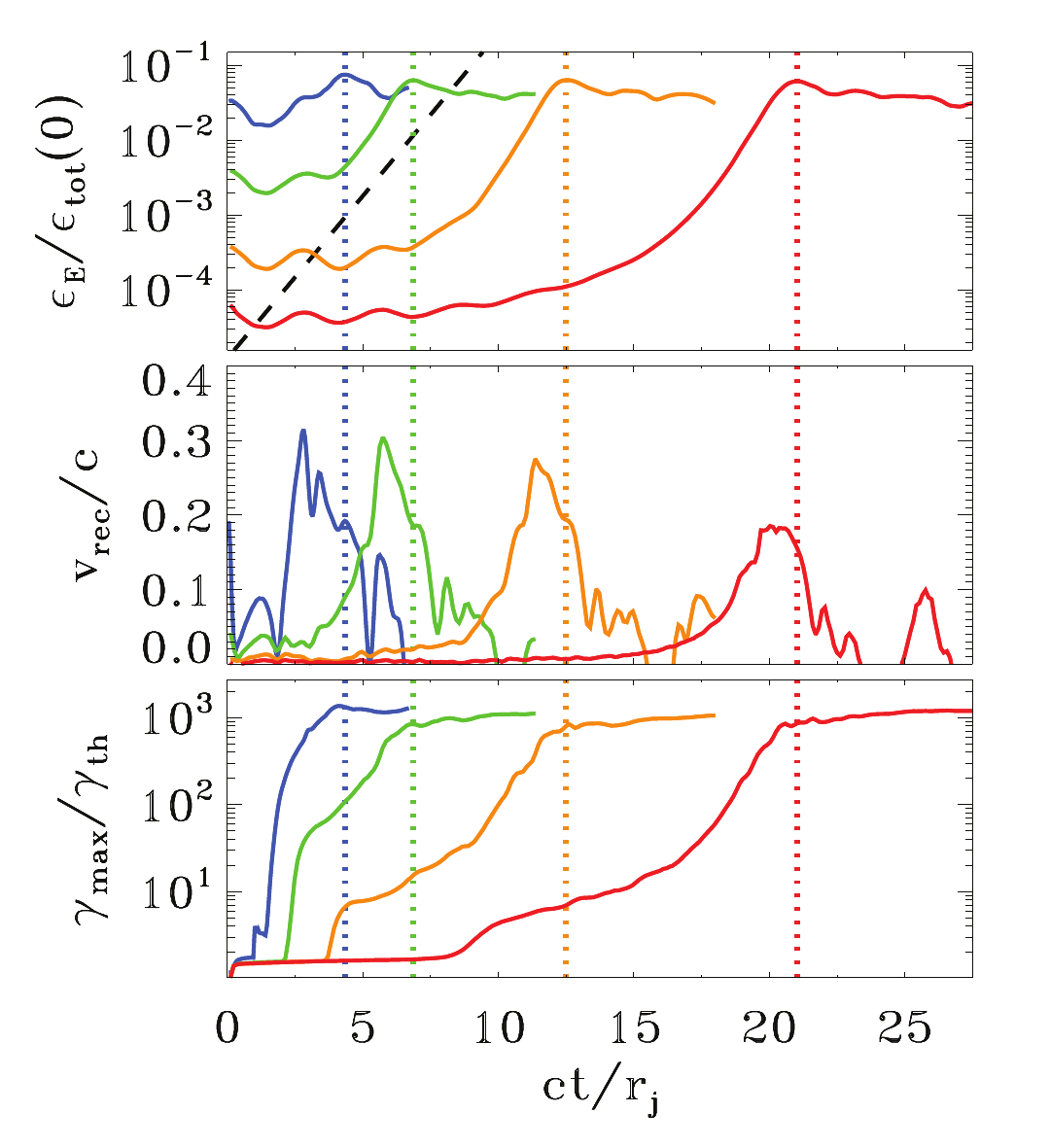} 
\caption{Temporal evolution of the electric energy (top panel; in units of the total initial energy), of the reconnection rate (middle panel; defined as the mean inflow velocity in a square of side $\rj/2$ centered at $x=y=0$) and of the maximum particle Lorentz factor (bottom panel; $\gammamax$ is defined in \eq{ggmax}, and it is normalized to the thermal Lorentz factor $\gamma_{\rm th}\simeq 1+(\hat{\gamma}-1)^{-1} kT/m c^2$), for a suite of four PIC simulations of Lundquist ropes with fixed $kT/mc^2=10^{-4}$, $\sigmain=43$ and $\rj/\rhot=61$, but different magnitudes of the initial velocity: $v_{\rm push}/c=3\times10^{-1}$ (blue), $v_{\rm push}/c=10^{-1}$ (green), $v_{\rm push}/c=3\times 10^{-2}$ (yellow) and $v_{\rm push}/c=10^{-2}$ (red). The dashed black line in the top panel shows that the electric energy grows exponentially as $\propto \exp{(ct/\rj)}$. The vertical dotted lines mark the time when the electric energy peaks (colors correspond to the four values of $v_{\rm push}/c$, as described above).}
\label{fig:lundveltime} 
\end{figure}
\begin{figure}
\centering
\includegraphics[width=.49\textwidth]{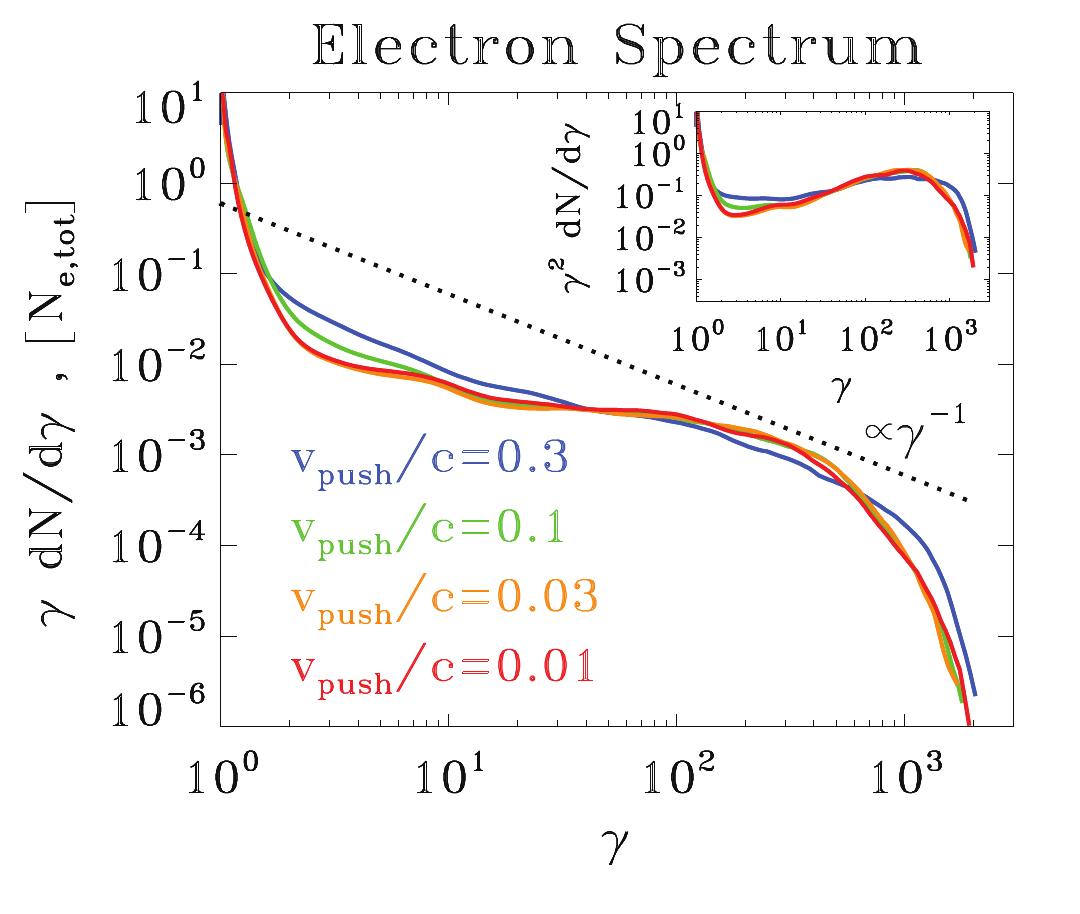} 
\caption{Particle spectrum at the time when the electric energy peaks, for a suite of four PIC simulations of Lundquist ropes with fixed $kT/mc^2=10^{-4}$, $\sigmain=43$ and $\rj/\rhot=61$, but different magnitudes of the initial velocity (same runs as in \fig{lundveltime}): $v_{\rm push}/c=3\times10^{-1}$ (blue), $v_{\rm push}/c=10^{-1}$ (green), $v_{\rm push}/c=3\times 10^{-2}$ (yellow) and $v_{\rm push}/c=10^{-2}$ (red). The main plot shows $\gamma dN/d\gamma$ to emphasize the particle content, whereas the inset presents $\gamma^2 dN/d\gamma$ to highlight the energy census. The dotted black line is a power law $\gamma dN/d\gamma\propto \gamma^{-1}$, corresponding to equal energy content per decade (which would result in a flat distribution in the inset). The particle spectrum is remarkably independent from the initial $v_{\rm push}$.}
\label{fig:lundvelspec} 
\end{figure}
\subsubsection{Dependence on the collision speed}
So far, we have investigated the merger of Lundquist flux ropes with a prescribed collision speed $v_{\rm push}/c=0.1$. In Figs.~\fign{lundveltime} and \fign{lundvelspec}, we explore the effect of different values of the driving speed on the temporal evolution of the merger and the resulting particle spectrum, for a suite of four PIC simulations with fixed $kT/mc^2=10^{-4}$, $\sigmain=43$ and $\rj/\rhot=61$, but different magnitudes of $v_{\rm push}$: $v_{\rm push}/c=3\times10^{-1}$ (blue), $v_{\rm push}/c=10^{-1}$ (green), $v_{\rm push}/c=3\times 10^{-2}$ (yellow) and $v_{\rm push}/c=10^{-2}$ (red).

\fig{lundveltime} shows the evolution of the electric energy (top panel, in units of the total initial energy). The initial value of the electric energy  scales as $\vpush^2$, just as a consequence of the electric field $-\bmath{v_{\rm push}}\times \bmath{B}/c$ that we initialize in the magnetic ropes. However, the subsequent evolution is remarkably independent of $\vpush$, apart from an overall  shift in the onset time. In all the cases, the electric energy grows exponentially as $\propto \exp{(ct/\rj)}$ (compare with the black dashed line) until it peaks at $\sim 5-10\%$ of the total initial energy (the time when the electric energy peaks is indicated with the vertical colored dotted lines). The peak of electric energy corresponds to the most violent phase of merger, and it shortly follows the peak time of the reconnection rate. As shown in the middle panel, the reconnection speed peaks at $v_{\rm rec}/c\sim 0.2-0.3$, with only a weak dependence on the driving speed $v_{\rm push}$.
Driven by fast reconnection during the dynamical merger event, the maximum particle energy $\gammamax$ grows explosively (bottom panel in \fig{lundveltime}), ultimately reaching $\gammamax/\gamma_{\rm th}\sim10^3$ regardless of the initial driving speed $v_{\rm push}$. The initial phase of growth of $\gammamax$ is sensitive to the value of $v_{\rm push}$, reaching higher values for larger $v_{\rm push}$ (compare the plateau at $\gammamax//\gamma_{\rm th}\sim 10^2$ in the green line at $ct/\rj\sim 3$ with the corresponding plateau at $\gammamax//\gamma_{\rm th}\sim 5$ in the red line at $ct/\rj\sim 12$). In contrast, the temporal profile of $\gammamax$ during the merger (i.e., shortly before the time indicated by the vertical dotted lines) is nearly indentical in all the cases (apart from an overall time shift). 
Overall, the similarity of the different curves in \fig{lundveltime} confirms that the evolution during the dynamical merger event is driven by self-consistent large-scale electromagnetic stresses, independently of the initial driving speed. 

As a consequence, it is not surprising that the particle spectrum measured at the time when the electric energy peaks (as indicated by the vertical colored lines in \fig{lundveltime}) bears no memory of the driving speed $\vpush$. In fact, the four curves in \fig{lundvelspec} nearly overlap (with the only marginal exception of the case with the largest $v_{\rm push}/c=0.3$, in blue).

%%%%%%%%%%%%%%%%%%%%%%%%%%%%%%%%%%%%
\begin{figure}
\centering
\includegraphics[width=.34\textwidth]{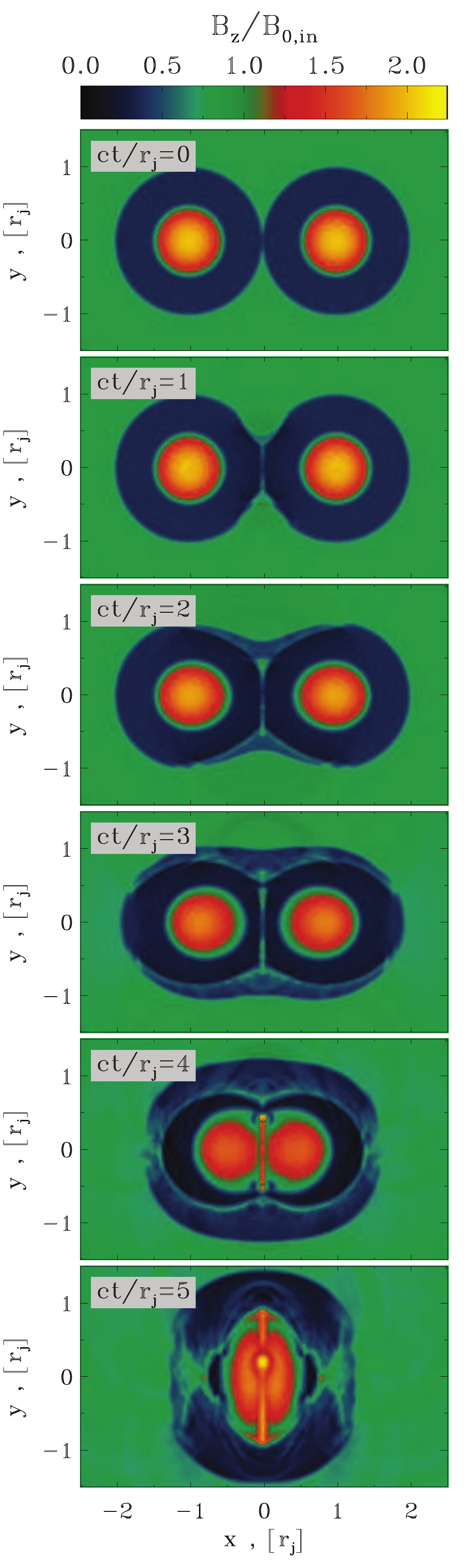} 
\includegraphics[width=.34\textwidth]{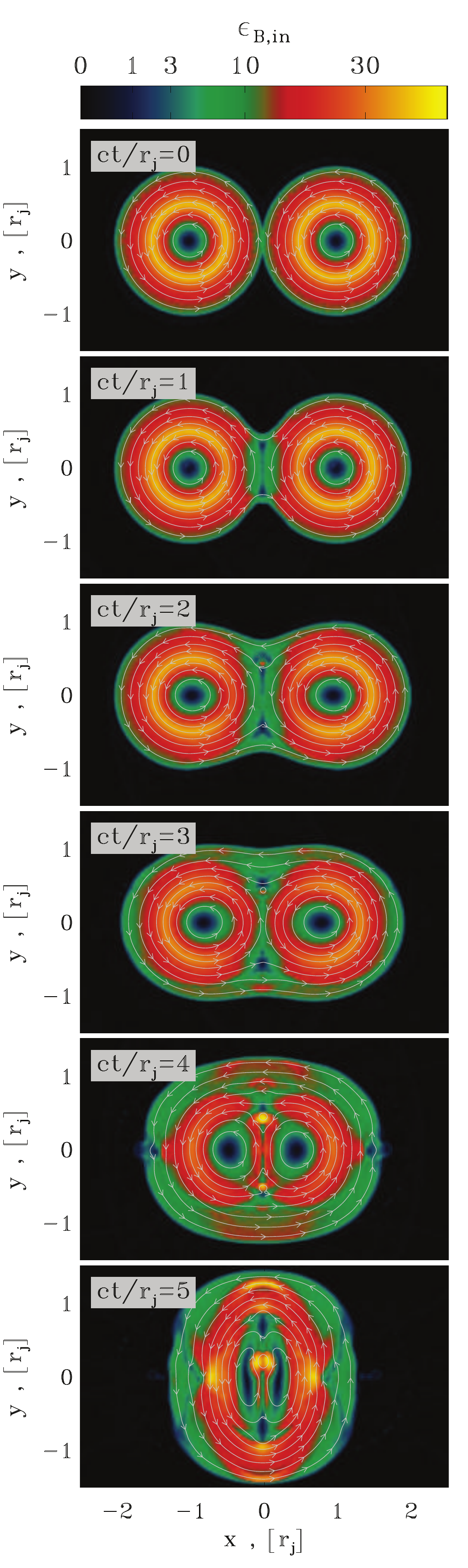} 
\caption{Temporal evolution of 2D core-envelope ropes (time is measured in $c/\rj$ and indicated in the grey box of each panel, increasing from top to bottom). The plot presents the 2D pattern of the out-of-plane field $B_z$ (left column; in units of $B_{0,\rm in}$) and of the in-plane magnetic energy fraction $\epsilon_{B,\rm in}=(B_x^2+B_y^2)/8 \pi n m c^2$ (right column; with superimposed magnetic field lines), from a PIC simulation with $kT/mc^2=10^{-4}$, $\sigmain=43$ and $\rj=61\,\rhot$, performed within a rectangular domain of size $10\rj\times 6\rj$ (but we only show a small region around the center). Spontaneous reconnection at the interface between the two flux ropes (i.e., around $x=y=0$) creates an envelope of field lines engulfing the two islands, whose tension force causes them to merge on a dynamical time. This figure should be compared with the force-free result in Fig. \protect\ref{fig:CEOverview}.}
\label{fig:corefluid} 
\end{figure}
\begin{figure}
\centering
\includegraphics[width=.49\textwidth]{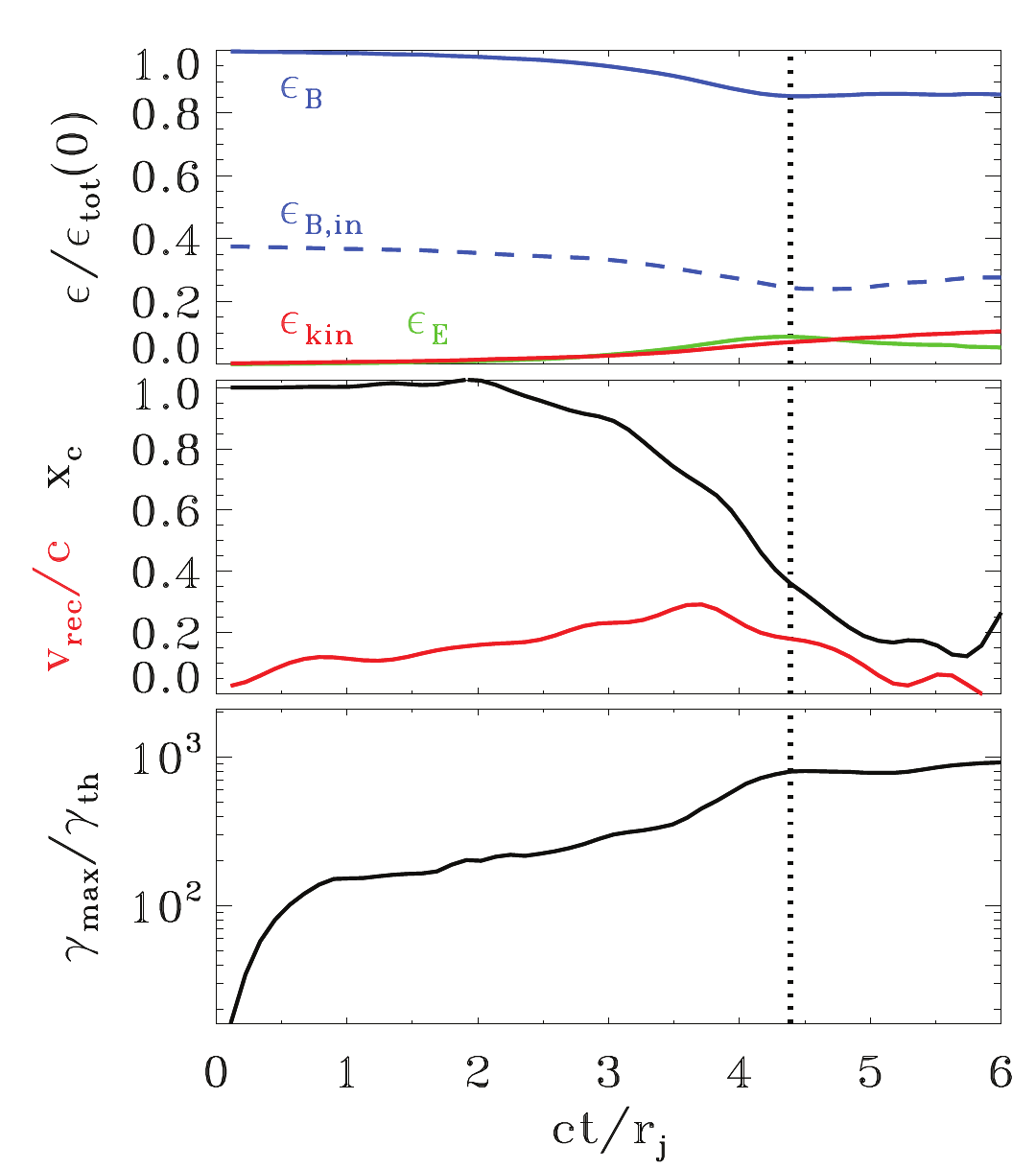} 
\caption{Temporal evolution of various quantities, from a 2D PIC simulation of core-envelope ropes with $kT/mc^2=10^{-4}$, $\sigmain=43$ and $\rj=61\,\rhot$ (the same as in \fig{corefluid}), performed within a rectangular domain of size $5\rj\times 3\rj$. Top panel: fraction of energy in magnetic fields (solid blue), in-plane magnetic fields (dashed blue), electric fields (green) and particles (red; excluding the rest mass energy), in units of the total initial energy. Middle panel: reconnection rate $v_{\rm rec}/c$ (red), defined as the  inflow speed along the $x$ direction averaged over a square of side equal to $\rj/2$ centered at $x=y=0$; and location $x_{\rm c}$ of the core of the rightmost flux rope (black), in units of $\rj$.
Bottom panel: evolution of the maximum Lorentz factor $\gammamax$, as defined in \eq{ggmax}, relative to the thermal Lorentz factor $\gamma_{\rm th}\simeq 1+(\hat{\gamma}-1)^{-1} kT/m c^2$, which for our case is $\gamma_{\rm th}\simeq 1$. Spontaneous reconnection at the interface between the two islands drives the early increase in $\gammamax$ up to $\gammamax/\gamma_{\rm th}\sim 20$, before stalling. At this stage, the cores of the two islands have not significantly moved (black line in the middle panel). At $ct/\rj\sim 3$, the tension force of the common envelope of field lines starts pushing the two islands toward each other (and the $x_{\rm c}$ location of the rightmost rope decreases, see the middle panel), resulting in a merger event occurring on a dynamical timescale. During the merger (at $ct/\rj\sim 4$), the reconnection rate peaks (red line in the middle panel), a fraction of the in-plane magnetic energy is transferred to the plasma (compare dashed blue and solid red lines in the top panel), and particles are quickly accelerated up to $\gammamax/\gamma_{\rm th}\sim 10^3$ (bottom panel). In all the panels, the vertical dotted black line indicates the time when the electric energy peaks, shortly after the most violent phase of the merger event.}
\label{fig:coretime} 
\end{figure}
\begin{figure}
\centering
\includegraphics[width=.49\textwidth]{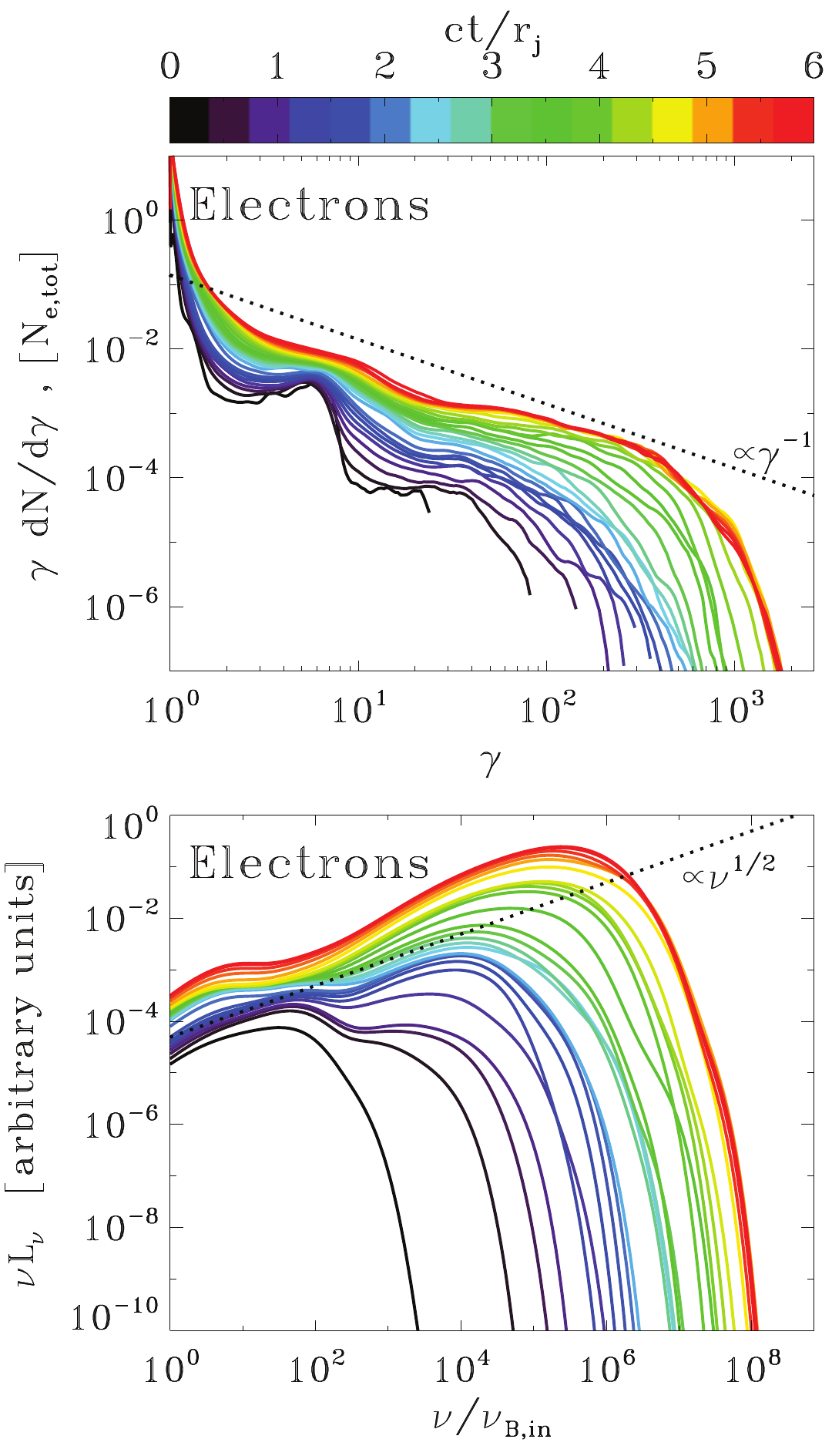} 
\caption{Particle energy spectrum and synchrotron spectrum from a 2D PIC simulation of core-envelope ropes with $kT/mc^2=10^{-4}$, $\sigmain=43$ and $\rj=61\,\rhot$ (the same as in \fig{corefluid} and \fig{coretime}), performed within a  domain of size $10\rj\times 6\rj$.  Time is measured in units of $\rj/c$, see the colorbar at the top. Top panel: evolution of the electron energy spectrum normalized to the total number of electrons. At late times, the high-energy spectrum approaches a distribution with $\gamma dN/d\gamma\propto \gamma^{-1}$, corresponding to equal energy content in each decade of $\gamma$ (compare with the dotted black line).  Bottom panel: evolution of the angle-averaged synchrotron spectrum emitted by electrons. The frequency on the horizontal axis is in units of $\nu_{B,\rm in}=\sqrt{\sigmain}\omega_{\rm p}/2\pi$. At late times, the synchrotron spectrum approaches a power law with $\nu L_\nu\propto \nu^{1/2}$ (compare with the dotted black line), as expected from an electron spectrum $\gamma dN/d\gamma\propto \gamma^{-1}$. 
}
\label{fig:corespec} 
\end{figure}
\begin{figure}
\centering
\includegraphics[width=.49\textwidth]{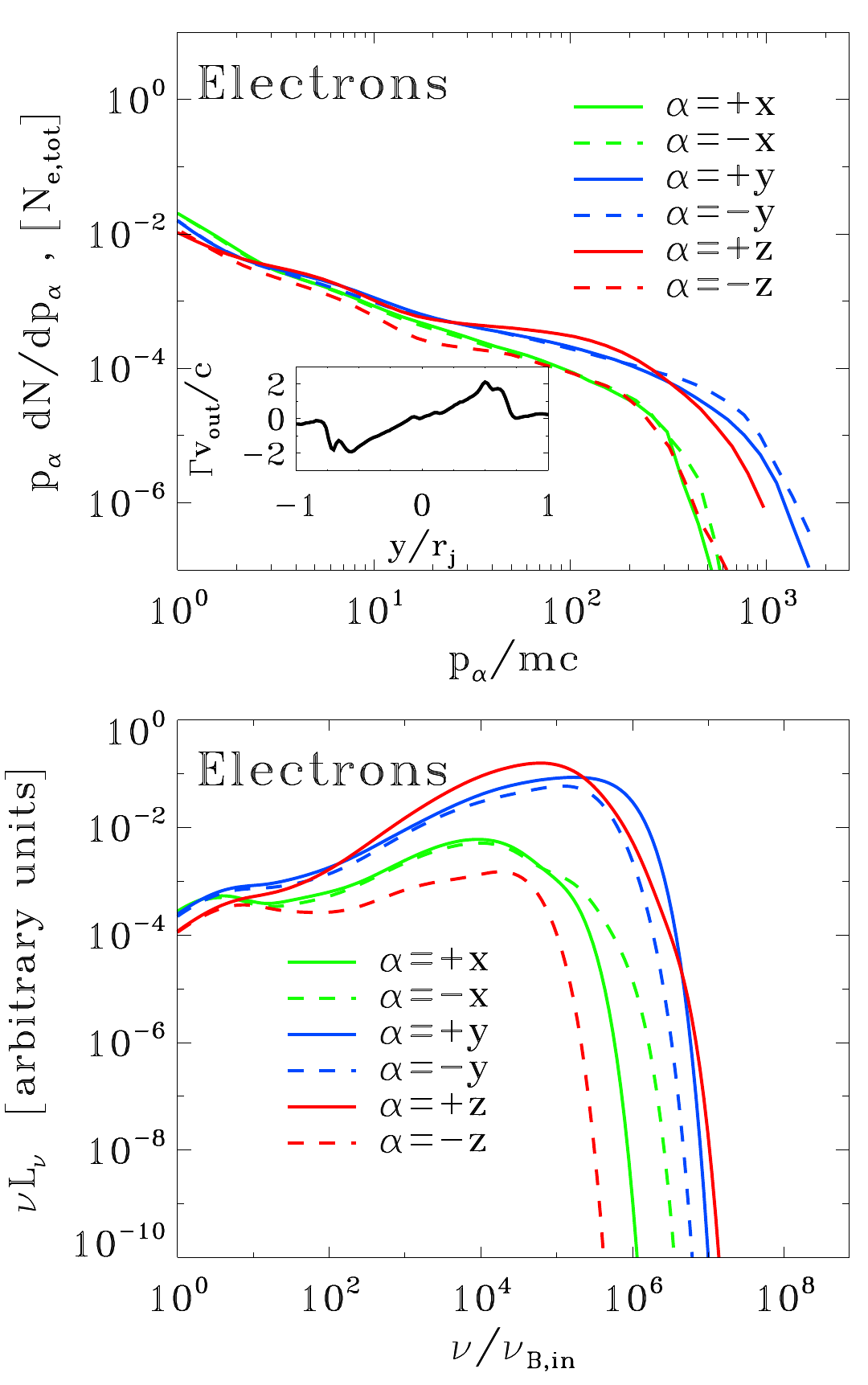} 
\caption{Particle momentum spectrum and anisotropy of the synchrotron emission from a 2D PIC simulation of core-envelope ropes with $kT/mc^2=10^{-4}$, $\sigmain=43$ and $\rj=61\,\rhot$ (the same as in \fig{corefluid}, \fig{coretime} and \fig{corespec}), performed within a  domain of size $10\rj\times 6\rj$. Top panel: electron momentum spectrum along different directions (as indicated in the legend), at the time when the electric energy peaks (as indicated by the vertical dotted black line in \fig{coretime}). The highest energy electrons are beamed along the direction $y$ of the reconnection outflow (blue lines) and along the direction $+z$ of the accelerating electric field (red solid line; positrons will be beamed along $-z$, due to the opposite charge).  The inset shows the 1D profile along $y$ of the bulk four-velocity in the outflow direction (i.e., along $y$), measured at $x=0$. 
Bottom panel: synchrotron spectrum at the time indicated in \fig{coretime} (vertical dotted black line) along different directions (within a solid angle of $\Delta \Omega/4\pi\sim 3\times10^{-3}$), as indicated in the legend. The resulting  anisotropy of the synchrotron emission is consistent with the particle anisotropy illustrated in the top panel.
}
\label{fig:corespecmom} 
\end{figure}
\subsection{PIC simulations of 2D flux tube mergers: core-envelope ropes}
The temporal evolution of the merger of two core-envelope ropes is shown in \fig{corefluid}, which should be compared with the force-free result in Fig. \ref{fig:CEOverview}.
The plot presents the 2D pattern of the out-of-plane field $B_z$ (left column; in units of $B_{0,\rm in}$) and of the in-plane magnetic energy fraction $\epsilon_{B,\rm in}=(B_x^2+B_y^2)/8 \pi n m c^2$ (right column; with superimposed magnetic field lines), from a PIC simulation with $kT/mc^2=10^{-4}$, $\sigmain=43$ and $\rj=61\,\rhot$. Time is measured in units of $\rj/c$ and indicated in the grey boxes within each panel. 

For the core-envelope geometry, the initial in-plane magnetic field is discontinuous at the interface $x=0$ between the two magnetic ropes. There, magnetic reconnection spontaneously starts since early times (in contrast to the case of Lundquist ropes, where the system needs to be driven by hand with a prescribed velocity push). The ongoing steady-state reconnection is manifested in \fig{corefluid} by the formation and subsequent ejection of small-scale plasmoids, as the current sheet at $x=0$ stretches up to a length $\sim 2\rj$. So far (i.e., until $ct/\rj\sim 2$), the cores of the two islands have not significantly moved (black line in the middle panel of \fig{coretime}, indicating the $x_{\rm c}$ location of the center of the rightmost island), the reconnection speed is small (around $v_{\rm rec}/c\sim 0.1$, red line in the middle panel of \fig{coretime}) and no significant energy exchange has occurred from the fields to the particles (compare the in-plane magnetic energy, shown by  the dashed blue line in the top panel of \fig{coretime}, with the particle kinetic energy, indicated with the red line).\footnote{As we have also pointed out for the case of Lundquist ropes, we remark that the energy balance described in the top panel of \fig{coretime} is dependent on the overall size of the box, everything else being the same. More specifically, for a larger box the relative ratios between $\epsilon_{\rm kin}$ (red line; or $\epsilon_{\rm E}$, green line) and $\epsilon_{B,\rm in}$ (dashed blue line) will not change, whereas they will all decrease as compared to the total magnetic energy $\epsilon_{B}$ (blue solid line).} The only significant evolution before $ct/\rj\sim 2$ is the reconnection-driven increase in the maximum particle Lorentz factor up to $\gammamax/\gamma_{\rm th}\sim 150$ occurring at $ct/\rj\sim 0.5$ (see the black line in the bottom panel of \fig{coretime}).

As a result of reconnection, an increasing number of field lines, that initially closed around one of the ropes, are now forming a common envelope around both magnetic islands. In analogy to the case of Lundquist ropes, the magnetic tension force causes the two islands to approach and merge on a quick (dynamical) timescale, starting at $ct/\rj\sim 3$ and ending at $ct/\rj\sim 5$ (see that the distance of the rightmost island from the center rapidly decreases, as indicated by the black line in the middle panel of \fig{coretime}). The tension force drives the particles in the flux ropes toward the center, with a fast reconnection speed peaking at $v_{\rm rec}/c\sim 0.3$ (red line in the middle panel of \fig{coretime}).\footnote{This value of the reconnection rate is roughly comparable to the results of Lundquist ropes.} In the reconnection layer at $x=0$, it is primarily the in-plane field that gets dissipated (compare the dashed and solid blue lines in the top panel of \fig{coretime}), driving an increase in the electric energy (green) and in the particle kinetic energy (red). In this phase of evolution, the fraction of initial energy released to the particles is small ($\epsilon_{\rm kin}/\epsilon_{\rm tot}(0)\sim 0.1$), but the particles advected into the central X-point experience a dramatic episode of acceleration. As shown in the bottom panel of \fig{coretime}, the cutoff Lorentz factor $\gammamax$ of the particle spectrum presents a dramatic evolution, increasing up to $\gammamax/\gamma_{\rm th}\sim 10^3$ within a couple of dynamical times. It is this phase of extremely fast particle acceleration that we associate with the generation of the Crab flares.

The two distinct evolutionary phases --- the early stage governed by steady-state reconnection at the central current sheet, and the subsequent dynamical merger driven by large-scale electromagnetic stresses --- are clearly apparent in the evolution of the particle energy spectrum (top panel in \fig{corespec}) and of the angle-averaged synchrotron emission (bottom panel in \fig{corespec}). Steady-state reconnection in the central current sheet leads to fast particle acceleration at early times (from black to light blue in the top panel).\footnote{We point out that the spectral bump at $\gamma\sim 5\sim\, 0.5\sigmain$, which is apparent since early times and later evolves up to $\gamma\sim 10$, is an artifact of the core-envelope solution. In this geometry, both the azimuthal field and the poloidal field are discontinuous at the boundary of the ropes. The discontinuity has to be sustained by the particles (everywhere around each rope), via their pressure and electric current. The system self-consistently builds up the pressure and current by energizing a few particles, thereby producing the bump at $\gamma\sim 5$. } The particle energy spectrum then freezes (see the clustering of the light blue and cyan lines, and compare with the phase at $1\lesssim ct/\rj\lesssim 2.5$ in the bottom panel of \fig{coretime}). Correspondingly, the angle-averaged synchrotron spectrum stops evolving (see the clustering of the light blue and cyan lines in the bottom panel). A second dramatic increase in the particle and emission spectral cutoff (even more dramatic than the initial growth) occurs between $ct/\rj\sim 3.5$ and $ct/\rj\sim 4.5$ (green to yellow curves in \fig{corespec}), and it directly corresponds to the dynamical merger of the two magnetic ropes, driven by large-scale stresses. The particle spectrum quickly extends up to $\gammamax\sim 10^3$ (yellow lines in the top panel), and correspondigly the peak of the $\nu L_\nu$ emission spectrum shifts up to $\sim \gammamax^2\nu_{B,\rm in}\sim10^6 \nu_{B,\rm in}$ (yellow lines in the bottom panel). The system does not show any sign of evolution at times later than $c/\rj\sim 5$ (see the clustering of the yellow to red lines). At late times, the high-energy spectrum approaches a distribution of the form $\gamma dN/d\gamma\propto \gamma^{-1}$ corresponding to equal energy content in each decade of $\gamma$ (compare with the dotted black line in the top panel). Consequently, the synchrotron spectrum approaches a power law with $\nu L_\nu\propto \nu^{1/2}$  (compare with the dotted black line in the bottom panel).

The particle distribution  is significantly anisotropic. In the top panel of \fig{corespecmom}, we plot the electron momentum spectrum at the time when the electric energy peaks (see the vertical black dotted line in \fig{coretime}) along different directions, as indicated in the legend. The highest energy electrons are beamed along the direction $y$ of the reconnection outflow (blue lines) and along the direction $+z$ anti-parallel to the accelerating electric field (red solid line; positrons will be beamed along $-z$, due to the opposite charge). This is consistent with our results for solitary X-point collapse in Paper I and for merger of Lundquist ropes presented above. Most of the anisotropy is to be attributed to the ``kinetic beaming'' discussed by \citet{2012ApJ...754L..33C}, rather than beaming associated with the bulk motion (which is only marginally relativistic, see the inset in the top panel of \fig{corespecmom}).  The   anisotropy of the synchrotron emission (bottom panel in \fig{corespecmom}) is consistent with the particle anisotropy illustrated in the top panel.

%%%%%%%%%%%%%%%%%%%%%%%%%%%%%
\begin{figure}
\centering
\includegraphics[width=.79\textwidth]{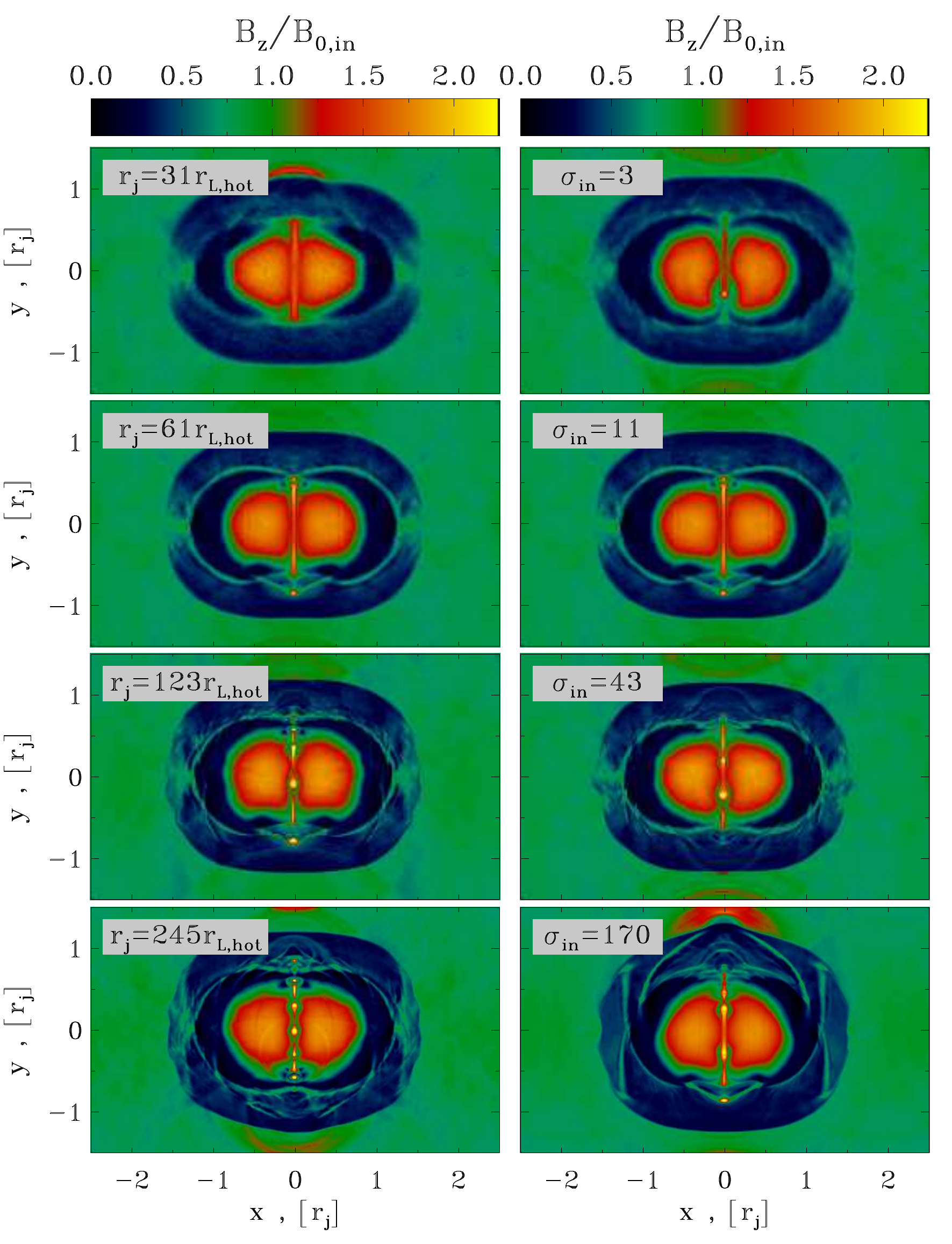} 
\caption{2D pattern of the out-of-plane field $B_z$ (in units of $B_{0,\rm in}$) at the most violent time of the merger of core-envelope ropes (i.e., when the electric energy peaks, as indicated by the vertical dotted lines in \fig{coretimecomp}) from a suite of PIC simulations. In the left column, we fix $kT/mc^2=10^{-4}$ and $\sigmain=11$ and we vary the ratio $\rj/\rhot$, from 31 to 245 (from top to bottom). In the right column, we fix $kT/mc^2=10^{-4}$ and $\rj/\rhot=61$ and we vary the magnetization $\sigmain$, from 3 to 170 (from top to bottom). In all cases, the simulation box is a rectangle of size $5\rj\times 3\rj$. The 2D structure of $B_z$ in all cases is quite similar, apart from the fact that larger $\rj/\rhot$ tend to lead to a more pronounced fragmentation of the current sheet.}
\label{fig:corefluidcomp} 
\end{figure}
\begin{figure}
\centering
\includegraphics[width=.79\textwidth]{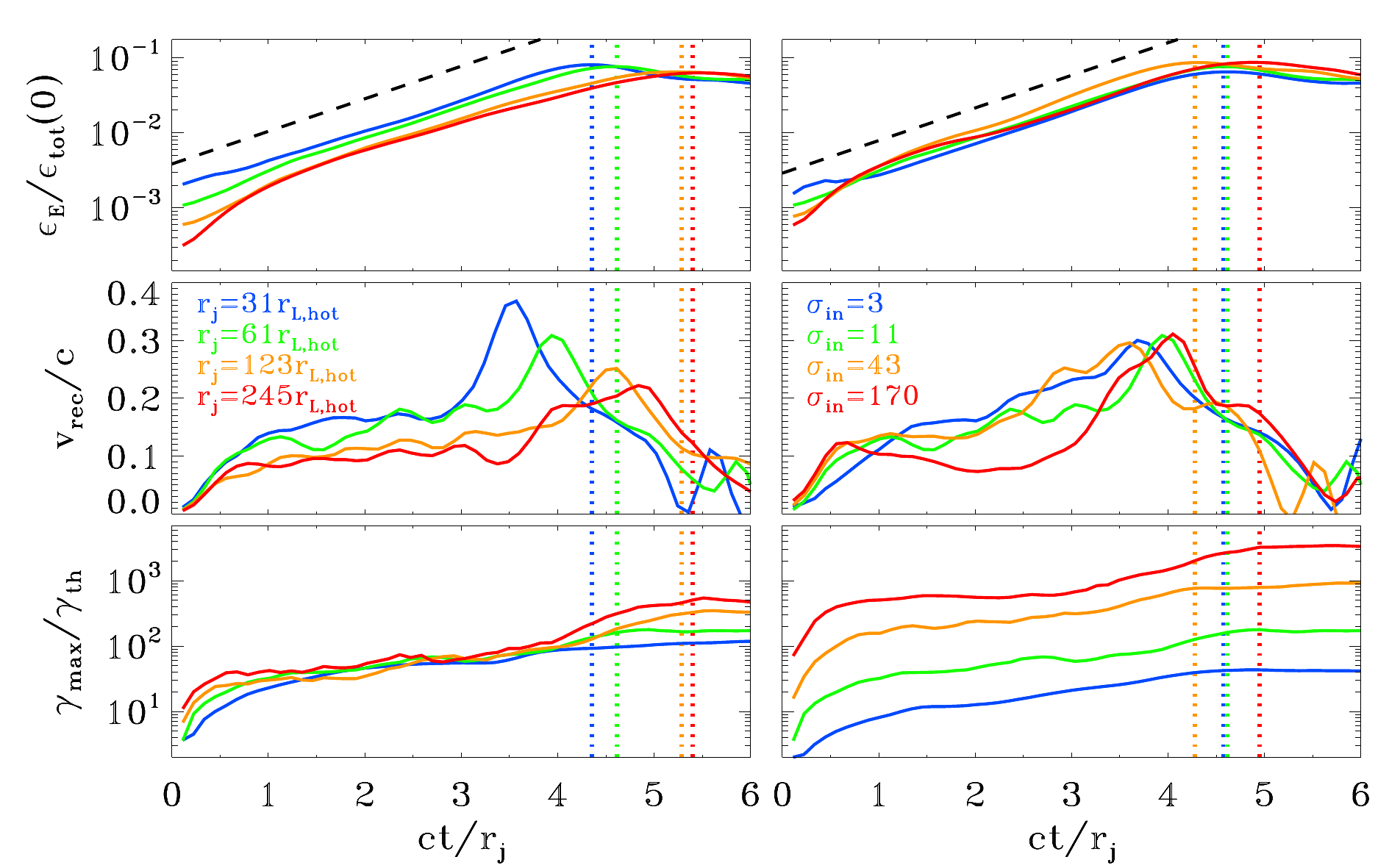} 
\caption{Temporal evolution of the electric energy (top panel; in units of the total initial energy), of the reconnection rate (middle panel; defined as the mean inflow velocity in a square of side $\rj/2$ centered at $x=y=0$) and of the maximum particle Lorentz factor (bottom panel; $\gammamax$ is defined in \eq{ggmax}, and it is normalized to the thermal Lorentz factor $\gamma_{\rm th}\simeq 1+(\hat{\gamma}-1)^{-1} kT/m c^2$), for a suite of PIC simulations of core-envelope ropes (same runs as in \fig{corefluidcomp}). In the left column, we fix $kT/mc^2=10^{-4}$ and $\sigmain=11$ and we vary the ratio $\rj/\rhot$ from 31 to 245 (from blue to red, as indicated in the legend). In the right column, we fix $kT/mc^2=10^{-4}$ and $\rj/\rhot=61$ and we vary the magnetization $\sigmain$ from 3 to 170 (from blue to red, as indicated in the legend). The maximum particle energy $\gammamax mc^2$ resulting from the merger increases for increasing $\rj/\rhot$ at fixed $\sigmain$ (left column) and for increasing $\sigmain$ at fixed $\rj/\rhot$.
The dashed black line in the top panel shows that the electric energy grows exponentially as $\propto \exp{(ct/\rj)}$. The vertical dotted lines mark the time when the electric energy peaks (colors as described above).}
\label{fig:coretimecomp} 
\end{figure}
\begin{figure}
\centering
\includegraphics[width=.79\textwidth]{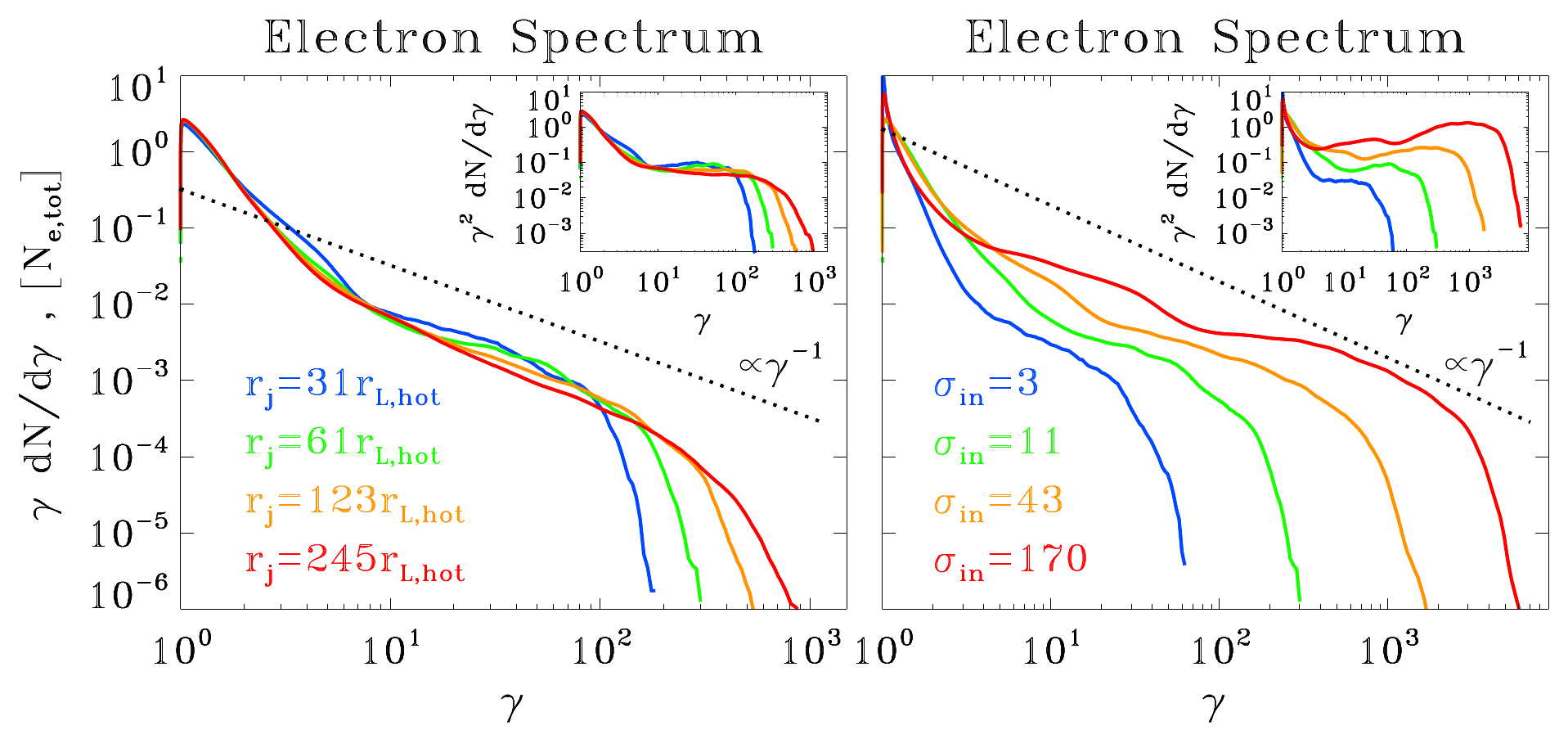} 
\caption{Particle spectrum at the time when the electric energy peaks, for a suite of PIC simulations of core-envelope ropes (same runs as in \fig{corefluidcomp} and \fig{coretimecomp}).  In the left column, we fix $kT/mc^2=10^{-4}$ and $\sigmain=11$ and we vary the ratio $\rj/\rhot$ from 31 to 245 (from blue to red, as indicated by the legend).  In the right column, we fix $kT/mc^2=10^{-4}$ and $\rj/\rhot=61$ and we vary the magnetization $\sigmain$ from 3 to 170 (from blue to red, as indicated by the legend). The main plot shows $\gamma dN/d\gamma$ to emphasize the particle content, whereas the inset presents $\gamma^2 dN/d\gamma$ to highlight the energy census. The dotted black line is a power law $\gamma dN/d\gamma\propto \gamma^{-1}$, corresponding to equal energy content per decade (which would result in a flat distribution in the insets). The spectral hardness is not a sensitive function of the ratio $\rj/\rhot$, but it is strongly dependent on $\sigmain$, with higher magnetizations giving harder spectra, up to the saturation slope of $-1$.}
\label{fig:corespeccomp} 
\end{figure}
\subsubsection{Dependence on the flow conditions}
We now investigate the dependence of our results on the magnetization $\sigmain$ and the ratio $\rj/\rhot$, where $\rhot=\sqrt{\sigmain}\comp$. In \fig{lundfluidcomp}, we present the 2D pattern of the out-of-plane field $B_z$ (in units of $B_{0,\rm in}$) during the most violent phase of rope merger (i.e., when the electric energy peaks, as indicated by the vertical dotted lines in \fig{coretimecomp}) from a suite of PIC simulations in a rectangular domain of size $5\rj\times 3\rj$. In the left column, we fix $kT/mc^2=10^{-4}$ and $\sigmain=11$ and we vary the ratio $\rj/\rhot$, from 31 to 245 (from top to bottom). In the right column, we fix $kT/mc^2=10^{-4}$ and $\rj/\rhot=61$ and we vary the magnetization $\sigmain$, from 3 to 170 (from top to bottom).  

The 2D pattern of $B_z$ presented in \fig{corefluidcomp} shows that the merger proceeds in a similar way in all the runs. The only difference is that  larger $\rj/\rhot$ lead to thinner current sheets, when fixing $\sigmain$ (left column in \fig{corefluidcomp}). In the right column, with $\rj/\rhot$ fixed, the thickness of the current sheet  ($\sim \rhot$) is a fixed fraction of the box size. In contrast, in the left column, the ratio of current sheet thickness to box size will scale as $\rhot/\rj$, as indeed it is observed. The tendency for fragmentation into secondary plasmoids is known to be an increasing function of the length-to-thickness ratio \citep[e.g.,][]{uzdensky_10,2016ApJ...816L...8W}. It follows that all the cases in the right column will display a similar tendency for fragmentation (and in particular, they do not appreciably fragment), whereas the likelihood of fragmentation is expected to increase from top to bottom in the left column. In fact, for the case with $\rj/\rhot=245$ (left bottom panel), a number of small-scale plasmoids appear in the current sheets. We find that as long as $\sigmain\gg1$, the secondary tearing mode discussed by \citet{uzdensky_10} --- that leads to current sheet fragmentation --- appears at $\rj/\rhot\gtrsim 100$, in the case of core-envelope ropes.\footnote{A similar result had been found for the case of ABC collapse, see Sect.\ref{unstr-latt-pic}, and for the merger of Lundquist ropes.}

In \fig{coretimecomp} we present the temporal evolution of the runs whose 2D structure is shown in \fig{corefluidcomp}. In the left column, we fix $kT/mc^2=10^{-4}$ and $\sigmain=11$ and we vary the ratio $\rj/\rhot$, from 31 to 245 (from blue to red, as indicated in the legend of the middle panel). In the right column, we fix $kT/mc^2=10^{-4}$ and $\rj/\rhot=61$ and we vary the magnetization $\sigmain$, from 3 to 170 (from blue to red, as indicated in the legend of the middle panel). 
 The top panels show that the evolution of the electric energy (in units of the total initial energy) is  similar for all the values of $\rj/\rhot$ and $\sigmain$ we explore. In particular, the electric energy grows approximately as $\propto \exp{(ct/\rj)}$ in all the cases (compare with the dashed black lines), and it peaks at $\sim 5-10\%$ of the total initial  energy. The onset time of the instability is also nearly independent of $\sigmain$ (top right panel). As regard to the dependence of the onset time on $\rj/\rhot$ at fixed $\sigmain$, the top left panel in \fig{coretimecomp} shows that larger values of $\rj/\rhot$ tend to grow later, but the variation is only moderate.

The peak reconnection rate in all the cases we have explored is around $v_{\rm rec}/c\sim 0.2-0.3$ (middle row in \fig{coretimecomp}). It is nearly insensitive to $\sigmain$ (middle right panel in \fig{coretimecomp}) and it marginally decreases with increasing $\rj/\rhot$ (but we have verified that it saturates at $v_{\rm rec}/c\sim 0.25$ in the limit $\rj/\rhot\gg1$, see the middle left panel). We had found similar values and trends for the peak reconnection rate in the case of ABC collapse, see Sect.\ref{unstr-latt-pic}, and in the merger of Lundquist ropes. This places our results on solid footing, since our main conclusions do not depend on the specific properties of a given field geometry.

In the evolution of the maximum particle Lorentz factor $\gammamax$ (bottom row in \fig{coretimecomp}), one can distinguish two phases. At early times ($ct/\rj\sim 3$), the increase in $\gammamax$ is moderate, when reconnection proceeds in a steady state fashion in the central region. At later times  ($ct/\rj\sim 4.5$), as the two magnetic ropes merge on a dynamical timescale, the maximum particle Lorentz factor grows explosively. Following the same argument detailed in Sect.\ref{unstr-latt-pic} for the ABC instability, we estimate that the high-energy cutoff of the particle spectrum at the end of the merger  event (which lasts for a few $\rj/c$) should scale as $\gammamax/\gamma_{\rm th}\propto v_{\rm rec}^2\sqrt{\sigmain} \rj\propto v_{\rm rec}^2\sigmain (\rj/\rhot)$. If the reconnection rate does not significantly depend on $\sigmain$, this implies that $\gammamax\propto \rj$ at fixed $\sigmain$. The trend for a steady increase of $\gammamax$ with $\rj$ at fixed $\sigmain$ is confirmed in the bottom left panel of \fig{coretimecomp}, both at the final time and at the peak time of the electric energy (which is slightly different among the four different cases, see the vertical dotted colored lines).\footnote{The fact that the dependence appears slightly sub-linear is due to the fact that the reconnection rate is slightly larger for smaller $\rj/\rhot$.} Similarly, if the reconnection rate does not significantly depend on $\rj/\rhot$, this implies that $\gammamax\propto \sigmain$ at fixed $\rj/\rhot$. This linear dependence of $\gammamax$ on $\sigmain$ is  confirmed in the bottom right panel of \fig{coretimecomp}. Once again, these conclusions parallel closely our findings for the instability of ABC structures and the merger of Lundquist ropes.

The dependence of the particle spectrum on $\rj/\rhot$ and $\sigmain$ is illustrated in \fig{corespeccomp} (left and right panel, respectively), where we present the particle energy distribution at the time when the electric energy peaks (as indicated by the colored vertical dotted lines in \fig{coretimecomp}). In the main panels we plot $\gamma dN/d\gamma$, to emphasize the particle content, whereas the insets show $\gamma^2 dN/d\gamma$, to highlight the energy census. 

 At the time when the electric energy peaks, most of the particles are still in the thermal component (at $\gamma\sim 1$), i.e., bulk heating is  negligible.\footnote{As we have explained above, the low-energy bump at $\gamma\sim\sigmain/2$ visible in the yellow and red curves on the right panel of \fig{corespeccomp} is due to hot particles at the boundaries of the ropes, where the initial fields are sharply discontinuous.} Yet, a dramatic event of particle acceleration is taking place, and the particle spectrum shows a pronounced non-thermal component, with a few lucky particles accelerated much beyond the mean energy per particle $\sim\gamma_{\rm th}\sigmain/2$ (for comparison, we point out that $\gamma_{\rm th}\sim 1$  and $\sigmain=11$ for the cases in the left panel). 
The spectral hardness is primarily controlled by the average in-plane magnetization $\sigmain$. The right panel in \fig{corespeccomp} shows that at fixed  $\rj/\rhot$ the spectrum becomes systematically harder with increasing $\sigmain$, approaching the asymptotic shape $\gamma dN/d\gamma\propto \rm const$ found for plane-parallel steady-state reconnection in the limit of high magnetizations \citep[][]{2014ApJ...783L..21S,2015ApJ...806..167G,2016ApJ...816L...8W}. At large $\rj/\rhot$, the hard spectrum of the high-$\sigmain$ cases will necessarily run into constraints of energy conservation (see \eq{ggmax}), unless the pressure feedback of the accelerated particles onto the flow structure ultimately leads to a spectral softening (in analogy to the case of cosmic ray modified shocks, see \citealt{amato_06}). This argument seems to be supported by the left panel in \fig{corespeccomp}. At fixed $\sigmain$, the left panel shows that larger systems (i.e., larger $\rj/\rhot$) lead to systematically steeper slopes, which possibly reconciles the increase in $\gammamax$ with the argument of energy conservation illustrated in \eq{ggmax}.

In application to the GeV flares of the Crab Nebula, which we attribute to the dynamical phase of rope merger (either Lundquist ropes or core-envelope ropes), we envision an optimal value of $\sigmain$ between $\sim 10$ and $\sim 100$. Based on our results, smaller $\sigmain\lesssim 10$ would correspond to smaller reconnection speeds (in units of the speed of light), and so weaker accelerating electric fields. On the other hand, $\sigmain\gtrsim 100$ would give hard spectra with slopes $s<2$, which would prohibit particle acceleration up to $\gammamax\gg \gamma_{\rm th}$ without violating energy conservation (for the sake of simplicity, here we ignore the potential spectral softening at high $\sigmain$ and large $\rj/\rhot$ discussed above).

% \clearpage
%%%%Lorenzo%%%%

\subsection{Merger of two flux tubes: conclusion}
\label{mergerconclusion}
In this section we have conducted a number of numerical simulations  of merging flux tubes with zero poloidal current. The key difference between this case and the 2D ABC structures, \S \ref{unstr-latt}, is that two zero-current flux tubes immersed either in external \Bf\ or external plasma represent a stable configuration, in contrast with the unstable 2D ABC structures. Two  barely touching flux tubes, basically, do not evolve -   there is no large scale stresses that force the islands to merge.  When the two flux tubes are pushed together, the initial evolution depends on the transient character of the initial conditions. 

In all the different configurations that we investigated the evolution proceeded according to the similar scenario: initially, perturbation lead to the reconnection of outer field lines and  the formation of common envelope. 
This   initial stage of the merger proceeds very slowly, driven by resistive effects.
With time the envelope grew in size and  a common magnetic envelope develops around the cores. 
 The dynamics changes when the two cores of the flux tube (which carry parallel currents) come into contact. Starting this moment the evolution of the two merging cores resembles the evolution of the   current-carrying flux tubes in the case of 2D ABC structures: the cores start to merge explosively and, similarly, later balanced by the forming current sheet. {\it Similar to the 2D ABC  case,  the fastest rate of particle acceleration occurs at this moment of fast dynamic merger.}
 
Thus, in the second stage of the flux tubes merger the dynamics is determined mostly by the properties of the cores, and not the details of the initial conditions (\eg how    flux tubes are pushed together). Also, particle acceleration proceeds here in a qualitatively similar way as in the case of 2D ABC structure - this is expected since the cores of merging flux tubes  do carry parallel currents that attract, similar to the 2D ABC case.

% \clearpage
%%%%Lorenzo%%%%

%%%%%%%Maxim%%%%%%%

\section{Magnetic island merger in highly magnetized plasma - analytical considerations}
\label{merger1}

\begin{figure}
\includegraphics[width=0.4\linewidth]{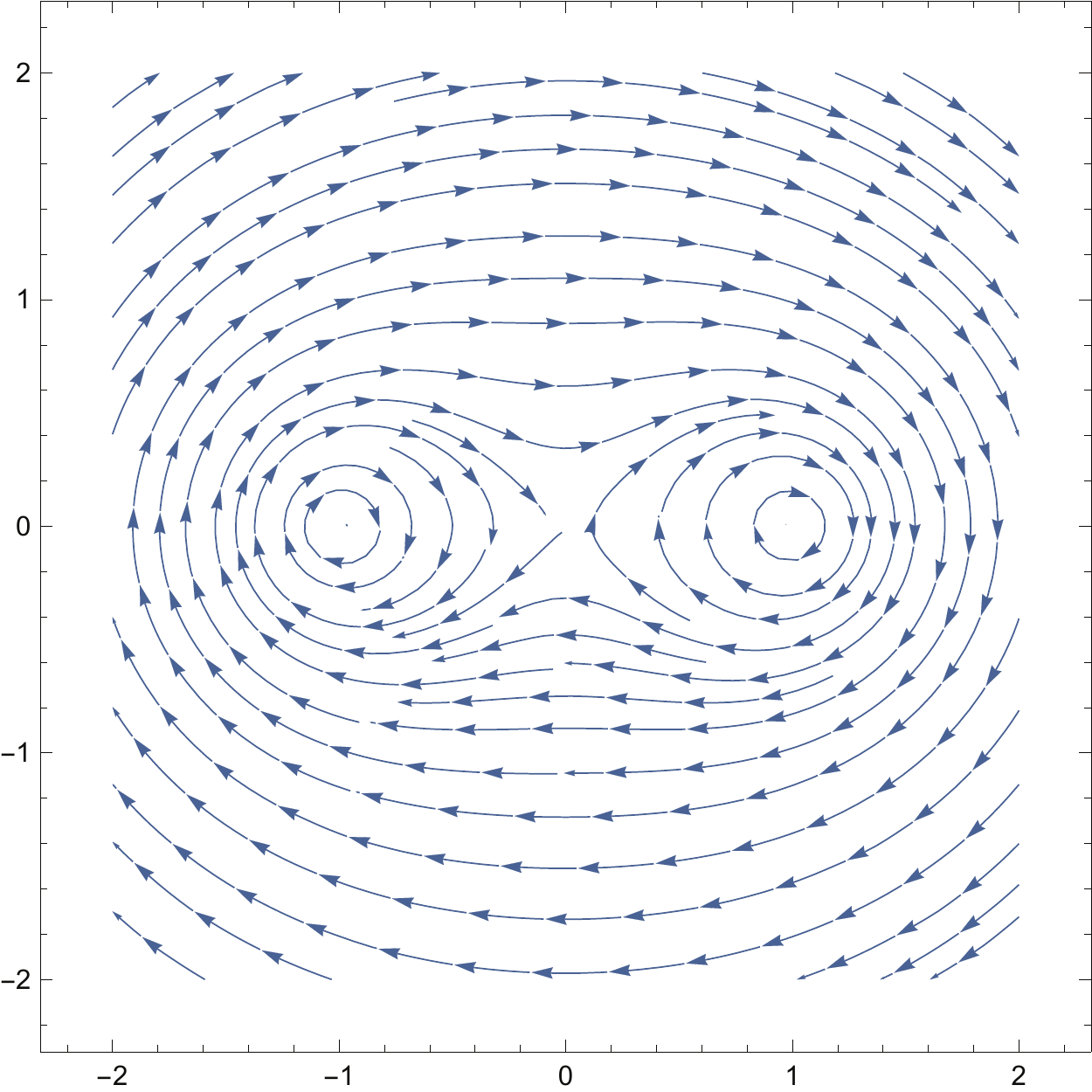}
\includegraphics[width=0.4\linewidth]{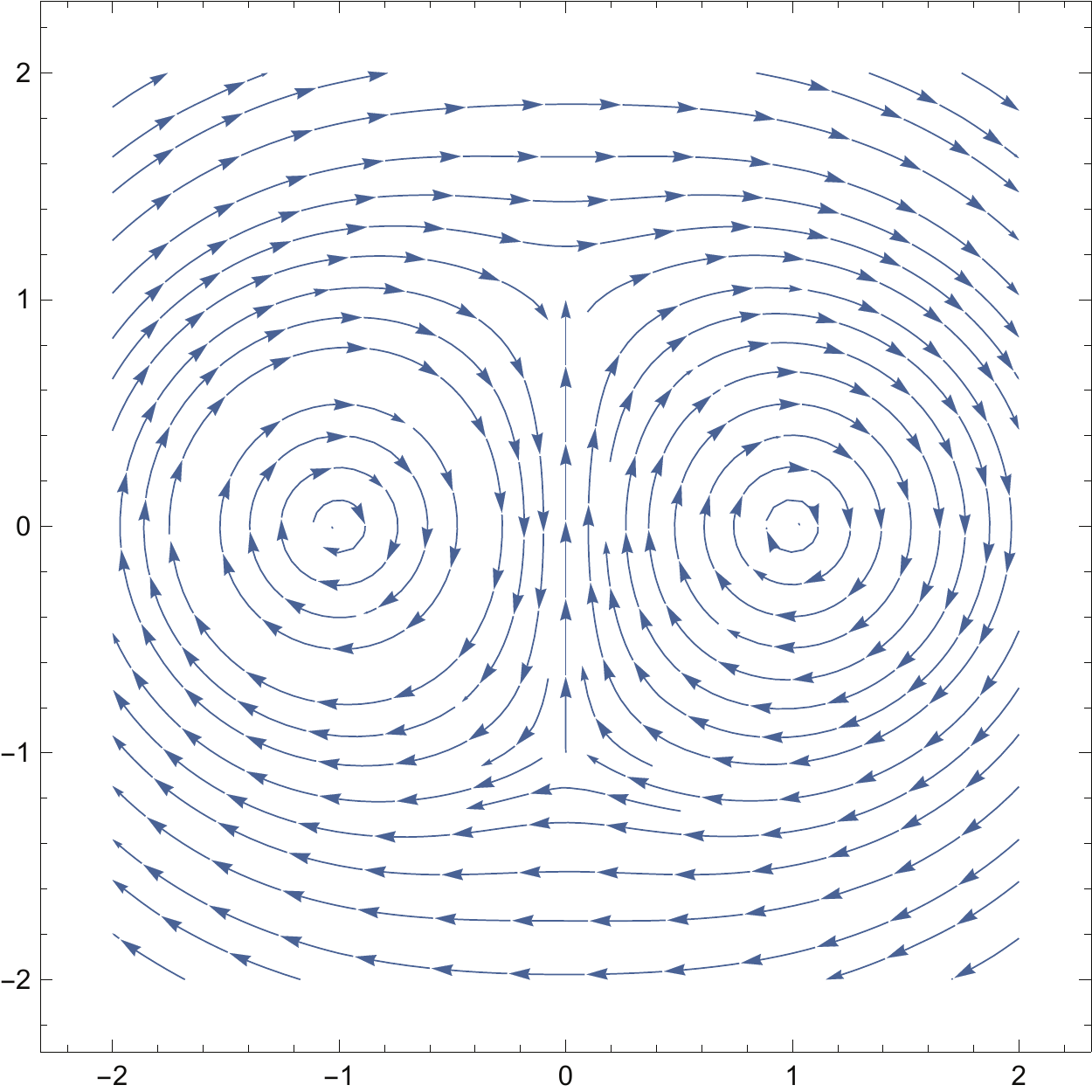}
\caption{Unstable X-point configuration (Left Panel) oscillates around   stable double Y-point configuration (Right  Panel) .
}
\label{IXpoint}
\end{figure}

Numerical simulations described above clearly show two stages of magnetic island merger - fast explosive stage and subsequent slower evolution. For intermediate times, the flux tubes show oscillations about some equilibrium position.  Let us next construct  an analytical model that captures the transition between the two stage. The  fast explosive stage is driven by large scale stresses of the type ``parallel currents attract". As an initial unstable configuration let us consider  X-point configuration of  two attracting  line currents  \citep{1965IAUS...22..398G}
\be
B_y + i B_x = { 4 I_0 z \over c( a_0^2 - z^2)}, \, z= x+ i y
\ee
where $a_0$ is the initial distance between the centers of the islands
 This non-equilibrium configuration will evolve into configuration  with Y-point along y-axis, $|y| < L$,  see Paper I  and Fig. \ref{IXpoint}, 
\be
 B_y + i B_x = { 4 I_0 z \over c} {a_0 \over \sqrt{a_0^2 +L^2} }\, {\sqrt{L^2+z^2} \over a_0^2 -z^2}
 \ee
For both configurations $\curl \B =0$.

 There is a current sheet at $x = 0 $ with surface current 
 \be
 g = 2 B_y (x=0)/(4 \pi)= { 2  \over \pi}  { \sqrt{ L^2 -y^2}  \over (a_0^2 + y^2) \sqrt{ a_0^2 +L^2 }}
 \ee
The surface current is {\it in the opposite direction} to the initial line currents. 
The repulsive force is 
\be
\int_{-L} ^L { 2 g I_0  \over \sqrt{a_0^2 +y^2}} {1\over \sqrt{a_0^2 +y^2}} dy = 
{2 I_0^2 L^2 \over c ( a_0^2 +L^2)}
\ee
The total force per unit length (it can also be obtained by integrating the stress over the $x=0$ surface is
\be
F = - {I_0^2 \over 2 c^2 a_0} \, {a_0^2-L^2 \over a_0^2 +L^2}
\ee
This balances the attractive force between the currents for 
$L = a_0$. Thus, attraction of parallel currents created a current sheet with the surface current flowing in the opposite direction; the repulsive force between each current and the current sheet balances exactly the attractive force between the currents. Evolution of the resulting equilibrium configuration will proceed on resistive time scales on the central current sheet.

We can also build a simple dynamic model of the current merger. Let us approximate the
effective \EM\ mass per unit length of a current-tube as
\be
m_{eff} c^2 = \int_{a_{min}}^{a_0} { 4I_0^2 \over 2 \pi} 2 \pi  r dr = (I_0/c)^2 \ln a_0 /a_{min}
\ee
Thus, $m_{eff} $ is nearly constant. Let us  next relate the current sheet length $L$ to the island separation $l$ assuming that for distance $2 l < 2 a_0$ ($ 2 a_0$ is the initial separation) 
\be
L^2 +l^2 = a_0^2
\ee
Thus, the motion of the flux tubes obeys 
\be
\ddot{l} = {I_0^2 \over m_{eff} c} \left(  {1 \over  l} -2  {l \over a_0^2} \right)
\ee
with initial conditions $l(0)= a_0$ and   $l'(0)= 0$. Solution $l(t)$ shows oscillations with equilibrium value $l=a_0/\sqrt{2}$, minimal value $l_{min} = 0.45$, period of oscillations is 
$4.39 a_0/c$.
 For small times $t \ll a_0 \sqrt{c} /I$,
\ba &&
l = a_0 - { c^2 t^2 \over 4 a_0 \ln (a_0/a_{min})}
\nn &&  
L = {c t \over \sqrt{ 2 \ln a_0/a_{min}}} 
\ea
For a distributed current, $\ln a_0/a_{min} \sim $ few.  Thus, the flux tubes oscillate around the equilibrium configuration, as the numerical simulations demonstrate.

Next, let us estimate the resulting \Ef\ and the electric potential during the initial nearly ideal stage of oscillations. Given the evolution of the \Bf\ described above, we can find a typical \Ef\ (by integrating $\partial_t \B + \curl \E=0$). We find at the point $x=y=0$
\ba && 
\dot{B} \approx {2 \sqrt{2} I_0 \over a_0^2 \sqrt{\ln a_0/a_{min}}}
\nn &&
E = {2 I_0 t \over a_0^2 \sqrt{\ln a_0/a_{min}} }  \ln \left( {2 \over \ln a_0/a_{min}} ( c t/a_0)^2 	\right) 
\ea
Thus, the electric field instantaneously becomes of the order of the \Bf; at the point $x=0, y=0$,
\be
E/B= { \ln \left( {2 \over \ln a_0/a_{min}} ( c t/a_0)^2 	\right)  \over  \sqrt{2}}
\ee
(value of \Bf\ at the point $x=0, y=0$ also increases linearly with time).

The model presented above explains the two stages of the island merger observed in numerical simulations. Initially, the merger is driven by the attraction of parallel currents. This stage of the instability is very fast, proceeding on dynamical (\Alfven) time scale. After the perturbation reaches non-linear stage, it takes about one dynamical time scale to reach  a new equilibrium consisting of two attracting current tubes and a repulsive  current sheet in between. During the initial stage \Efs\ of the order of the \Bfs\ develop. 
After the system reaches a new equilibrium the ensuing evolution proceeds on slower time-scales that depend on plasma resistivity.

% \clearpage
%%%%%%%
%%%%%%%%

\section{Conclusion}

In this paper we investigated dynamics and particle acceleration during merger of relativistic highly magnetized magnetic islands. We have considered the cases of
2D magnetic ABC equilibria (including a driven evolution) and a induced merger of zero-total-current flux tubes.

There is  a number of important basic plasma physics results.
\begin{description} 

\item {\it  Instability of 2D magnetic ABC structures.}  We have studied the instability of the 2D magnetic ABC structures  and  identified two main instability modes \citep[see also][]{2015PhRvL.115i5002E}.
  The instability is of the kind ``parallel currents attract''. \citep[][considered a  similar model problem for the magnetic field structure of the Solar corona and generation of Solar flares.]{1983ApJ...264..635P,1994ApJ...437..851L} 
We identified two stages of the instability - (i) the explosive stage, when the accelerating \Efs\ reach values close to the \Bfs, but little magnetic energy is dissipated; the most important process during this stage is the X-point collapse; (ii) slower, forced reconnection stage, whereas a large fraction of the initial \Bf\ energy is dissipated; at this stage the newly formed central current sheet repels the attracting currents.  
  Though the model is highly idealized, it illustrates two important concepts: (i) that ubiquitous  magnetic null lines  \cite[\eg][]{1999PhPl....6.4222A} serve as sites of current sheet formation, dissipation of magnetic energy and particle acceleration; (ii) that the evolution is driven by large-scale magnetic stresses.  The case of 2D ABC structures is different from the 3D ABC, which is stable \citep{1986JFM...166..359M}.

\item {\it  Triggered collapse of  2D  ABC structures.}  We have studied a number of set-ups that greatly accelerate the development of the instability of 2D ABC structures - either by a inhomogeneous large-scale compression or by a strong fast-mode wave. Large-scale perturbations can greatly speed-up the development of instability.

\item {\it  Collision of magnetic flux tubes.}  We have discussed the merger of  magnetic   flux tubes carrying no net electric current. In this case the tubes first develop a common envelope via resistive evolution and  then merge explosively. In comparison to two stages of the merger of  current-carrying flux tubes  the zero-current  flux tubes have in addition a slow initial stage of development of the common envelope.

\item {\it  Particle acceleration: different stages.}  We have discussed intensively the properties of particle acceleration during the $X$-point collapse,  during the development of the 2D ABC instability and during the merger of  flux tube with zero total current. These three different set-ups allow us to concentrate on somewhat different aspects of particle acceleration. In all the cases that we investigated the efficient particle acceleration always occurs in the region with $ E\geq B$  - by the charge-starved \Efs.  This stage is best probed with the $X$-point collapse simulations. In the case of ABC structures and flux tube mergers the
most efficiently particles are accelerated during the initial  explosive stage; during that stage not much of magnetic energy is dissipated. In case of 2D ABC system, this initial stage of  rapid acceleration is followed by a 
 forced reconnection stage; at this stage particles are further accelerated to higher energies, but the rate of acceleration is low. In case of the colliding/merging zero-current flux tubes, the fast dynamic stage is {\it preceded} by the slow resistive stage, when the outer field lines form an overlaying shroud that  pushes the parallel current-carrying cores together. When these parallel current-carrying cores come in contact the evolution proceeds similarly to the unstable ABC case. We stress that {\it the fastest particle acceleration occurs in the beginning of the dynamical stage of the merger} (right away in the X-point collapse simulations, in the initial stage of the instability of the ABC configuration, after the slow resistive evolution in case of colliding/merging flux tubes). 

\item {\it  Reconnection rates.}
 The key  advantage of the suggested model, if compared with  the  plane parallel case \citep[that mostly invokes tearing mode][]{2011ApJ...737L..40U,2012ApJ...746..148C,2012ApJ...754L..33C,2013ApJ...770..147C}, is that {\it macroscopic large scale magnetic stresses lead to much higher  reconnection rate and much faster particle acceleration. }
Specifically, the dynamics stage, associated with the  $X$-point collapse,
produces exceptionally high  reconnection/acceleration rates, as large as $\sim 0.8$; most importantly, this occurs over  macroscopic spacial scales. This high acceleration rate  is nearly an order of magnitude larger than is achieved in plane-parallel tearing mode-based models.

 \item {\it  Particle acceleration: particle spectra and bulk magnetization.} In all the cases we investigated the power-law slope of particle distribution depends on magnetization: higher $\sigma$ produce harder spectra. The critical case of $p =2$ corresponds approximately to $\sigma \leq 100$.  For $p\geq 2$ the maximal energy that  particles can achieve grows linearly  with the size of the acceleration region. In the regime $100 \leq \sigma \leq 1000$ particle spectra indices are $p< 2$, yet the maximal energy can still exceed $\sigma$ by orders of magnitude.   For very large $\sigma \geq 10^3$ the spectrum becomes hard, $p \rightarrow 1$, so that the maximal energy is limited by $\gamma_{p, max} \leq \sigma$.  
For small  $\sigma \leq  $ few the  acceleration rate becomes slow, non-relativistic \citep[see also][]{2016ApJ...816L...8W}.

\item {\it  Anisotropy of accelerated particles.} In all the cases we investigated the distribution of  accelerated particles, especially those with highest energy,  turned out to be highly anisotropic. Since the highest energy particles have large Larmor radii, this anisotropy is kinetic; qualitatively it resembles  beaming along the magnetic null line (and the direction of accelerating  \Ef). Importantly, this kinetic anisotropy shows   on macroscopic scales.

\end{description}

 Most importantly, both models - 2D ABC structures and magnetic flux tubes -  demonstrate a stage of explosive $X$-point collapse (Paper I) during which very fast particle acceleration occurs.  (In case of magnetic flux tubes this fast stage is preceded by resistive stage during which a common envelop is formed.)

We would like  to thank Roger Blandford,  Krzystof Nalewajko  Dmitri Uzdensky and Jonathan Zrake for discussions.

\acknowledgements
The simulations were performed on XSEDE resources under
contract No. TG-AST120010, and on NASA High-End Computing (HEC) resources through the NASA Advanced Supercomputing (NAS) Division at Ames Research Center. ML would like to thank   for hospitality Osservatorio Astrofisico di Arcetri and Institut de Ciencies de l'Espai, where large parts of this work were conducted. This work had been supported by NASA grant NNX12AF92G,  NSF  grant AST-1306672 and DoE grant DE-SC0016369.
OP is supported by the ERC Synergy Grant ``BlackHoleCam -- Imaging the Event Horizon of Black Holes'' (Grant 610058).

\clearpage
\bibliographystyle{jpp}
%\bibliographystyle{mn2e}
%  \bibliography{astro}
 \bibliography{/Users/maxim/Home/Research/BibTex} 

\appendix

%%%%%%%%%%%%%%%%%
 
\section{The inverse cascade}
\label{inverse}

The  Woltjer-Taylor relaxation principle \citep{Woltier:1958,1974PhRvL..33.1139T} states that magnetic  plasma configuration try to reach so-called constant-$\alpha$ configuration, conserving helicity in the process of relaxation. Such relaxation process is necessarily dissipative, though the hope typically is that it is weakly dissipative. In case of 2D magnetic ABC structures the 
 energy per helicity $ \propto  \alpha$; thus, according to the Woltjer-Taylor relaxation principle  the system will try to reach a state with smallest $\alpha$ consistent with the boundary condition. This, formally, constitutes an inverse-type cascade of magnetic energy to largest scales. On the other hand, such cascade is highly dissipative: conservation of helicity requires that $B \propto \alpha^{3/2}$, thus magnetic energy per flux tube  $B^2/\alpha^2 \propto \alpha$ decreases with the size of the tube ($\alpha$ is proportional to the inverse radius).  This implies that a large fraction of the magnetic energy is dissipated at each scale of the inverse cascade - cascade is highly non-inertial.  In contrast,   the Woltjer-Taylor relaxation principle assumes  a weakly resistive process.
Highly efficient dissipation of magnetic energy  is confirmed by our numerical results  that indicate that during merger approximately half of the magnetic energy is dissipated, \S \ref{unstr-latt}. As a results such cascade cannot lead to an efficient energy accumulation on the largest scales.  

To find the scaling of the cascade, consider  a helicity-conserving merger of two islands parametrized by $B_1 $ and $\alpha_1$. Conservation of helicity requires that in the final stage (subscript $2$) $H_2 \equiv B_2^2/ \alpha_2^3 =  2H_1 =
2 B_1^2 /\alpha_1^3$. From the conservation of area $\alpha_2 = \alpha_1/\sqrt{2}$; then $B_2 = B_1/\sqrt{2}$. The magnetic energy of the new state $E_2= \sqrt{2} B_1^2 /\alpha_1^2 < 2  B_1^2 /\alpha_1^2$.  Thus a fraction $(E_2 - 2 E_1)/(2 E_1) = 1-1/\sqrt{2}=0.29$ of the magnetic energy is dissipated in each step. 

Next, in each step the size of the island growth by $\sqrt{2}$. After $n$ steps the scale is $\propto \sqrt{2}^n$, while the energy  $\propto (1-1/\sqrt{2})^n$. The two power-laws are connected by the index $p = - \ln 2 /( 2  \ln (1-1/\sqrt{2}))= 3.54$. This is very close to the result of \cite{2014ApJ...794L..26Z} who concluded that the power-law index is close to $7/2$  \citep[the initial conditions used by][are different from ours]{2014ApJ...794L..26Z}. 

The efficient dissipation of the magnetic energy   is confirmed by our numerical results  that indicate that during merger a large fraction of the magnetic energy is dissipated. As a results such cascade cannot lead to an efficient energy accumulation on largest scales.  

\begin{figure}
\centering
\includegraphics[width=.35\textwidth]{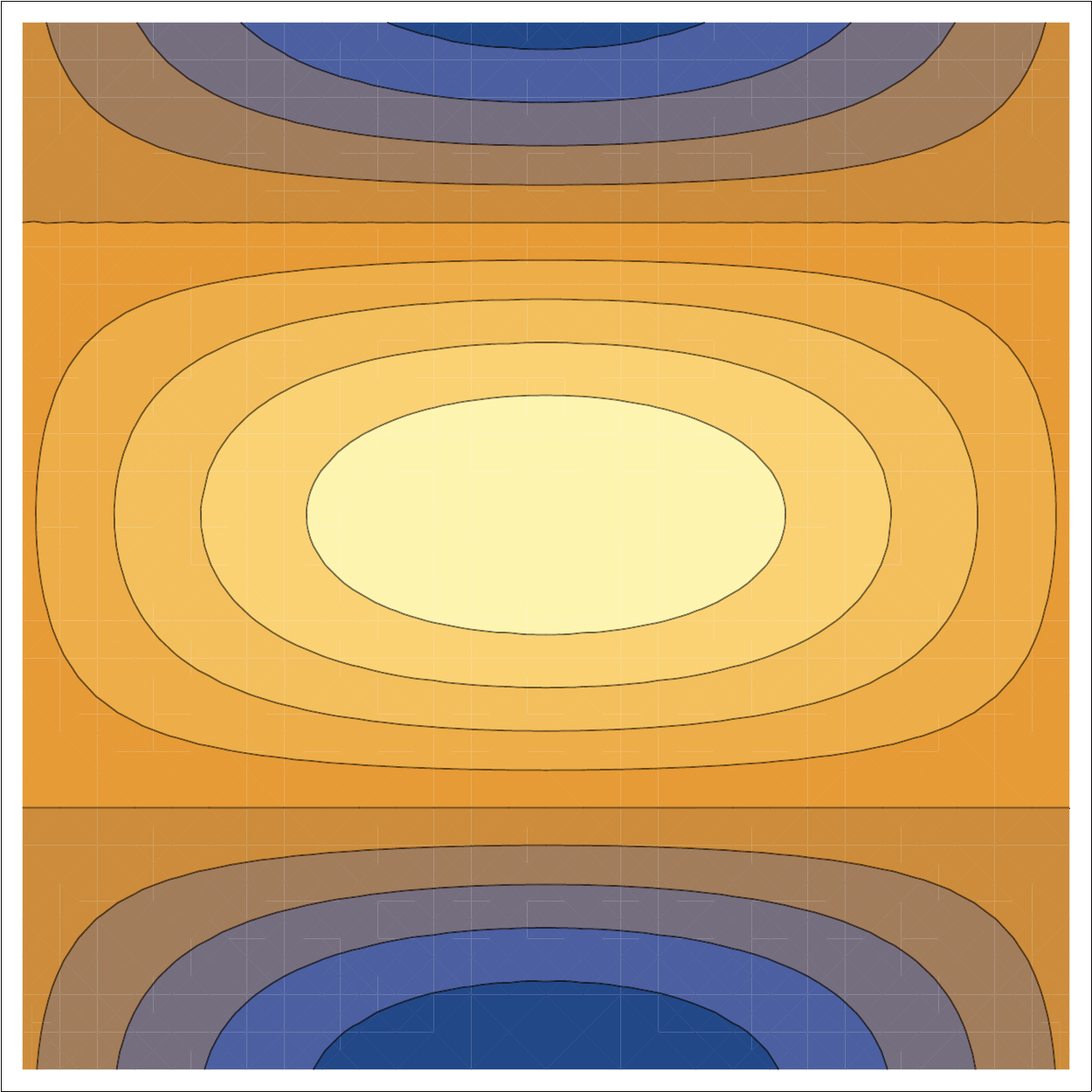}
\includegraphics[width=.35\textwidth]{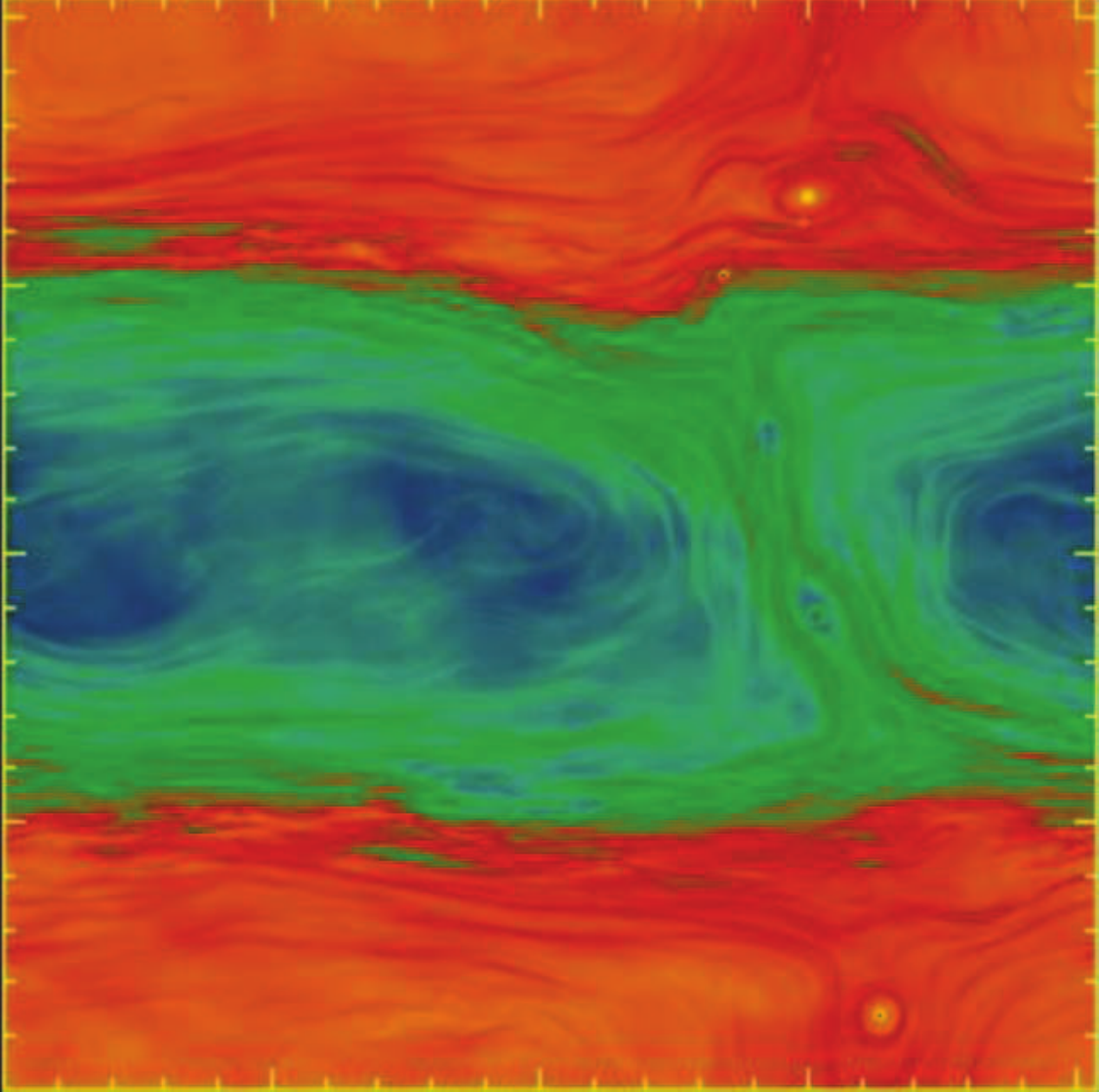}
\caption{Comparison of a 2:1 force-free state with the final structure of PIC simulations (at time  $19.01$). 
}
\label{final}
\end{figure}

In addition, we have verified that the final states are close to force-free. Due to square box size and periodic boundary conditions, the following family of force-free solutions is then appropriate:
\ba &&
A_z = \cos a x \sin \sqrt{\alpha^2 -a^2} y,
\nn  &&
\B = \nabla \times A_z + \alpha A_z {\bf e}_z
\ea
In particular, $A_z =\cos( x/\sqrt{5} )\sin (2 y\sqrt{5})$ produces two islands, Fig. \ref{final}.

Thus, we conclude that the inverse cascade of 2D magnetic ABC structures proceeds through (nearly) force-free self-similar configurations of ever increasing size. But the cascade is non-inertial  (highly dissipative) with the approximate index of $3.54$.

% \clearpage
%%%%%%%%%
%%%%%

   \end{document}